# Predictability can be dynamically constructed in deterministic systems

**Lars Koopmans[1], Elinor M. Kay[1,2,3], and Hyun Youk[1,2,3,4]**

[1]Department of Physics, University of Illinois Urbana-Champaign, Urbana IL, USA
[2]NSF Science and Technology Center for Quantitative Cell Biology, University of Illinois Urbana-Champaign, Urbana IL, USA
[3]Grainger College of Engineering, University of Illinois Urbana-Champaign, Urbana IL, USA
[4]Corresponding author: youk@illinois.edu

## ABSTRACT

A deterministic system's future can be fixed from the start yet remain unpredictable because of chaos or computational irreducibility. Here we reveal another source of unpredictability for non-chaotic deterministic systems: collective entities making the future legible are initially absent but dynamically constructed. In a generalized cellular automaton, disordered lattices evolved into static configurations, rectilinear waves, or spiral waves. Although initial configurations fixed each fate, machine-learning models did no better than randomly guessing fates from them. Recoding cell states geometrically revealed self-organization of prediction-enabling topological entities: vortices, non-contractible-loop strings, and, most generally, a winding field describing how same-state regions wrap the lattice. As the winding field self-organized, initially unpredictable fates became increasingly legible. Static and rectilinear-wave fates became progressively predictable, whereas spiral-wave fate became accurately predictable only near wave formation. These results establish that self-organization can build not merely order but entities that make certain futures visible, revealing a gap between determinism and practical predictability.

## INTRODUCTION

Complex self-organization can arise in systems whose components obey simple local rules. Yet when many discrete agents interact over long times, even fully deterministic dynamics can become difficult to predict or understand. The dimensionality of the system's state space, the accumulation of interactions across time, and the absence of obvious organizing principles can obscure how global order emerges from individual local interactions. A central question across physics, theoretical computer science, and biology is therefore how macroscopic organization arises step by step from microscopic interactions in high-dimensional discrete systems[1,2]. For deterministic systems, the long-standing and current view, as articulated by Laplace, is that the entire future is fixed by the initial configuration, so that one could predict the system's complete future given its exact initial configuration and the deterministic rules by which it evolves. For high-dimensional discrete systems that are non-chaotic and deterministic, this view remains uncontested. Here we ask whether the features that make macroscopic fate predictable can be absent from the initial configuration in any accessible form and, instead, emerge during the system's deterministic temporal evolution — the system self-organizing predictability itself.

Unpredictability arises in a chaotic deterministic system for two reasons acting together. First, we can never measure or store a system's initial configuration with infinite precision, so any real specification carries a small error. Second, in a chaotic system this error can grow exponentially in time: initial configurations that are arbitrarily close can evolve along diverging trajectories and ultimately yield very different outcomes[3]. We therefore cannot reliably forecast the exact long-term dynamics of chaotic systems in nature[4–6]. This familiar route to unpredictability does not arise in deterministic systems with a finite global state space built from discrete states and entities. This is for two reasons. First, a finite global state space necessitates that every initial configuration eventually yields a cyclic trajectory in state space. Second, discrete states and entities with deterministic interaction rules entail that we can, in principle, always exactly specify the initial configuration and the mapping from any state to its next state. For these two reasons, it is tempting to assume that, once the initial configuration is exactly determined, the long-term fate of such a non-chaotic deterministic system should be directly predictable from it, without tracing the full dynamics. Here we question this assumption — and show that it is untrue.

Here we show that a system's fate — its macroscopic configuration after a long time — may not yet be legible in the initial configuration, even when the system is exactly specifiable and deterministic. This runs against an implicit assumption of many machine-learning approaches[7,8], which have achieved remarkable success at predicting outcomes from inputs across complex systems: that a sufficiently expressive machine-learning model can predict long-term outcomes directly from the initial configuration because the fate-predictive structure is already organized into accessible, learnable features. If instead the structures that reveal macroscopic fate arise only during the system's temporal evolution, then prediction from the initial configuration may fail not because of insufficient model capacity or data, but because the relevant dynamical features have not yet formed. The problem of prediction in deterministic systems then becomes one of identifying when predictive structure becomes accessible and what physical objects carry it.

In this paper, we address this question by studying a class of generalized deterministic cellular automata, in which interactions extend beyond nearest neighbors over finite ranges, allowing discrete agents to influence one another through nonlocal interactions dictated by fully deterministic rules. As discrete dynamical systems with explicit update rules, generalized cellular automata allow microscopic interactions to be specified exactly while enabling direct observation of the collective behavior that emerges over long times[8–12]. One instance of such a system arises from secrete-and-sense cell–cell communication circuits[13,14], in which cells secrete diffusible molecules and respond to signals produced by their neighbors. These circuits are widespread in biological systems, including microbial quorum sensing and cAMP signaling during aggregation in *Dictyostelium discoideum*, where they generate large-scale spatiotemporal patterns[15–19]. In earlier work, we showed that a minimal circuit in which cells secrete and sense two signaling molecules produces persistent rectilinear and spiral waves across broad parameter regimes[20]. Although these patterns resemble waves familiar from reaction–diffusion systems[21–24], the mechanism here does not rely on diffusion-driven instabilities or continuous concentration fields[25,26], but instead arises in a discrete, deterministic cellular automaton governed by finite-range interactions[27–29].

To determine when and how predictive structure arises, we studied a generalized deterministic cellular automaton on a toroidal lattice in which disordered initial configurations evolved into static configurations, rectilinear waves, or spiral waves. Given an initial configuration, none of the machine-learning models we examined — including transformers, neural cellular automata, convolutional neural networks, and hidden Markov models — could predict which of the three fates would occur more accurately than random guessing, even though the initial configuration fully fixed the fate and the ensuing dynamics were non-chaotic and deterministic. Yet the same machine-learning models predicted fate more accurately from later configurations, revealing that prediction-enabling structures emerged as the dynamics unfolded.

Recoding cell states as a discrete phase field revealed discretized vortices that emerged from disorder, moved, interacted, and annihilated. Same-state cells formed "strings"—paths joining pairs of vortices—that could wrap the torus to form non-contractible loops (NCL strings). NCL strings sparsely represented the same wrapping topology captured more generally by the winding field, which assigns each cell a vector describing how its connected same-state region wraps the torus. The winding field carried nearly all the predictive signal available in the full configuration. NCL strings and regions with zero winding gradually vanished along trajectories ending in static configurations or rectilinear waves but persisted along spiral-wave trajectories. Their disappearance progressively forecast the first two fates to an observer following the automaton forward in time. Their continued presence, however, did not forecast spiral-wave fate: for most of the run, trajectories in which they would ultimately persist resembled those in which they would vanish. Only late, when winding-field size and expansion rate diverged among the three fates, did spiral-wave fate — and therefore all three fates — become reliably predictable. These results establish that even without chaos and without invoking computational irreducibility, being formally determined is not the same as being predictable in practice: a future can be completely fixed from the start yet practically unpredictable because the structures that make the future legible are absent and must be built later by the self-organizing dynamics.

## RESULTS

### Deterministic dynamics with unpredictable macroscopic outcomes

We studied a fully deterministic, generalized cellular automaton composed of hundreds of identical secrete-and-sense cells on a triangular lattice. Every cell secretes two diffusible molecules at either a low or high rate, governed by a biologically ubiquitous gene-regulatory circuit[1,30,31]: sensing molecule 1 promotes its own secretion while repressing secretion of molecule 2, whereas sensing molecule 2 pro-motes secretion of molecule 1 (Fig. 1a). Two genes—each expressed at either a high or low level—mirror these secretion rates, so that every cell occupies one of four discrete expression states, represented by four colors (Fig. 1b).

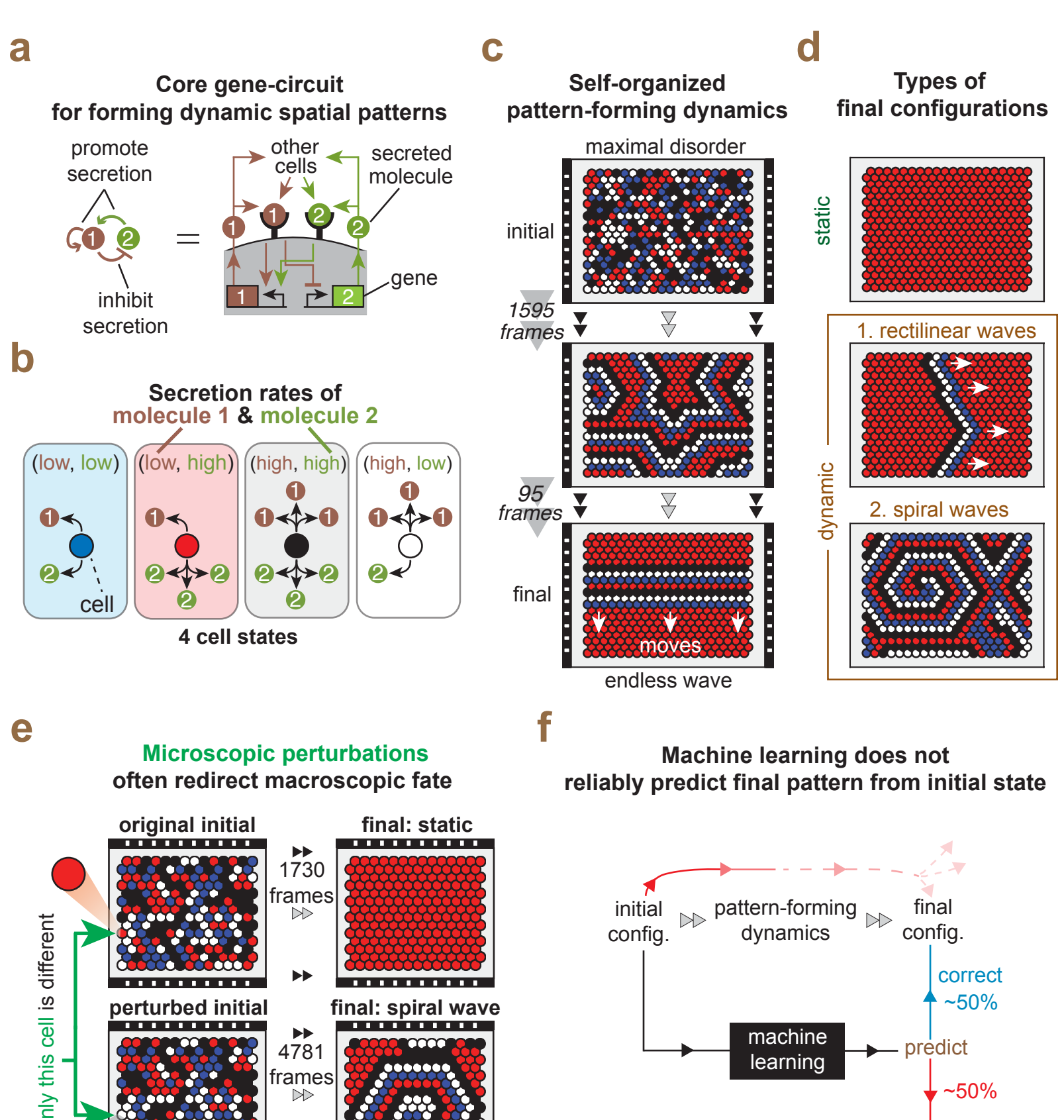

**Fig. 1 | Initial configurations fix pattern fate but do not make it predictable.**
**(a)** Regulatory circuit encoded in each cell. Molecule 1 (brown) activates its own secretion and represses secretion of molecule 2 (green); molecule 2 activates secretion of molecule 1. Diffusion-mediated interactions decay with distance and extend beyond nearest neighbors (Supplementary note 1).
**(b)** Four cell states defined by low or high secretion rates of molecule 1 and molecule 2. Each cell acts as a nearly perfect oscillator, cycling nearly unidirectionally through the four states at each timestep (Markov transitions in Supplementary fig. 1).
**(c)** Example trajectory from a disordered initial configuration to traveling rectilinear waves.
**(d)** Three final configurations generated with identical rules and parameters: static, rectilinear waves, and spiral waves (Supplementary fig. 2).
**(e)** Sensitivity to microscopic perturbations. Changing the state of exactly one cell in the initial configuration of a 14 × 14 lattice can redirect the system to a different macroscopic fate. ~50% of single-cell perturbations changed the final pattern class (Supplementary fig. 3).
**(f)** Prediction of static versus non-static fate from initial configurations. Logistic regression, tree ensembles, multilayer perceptrons, and convolutional neural networks trained on up to 850,000 trajectories yielded ~50% balanced accuracy (Supplementary figs. 4–9 and Supplementary note 2). A complementary, model-agnostic analysis calculated normalized mutual information (NMI) at each lattice site between that cell's initial state and the final binary fate. NMI estimates approached zero and matched label-shuffled controls as dataset size increased (Supplementary fig. 10).

At each timestep, every cell senses the quasi–steady-state concentrations generated by itself and by all other cells. Interactions decay exponentially with distance yet extend across several lattice shells—well beyond second-nearest neighbors—so that a substantial fraction of the sensed signal originates from spatially nonlocal cells (see details of setup in Supplementary note 1). When sensed concentrations cross fixed molecule-specific thresholds, cells synchronously update their state. Under these rules, we found that each cell behaved as a nearly perfect oscillator, cycling unidirectionally through the four states with only rare pauses (Supplementary fig. 1).

At the start of each simulation, we randomly assigned a state to every cell, ensuring maximal disorder with no spatial correlations and roughly equal representation of all four states (Supplementary note 1). Previous work showed that such initially disordered fields robustly self-organize into dynamic spatial patterns—persistent rectilinear or spiral waves of various shapes—across wide parameter ranges[20] (Fig. 1c). Reproducing these findings, we observed that identical update rules and parameters nonetheless produced distinct final pattern types—static configurations, rectilinear waves, or spiral waves—depending solely on the initial configuration within a large state space (already $4^{196}$ for a 14 x 14 lattice; Fig. 1d). The time required for pattern formation ranged from hundreds to tens of thousands of timesteps (Supplementary fig. 2).

To probe the sensitivity of this deterministic system, we randomly selected a reference initial configuration on a $14 \times 14$ lattice and ran the automaton to completion. We then repeated the simulation 588 times, each time starting from the same reference initial configuration, selecting one cell at random, and changing its state to one of its three alternative states chosen at random while keeping all other cells unchanged. We chose 588 perturbation runs because this equals the total number of distinct one-cell perturbations possible on a 14 x 14 lattice (14 x 14 x 3). We found that this minimal, one-cell perturbation frequently redirected the system toward a qualitatively different macroscopic outcome and substantially altered the time required for that outcome to emerge. Approximately 50% of the single-cell perturbations changed the final pattern class (Fig. 1e; Supplementary fig. 3). Across perturbed runs, static configurations, rectilinear waves, and spiral waves all occurred with comparable frequency. Even when the perturbation preserved the final pattern class, it often changed the detailed geometry of the pattern—for example, converting a horizontally traveling wave into a vertically traveling one—and altered the time required to reach that pattern by thousands of timesteps (Fig. 1e; Supplementary fig. 3).

These results demonstrate extreme sensitivity to microscopic perturbations in a fully deterministic discrete system, suggesting that macroscopic outcomes and temporal trajectories are not reliably inferable from the initial configuration alone.

## Limits of machine learning in predicting deterministic self-organization

The extreme sensitivity of final fate to single-cell perturbations raised a natural question: despite the apparent unpredictability of individual trajectories, does the initial configuration nonetheless contain statistical regularities that allow the final pattern type to be predicted directly from the initial configuration? To address this, we used machine learning as a systematic probe for predictive structure in the initial configuration[8].

To give prediction the best possible chance to succeed, we framed the task in its simplest form: binary classification of whether a random initial configuration ultimately relaxed into a static configuration or instead produced any dynamic spatial pattern. Using one million independently generated simulations, we trained a broad range of models—from linear classifiers and tree ensembles to deep neural networks—on datasets reaching up to 850,000 examples, with performance always evaluated on a large, frozen test set (schematized in Supplementary fig. 4 & Supplementary note 2).

Across all model classes and all training sizes, prediction performance remained indistinguishable from chance (~50% correct classification, equivalent to random guessing for this binary task). Neither increasing the amount of data nor increasing model expressivity improved generalization: linear models, boosted trees, multilayer perceptrons, and convolutional neural networks all failed to achieve balanced accuracy meaningfully above chance (Supplementary figs. 5–9).

Together, these results revealed a sharp distinction between being formally determined and predictable in practice. The initial configuration completely fixed the final fate, yet none of the tested machine-learning models — from linear and tree-based methods to deep and convolutional networks — could predict that fate more accurately than random guessing. A separate mutual-information analysis likewise found that no individual cell's initial state was predictive on its own (Supplementary fig. 10). Moreover, constructing a complete lookup table pairing every possible input configuration with its fate is not realistic: a 14 x 14 lattice with four cell states has $4^{196}$ possible initial configurations, far exceeding the approximately $10^{80}$ atoms in the observable universe. Taken together, we concluded that the fate was fixed but not readable from the initial configuration by any practical method we examined. We therefore shifted our focus from attempting to predict fate directly from initial configurations to examining the dynamics themselves.

## Emergence of three classes of discrete vortices

When we recast each cell state as a directional phase vector rather than as a color—right (0), up ($\pi/2$), left ($\pi$), and down ($3\pi/2$)—a previously hidden structure suddenly became visible (Fig. 2a). Neighboring cells that differed by a $\pi$ phase formed localized "cores," around which surrounding cells traced closed contours of rotating vectors. To determine whether these cores reflected systematic organization rather than coincidence, we developed an algorithm that traversed each contour and computed the net phase accumulated along one complete circuit (see Methods and Supplementary Note 3). We found that every core fell into one of three sharply defined classes. If the contour wound counterclockwise—accumulating $+2\pi$ or more—we classified it as a +1 vortex; if it wound clockwise—accumulating $-2\pi$ or more—as a $-1$ vortex; and if the net winding magnitude was less than $2\pi$, as a 0 vortex. We define the sign of this winding as the vortex's topological charge[32,33]. Regardless of the number of full rotations, we grouped vortices solely by the sign of the total winding. Each core and its contour therefore constituted a discrete topological object entirely invisible in the original color representation. Applying this classification across tens of thousands of automaton runs spanning multiple lattice sizes, we found that vortices emerged spontaneously in every run (Supplementary fig. 11).

### Universal three-stage dynamics

When we tracked vortices over time, we discovered that every run followed the same three-stage progression (Fig. 2b). First came a rapid vortex-creation phase: within ~20 timesteps, all three vortex types formed spontaneously in lattices that initially contained none, and the total vortex number reached a peak value (Fig. 2c & Supplementary fig. 12). Thus from maximal disorder, a dense population of vortices emerged almost immediately. Thereafter, vortices moved and progressively disappeared, causing the total vortex number to decrease in discrete, irreversible steps—a descending staircase (Fig. 2c & Supplementary fig. 13). We observed this staircase across lattice sizes and across tens of thousands of independent runs. We discovered that the final pattern was determined by whether any vortices survived: when all vortices vanished, either a static configuration or rectilinear waves always emerged, typically on the order of 100 timesteps after the last vortex disappeared (Fig. 2b & Supplementary fig. 14), whereas whenever vortices persisted, spiral waves always and eventually emerged (Supplementary fig. 15). A closer inspection of the staircase, however, revealed a hidden constraint.

### Hierarchical lifetimes and a hidden constraint

To probe the staircase more closely, we examined vortex abundance by type (Fig. 2d). We found that 0 vortices behaved differently from the others: they appeared prominently during the vortex-creation phase, but their abundance rapidly collapsed thereafter. In contrast, +1 and −1 vortices persisted far longer. Quantifying lifetimes confirmed this separation (Fig. 2e). 0 vortices survived on average ~10 timesteps after formation, whereas vortices with nonzero charge frequently persisted for hundreds to thousands of timesteps. Annihilations did not occur at uniform intervals. Instead, successive surviving charged vortices lived progressively longer than their predecessors. Each drop in the staircase occurred after a longer waiting time than the one before it (Figs. 2d-e & Supplementary figs. 13 & 16). As vortex abundance decreased, the interval to the next disappearance lengthened. Most strikingly, we observed that every disappearance of a +1 vortex was accompanied by the simultaneous disappearance of a −1 vortex (Fig. 2d). In runs that ultimately formed spiral waves, almost always exactly one such +1/-1 pair survived until the end; fewer than 5% of spiral-forming runs retained more than one pair, and none had zero vortices remaining (Supplementary fig. 16).

### Pairwise annihilation and emergent charge conservation

To understand the synchrony, we followed individual vortices in space and time (Fig. 2f). We found that +1 and −1 vortices moved without detectable attraction or repulsion until their cores approached within approximately two cell spacings. When their enclosing contours overlapped, the pair collapsed into a transient 0 vortex that decayed within ~10 timesteps. Annihilation therefore required local contour overlap and proceeded through a short-lived neutral intermediate. +1 and −1 vortices never disappeared independently—they always merged and vanished together. This microscopic mechanism explained the synchronized drops in vortex counts. Strikingly, this pairing reflected a deeper global constraint: at every timestep in every run, the total topological charge—the sum of all vortex charges—remained exactly zero (Supplementary fig. 17). No term in the local update rules enforced charge balance, yet we found that the dynamics respected it perfectly in every run we examined. This revealed an emergent conservation law: global topological charge remained invariant throughout the dynamics, despite being nowhere encoded in the update rules.

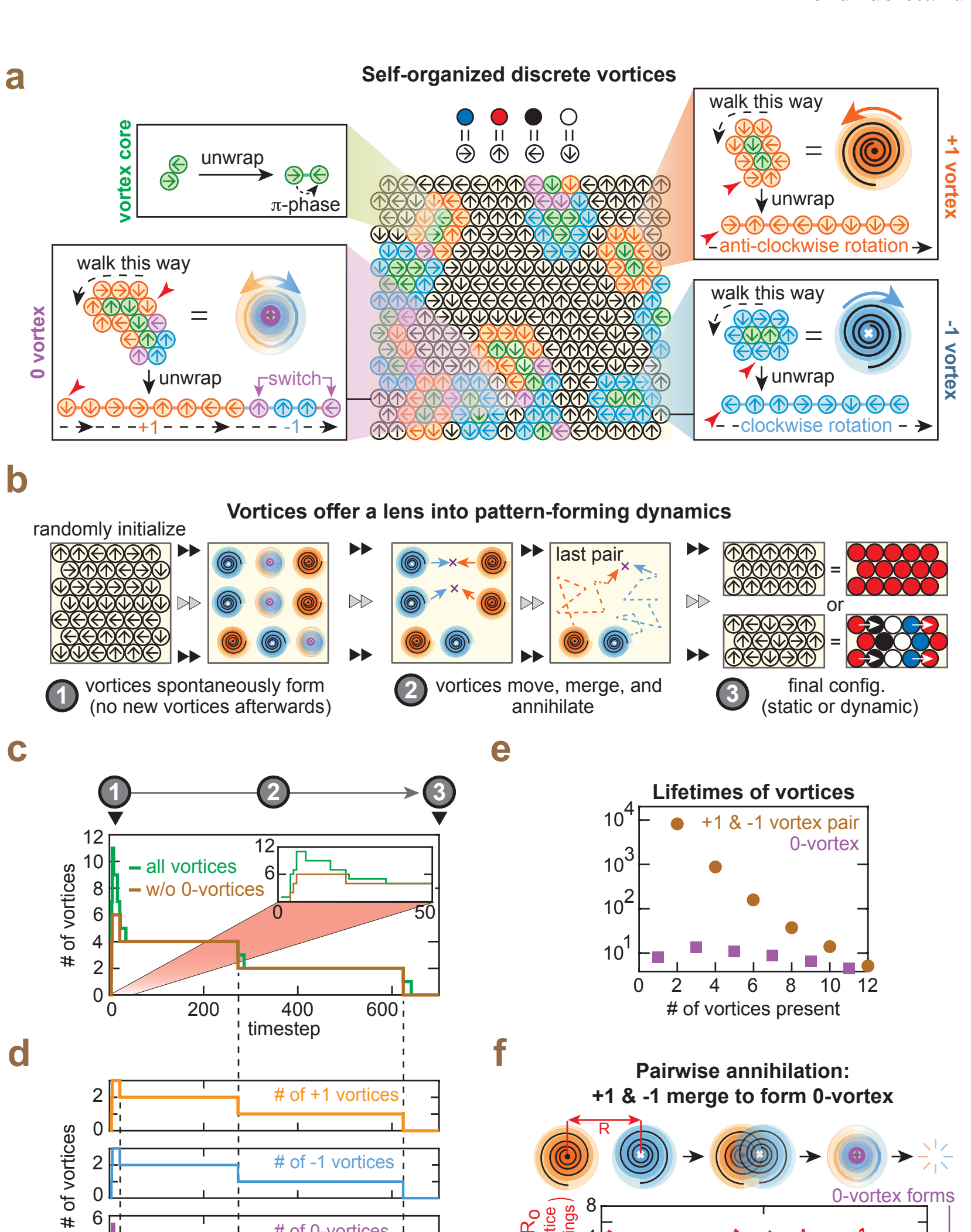

**Fig. 2 | Discrete vortices emerge spontaneously and offer a lens into self-organization dynamics.**
**(a)** Recoding the four cell states as discrete phase vectors (right = 0, up = $\pi/2$, left = $\pi$, down = $3\pi/2$). Neighboring cells differing by a $\pi$-phase form cores (green) around which surrounding vectors trace closed contours: +1 vortices (orange) wind counter-clockwise, −1 vortices (blue) wind clockwise, and 0 vortices (purple) have no net helicity (Supplementary note 3).
**(b)** Three-stage dynamics: vortex formation, vortex motion and annihilation, and pattern formation (static or rectilinear waves if all vortices annihilate; spiral wave otherwise).
**(c)** Total number of vortices (green) and charged vortices only (brown) versus time in a representative 16 x 16 lattice. Inset: first 50 timesteps (Supplementary figs. 12–13).
**(d)** Numbers of +1 (orange), −1 (blue), and 0 (purple) vortices for the run in (c). Dashed lines mark vortex annihilation events (Supplementary fig. 13).
**(e)** Vortex lifetime after formation versus number of vortices present, for +1/−1 pairs (brown circles) and 0 vortices (purple squares). 16 × 16 lattice; 1,000 runs (Supplementary fig. 16).
**(f)** Merger of a +1/-1 pair into a transient 0-vortex, followed by annihilation. R is core-to-core distance; $R_0$ is the core-to-core distance at the moment of annihilation. *Bottom:* normalized core-to-core distance $R - R_0$ measured in units of lattice spacing versus timesteps before merger for a representative run (16 × 16 lattice).

## Brownian motion recapitulates vortex dynamics

Having established that +1 and –1 vortices move freely and annihilate upon contact, we asked whether vortices themselves could be treated as effective particles undergoing Brownian motion—executing unbiased diffusive motion (Fig. 3a). If so, the relative velocity—the change in distance between the cores of each +1/–1 vortex pair at successive timesteps—would follow a normal distribution. We tested this prediction using our vortex-tracking algorithm applied to thousands of cellular automaton runs. The resulting distribution closely matched a normal distribution (Fig. 3b), indicating that vortex cores diffuse as Brownian particles without drift until collision and annihilation.

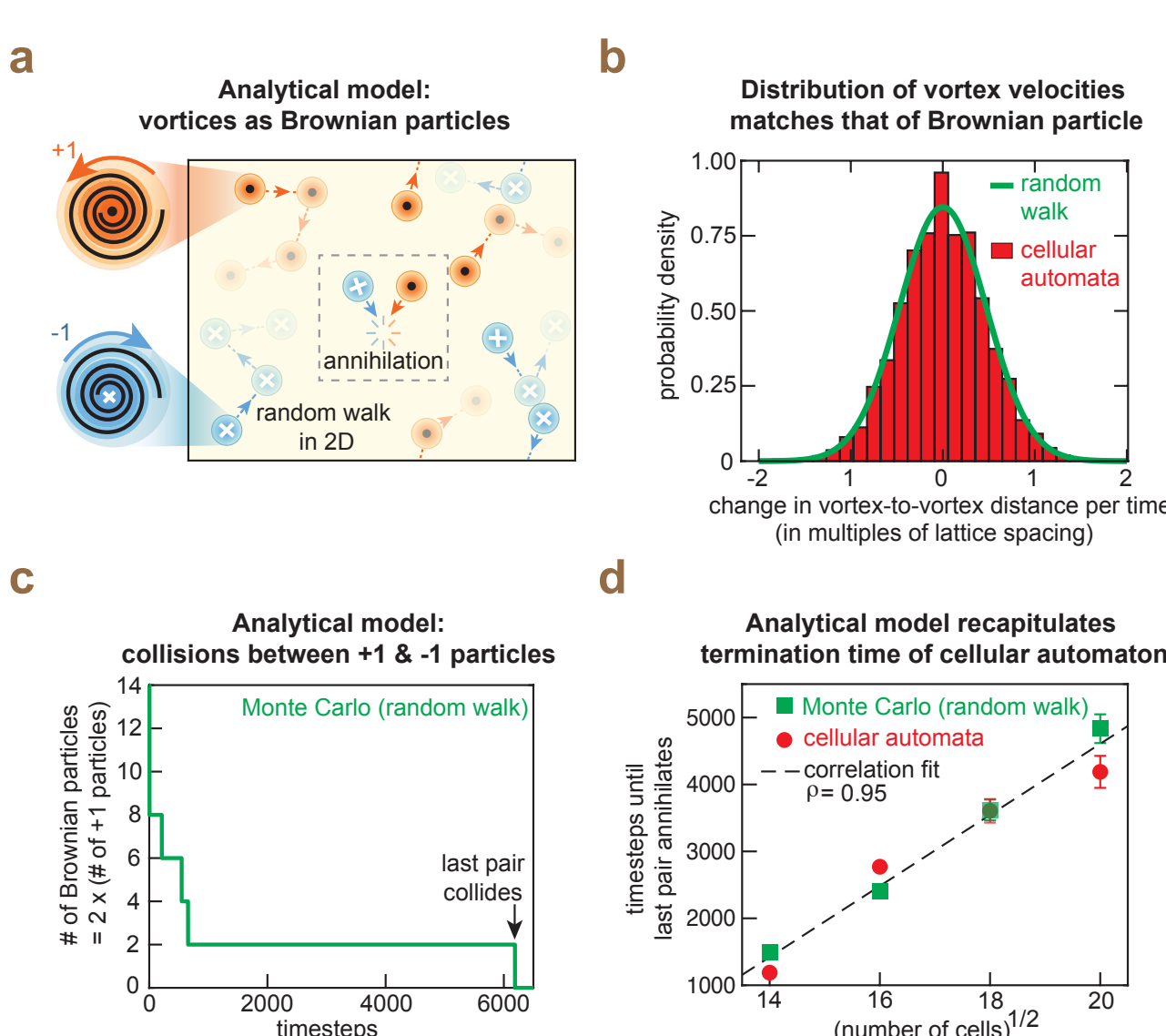

**Fig. 3 | Vortices behave as Brownian particles.**
**(a)** Brownian particle model. N vortices of each charge undergo independent two-dimensional random walks and annihilate upon contact (full details in Supplementary note 4).
**(b)** Distribution of relative vortex velocities — change in pair separation per timestep — from 1,000 cellular automaton runs (red), with normal-distribution fit (green). 16 x 16 lattice.
**(c)** Number of Brownian particle pairs versus time in a representative Monte Carlo simulation.
**(d)** Time until annihilation of the final vortex pair versus lattice size for the cellular automaton (red circles) and Monte Carlo model (green squares). Mean ± s.e.m.; 1,000 runs per lattice size and model. Pearson ρ = 0.95 (dashed line).

To test whether this particle-based description extends beyond single-step statistics, we constructed a minimal model of vortex dynamics: 2N Brownian particles—N of each type—diffusing on a lattice and annihilating upon contact (Supplementary note 4). Remarkably, this simple Monte Carlo model recapitulated the full temporal structure of the cellular automaton. As in the automaton, the number of vortex pairs decreased in a staircase-like fashion with progressively longer plateaus (Fig. 3c), and the time until the final pair annihilated—effectively setting the total run time—closely matched that of the cellular automaton across the lattice sizes examined (Fig. 3d; Pearson ρ = 0.95).

Together, these results show that, over the lattice sizes and time scales examined, the complex many-body vortex dynamics emerging from the cellular automaton are well approximated by an unexpectedly simple description: particles performing random walks with annihilation. This Brownian, particle-level description captures both vortex lifetimes and termination times, providing a minimal and intuitive framework for understanding the temporal structure of the pattern-forming dynamics. While we did not derive Brownian motion from first principles, the agreement between the particle model and the automaton suggests that vortex interactions effectively map the microscopic rules of cell-cell interaction into diffusive dynamics.

## Initial disorder seeds vortex creation

Having established an effective particle description of vortex motion and annihilation, we next asked what causes their spontaneous formation. Despite each automaton run starting from a distinct, maximally disordered configuration, we found a striking regularity across thousands of runs: the initial number of neighboring cell pairs differing by a $\pi$ phase was nearly constant, varying by only ~10% across runs (Fig. 4a; Supplementary fig. 18). This quantity rapidly declined over time, falling to roughly one-tenth of its initial value within ~50 timesteps on a $14 \times 14$ lattice (black curve, Fig. 4a). In parallel, the total number of vortices rose sharply, peaking within the first ~10 timesteps—marking the end of the vortex-creation phase—before monotonically declining thereafter (red curve, Fig. 4a).

The inverse temporal relationship between the initial $\pi$-phase differences and vortex number suggested that local $\pi$ mismatches act as seeds for vortex formation. To test this directly, we initialized lattices with controlled levels of disorder by varying the proportions of the four cell states, thereby tuning the initial number of $\pi$-phase-difference cells. As predicted, increasing initial $\pi$ mismatches increased the number of vortices created (Fig. 4b; Supplementary fig. 19).

This relationship saturated beyond a threshold: adding further disorder did not yield additional vortices. Moreover, the number of vortices formed was always smaller than the initial number of $\pi$ mismatches (Fig. 4b), indicating that many mismatches relax without producing vortices. Together, these results show that initial disorder seeds vortex creation, but only a subset of mismatches nucleate stable topological defects.

## Why +1 and -1 vortices must arise and disappear as pairs

While initial disorder seeds vortex formation, the strict pairing of +1 and –1 vortices arises not from the update rules, but from the topology of the lattice itself. The periodic boundary conditions render the lattice topologically equivalent to a torus—a closed surface on which isolated topological charge cannot exist. On such a surface, a lone +1 vortex cannot be embedded without introducing a compensat-ing −1 vortex elsewhere. To see why, consider imposing a single +1 vortex. Its defining contour—along which neighboring vectors rotate counter-clockwise by at most $+\pi/2$—must eventually wrap around the periodic boundaries and re-enter the lattice (Fig. 4c). Along this closed path, the rotation must reverse at some point, generating a local $\pi$-phase difference corresponding to a −1 vortex. Thus, +1 and −1 vortices are not independently created or destroyed by the cellular automaton's update rules; rather, they are globally constrained by lattice topology to appear and annihilate in pairs (see Supplementary note 5). This topological constraint acts as an effective conservation law for vortex dynamics.

## Non-contractible loops as hidden topological structures of vortex pairs

While examining how contours extend from vortices, we noticed an unexpected feature: starting from a +1 vortex core, one can reach a -1 vortex core by following a path composed entirely of cells in the same state (e.g., all 0-phase cells). Such a path, which we call a "string", can exist for each of the four cell states (Fig. 4d; Supplementary fig. 20), and thus constitutes an intrinsic structure of every vortex pair.

Because the lattice is a torus, closed paths of same-state cells may also wrap fully around the lattice—traversing the full horizontal and/or vertical extent of the lattice before returning to their starting point. We refer to these closed paths as "non-contractible loops" (NCLs) (Fig. 4d). An NCL cannot shrink to a point unless it is "cut"—that is, unless the cellular automaton update rule changes the state of at least one cell along the loop, which it can and does—meaning the term "non-contractible" refers to the topological property of the path on the torus, not to a constraint on the dynamics. Importantly, NCLs are global topological objects of the lattice and need not be associated with vortex cores. However, in some cases, a string connecting a +1/-1 vortex pair can itself constitute an NCL.

This distinction suggested a question: do vortex pairs behave differently when they are threaded by NCL strings? Since spiral waves arise only when at least one vortex pair survives indefinitely—whereas rectilinear waves and static configurations require complete vortex annihilation—we wondered whether the presence and dynamics of NCL strings might foreshadow which runs ultimately form spiral waves. To address this question, we monitored NCL strings over time, with particular focus on the late-stage regime in which only one vortex pair remains.

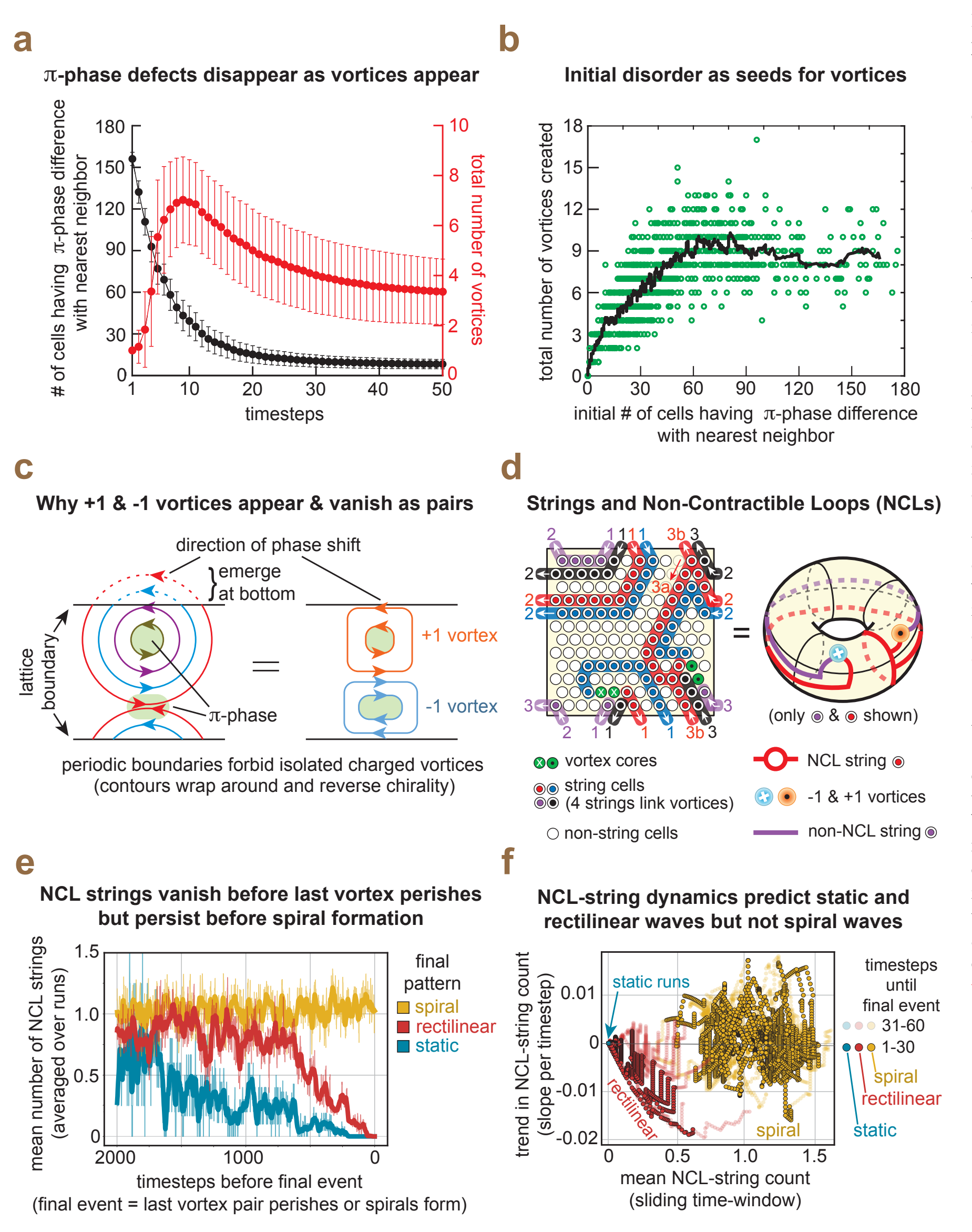

**Fig. 4 | Topology shapes vortex birth, pairing, and predictability.**
**(a)** Numbers of $\pi$-phase defects (black) and vortices (red) versus time from maximally disordered initial configurations. Mean ± s.d.; 14 x 14 lattice. See also Supplementary fig. 18.
**(b)** Total vortices created versus initial $\pi$ -phase-defect count in $16 \times 16$ lattices initialized with controlled cell-state proportions. Individual runs (green); running mean (black). See also Supplementary fig. 19.
**(c)** Periodic boundary conditions force +1 and −1 vortices to form and vanish in pairs; contours wrapping the torus must reverse chirality, topologically forbidding isolated charged vortices (proof in Supplementary note 5).
**(d)** *Left:* four state-specific strings connecting a +1/-1 pair in a 12 x 12 lattice. The red string (3a/3b) is the only non-contractible loop (NCL) string. The other three are contractible. White state shown as purple. Right: torus representations of NCL string (red) and a contractible string (purple) (See also Supplementary fig. 20).
**(e)** Mean NCL-string count versus timesteps before the final event — annihilation of the last vortex pair or spiral-wave formation — for runs on 26 x 26 lattices with one remaining vortex pair. Thin curves: ensemble means; Thick curves: 20-timestep sliding averages. Teal, static; red, rectilinear waves; gold, spiral waves (See also Supplementary figs. 21–22).
**(f)** Each point represents one timestep from one run (26 x 26 lattice, one vortex pair remaining), plotted in the space of mean NCL-string count versus fitted slope (both computed over a sliding time window). Opacity encodes proximity to the final event (opaque: 1-30 timesteps before; semi-transparent: 31-60 timesteps before). Spiral-wave runs (gold; 145 runs subsampled from 1,214) clustered near mean count of 1 and slope of 0 throughout. Rectilinear-wave runs (red; 145 runs) migrated toward lower mean and negative slope near the final event. Static runs (teal) had already lost all NCL strings before the plotted window and appear at the origin (all collapsed to one point). See also Supplementary fig. 23.

## Non-contractible loops predict final pattern types

Starting from the moment when only the final vortex pair remained, the number of NCL strings fluctuated for hundreds of timesteps around a mean near one, regardless of the final configuration. However, at later times—during the hundreds to thousands of timesteps preceding the final event (either annihilation of the last vortex pair or the onset of a spiral wave)—runs diverged into three late-stage behaviors, corresponding to the three final configuration types (Fig. 4e; Supplementary fig. 21).

In runs that yielded either rectilinear waves or static configurations, for which the final vortex pair eventually annihilated, the number of NCL strings increasingly spent time at zero (Supplementary figs. 21-22). Although the NCL-string count continued to fluctuate between zero and non-zero values, zeros occurred more frequently within a sliding time window as that window approached the annihilation event (Supplementary figs. 21-22). Consequently, the time-averaged number of NCL strings within the window gradually decreased toward zero as the system approached the annihilation event (Fig. 4e; Supplementary fig. 22). In runs yielding static configurations, this averaged value reached zero hundreds of timesteps before the annihilation (green curve in Fig. 4e), whereas in runs yielding rectilinear waves it typically reached zero only tens of timesteps before annihilation (red curve in Fig. 4e). In contrast, for runs yielding spiral waves, the number of NCL strings fluctuated around an average of approximately one until the spiral wave arose (yellow curve in Fig. 4e & Supplementary fig. 22).

These observations establish that an observer, without knowing how many timesteps remain in a run, can use the forward-time dynamics of NCL strings to predict that a spiral wave will not form and—conditional on that—whether the run will yield a static configuration or a rectilinear wave. This is achieved by tracking, within a 100-timestep sliding time window, both the mean number of NCL strings and the temporal trend of the NCL-string count within that window (Fig. 4f & Supplementary figs. 21 & 23). We computed the temporal trend as the slope of a linear least-squares fit to the NCL-string count versus time within each window. A consistently negative trend reflects an increasing dominance of zero-NCL-string timesteps and signals imminent vortex annihilation. Hence, once only a single vortex pair remains, the dynamics of NCL strings reveal whether vortex annihilation is imminent and, conditional on that, distinguish static from rectilinear-wave outcomes, but provide no analogous anticipatory signature for spiral-wave formation.

Yet the NCL-string analysis requires knowing in advance which topological construct to track. Having uncovered this mechanism, we asked a sharper question: even setting aside the full topological description, does the total vortex core size—a single scalar that compresses the entire lattice state into one number—at each timestep still reveal when fate-dependent predictive structure first emerges?

## Fate-dependent microstructures emerge in the final timesteps

To probe this, we examined a simple trajectory-level observable: the total size of all vortex cores at each timestep. This quantity counts how many cells exhibit a $\pi$-phase difference with at least one nearest-neighbor and thus tracks the overall level of vortex activity without resolving individual vortices or strings. We computed this time series for tens of thousands of simulations, each initialized from a distinct maximally disordered configuration and ending in one of the three final pattern types. We refer to each simulation as a trajectory.

When plotting this quantity over time for individual trajectories, we discovered that runs destined for completely different fates were virtually indistinguishable at the level of this observable for thousands of timesteps. Trajectories diverged only within the few dozen timesteps preceding final-pattern formation (Fig. 5a), suggesting that any reliable predictors of fate based on this observable — if present at all — must emerge extremely late in the dynamics.

To examine this systematically, we visualized 2,000 trajectories as a heat map, with 19 additional sets of 2,000 trajectories analyzed similarly (Supplementary fig. 24). Each row represents a trajectory, each column a timestep counted backward from the final configuration (time flows right to left), and each entry encodes the total number of cells belonging to vortex cores (Fig. 5b). Most of the heat map shows no shared structure across trajectories at the level of this observable. Consistent with this, the three final pattern types remain visually intermixed across most rows, with no obvious segregation by run duration or outcome. However, zooming into the final ~20 timesteps reveals barcode-like microstructures — dark horizontal bands — that are not apparent earlier in the trajectories. This qualitative observation motivated us to test whether similarities among these late-stage dynamics encoded final fate.

## Clustering late-stage dynamics reveals pattern fate

Motivated by this observation, we asked whether these late-stage microstructures encoded a sufficiently predictive signature to identify pattern fate. We applied a graph-based, unsupervised clustering approach using only the vortex-core-size time series, thereby treating each trajectory as a single object (see Supplementary note 6). We represented each trajectory as a node, with edges connecting trajectories whose time series were sufficiently similar.

We performed this analysis across sliding backward-time windows $[1, \tau]$, where $\tau$ is the number of timesteps before the final configuration formed. For large $\tau$ (e.g., 12,000), nearly all trajectories were isolated, forming no meaningful clusters (Fig. 5c, right). As $\tau$ decreased, trajectories—treated as single objects—were progressively absorbed into larger clusters. The total number of connected components fell reproducibly across 19 independent trajectory sets, while the fraction of trajectories remaining in singleton clusters dropped toward zero and the largest connected component grew in a stepwise manner (Supplementary fig. 25). By $\tau = 20$, the 2,000 trajectories collapsed into just 23 well-separated clusters (Fig. 5c, left), a result reproduced across additional trajectory sets (Supplementary fig. 25).

Within this reduced description, we discovered that fate became legible only after trajectories had already merged into these late-time clusters. Specifically, at earlier times, trajectories leading to different final configurations remained too fragmented for similarity-based prediction. A large fraction of trajectories persisted as isolated singletons, and the few clusters that did form were small, leaving no robust basis for inferring fate from resemblance to previously observed trajectories. Only near the end (small $\tau$) did the clustering reorganize the ensemble into larger groups segregated by the three final pattern types — static configuration, rectilinear wave, and spiral wave. Thus, although fate was unreadable from this observable for most of the trajectory, the final ~20 timesteps contained a sufficient signature for outcome-specific trajectory groups to emerge clearly.

## Fate signatures crystallize at late times

Tracking how clusters evolved as τ decreased revealed an uneven process: cluster sizes evolved in a stepwise fashion, with periods of approximate stability punctuated by rapid growth that accelerated at small τ (Fig. 5d). This behavior indicates that fate-defining features do not accumulate at a uniform rate but instead crystallize most fully in the final phase of each trajectory. Consistent with this picture, many trajectories became not merely similar but identical at late times. Among the 2,000 trajectories analyzed, only 180 distinct vortex-core-size profiles remained when considering the final 20 timesteps (Fig. 5e). Importantly, while the reduction in trajectory diversity accumulated across the full range of τ, the final ~20 timesteps produced the most compressed representation, collapsing 2,000 globally distinct trajectories into only 180 distinct profiles. We observed the same late-time convergence across all other trajectory sets (Supplementary fig. 26). Even when we pooled 38,000 trajectories from 19 additional trajectory sets, these globally distinct histories collapsed by the final 20 timesteps to only 977 exact vortex-core-size profiles—a 97.4% reduction in trajectory diversity (Supplementary fig. 27). Extending this analysis to smaller backward-time windows (smaller τ) revealed a striking asymmetry between outcomes (Supplementary fig. 28). Static-destined runs collapsed to a single exact terminal scalar sequence for the final four timesteps, rectilinear-wave runs collapsed to only a few terminal scalar sequences, whereas spiral-wave-destined runs remained far more diverse even at τ = 1. Thus, unlike annihilation-bound outcomes, spiral-wave formation did not converge onto a comparably compressed set of terminal scalar trajectories.

## Machine learning's ability to predict progressively emerges as dynamics unfold

Given that many of the same-fate trajectories, in the reduced-dimensional description of vortex-core areas, became more similar to each other as they evolved, we considered the possibility that a single lattice snapshot taken at a sufficiently late timestep could contain a legible structure that would allow a machine-learning model to accurately predict the final pattern fate. This is not precluded by our finding that none of the machine-learning models we examined predicted the pattern fate better than random guessing could from initial configurations (Fig. 1f). In fact, the NCL strings are late-time predictive structures, and we reasoned that some machine-learning models could independently discover at least the NCL strings as we did. To investigate this possibility, we handed a convolutional neural network (CNN) one lattice snapshot taken partway through a trajectory and asked it to predict which of the three patterns will form (Fig. 6a). This is a harder task than the binary static-versus-dynamic prediction we posed earlier with the initial configuration. We discovered that as we fed progressively later snapshots to the CNN, its three-fate balanced accuracy climbed steadily from chance (1/3) — equivalent to randomly guessing — to near-perfect prediction at the pattern-formation moment (Fig. 6b). The same progression occurred in the binary static-versus-dynamic prediction task (Supplementary fig. 29).

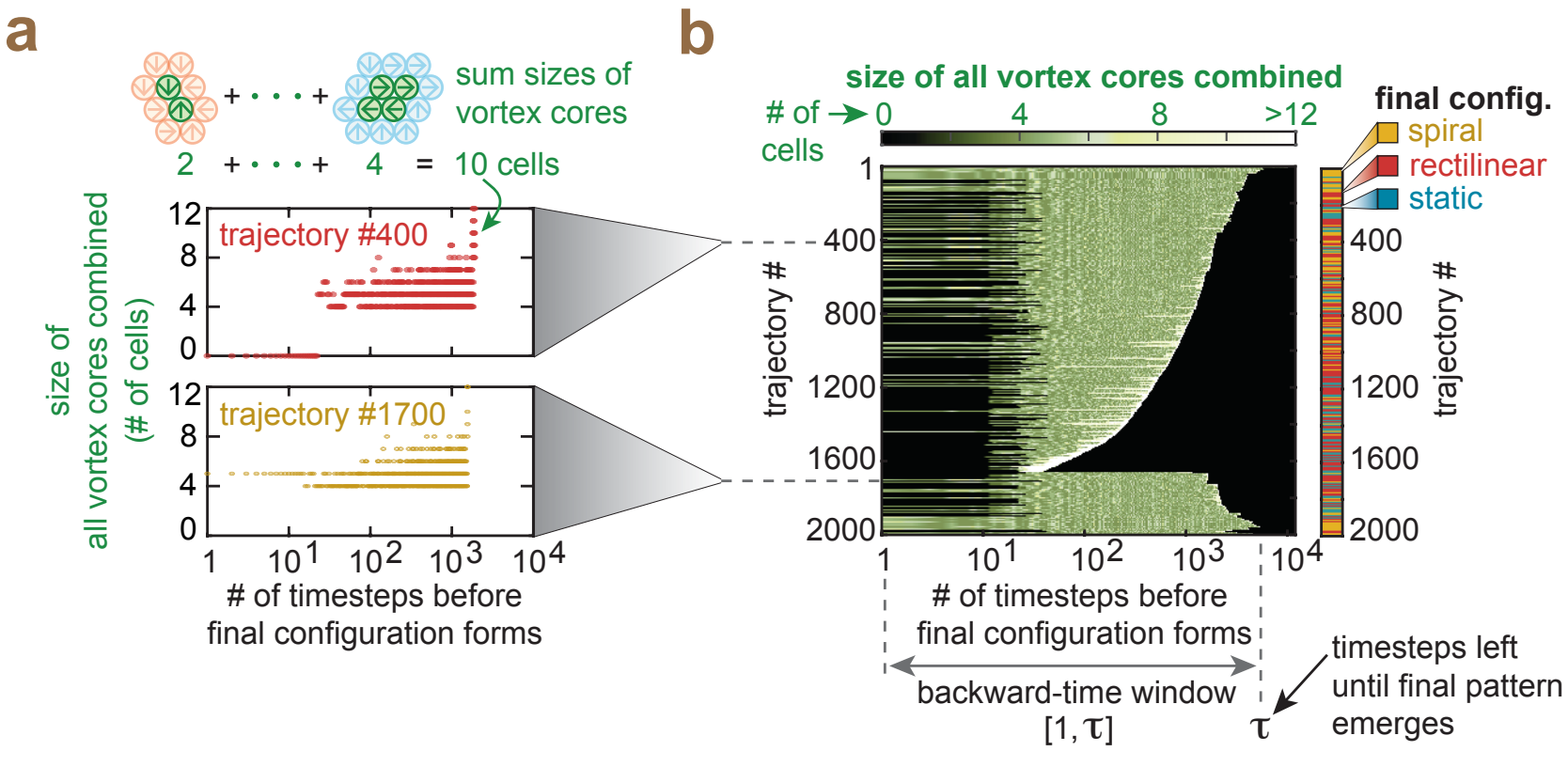

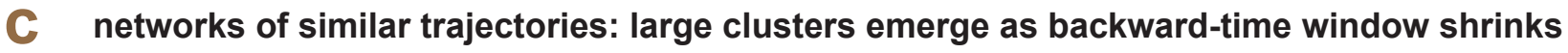

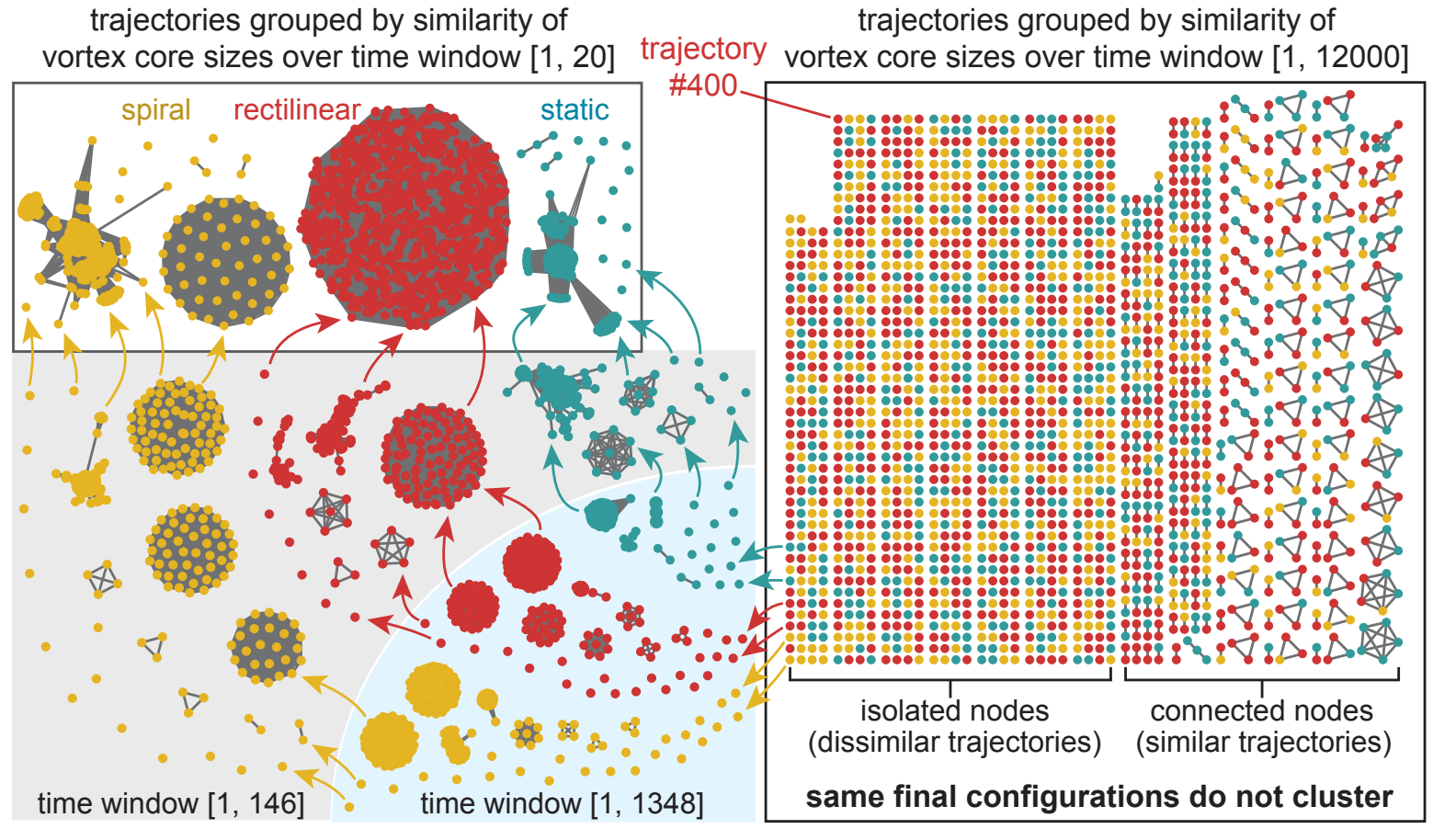

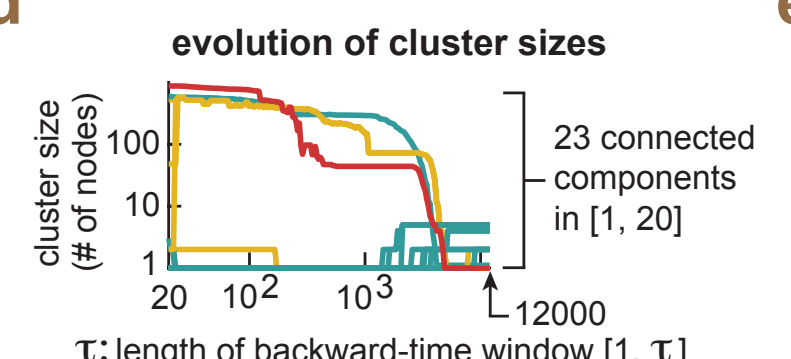

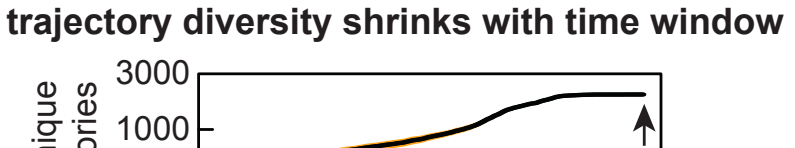

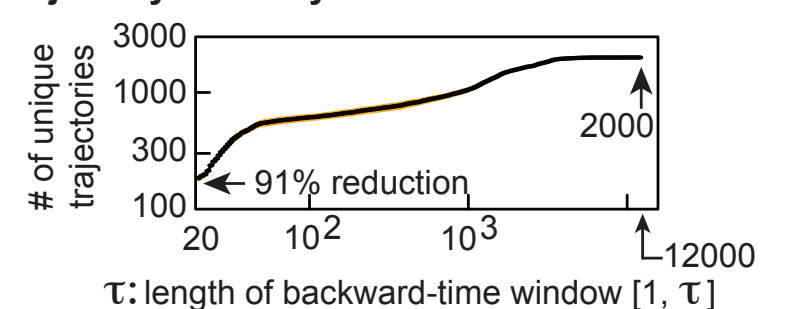

**Fig. 5 | A trajectory-similarity graph recovers fate-dependent clustering, but only in late times.**
**(a)** Total vortex-core size, defined as the number of cell belonging to all vortex cores combined, for representative trajectories forming a spiral wave (gold) or rectilinear wave (red). 14 x 14 lattices.
**(b)** Total vortex-core size for 2,000 trajectories from distinct disordered initial configurations. Rows denote trajectories; columns denote backward time τ, with τ = 1 one timestep before final-pattern formation. Rows were ordered greedily by nearest-neighbor Euclidean similarity between complete time series, beginning with the longest trajectory. Black denotes timepoints beyond a trajectory's duration; the right bar enotes final pattern. The late-time heat-map structure recurred in 19 additional trajectory sets (Supplementary fig. 24).
**(c)** Graph-based clustering using vortex-core-size time series over backward-time windows [1, τ]. Nodes denote trajectories. Edges connect sufficiently similar time series (Supplementary note 6). At large τ = 12,000, most nodes were isolated. At τ = 20 (left panel), the trajectories formed 23 connected components segregated by final pattern. Intermediate windows show the reorganization (Supplementary fig. 25).
**(d)** Connected-component sizes as τ decreases, showing extended plateaus and discrete increases.
**(e)** Number of unique scalar trajectories, defined by exact step-for-step matches, versus τ. The 2,000 trajectories at τ = 12,000 reduced to 180 profiles at τ = 20 (91% reduction). This analysis was repeated across 19 additional sets (Supplementary fig. 26). Pooling all 38,000 trajectories yielded a 97.4% reduction (Supplementary fig. 27).

We then asked whether this success was merely memorization — whether same-fate trajectories had simply merged onto configurations the CNN had already seen in training, so that it was recognizing stored snapshots rather than reading structures within a snapshot it had never seen. At the midpoint of the trajectories, only ~2% of the snapshots tested had an exact match anywhere in the training set, yet the CNN correctly predicted the pattern fate for far more than that (Supplementary fig. 30 & Supplementary note 7). The CNN was therefore predicting from configurations it had never encountered by generalizing — exact configuration-lookup could not explain its performance.

A CNN, which knows which cells are the nearest neighbors, matched the best accuracy of a fully connected multilayer perceptron (MLP), which does not, despite using ~8 times fewer training examples (Supplementary fig. 31 & Supplementary note 7). This suggested that a machine-learning model (network) that exploited locality could predict just as well with far fewer training examples. This hinted at the predictive signal living in the spatial arrangements of cells. Telling the CNN that the lattice was a torus changed little: wrapping CNN's convolutions around the periodic boundaries, rather than padding the lattice as a flat sheet, left its prediction accuracy nearly unchanged (Supplementary fig. 31). Hence, the CNN did not require convolutions that explicitly encoded the periodic boundary conditions to read the prediction-enabling structures. Finally, neither the vortex-connecting string mesh nor the NCL-string count explained the CNN's performance: each predicted fate poorly on its own for most of the trajectory, and deleting the entire string mesh from the configuration barely changed the CNN's accuracy (Supplementary fig. 32). These findings showed that none of the features we identified so far could, on its own, explain the CNN's performance, suggesting that the full configuration contained additional predictive structure.

Machine-learning models of widely distinct architectures — transformers, neural cellular automata (NCA), hidden Markov models (HMM), MLPs, and gradient-boosted decision trees (XGBoost) — likewise predicted the final fate no better than random guessing from initial configurations, but all performed better on lattice configurations taken from later parts of trajectories (Supplementary figs. 33-36). None outperformed the CNN at any point. Insufficient model depth did not explain the failure: for the CNN, NCA, and transformer, performance saturated at shallow depth, and increasing depth never improved prediction from initial configurations (Supplementary fig. 37). Together, these results show that the prediction failure from initial configurations was neither specific to one model class nor caused simply by insufficient computational depth, further supporting that the maximally disordered initial configurations contain no practically accessible structure that forecasts the final fate.

Despite the CNN's gradually increasing prediction accuracy, prediction remained genuinely difficult through much of the trajectory. Even at midpoint of trajectories, the CNN performed above chance but far from reliably (~65% balanced accuracy) (Fig. 6b). This difficulty was not shared equally among the three fates. The CNN could predict the fates for static-bound and rectilinear-wave-bound trajectories with at least 50% accuracy by ~40% of the runtime, whereas spiral-wave-bound trajectories stayed hard to predict until close to formation — the prediction accuracy remained only ~56% even by ~80% of the runtime and climbed steeply thereafter (Fig. 6c). The CNN thus showed the same asymmetry we had seen in the late-time clustering and in the NCL strings: vortex-annihilation-bound fates announce themselves early, through progressive topological simplification, whereas spiral-wave-bound fates carry no comparable precursor and resolve only as the pattern forms. We found the same asymmetry in the NCA and transformer at every training depth we examined (Supplementary fig. 37).

Taken together, these results place the NCL strings in context. They are genuine predictors, but only very late, when a single vortex pair remains. Earlier in each trajectory, the full configuration contained predictive structure beyond what the NCL strings alone captured—structure that was absent initially and self-organized over time. We next sought to identify it and to determine whether it too is topological.

### An emergent topological field carries the predictability

Motivated by the NCL strings, we considered their generalization: how a connected region of same-state cells wraps the torus. Cells of identical state could form connected regions across the lattice. On a torus, such a region could close on itself locally, or it could wrap all the way around — once, twice, or more — in either of the two directions (vertical or horizontal). We labeled each cell solely by how many times its region wraps in each direction — the same "winding vector" $(W_x, W_y)$ for every cell in the same connected region, where $W_x$ and $W_y$ are the horizontal and vertical winding numbers respectively. As a new description of the lattice, the winding field — we define it as the collection of the per-cell winding vectors of every cell — discards the cell's original state (phase) entirely. It tells us only the wrapping topology of the region a cell sits in, from which we cannot recover each cell's phase (Fig. 6d). Crucially, two distinct regions in the phase representation — one connected region of $\pi$-phase cells and another of $\pi/2$-phase cells — could be adjacent and have the same winding vector, thereby appearing as a single connected region in the winding-field representation (Fig. 6d - film strip; Supplementary note 7).

The winding vector was initially zero in every cell of virtually every initial configuration: any connected same-state region was initially tiny and almost never wrapped the torus. By counting both the percentage of cells with a non-zero winding vector and the total number of vortices in nearly a million trajectories, we found that the winding field switched on — a tiny fraction of cells first acquired non-zero winding vectors — during the same early time window (first ~1% of the runtime) in which the vortices formed (Fig. 6e & Supplementary fig. 38). In this same time window, the phase field (cell states) ordered first — its spatial order rose and nearly saturated within the first several timesteps — and only afterward did the winding field begin to order (Supplementary fig. 38). After the vortex-creation phase, the winding field further self-organized, slowly spreading to more cells as the vortices moved and annihilated. Eventually, as the last vortex pair annihilated and the static configuration or rectilinear wave was about to form, the winding field suddenly and rapidly expanded to fill the entire lattice (Fig. 6e & Supplementary fig. 39). Spiral-wave trajectories — the one fate with surviving vortices — exhibited no such jump; their winding field kept expanding slowly, reaching only ~60% of the lattice on average at wave formation (Fig. 6e; Supplementary fig. 39). For most of the dynamics, the three fates were alike in both the average size of their winding field and how that average size evolved over time, separating markedly only as they neared formation (Fig. 6e).

The separations, however, are averages over many trajectories, and they are laid out in scaled time, which presupposes knowing when each trajectory ends. To test whether the winding field carries the fate of each individual trajectory, and not merely on average, we returned to machine learning. Whether a same-state region wraps the torus is a global property: no bounded local neighborhood fixes a cell's winding number, so a network can determine it only by integrating spatial relationships across the entire lattice. Although computed deterministically from the configuration, the winding field makes the global topology directly available as a per-cell label, sparing the network from having to learn this global transformation. We fed the CNN the winding field alone and retrained it across different timepoints of trajectories. Strikingly, at every timestep of the trajectory, the winding field enabled the CNN to predict the fate nearly as accurately as the full lattice configuration did (Fig. 6f & Supplementary fig. 40), indicating that the winding field exposes nearly all of the fate-predicting signal accessible to the full configuration-fed CNN. Adding the full vortex description on top — the size, location, and chirality of every vortex — increased only slightly the accuracy (Fig. 6f). These results establish that the winding field is a general topological structure that carries the predictability, of which the NCL strings are the sparse, late-time instance.

We next asked how much detail in the winding field was needed for prediction. We found that the sign, or orientation, of each winding number — a matter of convention — did not matter: replacing positive and negative winding numbers with their magnitudes left the CNN's accuracy unchanged (Supplementary fig. 41). Moreover, same-state regions almost never wrapped the torus more than once. We could therefore reduce the winding field to a single yes-or-no label for each cell — whether its region wrapped the torus at all — and still predict fate nearly as accurately as with the complete winding field (Supplementary fig. 41). The exact number of wraps became informative only at the pattern-formation moment, and even then only helped distinguish rectilinear waves from the other two pattern types (Supplementary fig. 42).

Taken together, these findings establish that the predictive signal lives not in any one cell state but in the connectivity of same-state cells, so that each of the four states (phases) is a nearly equivalent window onto the same predictive structure. Hence, we have uncovered a system in which predictability is dynamically constructed: the emergent topological field that makes fate legible is absent at the start and takes shape, step by step, only as the system evolves.

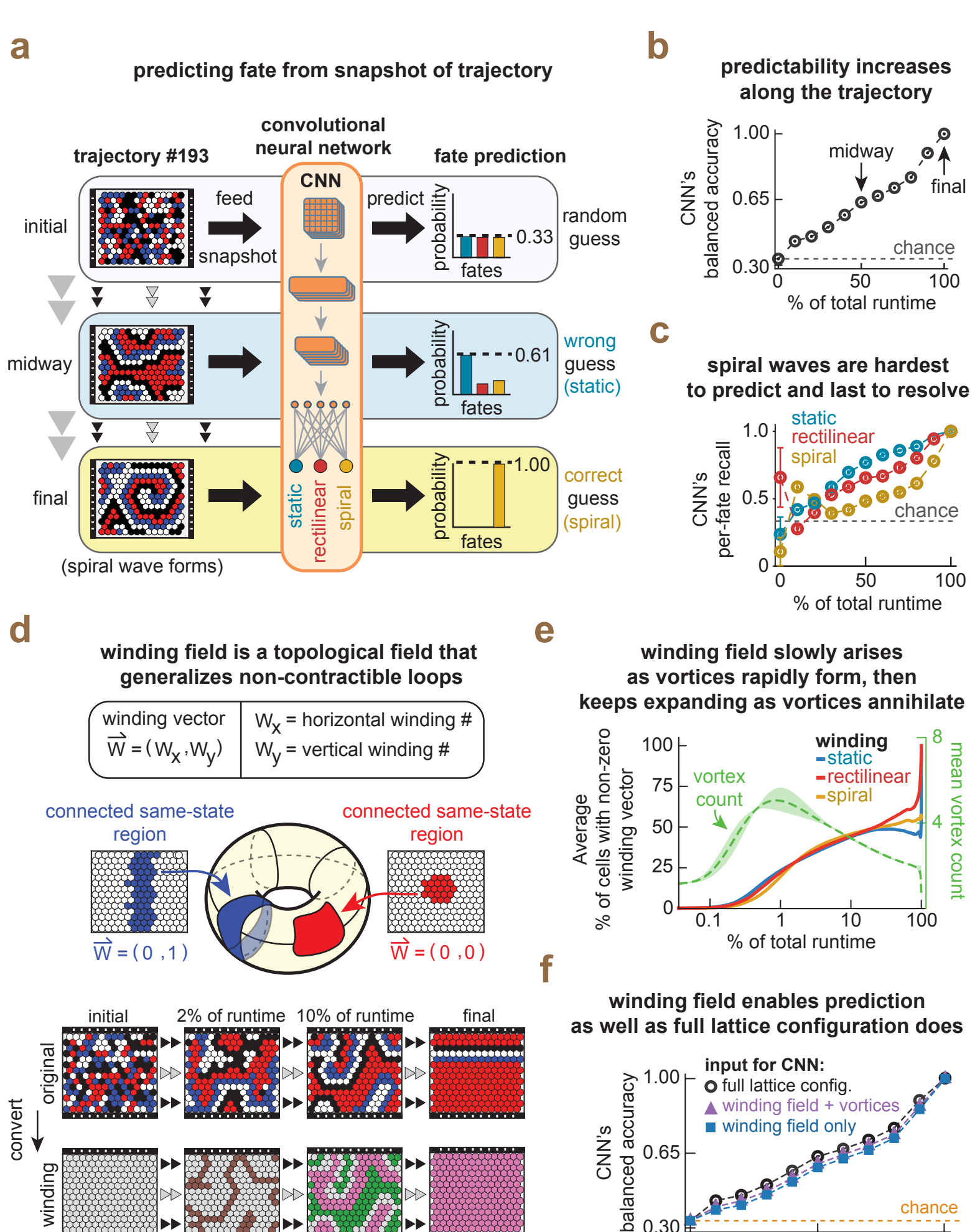

**Fig. 6 | A dynamically constructed topological field carries nearly all fate-predictive signal.**
**(a)** A CNN trained and tested on lattice snapshots sampled at specified percentages of total runtime. The task was to predict one of the three pattern fates. Details in Supplementary note 7.
**(b)** Balanced accuracy versus percentage of total runtime. Dotted line, three-class chance level (1/3). Means of three independent training runs; 850,000 training configurations and 100,000 held-out test trajectories per run. Maximum s.e.m. = 0.002, smaller than the data symbols.
**(c)** Per-fate recall of the CNN in **b**, defined as the fraction of trajectories of each fate classified correctly. Test set contained 26,784 static, 29,812 spiral-wave, and 43,404 rectilinear-wave trajectories. Means of three independent training runs; error bars, s.e.m; dotted line, 1/3. On initial configurations, each training run favored different fates, producing large variation in per-fate recall (static, 0 - 0.44; rectilinear wave, 0.25 - 1), although balanced accuracy remained at chance (**b**).
**(d)** Determining the winding vector of every cell. To the left of the torus, all blue cells have the same winding vector, **W** = (0, 1). To the right, all red cells have **W** = (0,0). Film strips show lattice snapshots in the original representation (top row) and the corresponding winding-field representation (bottom row).
**(e)** Percentage of cells with non-zero winding vector (left axis), separated by final fate, and total vortex count (right axis). Both are averages over one million trajectories. Maximum s.e.m. of the winding-fraction curves = 0.06%, smaller than the line widths. Green band, range of mean vortex counts across the three fates; dashed green line, overall mean. The starts of trajectories are omitted. See also Supplementary figs. 38 & 39.
**(f)** Balanced accuracy versus percentage of total runtime for CNNs receiving the full lattice configuration (original representation), the winding field alone, or the winding field plus vortex sizes, locations, and charges. Means of three independent training runs; 850,000 training configurations and 100,000 held-out test trajectories per run. Maximum s.e.m. = 0.005, smaller than the data symbols. See also Supplementary fig. 40.

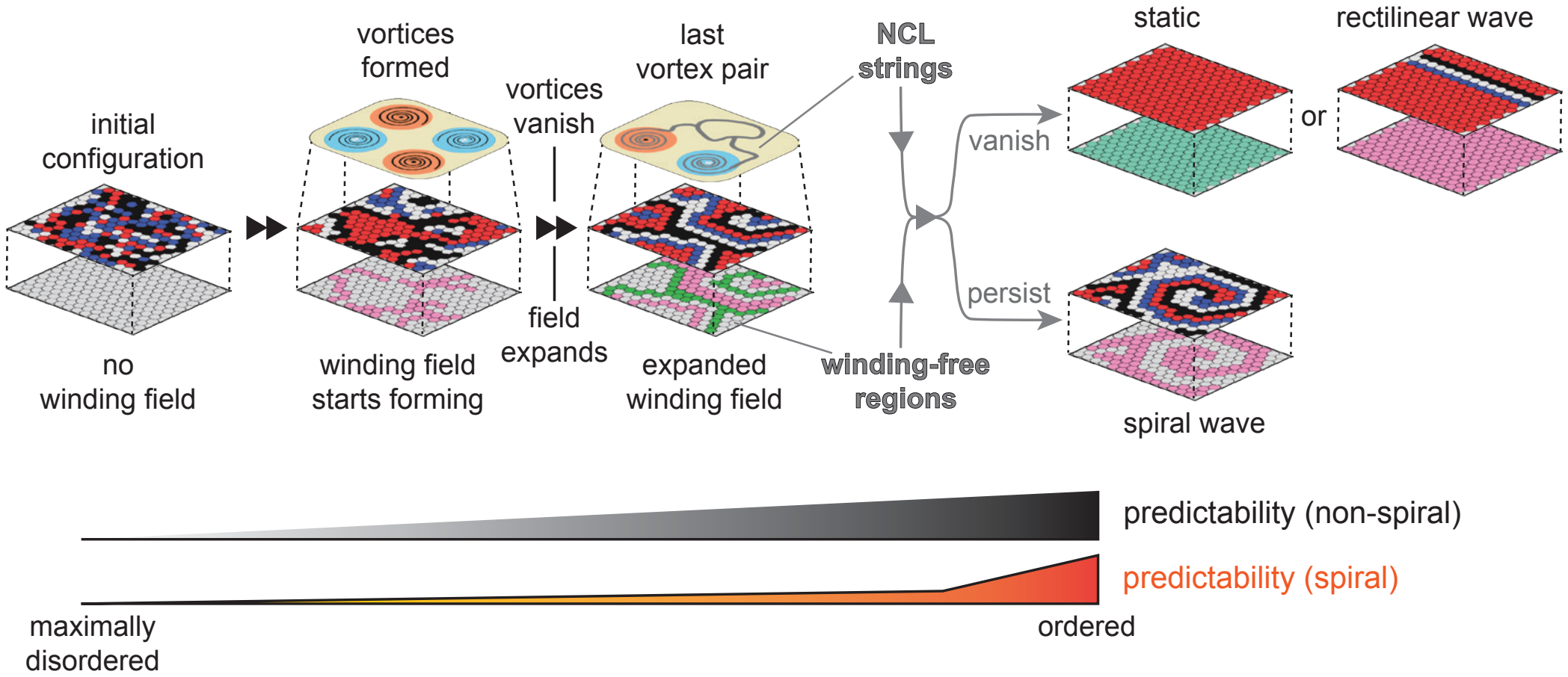

**Fig. 7 | Predictability is constructed gradually and asymmetrically through deterministic dynamics.**
Starting from a maximally disordered configuration without a winding field, vortices and the winding field emerge. The winding field then expands gradually as vortices annihilate, eventually filling the entire lattice only along trajectories in which all vortices annihilate to yield a static configuration or rectilinear wave. The dynamics of the non-contractible-loop (NCL) strings threading the final vortex pair can reliably forecast only non-spiral-wave fates, whereas the dynamics of winding-field expansion allow prediction of all three pattern fates. Annihilation of the NCL strings and vortices, accompanied by the disappearance of winding-free regions, announces a static or rectilinear-wave fate. By contrast, the persistence of winding-free regions provides no analogous early precursor to spiral-wave formation, which becomes reliably predictable only as spiral-wave emergence nears.

## DISCUSSION

Here we uncovered a distinction between determinism and predictability in a non-chaotic, deterministic system. This work establishes that the entire future can be fixed from the start and unfold through non-chaotic deterministic dynamics, yet it can remain unpredictable in practice. We found that predictability can instead be constructed gradually along the dynamics: the system's future becomes predictable only as its trajectory evolves. We then showed that prediction from an initial state can fail not because the dynamics are stochastic or chaotic, but because the collective variables that make the macroscopic fate legible can be emergent, taking shape only later during the system's evolution.

In the generalized cellular automaton we studied, disordered initial configurations evolved into static configurations, rectilinear waves, or spiral waves that could not be forecasted from the initial configurations, even by expressive machine-learning models trained on $10^6$ simulations. As global wrapping topology self-organized, the same machine-learning models became progressively more accurate. A model given only the winding field — a per-cell label recording how a connected same-state region wraps the torus — recovered most of the accuracy it achieved from the full configuration throughout the trajectory. The NCL strings threading the final vortex pair — a sparse representation of the same wrapping topology encoded by the winding field — and the winding-free regions both vanished along trajectories ending in static configurations or rectilinear waves but persisted along spiral-wave-bound trajectories (Fig. 7). This correspondence had asymmetric consequences: vanishing is a change that a forward-time observer can detect without knowing when the trajectory will end, whereas persistence alone cannot announce a spiral-wave fate. Only as pattern formation neared did winding-field expansion accelerate enough to provide an observable change, making spiral-wave fate reliably readable from its expansion dynamics. Crucially, the prediction-carrying winding field was absent from virtually every initial configuration and therefore unavailable for any model to read, strongly suggesting that this topological absence is the reason that fate prediction from an initial configuration is practically unachievable.

These results differ from existing accounts of unpredictability in deterministic systems. Wolfram's classical results[9] concern one-dimensional cellular automata that, like ours, are discrete in both time and state. The relevant difference here is the size of global state space. Such automata are typically defined on infinite lattices, giving them infinite global configuration spaces and allowing computational resources to grow without bound. Within this class, computational universality makes some questions about long-term behavior formally undecidable[34], while other automata exhibit chaos on infinite lattices. Our finite lattice instead has a finite global configuration space and eventually periodic dynamics. The unpredictability we discovered is therefore neither chaos nor formal undecidability. Later work showed that Wolfram-type automata can nonetheless be coarse-grained into predictable effective descriptions[10,11], and, in a continuous and often chaotic setting, other work estimates the predictability of individual states from trajectory statistics[35]. Our contribution is to identify when predictive structure appears, what physical objects carry it, why it emerges asymmetrically, and how it self-organizes along the dynamics — treating predictability itself as an entity embedded in physical structure, and tying a disorder-to-order transition to an unpredictable-to-predictable one.

In calling a system unpredictable, we distinguish being "formally determined" from being "predictable in practice." Established frameworks ask different questions about prediction over ensembles of trajectories. Predictive information asks how much observing the past reduces uncertainty about the future, statistical complexity asks how much memory the simplest predictive model needs, and transition-path theory asks how likely a state is to reach one outcome rather than another under stochastic dynamics[36–38]. Our question was different. In our deterministic system, the exact configuration does not merely make one fate more likely than another but fixes one fate uniquely at every time. Thus, how completely the state determines fate does not change. What changes is whether a practical predictor can read that fate from the state. Practical predictability therefore depends not only on the system but also on the predictor used. Holding the predictor classes and training procedure fixed across time isolates what changed: later configurations made each trajectory's already-fixed fate progressively easier for those predictors to read. The winding field makes this distinction concrete: it introduces no independent input, yet as global wrapping topology self-organizes, it becomes an increasingly fate-predictive representation. We showed that even when the future is fully determined, the structure that makes it practically predictable may have to be dynamically constructed.

Although we studied a single generalized cellular automaton, three ingredients help explain the topological objects we found. First, cyclic internal states let the system be recast as a discrete phase field with topological defects of discretized winding. Second, spatial coupling lets those defects act as coherent, interacting objects. Third, periodic boundary conditions enforce net topological-charge neutrality, while the toroidal domain permits non-contractible loops. In our system, defects consequently appeared and vanished in correlated pairs. These ingredients, or close analogues of them, recur across nonequilibrium systems — excitable media[39–41], coupled oscillator networks[42,43], and lattice models with discrete rotational symmetry[42,44–48] — making such systems natural settings in which to test whether related topological structures can render macroscopic outcomes legible (Supplementary note 8). Our findings raise the broader possibility that a disorder-to-order transition can coincide with an unpredictable-to-predictable one whenever the structure that orders the dynamics is also what renders the outcome legible. Demonstrating this in one system, and identifying these ingredients, gives a concrete basis for testing whether the same dynamics arise elsewhere[49].

More broadly, our findings suggest that in a class of non-chaotic deterministic systems, mechanistic understanding may reside not in the initial state or the final fate alone, but in collective entities that arise during the dynamics. In this view, self-organization is not merely the unfolding of a pre-encoded outcome, but a dynamical process that constructs the structure that makes prediction possible.

## METHODS

**Initialization of cellular states.** We initialized each simulation from a spatially disordered configuration designed to have negligible spatial correlations while maintaining controlled fractions of cells expressing each gene. Each cell contains two genes, and each gene can be either ON (representing 'high' secretion rate) or OFF (representing 'low' secretion rate). We represent these states by binary variables $X_m^{(z)} \in \{0,1\}$, where $m$ indexes the cell and $z \in \{1,2\}$ denotes the gene. The four possible combinations $(X_m^{(1)}, X_m^{(2)})$ correspond to the four cellular states shown in Fig. 1b. We initialized the two genes independently. For each gene z, we first generated a binary lattice in which a fraction $p_z$ of cells were assigned $X_m^{(z)} = 1$ and the remaining cells were assigned $X_m^{(z)} = 0$. We quantified spatial correlations in these initial gene-expression patterns using modified Moran's $I$ statistic,

$$I_z = N / (\Sigma_m \Sigma_{j \neq m} w_{mj}) \times [\Sigma_m \Sigma_{j \neq m} w_{mj} (X_m^{(z)} - \bar{X}^{(z)}) (X_j^{(z)} - \bar{X}^{(z)})] / [\Sigma_m (X_m^{(z)} - \bar{X}^{(z)})^2],$$

where $N$ is the total number of cells, $w_{mj}$ equals 1 for nearest neighbors and 0 otherwise, and $\bar{X}^{(z)}$ denotes the mean value of $X^{(z)}$ across the lattice. To obtain configurations with negligible spatial correlations ($I_z \approx 0$), we iteratively swapped the states of randomly selected pairs of cells while preserving the total number of ON and OFF cells. After each proposed swap, we recomputed $I_z$ and accepted the swap if it moved the system closer to the target value. Repeating this procedure generated binary patterns with $I_z$ very close to zero while maintaining the prescribed fraction $p_z$. In the simulations reported in this study, we set $p_1 \approx 0.5$ and $p_2 \approx 0.5$ and generated configurations with $I_1 \approx 0$ and $I_2 \approx 0$. These conditions produce spatially disordered lattices in which each gene is expressed in roughly half of the cells and exhibits no detectable spatial clustering. Because the two genes were initialized independently, the four cellular states occurred in approximately equal proportions across the lattice. All subsequent automaton dynamics evolved from these initial conditions. At each timestep, every cell sensed the concentrations generated by all cells and updated its state simultaneously according to the regulatory rules described in Supplementary note 1.

**Machine-learning analysis on initial configurations.** To determine whether the final macroscopic fate of the automaton could be predicted from the initial configurations alone by tractable supervised learning approaches, we designed an extensive machine-learning (ML) analysis that was deliberately biased in favor of finding predictive signal (Supplementary fig. 4 and Supplementary note 2). We sought to ask the most forgiving version of the question: can any supervised learning approach predict fate from the initial state, given virtually unlimited training data? We therefore reduced the classification problem to its simplest binary form—predicting whether a run would end in a static configuration (label 1) or any dynamic pattern (rectilinear or spiral waves; label 0)—so that the prediction task was strictly easier than distinguishing all three final outcomes. We generated 1,000,000 independent simulations on a 14 × 14 triangular lattice, each starting from a randomly sampled, maximally disordered initial configuration. For storage, we saved each 14 x 14 lattice configuration as a flattened vector of 196 cell states encoded as integers {1, 2, 3, 4}, together with a binary fate label determined by evolving the automaton until its final pattern type was established. Before being supplied to the CNN as an input, we reshaped each vector as a 14 × 14 lattice and one-hot encoded it as a four-channel two-dimensional image, with one channel for each cellular state. Thus, each CNN input had shape 4 × 14 × 14; only the stored data were flattened. We split these data into a training pool of 850,000 samples, a validation set of 50,000 samples, and a frozen test set of 100,000 samples that was never used during training or model selection. Approximately 26.6% of runs yielded static outcomes and 73.4% yielded dynamic outcomes. We trained and evaluated six model classes spanning a wide range of inductive biases and representational capacities: logistic regression with one-hot encoding, Extremely Randomized Trees (ExtraTrees), Histogram-based Gradient Boosting (HGB), XGBoost, a multilayer perceptron (MLP), and a convolutional neural network (CNN). For each model class, we trained on progressively larger subsets of the training pool—from $10^3$ to $8.5 \times 10^5$ examples—and evaluated performance on the frozen test set at each scale (Supplementary figs. 5–9). We assessed performance using balanced accuracy and ROC–AUC, both of which equal 0.5 under chance-level prediction, and we calibrated decision thresholds on the validation set to maximize balanced accuracy before evaluating on the test set (Supplementary fig. 6). Across all model classes and all training set sizes, we discovered that performance saturated at chance level: balanced accuracy ≈ 0.5 and ROC–AUC ≈ 0.5 with no systematic improvement as training data grew by nearly three orders of magnitude. Increasing model expressivity—from linear classifiers to CNNs with spatial inductive biases—likewise failed to rescue predictive performance. These results indicate that, across the model classes and data scales examined here, the failure of prediction is not explained by model capacity or data quantity but instead reflects the absence of detectable predictive structure in the initial configuration. The CNN read the triangular lattice through a square-array embedding. All six triangular-lattice nearest neighbors of each cell lay within the CNN's local square receptive field, so local structure was preserved; however, that receptive field also included array positions that were not triangular-lattice nearest neighbors. The embedding therefore approximated rather than exactly encoded the six-neighbor triangular connectivity. A geometry-matched architecture would make the exact connectivity more explicit, while the fully connected network, which makes no spatial assumptions, also found no signal (Supplementary note 2).

**Normalized mutual information analysis on initial configurations.** As a model-agnostic complement to the ML analysis, we quantified the normalized mutual information (NMI) between each cell's initial four-state identity, considered individually, and the binary fate label (Supplementary fig. 10 and Supplementary note 2). We computed NMI separately for every lattice cell across datasets of increasing size. We found that the single-cell NMI values collapsed onto the shuffle baseline at all dataset sizes and decreased toward zero as sample size $N$ grew, consistent with the learning-curve results. The small positive values observed at small $N$ reflect finite-sample estimation noise rather than detectable statistical dependence between an individual cell's initial state and macroscopic fate. Their decrease with increasing $N$ is consistent with increasingly accurate estimates approaching zero, not with the emergence of a weak signal. Because NMI was computed separately for each site, this analysis does not test predictive information encoded jointly across multiple sites. Together with the ML learning curves, this analysis shows that no individual lattice site carried detectable fate-predictive information on its own, while the tested multivariate model classes detected no accessible predictive signal from the complete initial configurations.

**Vortex core identification.** We identified vortex cores as connected regions of cells for which at least one nearest neighbor exhibited a $\pi$ -phase difference. Candidate core cells were first marked by detecting nearest-neighbor phase differences of magnitude $\pi$. We then applied a connected-component labeling algorithm adapted to the triangular lattice with periodic boundary conditions. In a first pass, provisional labels were assigned to all candidate core cells. For each such cell, we examined its nearest neighbors: if any neighbor already carried a label, we assigned the smallest neighboring label to the current cell; otherwise, we assigned a new unique label. In a second pass, equivalent labels were merged so that each spatially connected core region received a single consistent label. In a final pass, labels were renumbered consecutively to remove gaps arising from merges. Each connected labeled region was treated as one vortex core.

**Vortex winding calculation.** For each labeled core, we defined its contour as the set of cells immediately adjacent (distance one) to any core cell. To account for periodic boundary conditions on the toroidal lattice, we computed the center of mass of the core using a periodic-aware algorithm. Contour cells were sorted by their polar angle around the core's center of mass using arctan2. Traversing the contour counterclockwise, we computed the phase difference $\Delta\varphi$ between each consecutive pair of contour cells. Phase wraparound (e.g., transitions between $3\pi/2$ and 0) was corrected so that $\Delta\varphi$ lay in the interval $(-\pi, \pi]$. The total winding number was then computed as:

$$W = (1/2\pi)\, \Sigma\, \Delta\varphi.$$

**Vortex classification and topological charge.** Vortices were classified according to the sign of their total winding number $W$:

- $W \geq 1$: classified as +1 vortex (counterclockwise winding),
- $W \leq -1$: classified as −1 vortex (clockwise winding),
- $|W| < 1$: classified as 0 vortex.

We refer to this signed winding classification as the vortex's topological charge. In rare cases, contours yielded $|W| \geq 2$, corresponding to multiple complete windings. Such vortices were grouped according to the sign of $W$ and counted as ±1 vortices in our analysis. Thus, vortices were classified by chirality (sign of winding), not by winding magnitude. Because the lattice has periodic boundary conditions (topologically equivalent to a torus), the total topological charge across the lattice must sum to zero. As shown in the Results, this constraint was dynamically respected in every simulation. Occasional single-timestep fluctuations in vortex counts arising from contour-labeling ambiguities were detected algorithmically and excluded in post-processing (Supplementary note 3). These events occurred at rates $< 10^{-4}$ per timestep and did not affect the dynamical conclusions.

**Brownian particle model of vortex dynamics.** Tracking individual vortices over time, we discovered that charged vortices moved through the lattice and annihilated upon encounter with an oppositely charged partner, suggesting that their collective dynamics might admit a coarse-grained description as a gas of diffusing particles. To test this, we sought to determine whether a minimal stochastic model could reproduce the statistical structure of vortex lifetimes and annihilation events observed in the automaton (Supplementary note 4 and Fig. 3d). We first characterized vortex motion empirically. Using vortex centroid trajectories extracted from automaton simulations, we computed, for each nearest-neighbor +1/−1 vortex pair, the scalar separation distance $r(t)$ and its discrete increments $\Delta r(t)$. We found that the distribution of $\Delta r$ was approximately symmetric and well described by a normal distribution centered near zero, indicating that vortex motion lacked a detectable drift component and was consistent with unbiased diffusive motion. We also observed intermittent pauses in vortex displacement, with waiting intervals that decayed rapidly in probability and rarely exceeded four timesteps. Based on these observations, we constructed a minimal Brownian particle model in which each vortex was represented as a point particle undergoing isotropic two-dimensional diffusion according to the Euler–Maruyama discretization

$$X_{t+1} = X_t + \sigma\, \eta_x, \quad Y_{t+1} = Y_t + \sigma\, \eta_y,$$

where $\eta_x$ and $\eta_y$ are independent standard normal samples and $\sigma$ is the diffusion parameter. We inferred $\sigma$ from the empirical distribution of vortex velocity increments, using differences taken over four timesteps to avoid contamination from waiting intervals. An annihilation threshold distance $r_{threshold}$ was estimated from the distribution of pairwise vortex separations at the moment of annihilation in the cellular automaton. Each model simulation began with $2N$ particles at random positions, with equal numbers assigned charges +1 and −1; whenever two oppositely charged particles came within $r_{threshold}$ of each other, the pair was removed. We found that this minimal model reproduced the same qualitative staircase structure in vortex-pair number observed in the automaton (Fig. 3c) and generated annihilation timescales of the same order of magnitude, with predicted and observed termination times strongly correlated across lattice sizes (Fig. 3d; Supplementary figs. 16–18).

**Detection of non-contractible loops (NCLs).** We detected non-contractible loops (NCLs) by analyzing connected clusters of cells in identical states on the periodic lattice. Because the lattice uses periodic boundary conditions, it is topologically equivalent to a torus, on which loops can be either contractible (shrinkable to a point) or non-contractible (wrapping around the torus in at least one periodic direction). We first identified clusters of connected same-state cells using a connected-component labeling procedure. For each of the four cellular states, we labeled all cells that were connected through nearest-neighbor adjacency on the triangular lattice while respecting periodic boundary conditions. For each labeled cluster, we searched for closed paths using a depth-first search. Starting from each cell in the cluster, the algorithm recursively followed neighboring cells belonging to the same cluster. During this traversal we recorded visited nodes and parent relationships in the search tree. Whenever the search encountered a previously visited node that was not the immediate parent of the current node, we identified a back edge, indicating the presence of a closed loop within the cluster. For every detected loop we computed its net displacement across the periodic lattice. As the loop was traversed, we accumulated the displacement of each step in the x and y lattice directions while accounting for periodic wrapping across the lattice boundaries. We then calculated the loop's winding numbers as

$$w_x = \Delta x / L_x, \qquad w_y = \Delta y / L_y,$$

where $\Delta x$ and $\Delta y$ are the total displacements along the two periodic directions and $L_x$ and $L_y$ denote the lattice dimensions. A loop was classified as non-contractible if either $w_x$ or $w_y$ was non-zero, indicating that the loop wrapped around the torus in at least one periodic direction. We removed duplicate detections arising from different starting points by comparing loop node sets and retaining only unique loops. This procedure identified all NCLs present in the lattice regardless of whether they connected vortex cores.

**Identification of NCL strings connecting vortices.** We next determined which of the strings connecting vortex pairs corresponded to non-contractible loops. In our system, a string is defined as a path of same-state cells that connects two vortex cores. We first identified vortex-connected strings by analyzing the clusters of same-state cells that intersect vortex contours. For each cluster we determined the shortest path connecting each pair of vortices through that cluster using a graph search restricted to the cluster nodes. In most configurations this procedure produced four distinct vortex-connecting strings, corresponding to the four cellular states. Rarely, additional clusters produced an extra string. For every vortex pair we stored the set of nodes belonging to each string. We then determined whether a given string coincided with an NCL by comparing the nodes of the string with the nodes belonging to previously identified non-contractible loops. Specifically, we computed the set intersection between each string's nodes and the nodes of each detected NCL. If a string shared one or more nodes with an NCL, we classified that string as an NCL string. Finally, we calculated the fraction of vortex-connecting strings that were NCLs. To avoid counting an excessive number of combinatorial paths within large clusters, we restricted this analysis to the shortest string associated with each cluster. Under typical conditions this yielded four strings per vortex pair, allowing a well-defined measure of the proportion of strings that belonged to non-contractible loops.

**Quantification of NCL-string abundance over sliding time windows.** To quantify the late-stage dynamics of non-contractible-loop (NCL) strings, we analyzed the number of NCL strings present at each timestep of every run. For each trajectory, we first identified the time interval beginning when only a single vortex pair remained on the lattice and ending at the final event, defined as either annihilation of the final vortex pair or the onset of spiral-wave formation. Within this interval we computed the abundance of NCL strings using a sliding time window of fixed duration. Let N(t) denote the number of NCL strings present at timestep t. For each timestep t within the analyzed interval, we defined a sliding window of length W timesteps spanning the interval $[t - W + 1, t]$. When fewer than W timesteps remained before the final event, the window included all available timesteps. Within each window we calculated the temporally averaged NCL-string abundance,

$$\langle N \rangle_t = (1/W_t)\ \Sigma_{k=t-Wt+1}^{\ t} N(k),$$

where $W_t = \min(W, t)$ denotes the number of timesteps included in the window. Each window therefore produced a single value representing the mean NCL-string abundance over the preceding interval. Repeating this procedure for all timesteps generated a time series of averaged NCL-string abundances associated with overlapping windows along each trajectory.

**Estimation of temporal trends in NCL-string abundance.** To determine whether the number of NCL strings N(t) increased, decreased, or remained approximately constant within each sliding time-window, we estimated the temporal trend of N(t) using linear regression. For every sliding window ending at timestep t, we fitted the NCL-string counts within that window to a linear model,

$$N(k) = a_t + b_t k,$$

where $k$ indexes the timesteps contained in the window. The parameter $b_t$ represents the slope of the best-fit line obtained by ordinary least-squares regression and therefore quantifies the temporal trend in NCL-string abundance over the window. Positive values of $b_t$ indicate increasing NCL-string counts, negative values indicate decreasing counts, and values near zero indicate approximately stationary behavior. Each sliding window thus yielded two quantities: the mean NCL-string abundance $\langle N \rangle_t$ and the corresponding slope $b_t$. We used these two quantities to construct the phase-space representation shown in Fig. 4f and Supplementary figs. 21 and 23, in which each point corresponds to a single sliding window from an individual trajectory. This representation allowed us to compare the dynamical evolution of NCL strings across many runs without requiring knowledge of how many timesteps remained before the final event.

**Graph-based clustering of trajectories.** To determine whether trajectories with similar late-stage vortex dynamics grouped together before becoming identical, we performed a graph-based clustering analysis on the same total-vortex-core-size time series. For each backward-time window [1, τ], we represented trajectory $i$ by the truncated vector

$$x_i^{(\tau)} = (x_i(1), x_i(2), \ldots, x_i(\tau)),$$

where $x_i(t)$ denotes the total vortex-core size of trajectory $i$ at backward time $t$. For every pair of trajectories $i$ and $j$, we computed their Euclidean distance,

$$d(i, j) = \|x_i^{(\tau)} - x_j^{(\tau)}\|_2.$$

We then constructed an undirected proximity graph in which each node represents one trajectory and an edge connects nodes i and j if $d(i, j) \leq d_{thr}(\tau)$, with threshold $d_{thr}(\tau) = \sqrt{\tau}$

This threshold corresponds to allowing a root-mean-square difference of at most one vortex-core cell per timestep over the window. We identified the connected components of the resulting graph and treated each connected component as one trajectory cluster. For each τ, we recorded the total number of connected components, the fraction of trajectories belonging to singleton components, and the size of the largest connected component. We also visualized the clustering structure by drawing network layouts for selected τ values and by tracking how the sizes of late-time clusters evolved as τ varied. This analysis yielded the graph-based clustering results shown in Figs. 5c-d and Supplementary fig. 25.

**Unique-trajectory analysis.** To determine when distinct runs of the cellular automaton began to converge onto identical late-stage dynamics, we analyzed the time series of total vortex-core size for each trajectory. For every run, we indexed time backward from the final configuration so that $t = 1$ denotes the timestep immediately preceding final-pattern formation, $t = 2$ denotes two timesteps before the end, and so on. We represented each trajectory by a row vector of length $T_{max}$, where $T_{max}$ is the maximum trajectory length in the analyzed ensemble, and padded shorter trajectories beyond their actual duration so that all trajectories could be stored in a common matrix format. For a given backward-time window [1, τ], we extracted from each trajectory the sub-vector spanning the first τ columns and identified unique row patterns. Two trajectories were classified as identical within [1, τ] if their total-vortex-core-size time series matched exactly at every timestep in that window. We then counted the number of distinct row patterns as a function of τ. This procedure yielded the unique-trajectory count shown in Fig. 5e and Supplementary figs. 26-28. Because this criterion required exact agreement over the entire window, even a one-cell difference in vortex-core size at a single timestep caused two trajectories to be counted as distinct.

**Trajectory-resolved fate prediction.** This analysis extends the initial-configuration analysis above to configurations sampled later along a trajectory (Supplementary note 7 and Supplementary figs. 29 & 40). Because the automaton is deterministic and its configuration space for a 14 x 14 lattice is finite ($4^{196}$ states), every trajectory eventually revisits a configuration and then cycles. We marked pattern formation with two times: the first-recurrence time $T_{first}$, the first time-step at which the lattice repeats a configuration seen earlier in the same run, and the formation time $T_{true} = T_{first} - P$, the first timestep of the terminal cycle of period P. Elsewhere in this study we normalized time by $T_{first}$, because a forward-time observer cannot know the system has entered its terminal cycle until a configuration first recurs, and the terminal periods are often long. Here and in Fig. 6 we instead normalized by $T_{true}$, defining $f = t / T_{true}$ so that $f = 0$ and $f = 1$ mark the initial configuration and pattern formation for every trajectory regardless of period. Using the same 1,000,000 simulations and the same training, validation, and test splits as above, we took from each trajectory the single configuration at $t = \mathrm{round}(f \cdot T_{true})$ and trained both a multilayer perceptron (MLP) and a convolutional neural network (CNN) to classify it as one of the three fates, static, rectilinear, or spiral. For the CNN, we presented each lattice as a four-channel image, one channel per cellular state (a square-grid representation of the triangular lattice; see Supplementary note 7), and trained with early stopping on the validation set. We report balanced accuracy, the mean of the three per-fate recalls (chance = 1/3), averaged over three independent training runs. For the transformer, neural cellular automaton, and hidden Markov model, we trained on 200,000 configurations, evaluated on the same frozen test set, and in every case ran positive controls at $f = 0.5$ and $f = 1$ and a label-permuted null in which we scrambled the fate labels before training. Per-fate recall and the dependence on model depth for the convolutional network, the neural cellular automaton, and the transformer are shown in Supplementary fig. 37.

**Multilayer perceptron (MLP).** We trained a fully connected network that treated the 196 cells on a 14 x 14 lattice as an unordered list, with no built-in notion of spatial neighborhood. We supplied each configuration as a one-hot encoding of the four cellular states, flattened into a 784-element vector, and passed it through four hidden layers of 2048, 2048, 1024, and 512 units with ReLU activations and 25% dropout. We trained with a class-weighted cross-entropy loss and the AdamW optimizer for up to 50 epochs, selecting the stopping epoch by validation balanced accuracy. At each f, we applied logistic regression and gradient-boosted decision trees (XGBoost) to the same inputs, both of which likewise treat the cells as unordered features. Full details are in the caption of Supplementary fig. 33.

**Neural cellular automaton (NCA).** We built a learnable cellular automaton: a single convolutional update block, applied repeatedly to its own output, followed by a spatially pooled readout. We gave every cell on a 14 x 14 lattice 32 hidden channels as working space and used circular padding so that the network respected the periodic boundary conditions. Because the update block's weights were shared across every application, the parameter count stayed fixed at about 10,600 while the amount of computation grew, so the number of applications, which we call unrolls, isolated computation from model capacity. We trained a separate network for each unroll count from 1 to 256, and we also re-evaluated trained networks at unroll counts other than their own to test whether they had learned a time-invariant update rule. Full details are in the caption of Supplementary fig. 34.

**Transformer.** We treated each of the 196 cells on the 14 x 14 lattice as one token and allowed every cell to attend to every other cell directly, with no restriction on range. We made attention depend only on the toroidal displacement between two cells, which made the network exactly equivariant to lattice translations, and pooled over all tokens before classification, which made it exactly invariant to them. We initialized each attention head's displacement bias to decay with lattice distance, using a different decay length per head, because otherwise the network would treat all 196 cells as interchangeable at the start of training and converge poorly. We swept 2, 4, and 8 attention layers and three schemes for supplying spatial information, using 128-dimensional embeddings, four attention heads, and about 1.05 million parameters. Full details are in the caption of Supplementary fig. 35.

**Hidden Markov model (HMM).** Because a hidden Markov model describes sequences, we first read each 14 × 14 lattice out as a sequence of 196 symbols, using three reading orders: row-major, column-major, and a serpentine order that reverses every second row so that consecutive symbols are always lattice neighbors. We fitted one categorical model per fate by the Baum–Welch algorithm on 15,000 training sequences of that fate, sweeping the number of hidden states from 2 to 64. We classified each test configuration by whichever fate model assigned it the highest log-likelihood, using equal prior probabilities so that the decision rule matches balanced accuracy. Full details are in the caption of Supplementary fig. 36.

**Configuration diversity and exact-match coverage.** To check whether the CNN relied on memorizing training configurations, we characterized how configurations recur and overlap between the training and test sets (Supplementary note 7 & Supplementary fig. 30). For each fraction f, we computed the exact-match coverage, the fraction of held-out test configurations with an exact cell-by-cell match somewhere in the training pool, and we evaluated a lookup predictor that assigns each test configuration the fate of its exact training match when one exists and guesses otherwise. We also counted the number of distinct configurations of each fate as a function of f.

**Locality and boundary controls.** To test whether the predictive structure was local or global, we compared two networks as a function of training-set size, at the trajectory midpoint, on the binary static-versus-dynamic task: a convolutional neural network (CNN), whose early layers combine each cell only with its neighbors, and a multilayer perceptron (MLP), a fully connected network that treats the 196 cells as an unordered list with no built-in notion of spatial neighborhood (Supplementary note 7 & Supplementary fig. 31). To test whether the network needed the periodic boundary, we compared a convolutional network with circular (torus) padding against one with flat (zero) padding, using a single-state input.

**Object ablations of the configuration.** To test whether the prediction was carried by the objects we had already characterized, we added and removed them as network inputs (Supplementary note 7 & Supplementary fig. 32). We supplied individual objects on their own, either the string mesh connecting the vortices or the count of non-contractible-loop (NCL) strings. For the NCL-string count, we used logistic and gradient-boosted predictors. We also removed the string mesh from the full configuration and, as a control for the number of cells removed, removed a matched number of randomly chosen cells instead.

**Discrete winding field.** To test whether the predictive structure carried by non-contractible loops could be reduced to a purely topological, per-cell input, we constructed a winding field that labels every lattice site by the torus-winding class of its same-state cluster. Using the cluster-labeling and depth-first loop search described above, we enumerated the closed loops within each connected same-state cluster and computed each loop's winding numbers $w_x = \Delta x/L_x$ and $w_y = \Delta y/L_y$ from its wrap-corrected displacement around the periodic lattice. We summarized each cluster by selecting, independently for each component, the loop of largest absolute winding and retaining its signed value, and we stamped the resulting pair ($w_x$, $w_y$) on every cell of the cluster. Cells in clusters with no non-contractible loop received (0, 0). Because a cluster is connected, all of its cells share the same set of achievable loop windings, so ($w_x$, $w_y$) is a property of the cluster rather than of any single cell. The sign of each component records the loop-traversal orientation and is a gauge convention rather than a physical distinction. The field is global—whether a cell's cluster wraps the torus cannot be read from any bounded neighborhood—and dense, with roughly half of all cells carrying a non-zero winding at intermediate and late times. We computed this field for every configuration in the machine-learning splits, aligned it cell-for-cell with the lattice input, and supplied it to the CNN in place of the cell-state configuration. We also supplied the winding field together with the full vortex description, comprising the size, location, and chirality of every vortex, to test whether adding the vortices improved prediction beyond the winding field alone (Supplementary note 7 and Fig. 6f).

**Spatial index of the winding and cell-state fields.** To track how the winding field develops over time, we computed spatial order of the lattice, both in terms of the winding field and phases, across the full ensemble of trajectories (Supplementary note 7 & Supplementary fig. 38). The spatial index is a global autocorrelation function of time (weighted Moran's *I* in which every cell pair is weighed by the strength of communication between them, as defined in Supplementary note 1).

**Winding-field encodings and the wrap mask.** To determine which part of the winding field the network uses, we supplied it to the convolutional network under four encodings (Supplementary note 7 and Supplementary figs. 41 & 42): the signed pair ($w_x$, $w_y$); its magnitudes $|w_x|$ and $|w_y|$; a gauge-fixed canonical form; and a binary wrap mask, marking only whether each cell's region wraps the torus at all. We also measured the distribution of winding magnitude among the wrapping cells as a function of f.

**Software.** All cellular automaton simulations and the associated vortex, NCL, and pattern-timing analyses were implemented in Python 3.14 using NumPy, SciPy, and Matplotlib. Large simulation ensembles were executed in parallel using Python's multiprocessing module with configurable worker counts and chunk sizes. The machine-learning analyses used scikit-learn (logistic regression, ExtraTrees, and histogram-based gradient boosting), XGBoost, and PyTorch (multilayer perceptron, convolutional neural network, neural cellular automaton, and transformer). The graph-based clustering and unique-trajectory analyses were implemented in MATLAB (R2025a) and parallelized using MATLAB's Parallel Computing Toolbox with parfor loops. Visualization of inter-event time distributions from the vortex-annihilation analyses was performed in R using ggplot2, with NumPy data files loaded via the reticulate package.

## ACKNOWLEDGEMENTS

We thank Yiteng Dang, Diederik Laman Trip, Amir Mitchell, Sihai D Zhao, Mobolaji Williams, Ling-Wei Kong, Karin Dahmen, Sergei Maslov, and members of the Youk laboratory for insightful discussions or comments on the manuscript. This work was supported by the grant from the National Institutes of Health (NIH-NIGMS R35 Grant, GM147508). H.Y. was supported in part by the NSF Science and Technology Center for Quantitative Cell Biology (NSF DBI 2243257).

## DATA AVAILABILITY

Source Data are provided with this paper. The derived datasets supporting the machine-learning and winding-field analyses are available in Zenodo under DOI: dx.doi.org/10.5281/zenodo.21728102

## CODE AVAILABILITY

All code is available at https://github.com/youklab/GCA-predictability-2026 and archived in Zenodo under DOI 10.5281/zenodo.21754318.

## AUTHOR CONTRIBUTIONS

L.K., E.M.K., and H.Y. performed the computational experiments. L.K. developed the initial codes for the topological formulation of cellular automaton and the algorithms for analyzing vortex properties. L.K. and H.Y. analyzed the creation and annihilation dynamics of vortices. E.M.K. and H.Y. expanded the codebase and developed algorithms for studying strings, non-contractible loops (NCLs), and vortex properties. L.K. and H.Y. performed the graph-based clustering and the trajectory-collapse analyses. H.Y. designed and performed the machine-learning and winding-field analyses. All authors discussed the data and contributed to writing the manuscript. H.Y. conceived and supervised the project.

## COMPETING INTERESTS

The authors declare no competing interests.

## SUPPLEMENTARY INFORMATION

Supplementary Figs. 1-42 and Supplementary Notes 1-8.

# Supplementary Information

# Predictability can be dynamically constructed in deterministic systems

Lars Koopmans[1], Elinor M. Kay[1,2,3], and Hyun Youk[1,2,3,4]

[1]Department of Physics, University of Illinois Urbana-Champaign, Urbana IL, USA

[2]NSF Science and Technology Center for Quantitative Cell Biology, University of Illinois Urbana-Champaign, Urbana IL, USA

[3]Grainger College of Engineering, University of Illinois Urbana-Champaign, Urbana IL, USA

[4]Corresponding author: youk@illinois.edu

---

**Table of Contents:**

## Supplementary Figures

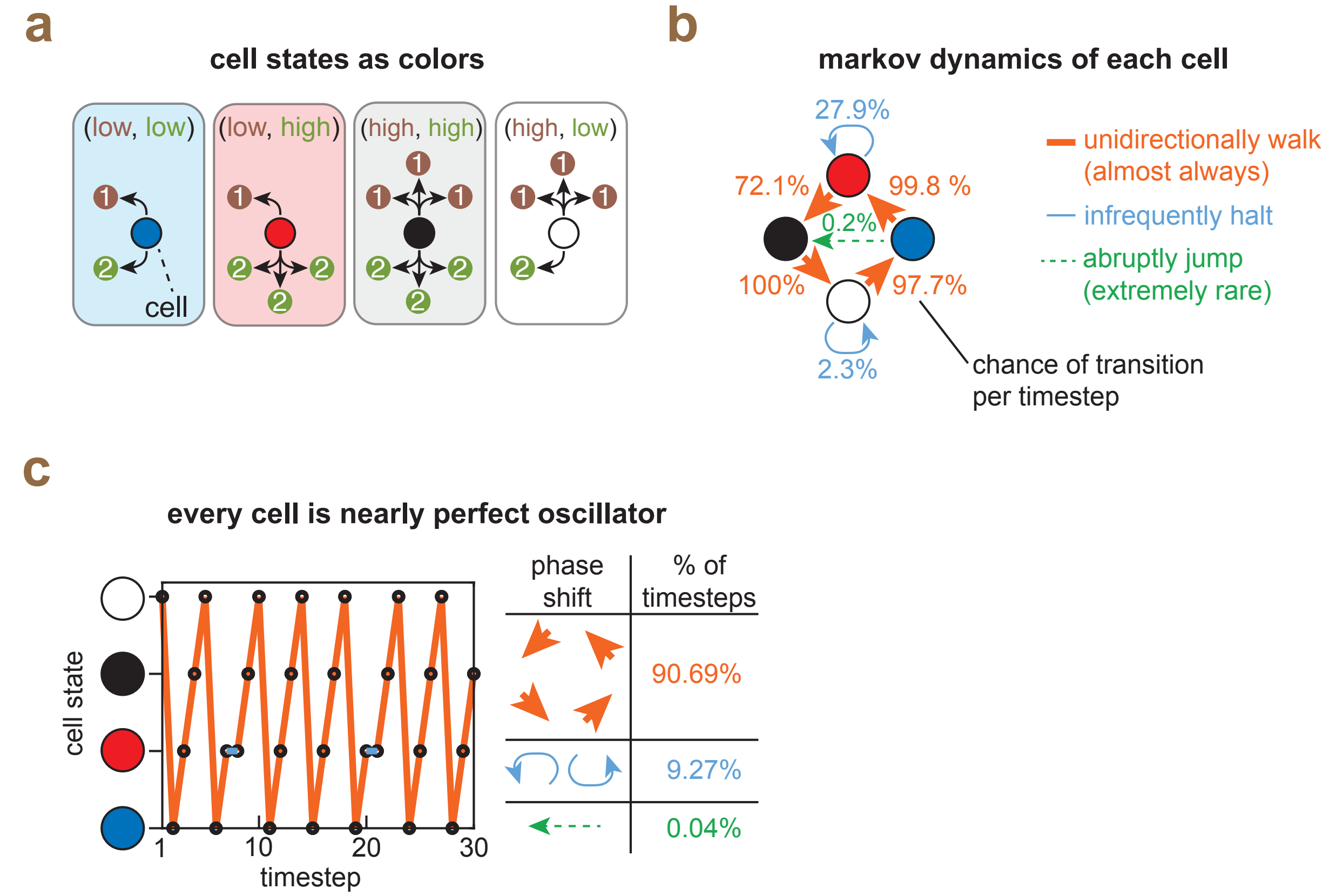

**Supplementary Figure 1: Individual cells behave as nearly perfect oscillators.**
**(a)** Each cell in the cellular automaton occupies one of four possible states determined by the secretion rates of two diffusible molecules, labeled molecule 1 and molecule 2. Each molecule can be secreted at either a low or high rate, yielding the four possible combinations: (low, low), (low, high), (high, high), and (high, low). We represented these four states by four colors as shown.

**(b)** To quantify how cells transitioned between the states, we analyzed cellular automaton runs on a 16 $\times$ 16 lattice (256 cells). Starting from maximally disordered initial configurations (see Supplementary note 1 for how we ensured such initialization), we ran the automaton 1000 independent times and recorded the state of every cell at every timestep. We determined all possible ways that a cell could transition between states, by observing the transition for every cell at every timestep. The resulting transition diagram shown in (b) summarizes the probability that a cell transitions from one state to another during a single timestep. Transitions overwhelmingly followed a unidirectional progression–counter-clockwise cycling–through the four states (orange arrows).
***(caption continues)***

**Supplementary Figure 1 *(continued)*:**
Specifically, most of the time, a cell transitioned from its current state to a state that was one Hamming distance away in the following, biased sense: blue (low, low) $\rightarrow$ red (low ,high) $\rightarrow$ black (high, high) $\rightarrow$ white (high, low) $\rightarrow$ blue (low, low). There were exceptions. Cells occasionally remained in the same state for one timestep (blue arrows), producing a temporary halt in the progression. Abrupt jumps that skipped intermediate states occurred extremely rarely (green dashed arrows).

**(c)** Tracking the time series of individual cells revealed that most cells repeatedly cycled through the four states in the same order. The typical transition corresponded to advancing to the immediate next state–a state that was just one Hamming distance away from its current state. This cycle then repeated. Quantitatively, cells advanced to the immediate next step, one Hamming distance away, in 90.69% of timesteps, remained in the same state in 9.27% of timesteps, and jumped non-sequentially between states in only 0.04% of timesteps. Thus, despite the complex collective behavior that emerges at the population level, the dynamics of each individual cell are highly regular and resemble those of a nearly deterministic oscillator that cycles through the four states.

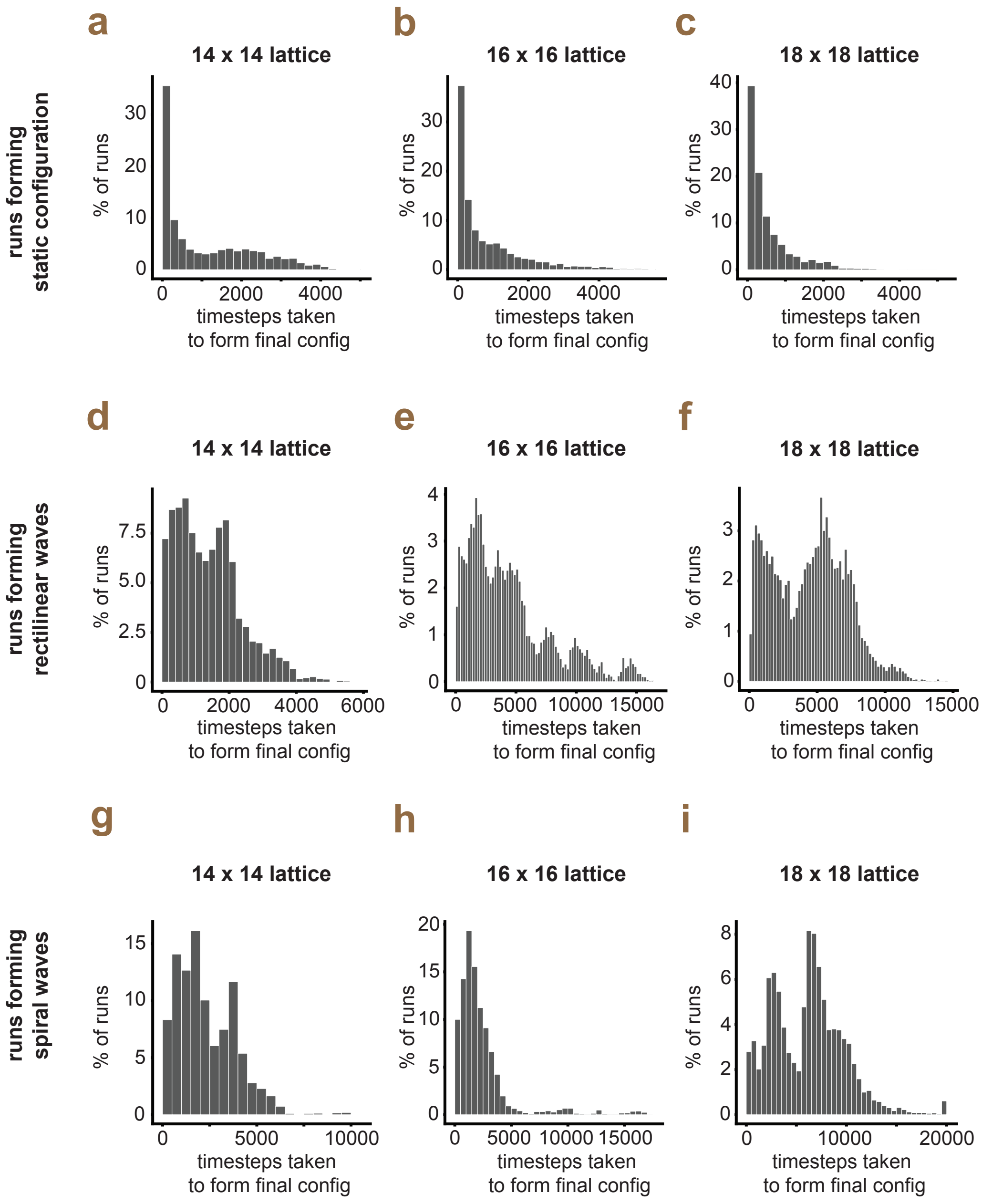

**Supplementary Figure 2: Distributions of pattern-formation times for each final pattern type.** Histograms show the time required for the cellular automaton to reach its final configuration for runs that yielded **(a–c)** static configurations, **(d–f)** rectilinear waves, or **(g–i)** spiral waves. Columns correspond to different lattice sizes: $14 \times 14$, $16 \times 16$, and $18 \times 18$. For each lattice size, we ran 20,000 simulations, each starting from a randomly chosen, maximally disordered configuration. For each run, measured the timestep at which the final pattern emerged. Histograms are plotted as the percentage of runs within each pattern type.

***(caption continues)***

**Supplementary Figure 2 *(continued)*:**

The numbers of runs contributing to each set of panels are: static configurations — $n$ = 7035, $n$ = 10639, and $n$ = 5661 for $14 \times 14$, $16 \times 16$, and $18 \times 18$ lattices, respectively; rectilinear waves — $n$ = 7241, $n$ = 5632, and $n$ = 3493; spiral waves — $n$ = 5724, $n$ = 3729, and $n$ = 10846.

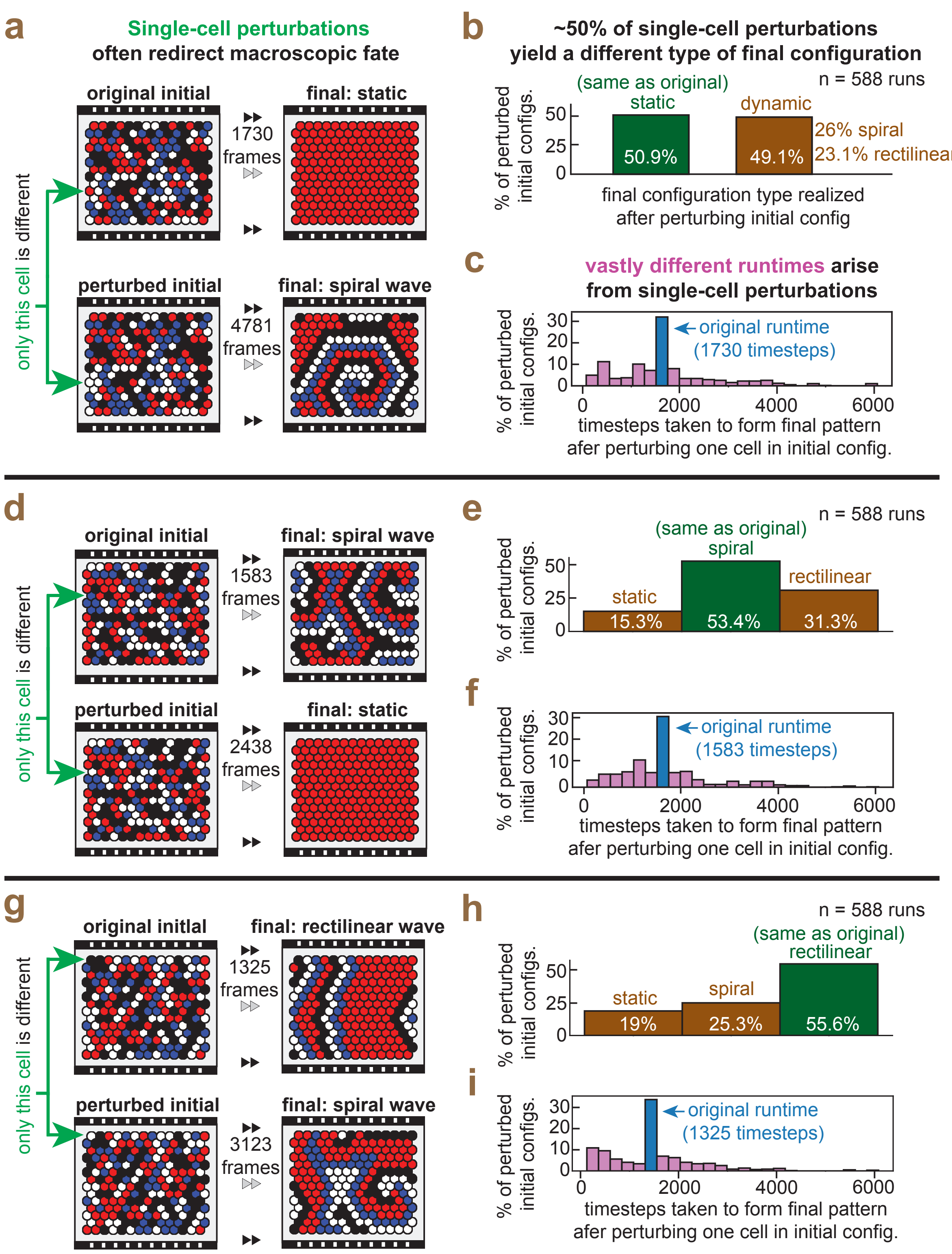

**Supplementary Figure 3: Exhaustive single-cell perturbations reveal that macroscopic fate and runtime are extremely sensitive to perturbing just one cell in the initial configuration.** ***(caption continues)***

**Supplementary Figure 3 *(continued)*:**
This figure provides a systematic analysis of how perturbing the state of a single cell in the initial configuration affects the final pattern realized by the cellular automaton. All simulations were performed on a $14 \times 14$ lattice (196 cells). For a given initial configuration, we generated every possible single-cell perturbation by selecting one cell and changing its state to one of the three alternative states. Because each of the 196 cells can be changed to three different states, this procedure produces $14 \times 14 \times 3 = 588$ distinct perturbed initial configurations. Each of these perturbed configurations was simulated independently until the final pattern formed.

**(a–c)** Example in which a single-cell perturbation redirects the macroscopic fate of the system. Panel (a) shows one initial configuration that evolved into a static final configuration after 1730 timesteps. As an example, changing the state of only one cell at $t = 0$ could yield a perturbed initial configuration that instead evolved into a spiral wave after 4781 timesteps (bottom row of panel (a)). Panel (b) summarizes the outcomes across all $n = 588$ single-cell perturbations derived from this initial configuration. Approximately half of the perturbations preserved the original outcome (static), whereas the other half redirected the system to dynamic spatial patterns (spiral or rectilinear waves). Panel (c) shows the distribution of runtimes required to reach the final configuration across all the single-cell perturbations. The runtime varied widely—from hundreds to several thousand timesteps—even though the initial configurations differ by only a single cell. The runtime of the unperturbed simulation is indicated in blue for reference.

**(d–f)** A second example in which the unperturbed configuration produced a spiral wave after 1583 timesteps. As an example, perturbing a single cell could redirect the system to a static configuration (bottom row of panel d). Panel (e) shows the distribution of final pattern types across the 588 perturbed configurations. Although the original outcome was a spiral wave, many of the single-cell perturbations produced rectilinear waves or static configurations. Panel (f) shows that the runtimes of perturbed simulations again spanned a wide range relative to the original runtime.

**(g–i)** A third example in which the unperturbed configuration yielded a rectilinear wave after 1325 timesteps. As an example, a single-cell perturbation could redirect the system to a spiral wave and substantially alter the time required for the pattern formation (bottom row of panel g). Panel (h) summarizes the distribution of final pattern types across all the single-cell perturbations, and panel (i) shows the corresponding distribution of runtimes.
***(caption continues)***

**Supplementary Figure 3 *(continued)*:**

Together, these results demonstrate that the smallest unit of cell-state perturbation—changing the state of just one cell in the initial configuration—frequently redirect the macroscopic fate of the system and strongly alter the time required for pattern formation. Even when the perturbation preserves the same pattern class (e.g., rectilinear waves), the detailed geometry of the pattern can differ (for example, waves traveling in different directions) and the runtime can vary by thousands of timesteps.

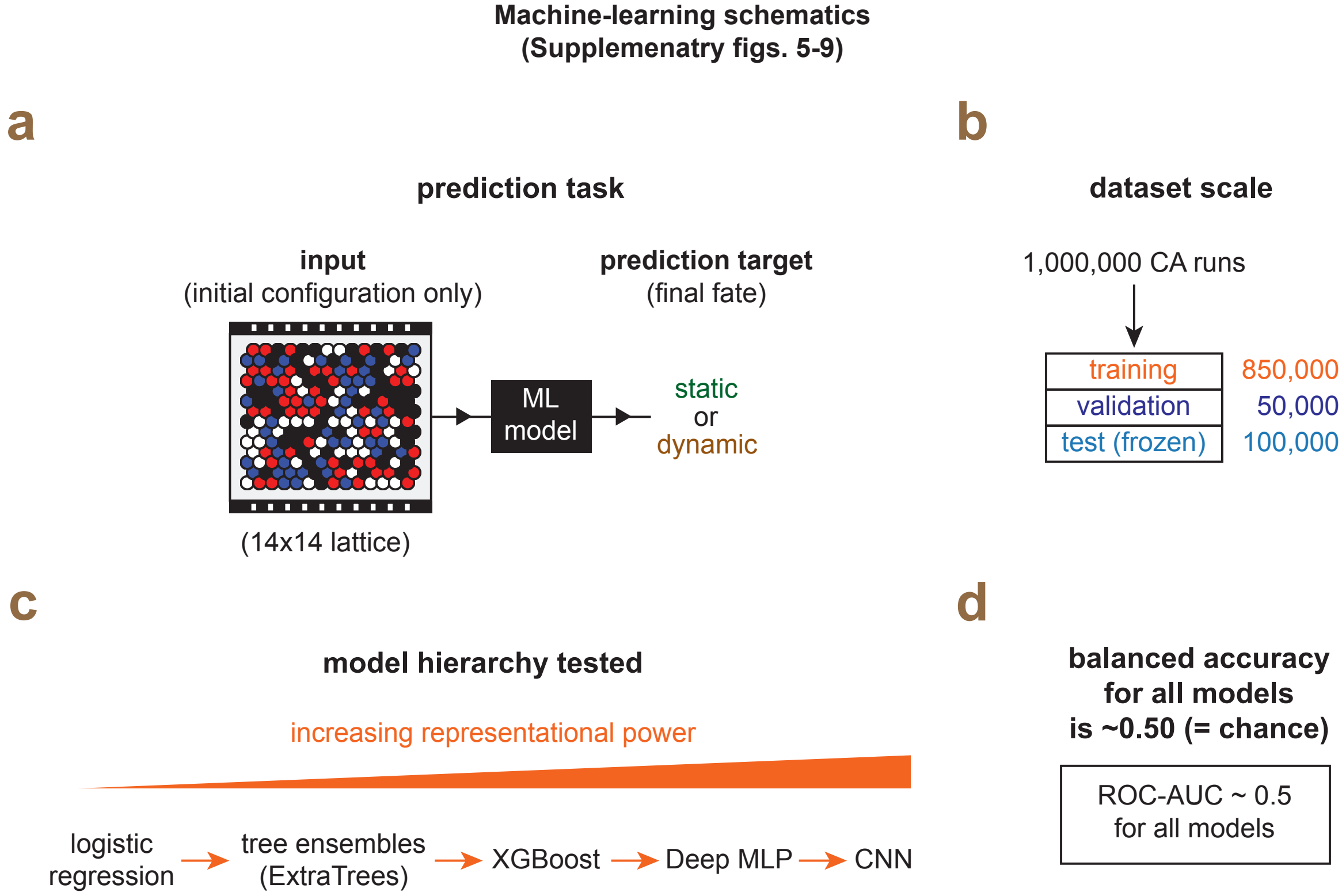

**Supplementary Figure 4: Systematic machine-learning tests of predictability from initial configurations.**

This schematic summarizes the machine-learning analyses used to test whether the final macroscopic fate of the cellular automaton (static or non-static (dynamic)) can be predicted from the initial configuration alone. The analyses evaluated whether predictive signal about the eventual outcome could be extracted from the initial configuration, before the deterministic dynamics unfolded.

**(a) Prediction task.** Each input consisted solely of the initial configuration of the cellular automaton: a $14 \times 14$ lattice (196 cells). The learning task was a simple binary classification: given this initial configuration, the model had to predict whether the deterministic simulation (the cellular automaton) will ultimately produce a static configuration or a non-static configuration (i.e., dynamic spatial pattern, which could be either a rectilinear wave or spiral wave). Importantly, the models received no temporal information from the system's evolution; they only observed the initial configuration.

**(b) Dataset scale.** To rigorously test predictability of the final pattern type, we generated a dataset of 1,000,000 independent deterministic cellular-automaton simulations starting from randomly sampled, maximally disordered initial configurations.

***(caption continues)***

**Supplementary Figure 4 *(continued)*:**
We partitioned these runs into a training pool of 850,000 simulations, a validation set of 50,000 simulations used exclusively for model selection and threshold calibration, and a frozen test set of 100,000 previously unseen simulations used only for final evaluation.

**(c) Model hierarchy tested.** We evaluated a hierarchy of supervised learning models with progressively greater representational flexibility in order to determine whether increasingly expressive algorithms could extract predictive structure from the initial configuration. The tested models include logistic regression (linear classifier), nonlinear tree ensembles (Extremely Randomized Trees), gradient-boosted decision trees (XGBoost), deep multilayer perceptrons, and convolutional neural networks that incorporate spatial inductive bias through local receptive fields. This progression spans standard tabular models, powerful nonlinear ensembles, and modern deep-learning architectures that explicitly exploit spatial structure.

**(d) Summary outcome.** Across all model classes and training set sizes, predictive performance remained at chance level (see redSupplementary figs. 5-9). Balanced accuracy on the frozen test set remained $\approx 0.50$, and ROC–AUC remained $\approx 0.5$ for all models, indicating that none of the algorithms could discriminate between static and dynamic outcomes based solely on the initial configuration. These results demonstrate that we detected no practically extractable predictive signal in the initial configuration about the system's final fate.

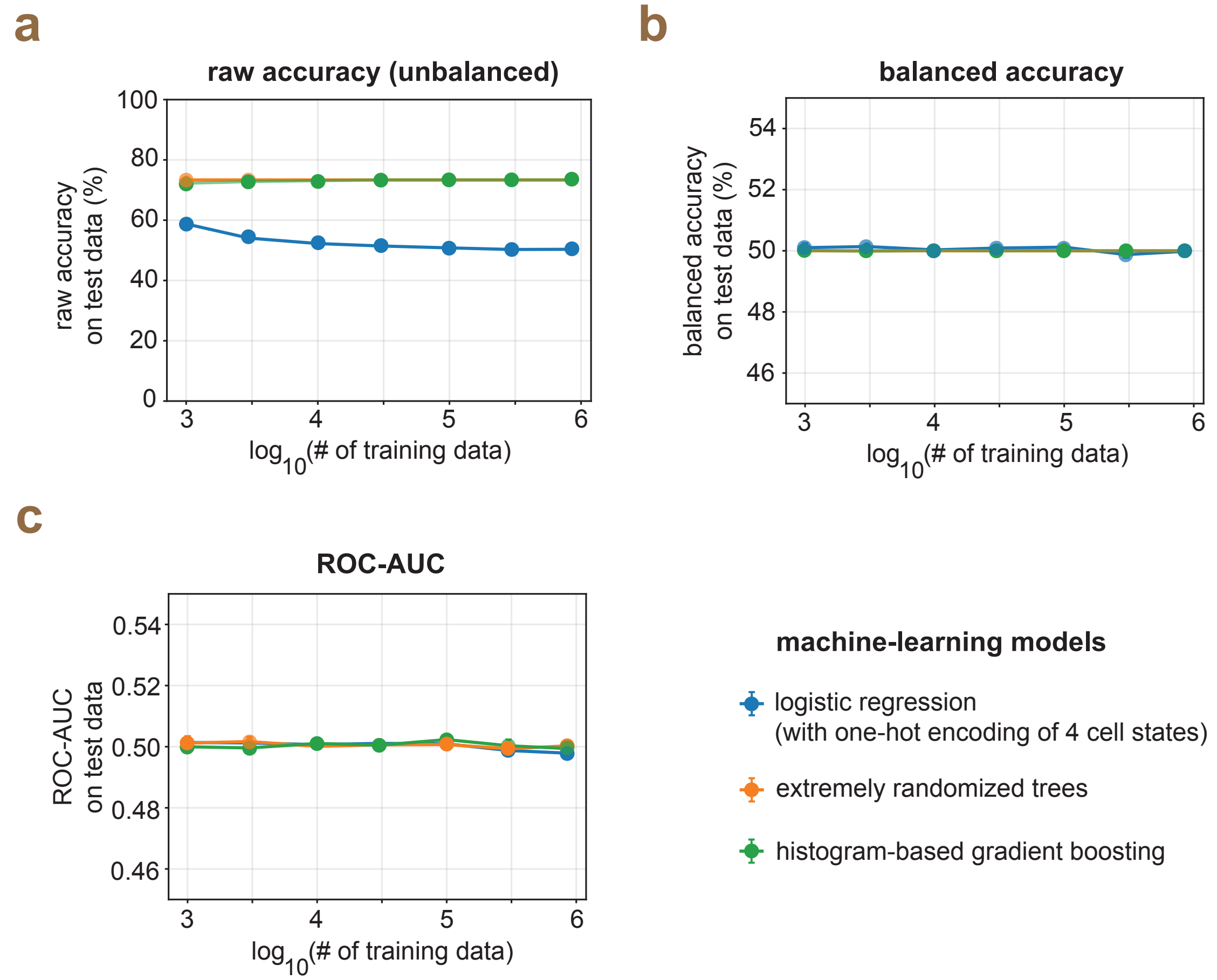

**Supplementary Figure 5: Learning curves for baseline static classifiers show chance-level predictability from the initial configuration.**
We examined whether standard supervised learning methods could predict the final macroscopic fate of the system–static or non-static (dynamic)–directly from the initial configuration as an input. The prediction task was a binary classification: whether a randomly sampled initial configuration evolves into a static configuration (label = 1) or into any dynamic spatial pattern (label = 0). Inputs consisted of flattened $14 \times 14$ lattices (196 cells).

We used three baseline classifiers that operated directly on the initial configuration: logistic regression with one-hot encoding of the four cell-states (blue), extremely randomized trees (orange), and histogram-based gradient boosting (green). We trained the models on subsets of increasing size drawn from a fixed training pool of 850,000 cellular automaton runs ($N = 10^3$, $3 \times 10^3$, $10^4$, $3 \times 10^4$, $10^5$, $3 \times 10^5$, and $8.5 \times 10^5$). For each training size, we used three independent random subsamples (seeds). We evaluated performance on a frozen test set of 100,000 previously unseen initial configurations, held fixed across all models and training sizes.
***(caption continues)***

**Supplementary Figure 5 *(continued)*:**
Panels show test-set performance measured by **(a)** raw accuracy, **(b)** balanced accuracy, and **(c)** ROC–AUC (area under the receiver operating characteristic curve). Balanced accuracy and ROC–AUC are insensitive to the moderate class imbalance in the dataset ($\approx$ 26.6% static, $\approx$ 73.4% dynamic), whereas raw accuracy can be inflated by majority-class bias and is shown here for completeness only. Points indicate mean $\pm$ standard deviation across seeds; error bars are typically smaller than the marker size and are therefore not visible.

Across all models and all training sizes, performance saturated near chance (balanced accuracy $\approx$ 0.5; ROC–AUC $\approx$ 0.5), with no systematic improvement as the amount of training data increased by nearly three orders of magnitude. The learning curves demonstrate that within these baseline linear and nonlinear tabular classifiers, we found no detectable predictive signal about final pattern fate in the initial configuration.

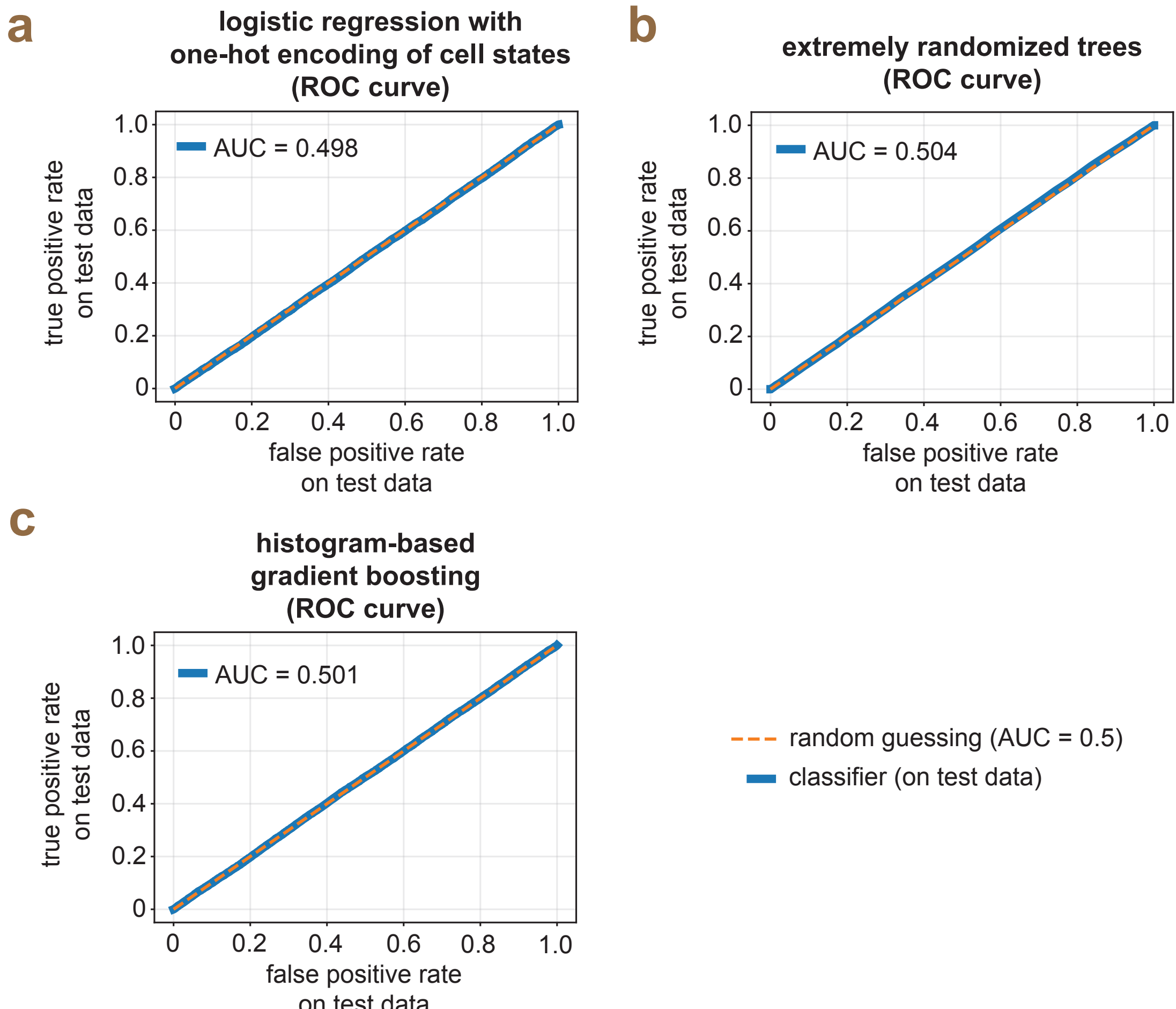

**Supplementary Figure 6: Validation-based threshold calibration does not improve predictability for baseline classifiers.**

Receiver operating characteristic (ROC) curves for three baseline static classifiers (Supplementary fig. 5) evaluated on a frozen test set of 100,000 previously unseen initial configurations: **(a)** logistic regression with one-hot encoding of the four cell states, **(b)** extremely randomized trees, and **(c)** histogram-based gradient boosting. We trained each classifier on the full training pool (850,000 cellular automaton runs) to produce a continuous score interpreted as the probability that an initial configuration evolves into a static outcome. We selected decision thresholds exclusively on a separate validation set (50,000 runs) to maximize balanced accuracy and then held them fixed when evaluating performance on the test set. In each panel, the blue curve shows the ROC curve on the test data, plotting the true positive rate (fraction of static outcomes correctly identified) against the false positive rate (fraction of dynamic outcomes incorrectly labeled static) as the decision threshold is varied. The dashed orange diagonal indicates random guessing, for which the area under the ROC curve (ROC–AUC) equals 0.5.

***(caption continues)***

**Supplementary Figure 6** ***(continued)*****:**
In all cases, the ROC curves lie essentially on the chance diagonal, with ROC–AUC $\approx 0.5$, indicating that the score rankings produced by these models contain no discriminative information about final pattern fate. Although different threshold choices can substantially change raw accuracy by biasing predictions toward one class or the other, validation-based threshold calibration did not improve test-set balanced accuracy.

These results show that, within these baseline classifiers, when score distributions do not separate outcome classes, optimizing decision thresholds does not yield improved predictability from the static initial configuration.

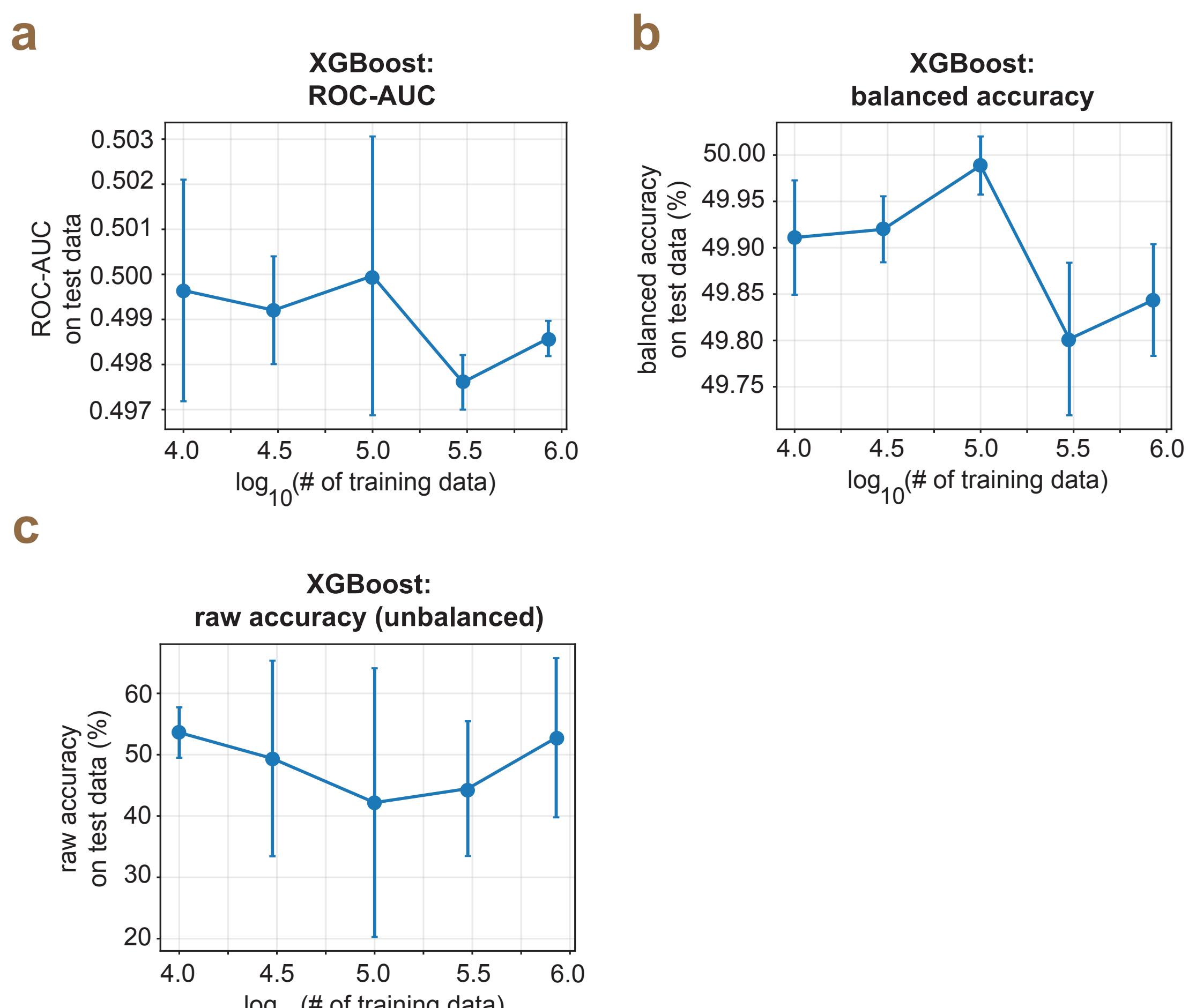

**Supplementary Figure 7: Learning curves for aggressively tuned XGBoost classifiers show chance-level predictability from the initial configuration.**

To test whether substantially more expressive nonlinear models ciykd uncover predictive signal missed by baseline methods, we trained gradient-boosted decision trees (XGBoost) on one-hot encoded initial configurations. Each cell's four-state identity was encoded as a 4-dimensional one-hot vector, yielding 784 binary input features per sample. Models were trained on the same sequence of training set sizes ($N = 10^4$, $3 \times 10^4$, $10^5$, $3 \times 10^5$, $8.5 \times 10^5$), with three independent random seeds per size. Training used aggressive hyperparameters (large maximum number of trees and a small learning rate) combined with early stopping based on validation-set ROC–AUC. Class imbalance was handled via `scale_pos_weight` computed from the training subset. All threshold selection was performed exclusively on the validation set, and final performance was evaluated on the same frozen test set used throughout. Despite its greater representational capacity and extensive training data, XGBoost achieves chance-level performance across all metrics: test-set ROC–AUC $\approx$ 0.5 and balanced accuracy $\approx$ 0.5 at all training sizes. Learning curves flatten rapidly, and increasing training data or model complexity does not yield systematic improvement.

***(caption continues)***

**Supplementary Figure 7** ***(continued)*****:**
These results show that, within this model class and training regime, even highly expressive boosted tree models did not detect predictive signal in the static initial configuration.

**(a) ROC–AUC on the held-out test set** as a function of training set size for an XGBoost classifier trained on one-hot encoded initial configurations (784 sparse features = 196 lattice sites $\times$ 4 states). Points show mean $\pm$ s.d. across three random subsamples per training size. ROC–AUC remains near 0.5 across all sample sizes up to 850,000 training examples, indicating no discriminative signal accessible to this model class from the initial configuration alone.

**(b) Balanced accuracy on the test set** using a decision threshold selected only on the validation set to maximize balanced accuracy. Balanced accuracy remains near 0.5 across all training sizes, consistent with chance-level prediction even after threshold calibration.

**(c) Plain accuracy on the test set** using the same validation-selected threshold. Accuracy varies substantially with training size because the dataset is class-imbalanced and the fitted models tend to output near-constant scores. Consequently, the chosen threshold can bias predictions toward one class without improving discrimination (cf. panels a–b).

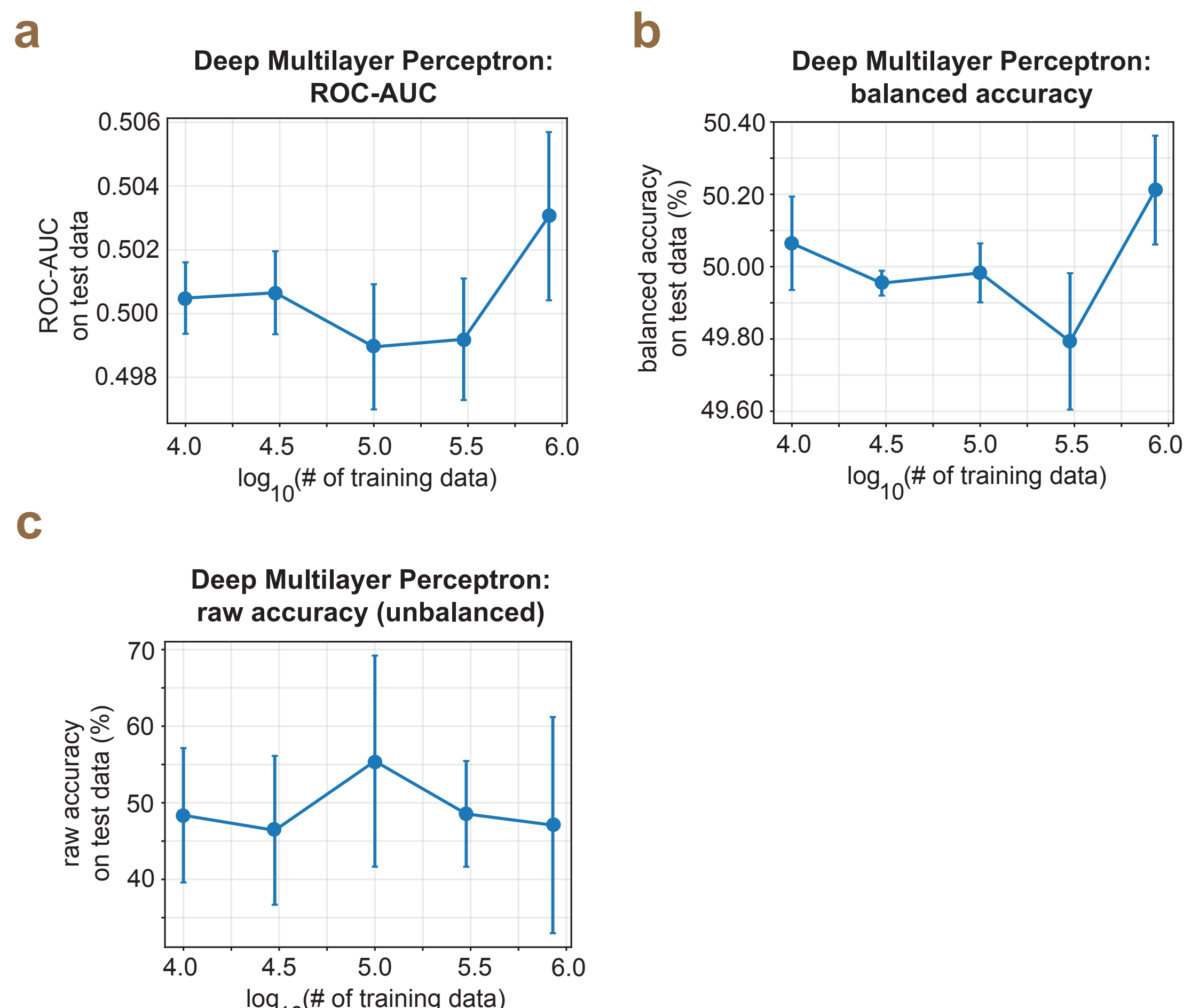

**Supplementary Figure 8: Deep neural networks (multilayer perceptrons) remain at chance level when predicting final fate from the initial configuration.**

We examined whether a high-capacity neural network could extract predictive signatures from the initial configuration alone. We trained a fully connected deep multilayer perceptron (MLP) implemented in PyTorch on one-hot encoded initial configurations of the cellular automaton. Each $14 \times 14$ lattice (196 cells) was represented by a sparse binary vector encoding the four possible cell states. The network contained multiple hidden layers with nonlinear activations. Training used stochastic gradient descent with early stopping based on validation-set ROC–AUC. We trained the models on random subsets of increasing size drawn from a fixed training pool (from $10^4$ up to $8.5\times10^5$ samples), with three independent random seeds per training size. For each run, we selected the epoch with the best validation ROC–AUC and chose a decision threshold on the validation set to maximize balanced accuracy. All reported metrics were then evaluated on the same frozen test set of 100,000 previously unseen initial configurations.

***(caption continues)***

**Supplementary Figure 8 *(continued)*:**

Panels show test-set performance as a function of training set size: **(a)** ROC–AUC (area under the receiver operating characteristic curve), a threshold-independent measure of discriminative ability; **(b)** balanced accuracy using the validation-selected threshold, which accounts for class imbalance; and **(c)** plain (unbalanced) accuracy, shown for completeness.

Across all training sizes, the MLP failed to achieve meaningful discrimination. Test-set ROC–AUC remained near chance ($\approx$ 0.5), and test-set balanced accuracy remained $\approx$ 0.5, indicating performance equivalent to random guessing. At small training sizes, training loss decreased while validation and test metrics remained flat, consistent with overfitting without generalization. Even at the largest training size, increasing model capacity and data did not improve predictability. These results show that, within this model class and training regime, deep neural networks did not detect predictive signal in the initial configuration alone.

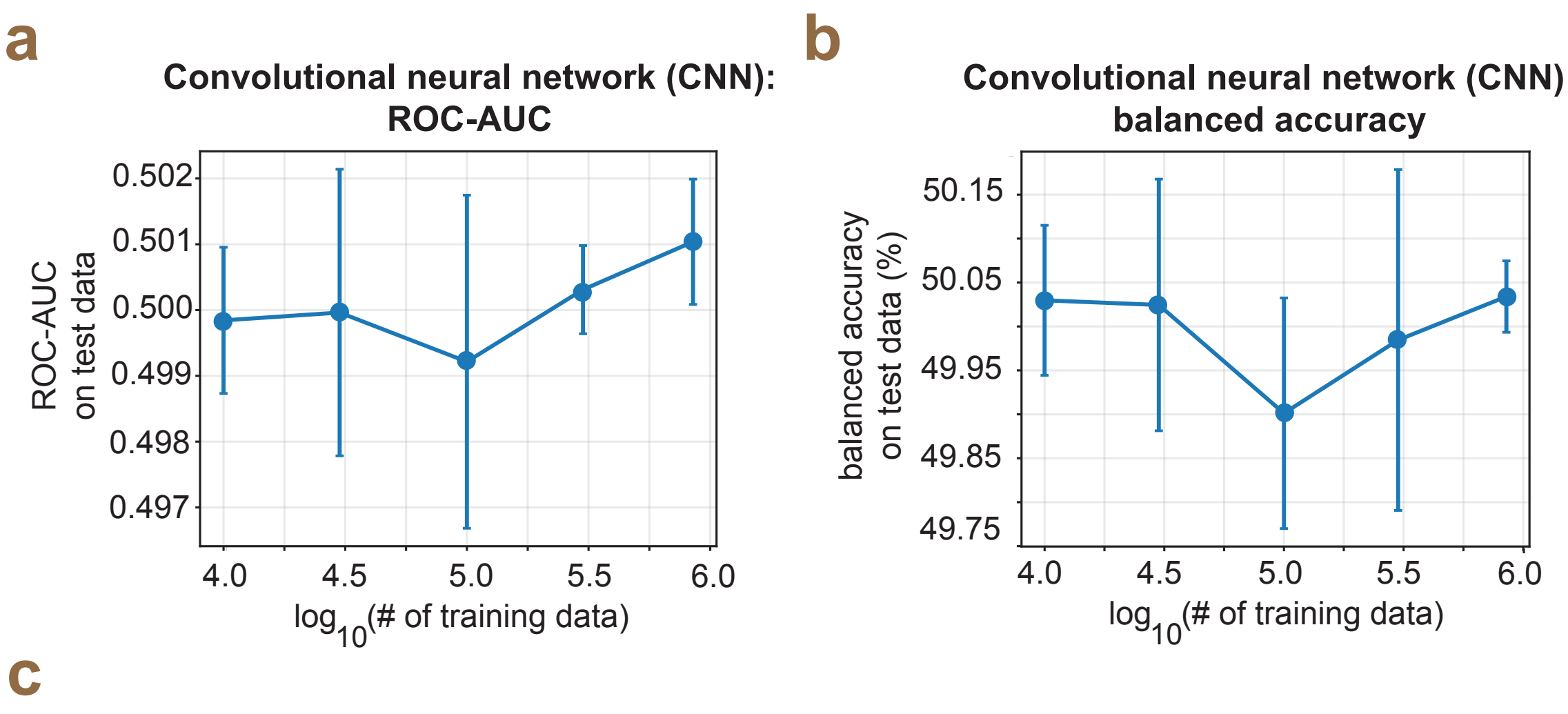

c

Convolutional neural network:
raw accuracy (unbalanced; shown for completeness)

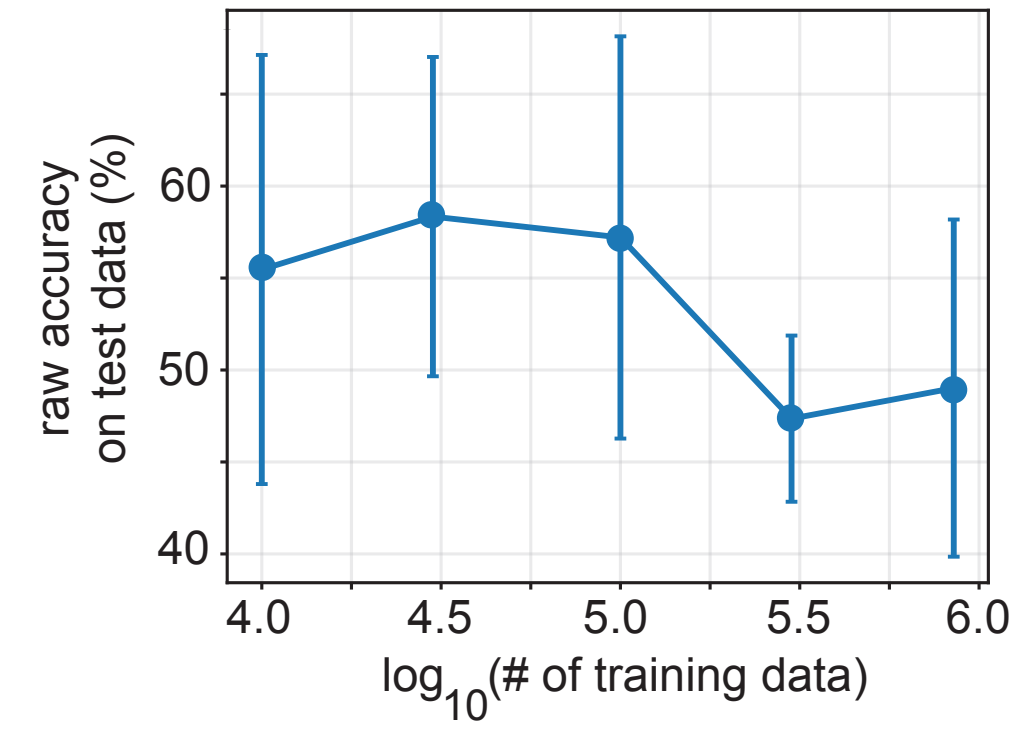

**Supplementary Figure 9: Spatially structured convolutional neural networks remain at chance level when predicting from the initial configuration.**

We tested whether explicitly incorporating the spatial geometry of the lattice enables machine learning to extract predictive information about the system's final macroscopic fate–guessing static versus non-static–from the initial configuration. Following the failure of tabular classifiers and fully connected neural networks (Supplementary Figs. 5–8), we trained a convolutional neural network (CNN) designed to exploit local spatial correlations in the initial lattice configuration. The prediction task was binary classification: whether a deterministic cellular-automaton run yielded a static configuration (label = 1) or any dynamic spatial pattern (label = 0). Each initial configuration consisted of a $14 \times 14$ lattice (196 cells). We encoded the configuration as a four-channel two-dimensional tensor, with one channel per discrete cell state. We trained CNN models using fixed data splits derived from $10^6$ independent cellular-automaton runs: a training pool of 850,000 runs, a validation set of 50,000 runs, and a frozen test set of 100,000 runs held out throughout all experiments. For each training size ($10^4$, $3 \times 10^4$, $10^5$, $3 \times 10^5$, $8.5 \times 10^5$), we trained three independent models using different random subsamples of the training pool and report mean $\pm$ s.d. across runs.

***(caption continues)***

**Supplementary Figure 9 *(continued)*:**
Training used PyTorch with AdamW optimization and early stopping based on validation ROC–AUC. Decision thresholds were selected exclusively on the validation set to maximize balanced accuracy and were then held fixed when evaluating the test set. Panels show test-set performance as a function of training set size: **(a)** ROC–AUC (area under the receiver operating characteristic curve), a threshold-independent measure of discriminative ability; **(b)** balanced accuracy computed using the validation-selected decision threshold, which corrects for class imbalance; and **(c)** raw (unbalanced) accuracy using the same threshold, shown for completeness. Across all training sizes, ROC–AUC remained indistinguishable from chance ($\approx 0.5$), and balanced accuracy remained $\approx 0.5$, indicating no measurable discrimination between static and dynamic outcomes from the initial configuration. Increasing the training dataset by nearly two orders of magnitude and using a spatially aware neural architecture did not improve performance. Variations in raw accuracy arise from class imbalance and threshold bias rather than genuine discriminative ability: because the learned scores contain little or no separation between outcome classes, small shifts in thresholding can bias predictions toward the majority class without improving class separation.

These results show that, within this model class and training regime, incorporating spatial structure in the initial configuration did not reveal predictive signal about final fate.

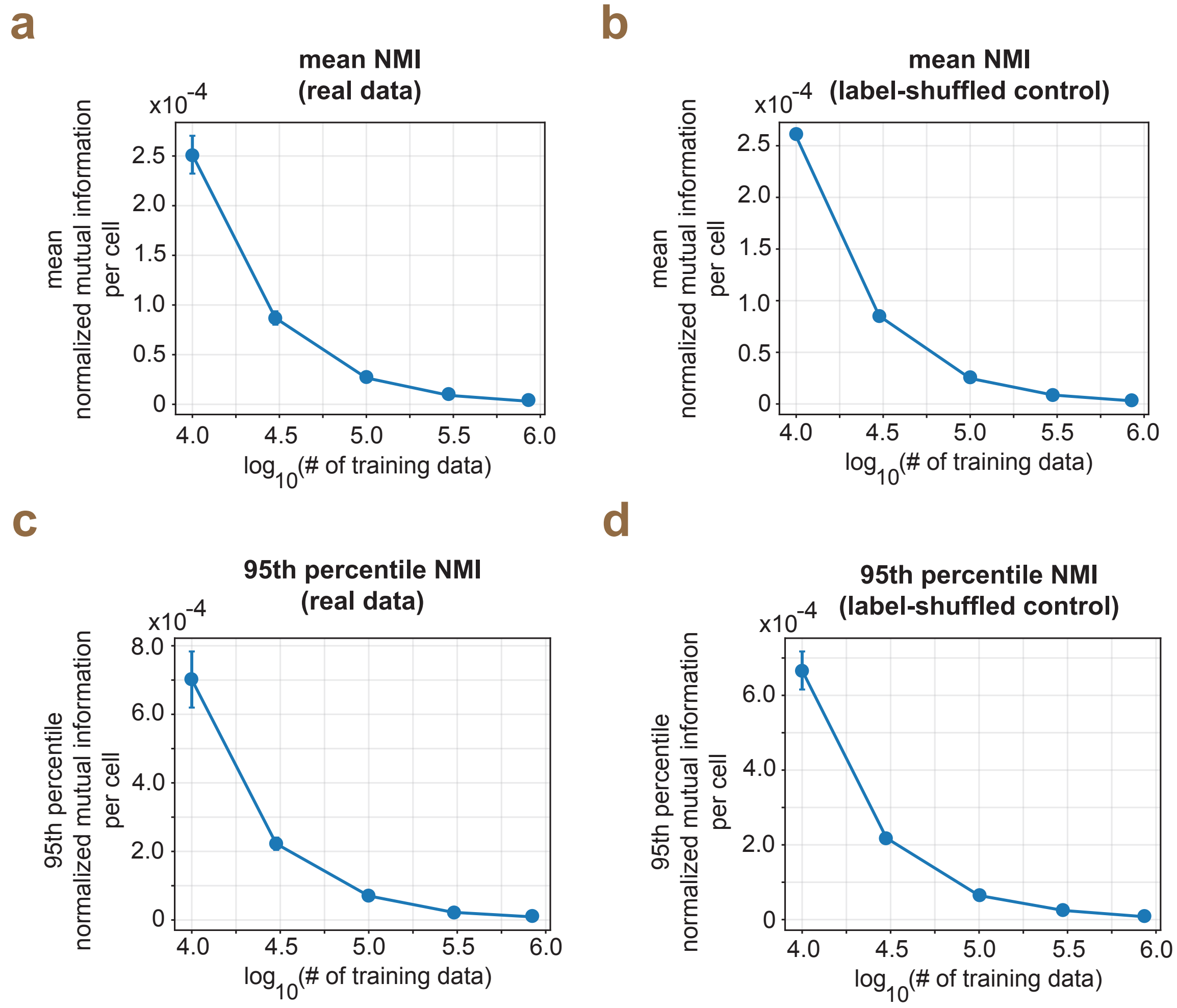

**Supplementary Figure 10: Normalized mutual information — No individual cell's initial state was detectably predictive of final fate.**

For each of the 196 lattice sites in a $14 \times 14$ lattice, we separately calculated the normalized mutual information (NMI) between the cell's initial state, which could be one of four states, and the final binary fate (static or dynamic). NMI was normalized by the entropy of the fate labels. For each dataset size $N$, we show the mean (**a**, **b**) and 95$^{th}$ percentile (**c**, **d**) of the NMI values across all sites. Results are shown for the true fate labels (**a**, **c**) and for a control in which the fate labels were randomly shuffled (**b**, **d**). Each point is the average over three independent random subsamples of size $N$. For dataset sizes up to $8.5 \times 10^5$, the NMI values for the true labels closely followed the shuffled-label baseline and decreased toward zero as $N$ increased. Thus, the small positive values at smaller $N$ are consistent with finite-sample effects rather than a detectable association between any individual cell's initial state and final fate.

***(caption continues)***

**Supplementary Figure 10** ***(continued)*****:** Because NMI was calculated separately for each cell, this analysis does not test predictive information encoded jointly across multiple cells or in their spatial arrangement. The result instead reveals an informative contrast with the perturbation experiment in Fig. 1e: changing one cell's state could redirect a trajectory to a different fate, yet that cell's initial state was not predictive of fate on its own. A cell can therefore be causally pivotal without being predictive in isolation, because the effect of changing it depends on the rest of the configuration.

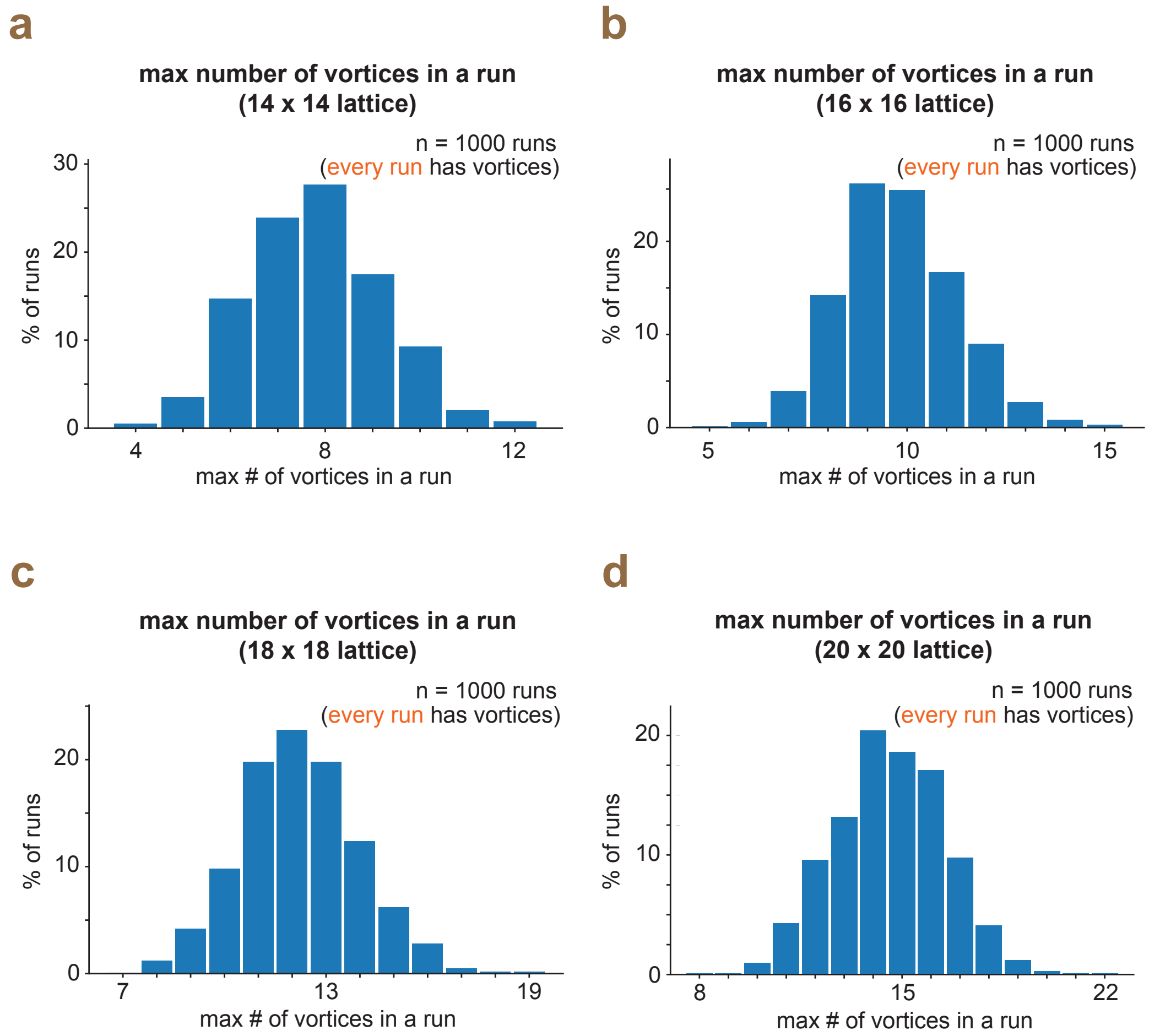

**Supplementary Figure 11: Distribution of the maximum number of vortices formed during a single run of the cellular automaton.**
For each lattice size, we simulated $n$ = 1000 independent runs of the cellular automaton and recorded the maximum number of vortices simultaneously present at any time during each run. Histograms show the distribution of this quantity for lattices of increasing size: **(a)** $14 \times 14$, **(b)** $16 \times 16$, **(c)** $18 \times 18$, and **(d)** $20 \times 20$ cells. The x-axis reports the maximum number of vortices observed within a run, while the y-axis shows the percentage of runs exhibiting that value. In every run, vortices appeared at some point during the dynamics (every run contained vortices). These measurements quantify the typical peak abundance of vortices spontaneously generated during the self-organization process before vortices subsequently move, merge, and annihilate as the system approaches its final configuration.

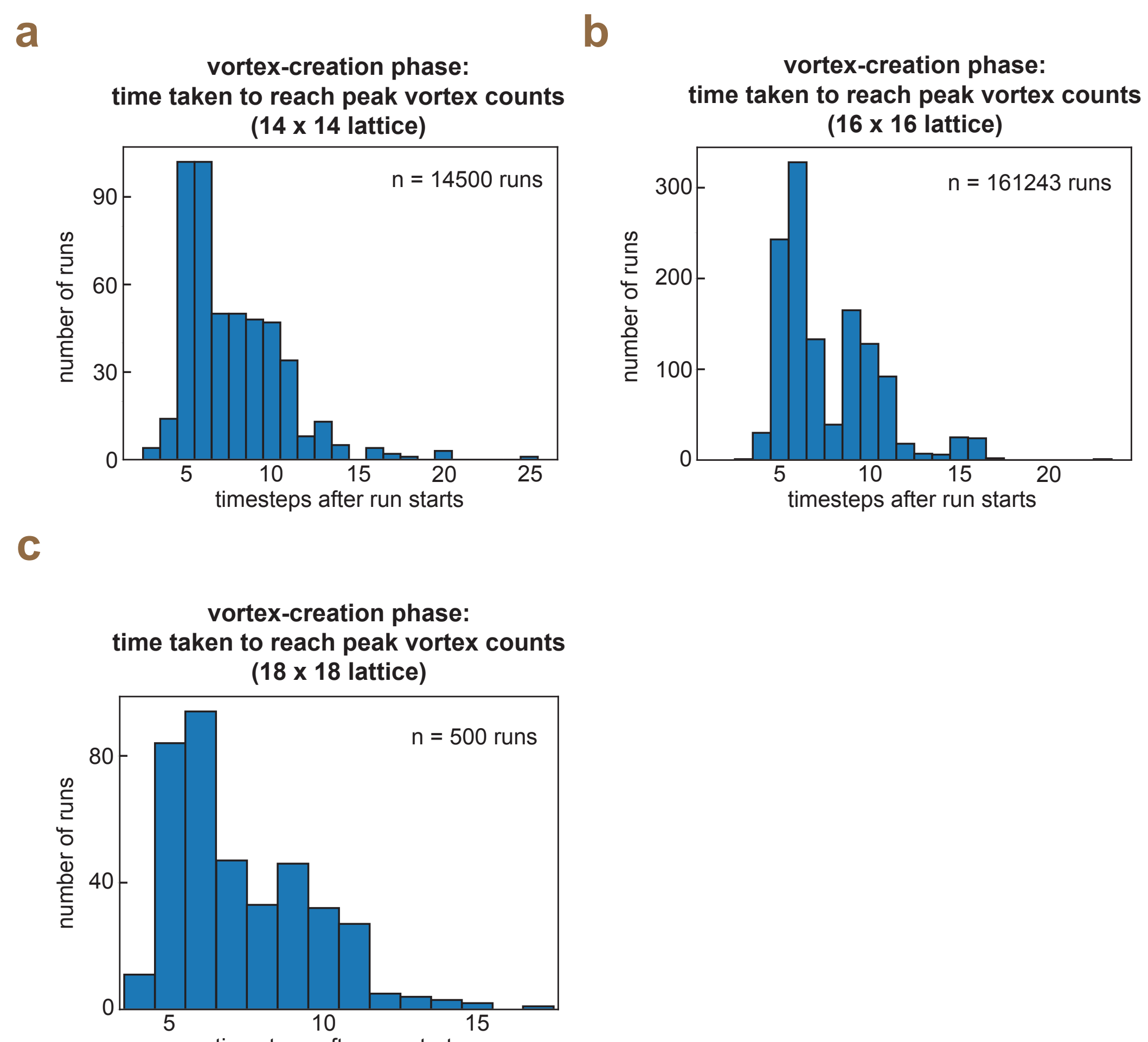

**Supplementary Figure 12: Time required to reach the peak number of vortices during the vortex-creation phase.**

For each run of the cellular automaton, we recorded the timestep at which the total number of vortices first reached its maximum value. Histograms show the distribution of this time for lattices of different sizes: **(a)** $14 \times 14$, **(b)** $16 \times 16$, and **(c)** $18 \times 18$ cells. The x-axis reports the number of timesteps elapsed after the start of the run, and the y-axis shows the number of runs exhibiting that value. Each panel indicates the total number of runs analyzed. Across all lattice sizes, the system typically reaches its peak vortex count within the first $\sim$ 5–10 timesteps, indicating that vortex creation occurs very early in the dynamics. After this brief vortex-creation phase, the number of vortices typically decreased as the charged vortices moved, merged, and annihilated during the subsequent evolution toward the final pattern.

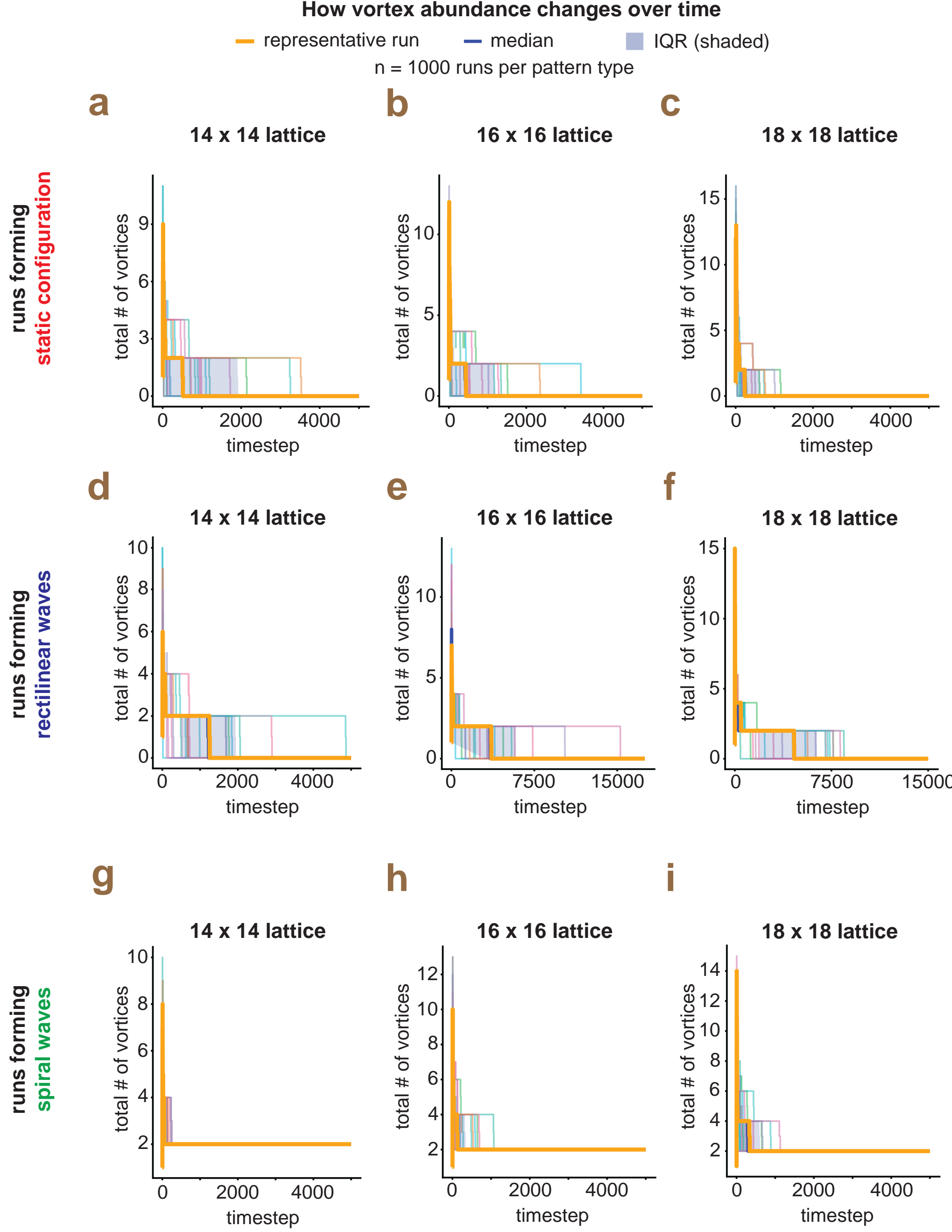

**Supplementary Figure 13: "Staircase" dynamics: declining vortex abundance across lattice sizes and final pattern types.**

This figure expands the vortex-abundance dynamics shown in Fig. 2c by quantifying how the total number of vortices changes over time across lattice sizes and final pattern types. Each panel shows the temporal trajectory of vortex abundance for runs that ultimately produced one of three final configurations: static configuration, rectilinear waves, or spiral waves. Every run yielded the characteristic "staircase" shapes. ***(caption continues)***

**Supplementary Figure 13 *(continued)*:**
Here, each column corresponds to a different lattice size: $14 \times 14$, $16 \times 16$, and $18 \times 18$ cells. For each final pattern type and lattice size, we analyzed 1,000 independent runs. Thin colored curves show representative individual trajectories sampled from these runs, illustrating the characteristic stepwise decreases in vortex number over time. The blue curve shows the median vortex count across runs at each timestep, and the shaded region indicates the interquartile range (IQR; $25^{th}$–$75^{th}$ percentiles). The orange curve highlights a representative run whose 'staircase' properties (vortex lifetime, number of annihilation events, and cumulative vortex abundance) lie closest to the median across runs.

Across all lattice sizes and pattern types, vortex abundance declined through discrete downward steps rather than continuously, reflecting successive pairwise annihilation events that removed vortices from the lattice. Runs that terminated in static configurations or rectilinear waves eventually lost all vortices. In contrast, runs that produced spiral waves retained at least—and almost always—two vortices indefinitely, leaving a persistent plateau corresponding to the surviving vortex pair that sustains the spiral pattern.

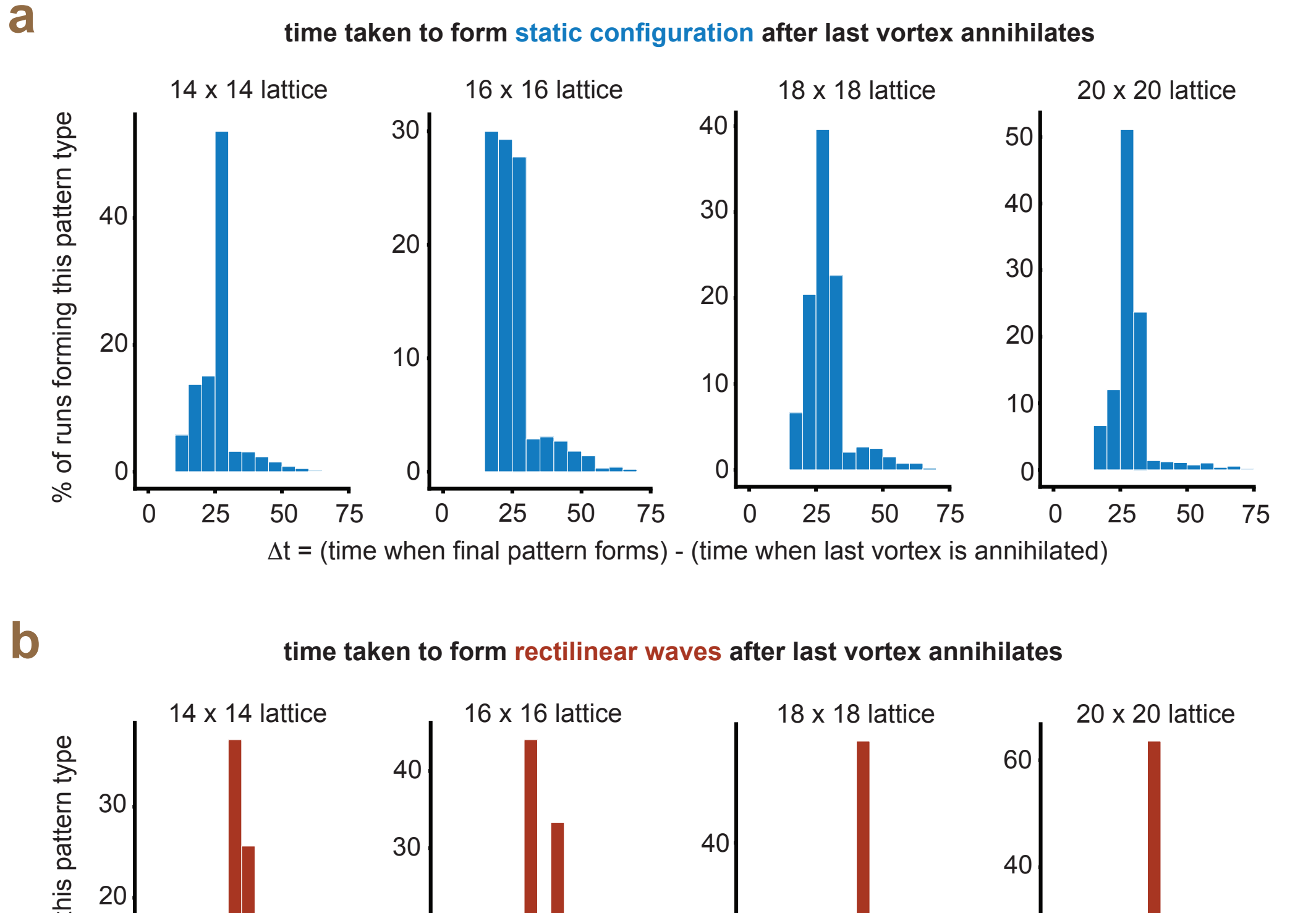

**Supplementary Figure 14: Pattern-formation times after vortex extinction (for runs forming static configuration or rectilinear waves) and spiral formation times.**

For each latice size ($14 \times 14$, $16 \times 16$, $18 \times 18$, $20 \times 20$), we ran 20,000 simulations and measured the 'run time' (i.e., time taken for a run to yield its final configuration). For runs that ended without any surviving vortices, we quantified $\Delta t = t_{\text{form}} - t_{\text{lastPairDisappear}}$, the delay between the disappearance of the last vortex pair and the time the final static configuration or rectilinear wave first appeared. Histograms show the distribution of $\Delta t$ within each type, plotted as the percentage of runs in that type and lattice size. Although most runs settled within $O(10^2)$ timesteps, rare runs exhibited substantially longer delays. For spiral-wave-forming runs, vortex pairs persisted through formation, so we instead plotted here the distribution of the total formation time $t_{\text{form}}$. Supplementary fig. 15 shows the distribution of the number of charged vortices remaining at the end of spiral-forming runs (vast majority have two, one +1 and one -1).

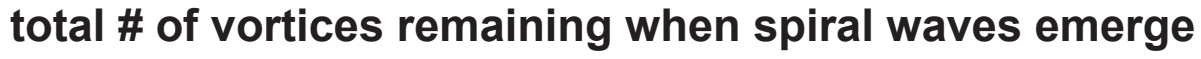

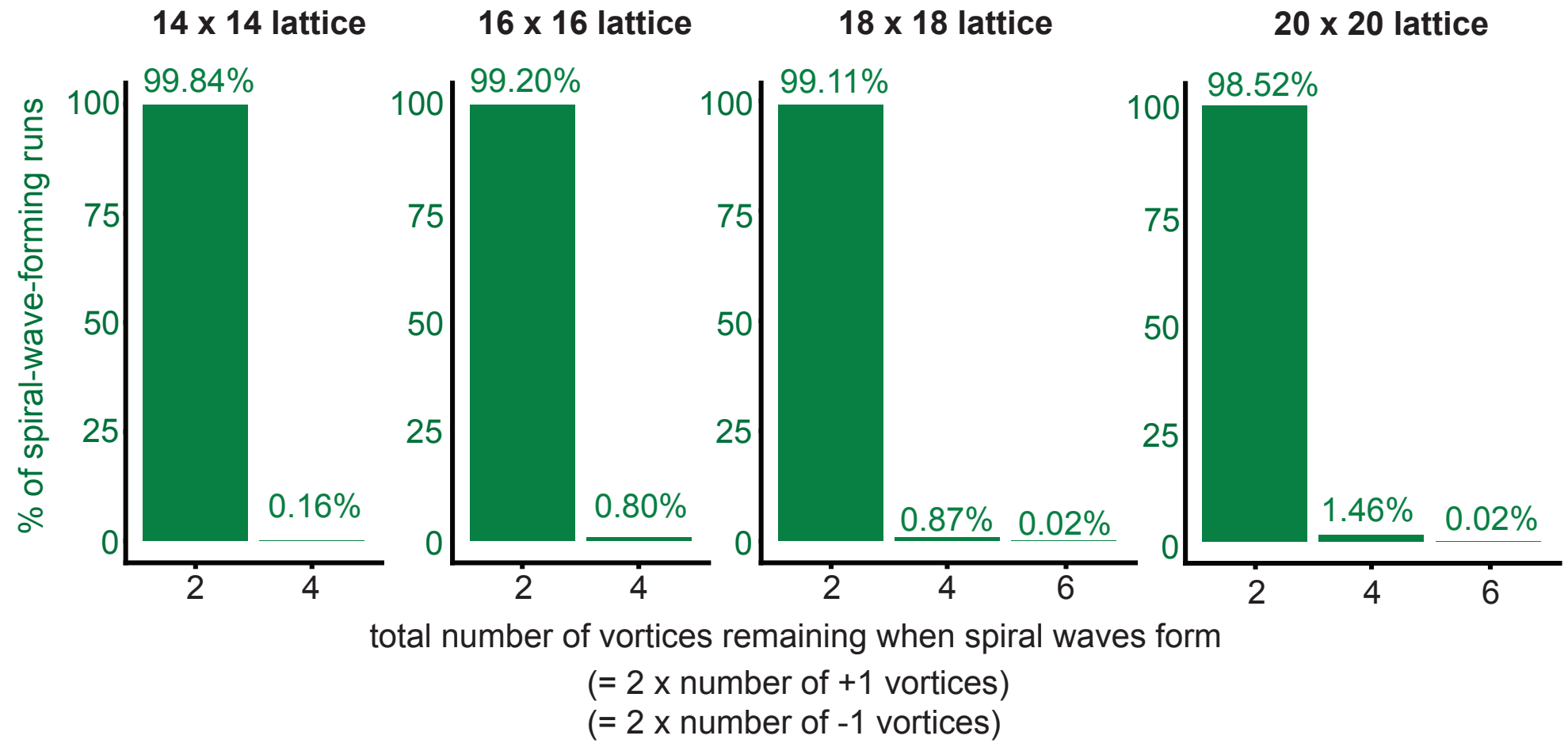

**Supplementary Figure 15: Virtually every spiral-wave-forming run retains exactly one vortex pair when the spiral wave emerges.**

For each lattice size, we ran 20,000 simulations of the cellular automaton and identified the subset of runs that formed spiral waves. Bar plots show the percentage of spiral-wave-forming runs that retained 2, 4, or 6 charged vortices at the moment the spiral wave formed, corresponding to 1, 2, or 3 surviving $+1/-1$ vortex pairs, respectively. The panels show results for $14 \times 14$ ($n$ = 5724 spiral-wave-forming runs), $16 \times 16$ ($n$ = 3729), $18 \times 18$ ($n$ = 10846), and $20 \times 20$ ($n$ = 12361) lattices. Across all lattice sizes, nearly all spiral-wave-forming runs retained exactly two charged vortices when the spiral pattern emerged, while only a small minority retained four vortices and a vanishingly small fraction retained six.

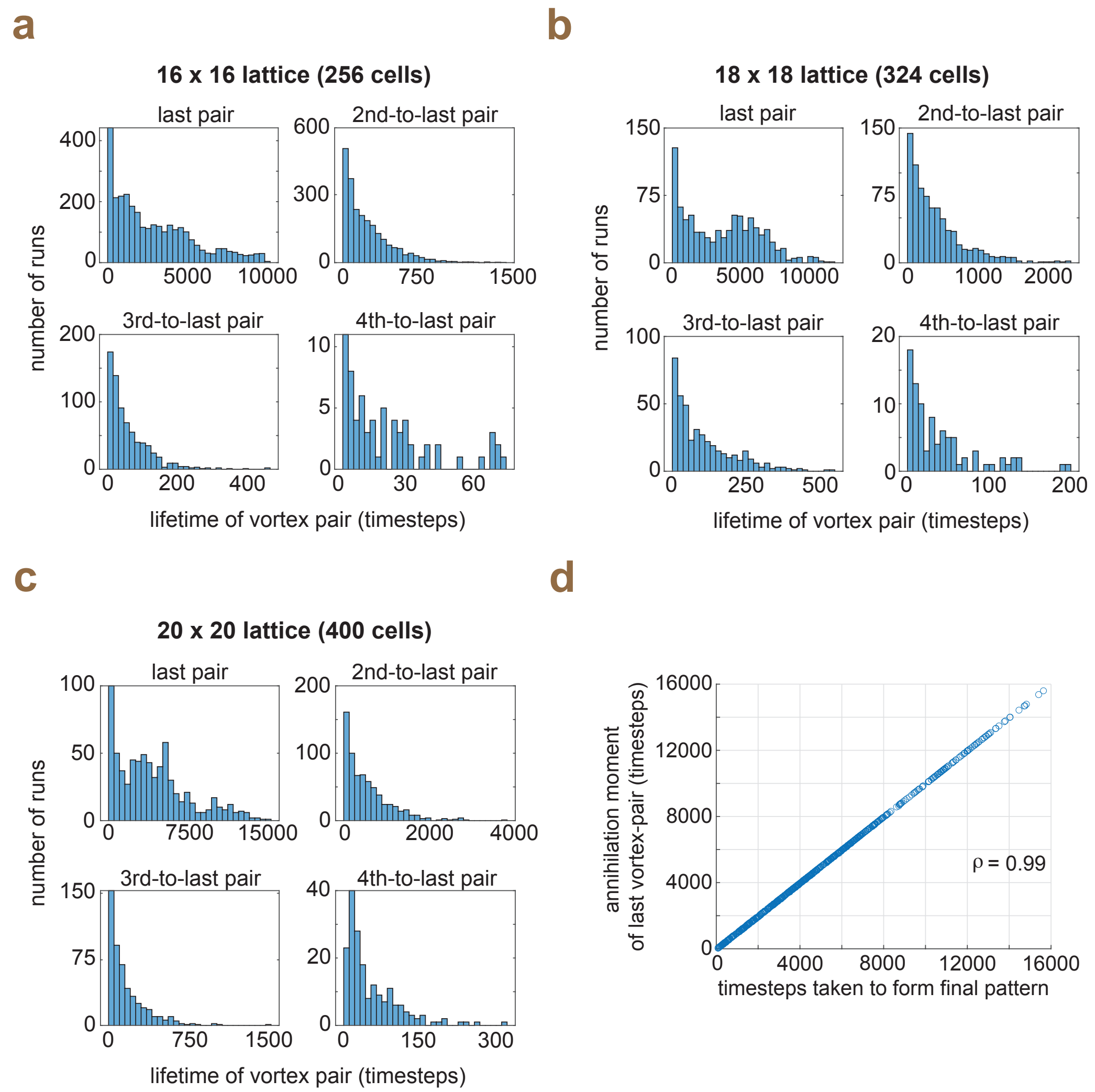

**Supplementary Figure 16: Lifetimes of the final vortex pairs dominate the duration of pattern formation.**

**(a–c)** Distributions of the lifetimes of the last four oppositely charged vortex pairs in cellular automaton runs for lattice sizes **(a)** $16 \times 16$, **(b)** $18 \times 18$, and **(c)** $20 \times 20$. For each run, we tracked successive annihilation events and measured the lifetime of the vortex pairs that disappeared last, second-to-last, third-to-last, and fourth-to-last. Histograms show the number of runs exhibiting each lifetime. Across lattice sizes, the lifetime distribution broadens dramatically for the final vortex pair compared with earlier pairs, indicating that once only a few vortices remain, the remaining pair can wander for long periods before annihilation.

***(caption continues)***

**Supplementary Figure 16 *(continued)*:**

**(d)** Correlation between the annihilation moment of the final vortex pair and the total run time of the cellular automaton (i.e., total time taken to form the final pattern, starting from a maximally disordered initial configuration). Each point corresponds to one simulation ($20 \times 20$ lattice, $N$ = 2048 runs). Because spiral-wave runs retain surviving vortices, they do not exhibit a final vortex-pair annihilation event and are therefore excluded from this panel. For the remaining runs, the near-perfect correlation ($\rho$ = 0.99) shows that the disappearance of the final vortex pair nearly coincides with the moment when the final pattern forms. Thus, the overall duration of these runs is largely determined by how long the last surviving vortex pair persists before annihilating.

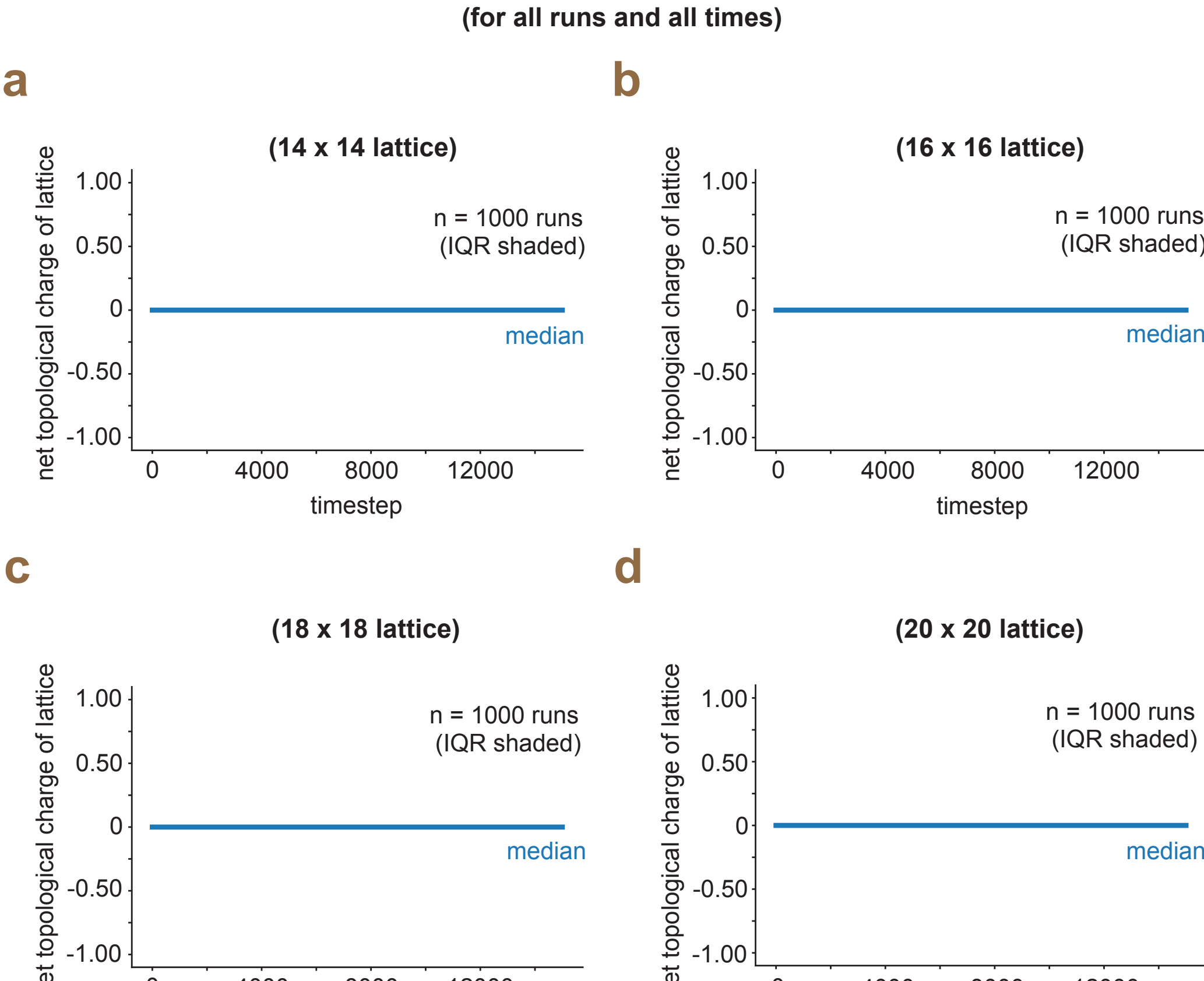

**Supplementary Figure 17: Conservation of total topological charge in every run.**
Time evolution of the net topological charge of the lattice during every run of the cellular automaton for lattice sizes **(a)** $14 \times 14$, **(b)** $16 \times 16$, **(c)** $18 \times 18$, and **(d)** $20 \times 20$. Each panel summarizes $n = 1000$ independent runs. Solid lines show the median net topological charge across runs, and the shaded regions indicate the interquartile range (IQR). At all timesteps and for all runs, the total topological charge of the lattice remained zero.

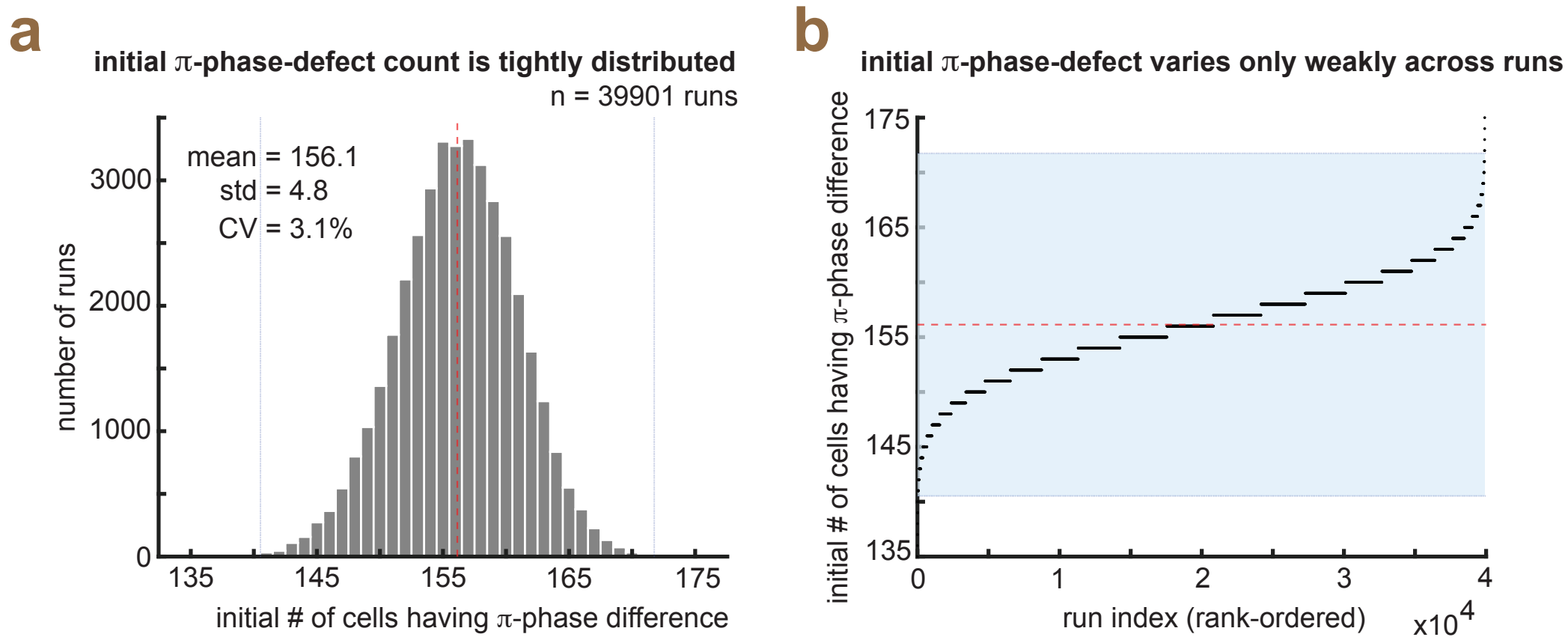

**Supplementary Figure 18: Initial $\pi$-phase-defect counts vary only weakly across disordered initial configurations.**

Distribution of the initial number of cells having a $\pi$-phase difference with at least one nearest neighbor across $n$ = 39,901 independent runs of the cellular automaton on a $14 \times 14$ lattice.

**(a)**, histogram of the initial $\pi$-phase-defect count for all runs. The dashed red line marks the mean value, and the blue dotted lines indicate $\pm$10% of the mean.

**(b)**, the same values plotted in rank-ordered form across runs. The shaded region again denotes the $\pm$10% interval around the mean. Despite each run beginning from a distinct maximally disordered configuration, the initial number of $\pi$-phase defects varies only weakly across runs (mean $\mu$ = 156.1, standard deviation $\sigma$ = 4.8, coefficient of variation = 3.1%). This near constancy reflects the statistical regularity of $\pi$-phase mismatches in randomly initialized states.

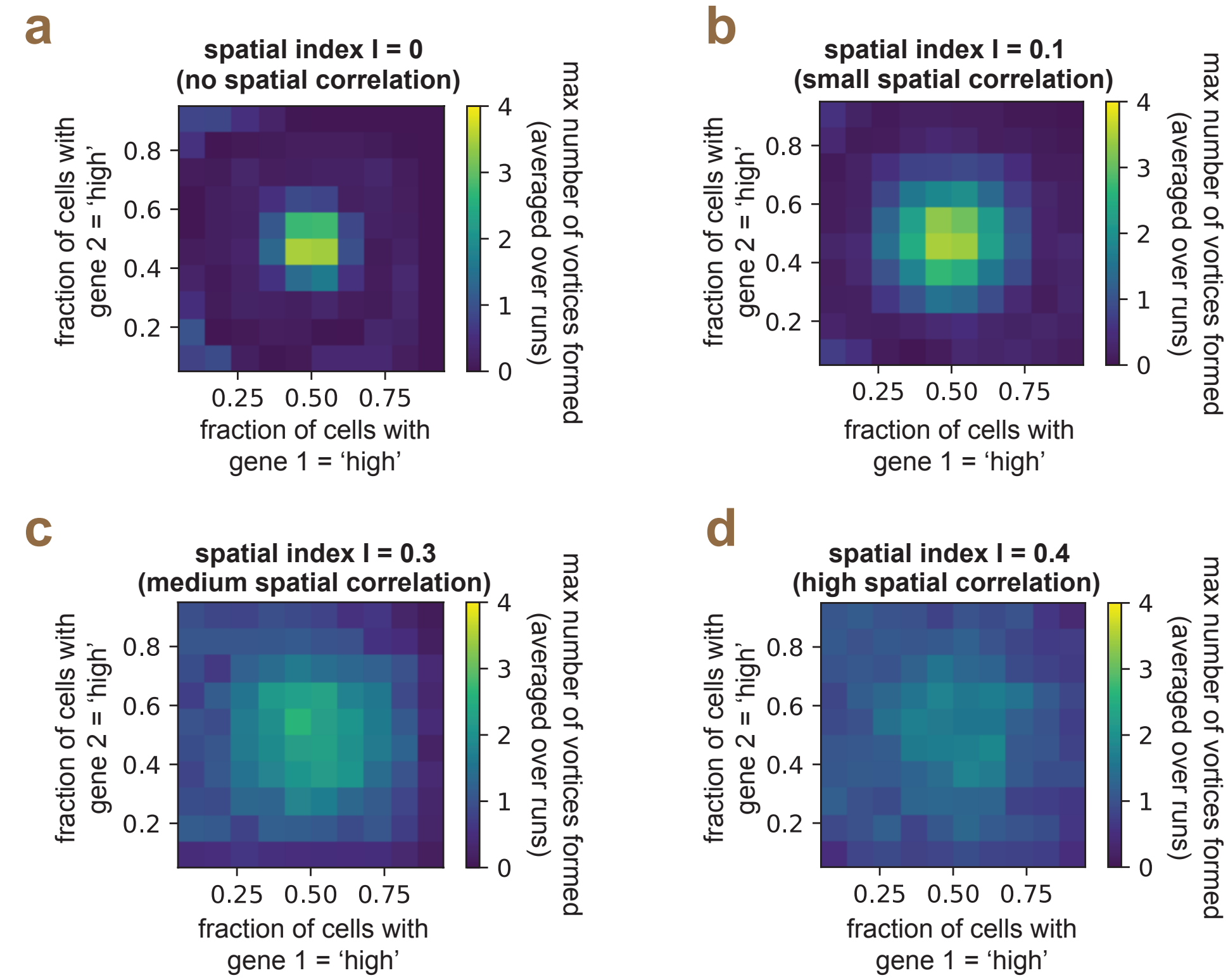

**Supplementary Figure 19: Initial disorder seeds vortex creation across spatially correlated initial states.**

Heat maps showing the maximum number of vortices formed during the early vortex-creation phase as a function of the initial fractions of cells with gene 1 in the high state ($p_1$) and gene 2 in the high state ($p_2$). Each panel corresponds to a different spatial index $I$ (defined in Supplementary Note 1), which measures spatial autocorrelation in the initial configuration (with $I_1 = I_2$): **a**, $I = 0$ (no spatial correlation); **b**, $I = 0.1$ (small spatial correlation); **c**, $I = 0.3$ (medium spatial correlation); **d**, $I = 0.4$ (high spatial correlation). The color scale indicates the maximum number of vortices created, averaged over runs with the same initial statistics. Across all spatial correlations, vortex creation peaks near $p_1 = p_2 = 0.5$, where the four cell states occur with approximately equal abundance and neighboring cells most frequently exhibit $\pi$-phase differences. Moving away from this point reduces the probability of $\pi$-phase mismatches and correspondingly lowers vortex production. Increasing spatial correlation further suppresses vortex formation by clustering identical states into larger domains, which reduces the density of domain boundaries where $\pi$-phase mismatches—and thus vortex nucleation events—can arise. These results show that the density of initial $\pi$-phase mismatches controls vortex nucleation, demonstrating that initial disorder seeds the creation of vortices.

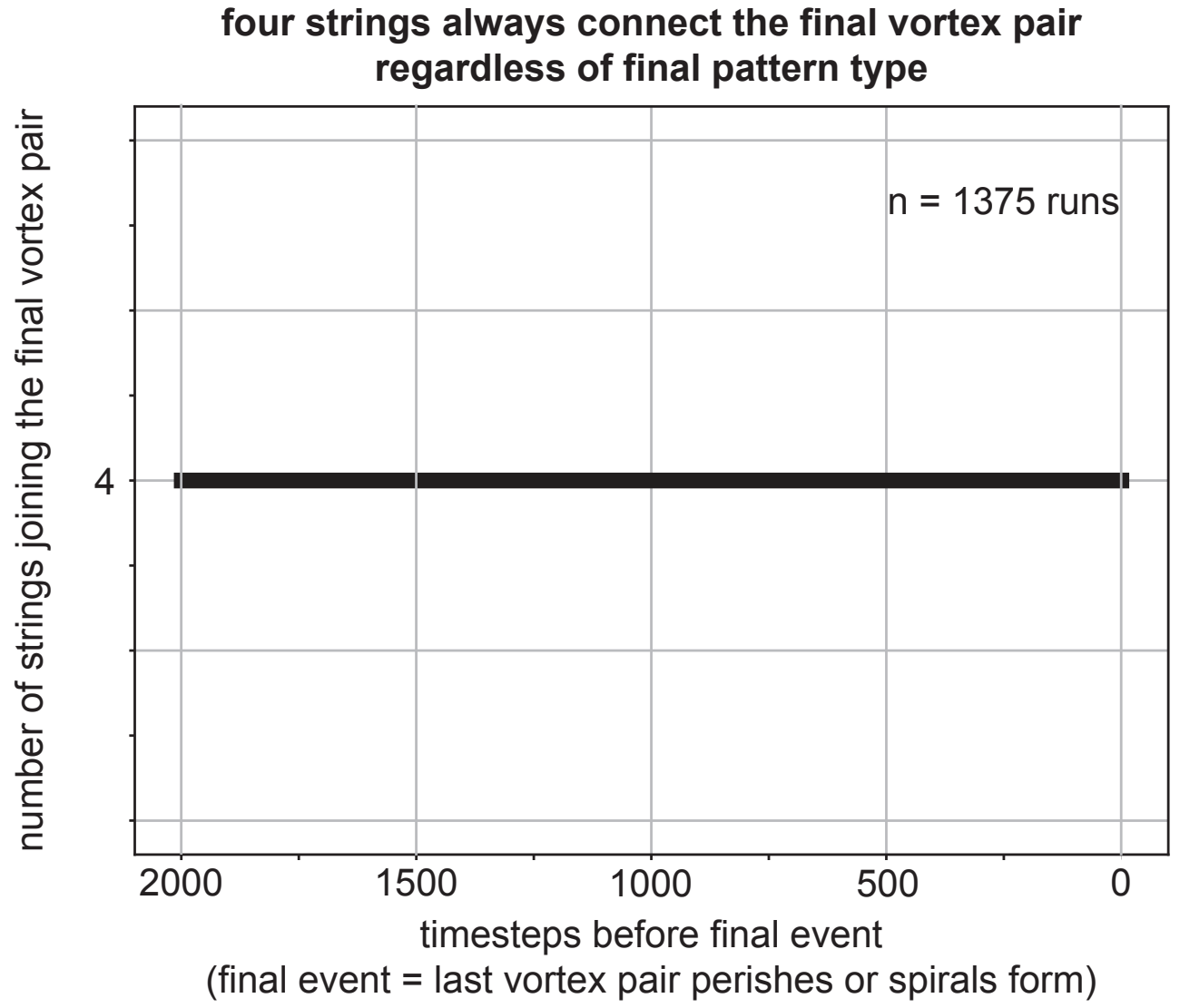

**Supplementary Figure 20: Four strings always connect the final vortex pair.**
We examined how many strings connect the final remaining vortex pair once only a single +1/1 vortex pair remains on the lattice. For each timestep after the system had entered this final-pair regime, we counted the number of strings composed of cells in identical states that connect the two vortex cores. Across all lattice sizes examined, the final vortex pair was always connected by exactly four strings—one for each of the four cell states. The plot shown here illustrates this result for 1,375 independent runs on a $26 \times 26$ lattice. The graph shows the raw (unsmoothed) number of strings as a function of time measured backward from the final event (defined as either annihilation of the last vortex pair, for runs forming static configurations or rectilinear waves, or the onset of spiral-wave recurrence). The four-string structure persisted throughout the entire late-stage regime in every trajectory. In 9 of the 1,375 runs (0.65%), transient deviations from four strings appeared as isolated single-timestep blips. Inspection revealed that these events arose from rare detection artifacts in the string-identification algorithm rather than genuine topological changes. After filtering these artifacts, the number of strings remained identically four at all times shown.

These results establish that once the system enters the final-vortex-pair regime, the two vortices are invariably connected by four strings, revealing a robust structural constraint of the lattice dynamics.

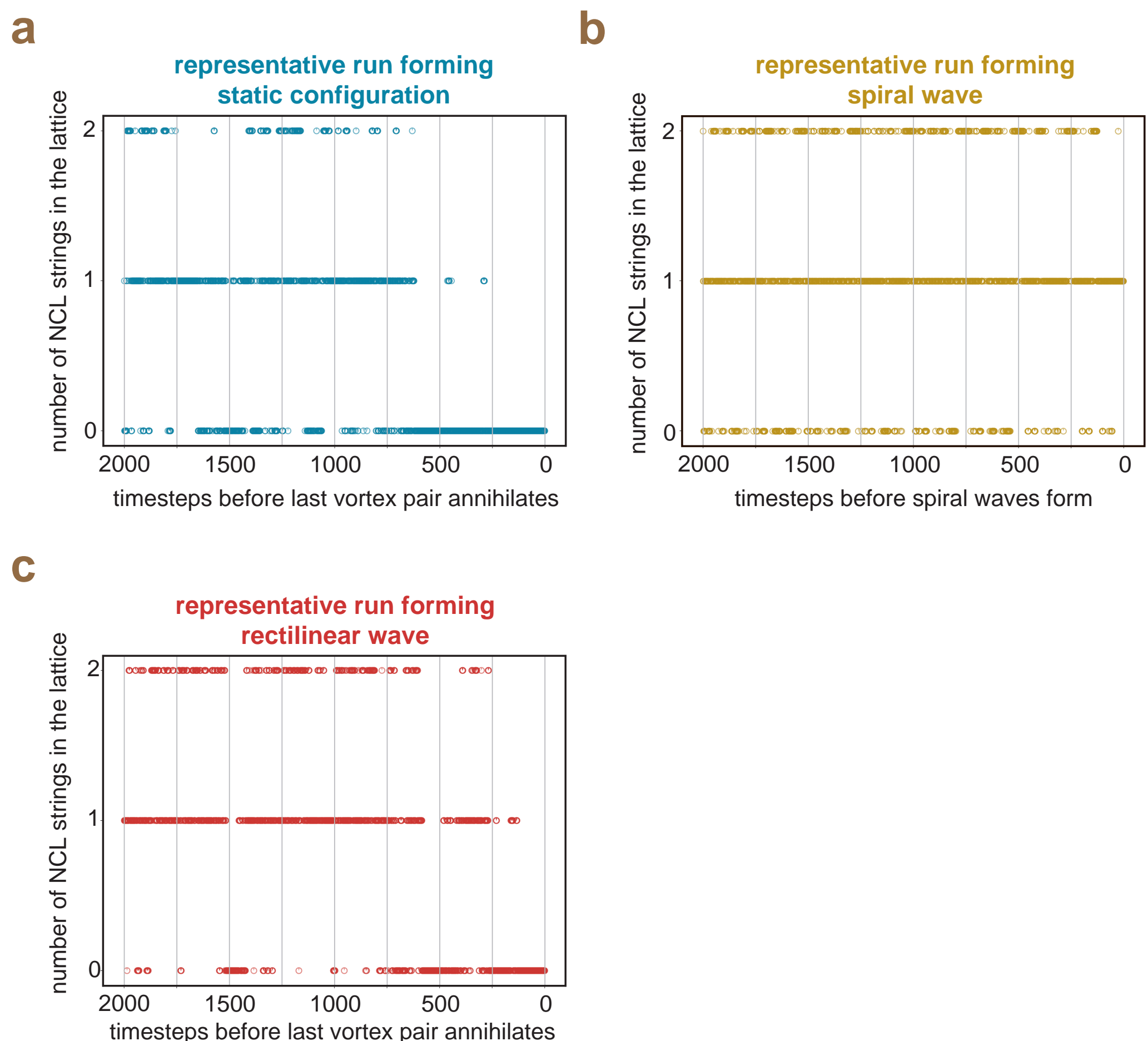

**Supplementary Figure 21: Raw dynamics of non-contractible-loop (NCL) strings connecting the final vortex pair.**

Each panel shows the raw number of NCL strings present in the lattice as a function of time during the late-stage regime in which only a single +1/-1 vortex pair remains. Each dot corresponds to one timestep. Time is measured backward from the final event, defined as either annihilation of the last vortex pair (runs forming static configurations or rectilinear waves) or the onset of spiral waves. Panels show representative runs that ultimately yielded **(a)** a static configuration, **(b)** a spiral wave, and **(c)** a rectilinear wave. Only timesteps during which exactly two vortices were present in the lattice are included. The NCL-string count fluctuates between 0, 1, and 2 over thousands of timesteps prior to the final event.

These raw trajectories illustrate the late-stage dynamics from which the sliding-window averages and temporal trends shown in Fig. 4e–f were computed. Notably, runs in which the final vortices eventually annihilate increasingly spent time without any NCL strings as the final event approaches, whereas spiral-forming runs continue to fluctuate around nonzero NCL-string counts.

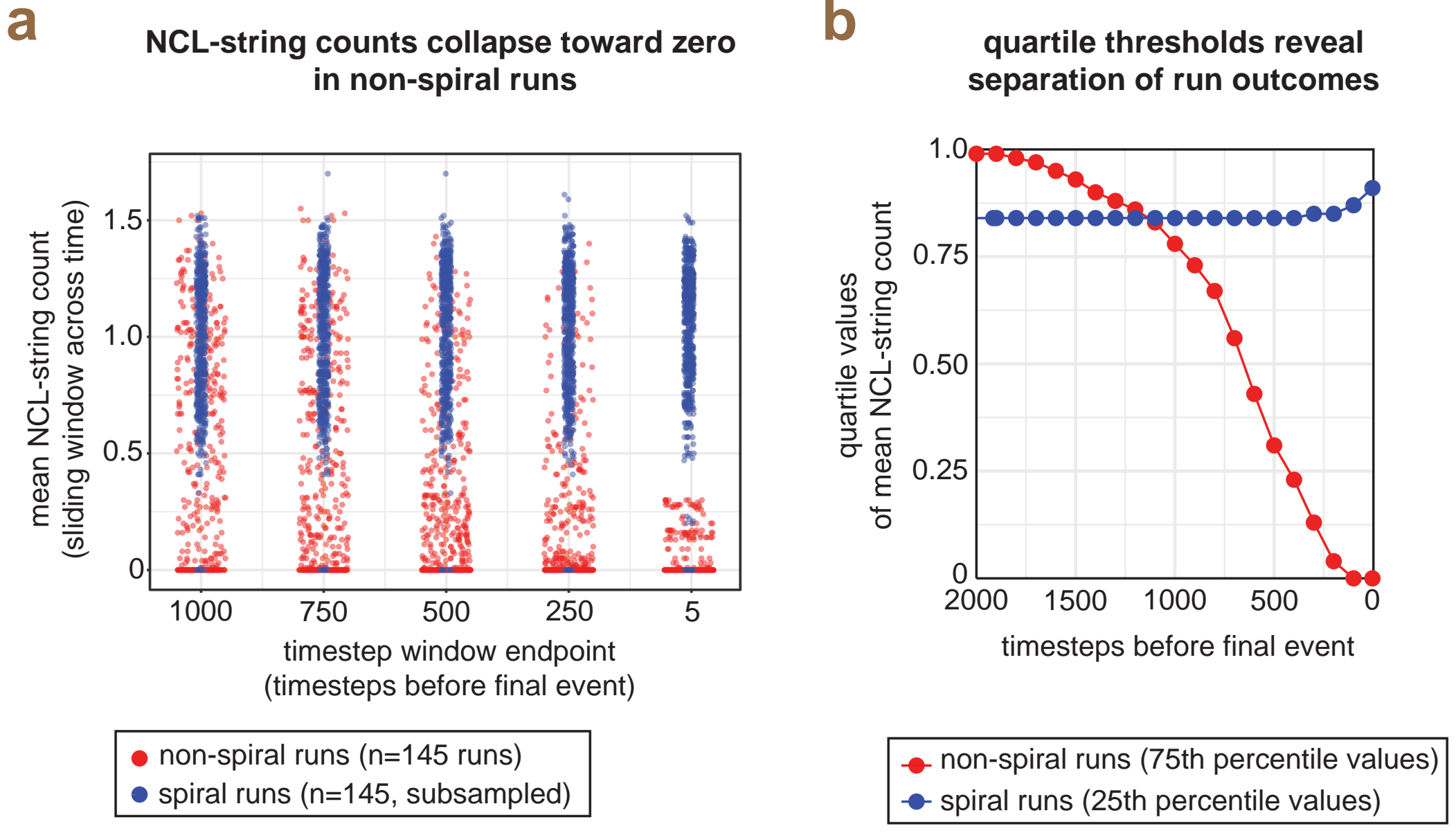

**Supplementary Figure 22: Distributions of temporally averaged NCL-string counts separate spiral and non-spiral runs.**

**(a)** Distribution of mean NCL-string counts computed within sliding windows for runs that ultimately form spiral waves (blue) or non-spiral outcomes (static configurations or rectilinear waves; red). Each dot represents the mean number of NCL strings within a sliding averaging window during the late-stage regime in which only a single vortex pair remains. The horizontal axis indicates the endpoint of the averaging window, measured as timesteps before the final event. For each labeled endpoint, the sliding window spans the preceding 100 timesteps toward the final event (i.e., toward 0). When fewer than 100 timesteps remain before the final event—for example, at endpoint = 5—the window includes all available timesteps. To reduce visual clutter, we plot values only at five approximately evenly spaced window endpoints between the labeled endpoint and the final event (e.g., for endpoint = 1000, points correspond to windows ending at 1000, 750, 500, 250, and 1). Spiral runs were randomly subsampled to 145 trajectories to match the number of non-spiral runs (145) and avoid imbalance in the distributions. As the final event approaches, the distribution for non-spiral runs progressively contracts toward zero NCL strings, whereas spiral-forming runs remain centered near one.

**(b)** Summary of the separation between the two outcome classes. Red points show the 75th percentile of the non-spiral distributions, while blue points show the 25th percentile of the spiral distributions. Already $\sim$1000 timesteps before the final event, the upper quartile of the non-spiral distribution falls below the lower quartile of the spiral distribution, revealing a clear statistical separation between the two trajectory classes that widens as the system approaches the final event.

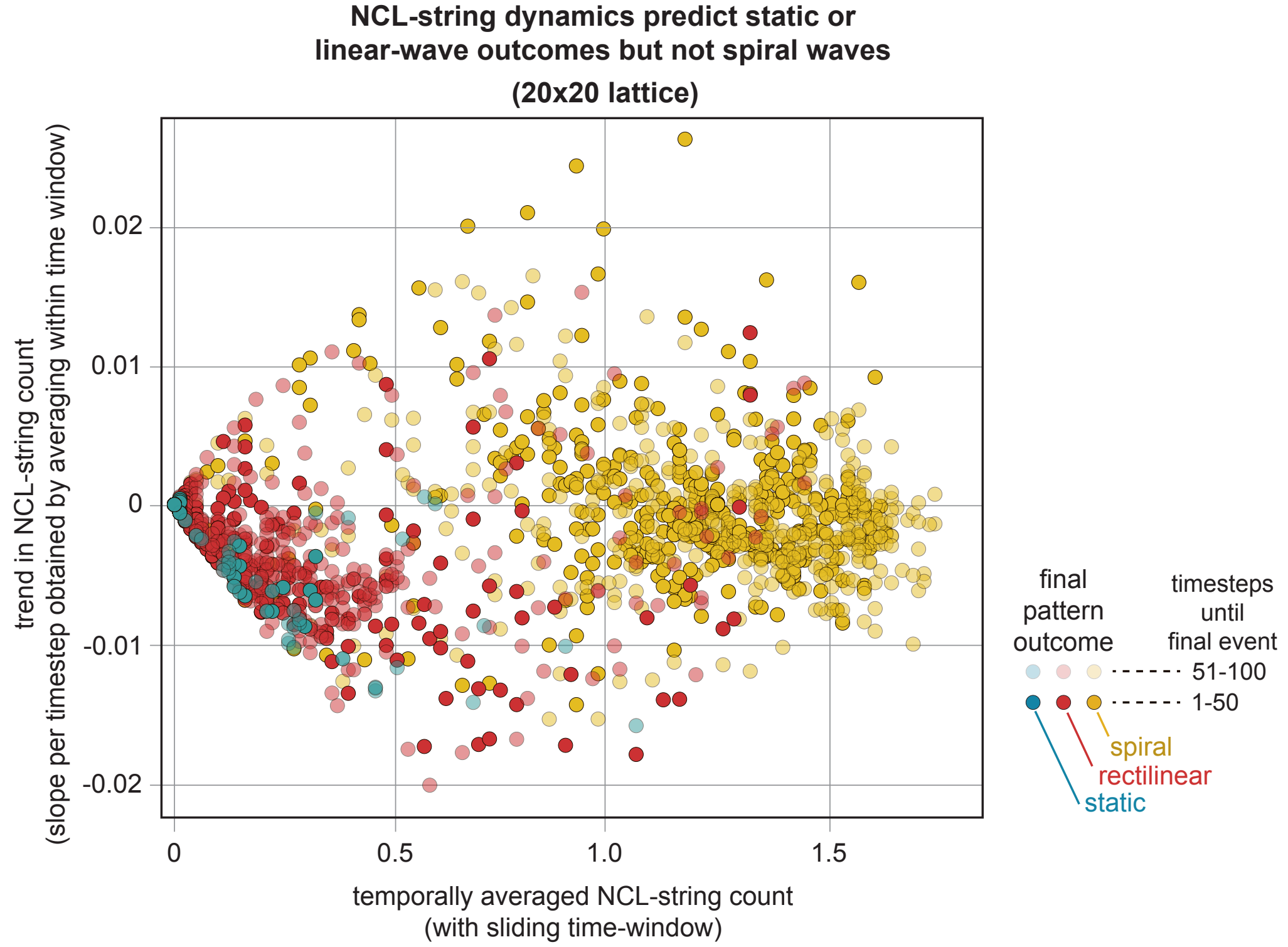

**Supplementary Figure 23: NCL-string dynamics on a smaller lattice show the same predictive structure as in Fig. 4f.**

Scatter plot of the temporal trend in NCL-string count versus the temporally averaged NCL-string count during the late-stage regime in which only a single vortex pair remains on a $20 \times 20$ lattice. Each point corresponds to a sliding time window along an individual trajectory. The horizontal axis shows the mean NCL-string count within the window, while the vertical axis shows the temporal trend (slope per timestep obtained from a linear least-squares fit of the NCL-string count versus time) within that same window. Each window spans up to 100 timesteps preceding the final event (defined as either annihilation of the last vortex pair for runs forming static configurations or rectilinear waves, or the onset of spiral-wave recurrence). When fewer than 100 timesteps remain before the final event, all available timesteps are used. Points are colored by the eventual pattern outcome: static configurations (cyan), rectilinear waves (red), and spiral waves (yellow). Filled points denote windows occurring within the final 50 timesteps before the final event, whereas lighter points correspond to windows occurring 51–100 timesteps before the final event. Each trajectory contributes multiple overlapping windows as the system approaches the final event.

***(caption continues)***

**Supplementary Figure 23 *(continued)*:**

Consistent with the behavior shown for the 26 $\times$ 26 lattice in Fig. 4f, runs that ultimately produce static configurations or rectilinear waves cluster at low mean NCL-string counts with negative temporal trends, whereas spiral-forming runs remain concentrated near approximately one NCL string with near-zero temporal trend. This separation indicates that NCL-string dynamics predict annihilating outcomes but not spiral-wave formation even on smaller lattices ($n$ = 211 independent runs).

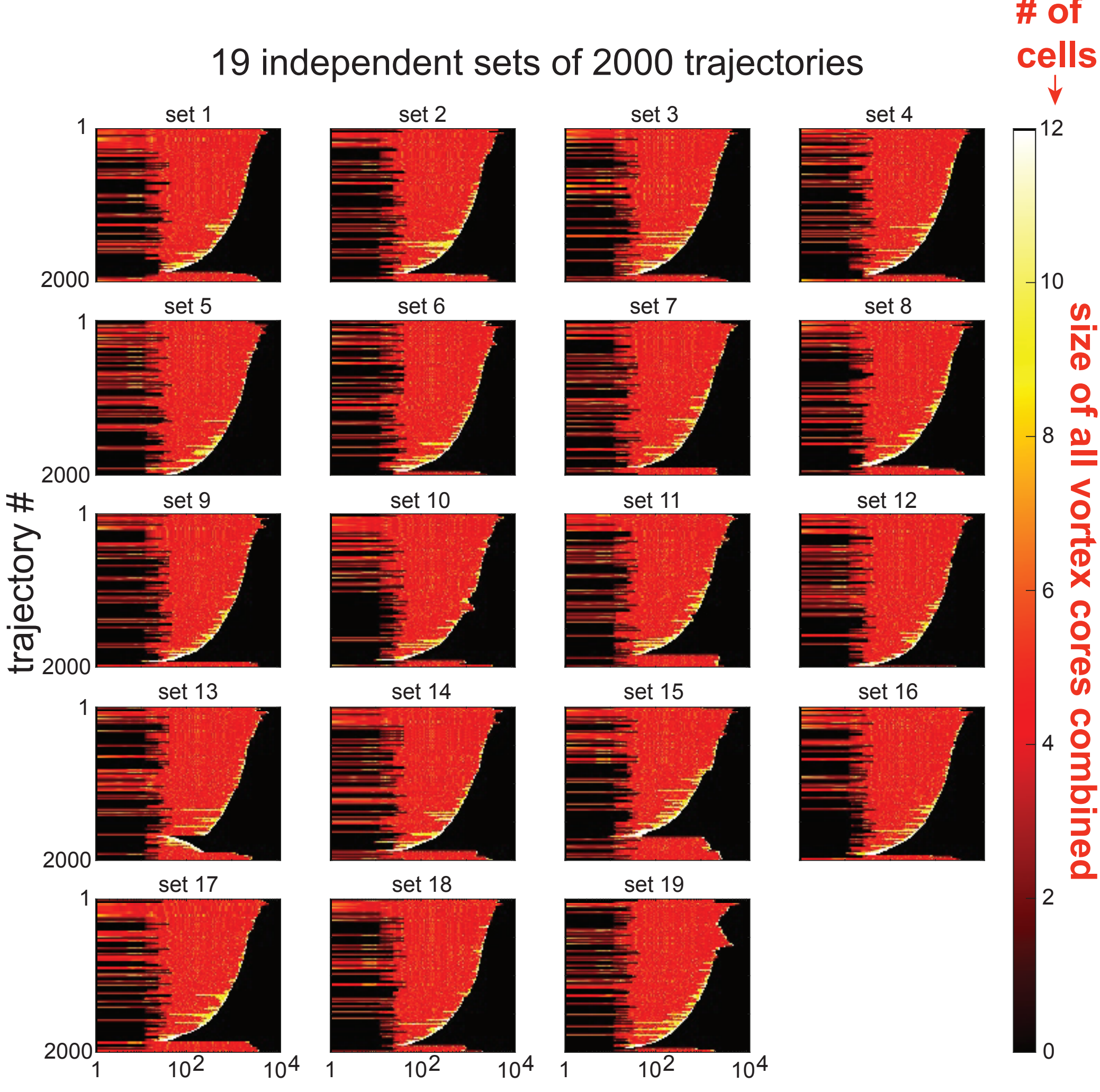

**Supplementary Figure 24: Heat-map structure analogous to Fig. 5b recurs across additional trajectory sets.**

For each of 19 additional trajectory sets, we computed the same scalar observable used in Fig. 5b: the total number of cells belonging to vortex cores at each backward-timestep (i.e., timesteps prior to either the last vortex pair annihilating or the spiral-wave emerging). Each panel shows one set, with trajectories reordered by greedy nearest-neighbor similarity to place visually similar scalar histories adjacent to one another, exactly as in the construction of Fig. 5b. Color encodes total vortex core size, capped at 12 as in Fig. 5b. In every set, most of the heat map remained visually structure-poor over the majority of the trajectory, indicating that trajectories leading to different final outcomes were broadly indistinguishable at the level of this scalar observable for most of their evolution.

***(caption continues)***

**Supplementary Figure 24 *(continued)*:**

By contrast, barcode-like horizontal microstructures emerged reproducibly only near the final timesteps (small $\tau$), showing that the late-time visual signature highlighted in Fig. 5b is not specific to the single trajectory set used there but recurs across additional independently analyzed trajectory sets.

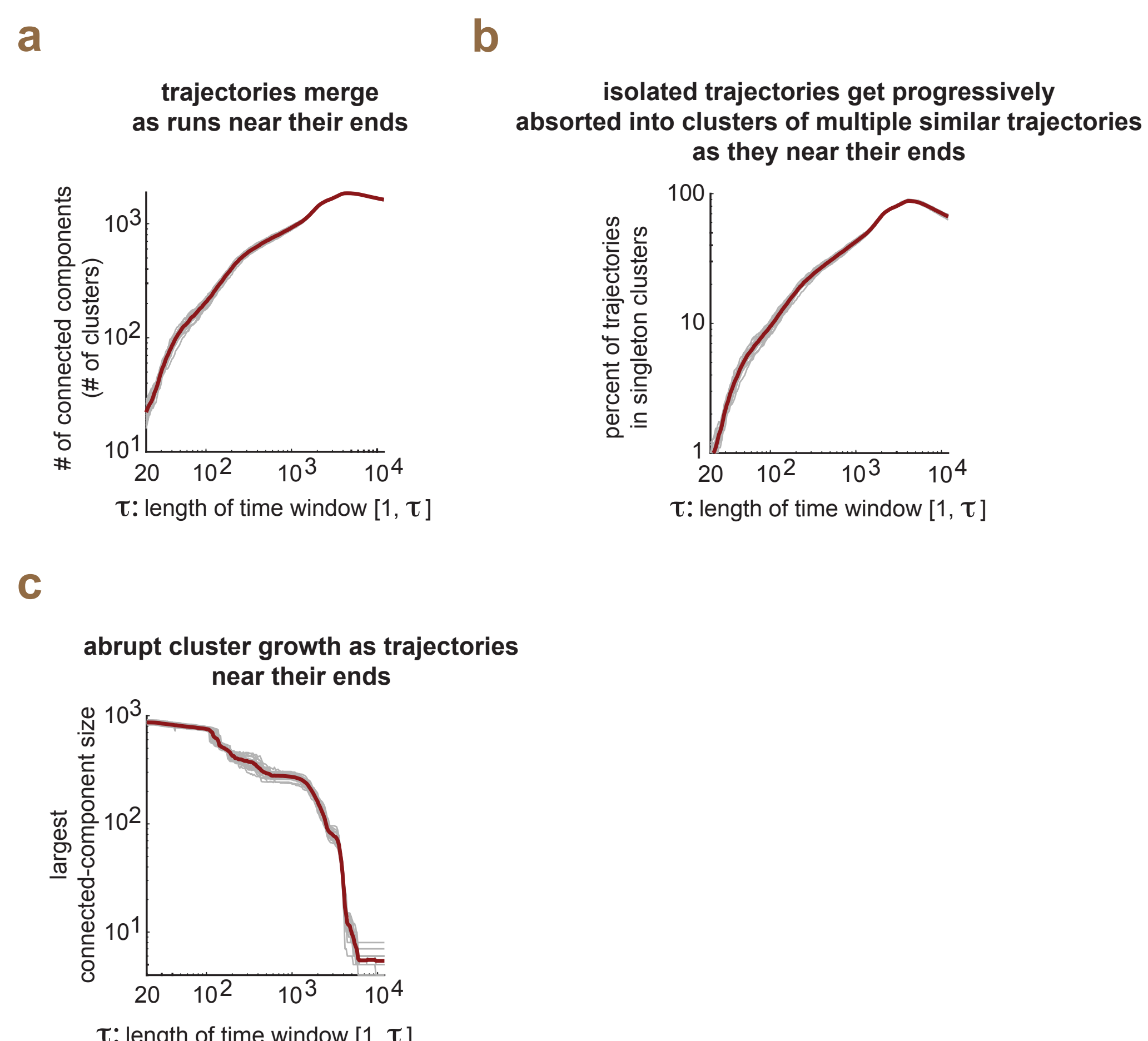

**Supplementary Figure 25: Graph-based clustering dynamics are reproducible across additional trajectory sets.**

We repeated the graph-based clustering analysis used in Fig. 5c,d on 19 additional trajectory sets, each containing 2,000 trajectories. For each backward-time window $[1, \tau]$, where $t = 1$ denotes the timestep immediately preceding final-pattern formation, we represented every trajectory by its total-vortex-core-size time series over that window, connected pairs of trajectories whose Euclidean distance fell below the same threshold used in the main analysis, and identified the connected components of the resulting graph. Thin gray curves show the 19 trajectory sets individually and the dark red curves show their mean.

(**a**) Total number of connected components as a function of $\tau$. In every set, the number of components remained large for long windows and then decreased reproducibly as the window shortened, showing that late-time trajectories merged into a smaller number of groups.

***(caption continues)***

**Supplementary Figure 25 *(continued)*:**

(**b**) Fraction of trajectories that belonged to singleton connected components. This quantity decreased toward small $\tau$, showing that isolated trajectories were progressively absorbed into multi-trajectory clusters as the final timesteps approached.

(**c**) Size of the largest connected component as a function of $\tau$. In every set, the largest component grew in a stepwise manner as $\tau$ decreased, quantitatively supporting the abrupt cluster-growth behavior shown in Fig. 5d.

Together, these panels show that the graph-based clustering dynamics reported in Fig. 5c,d are not specific to the trajectory set used there but recur reproducibly across additional trajectory sets.

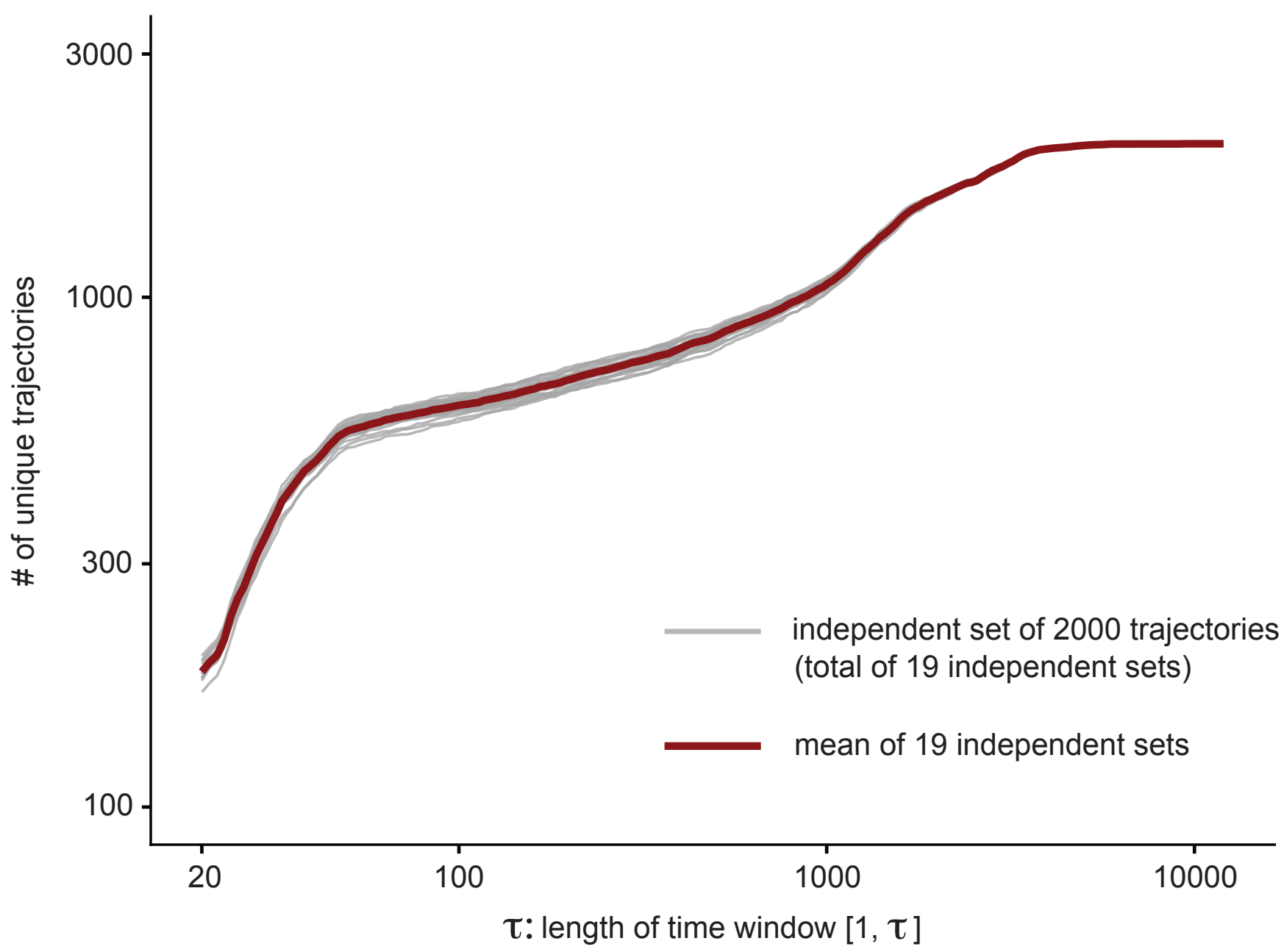

**Supplementary Figure 26: Unique-trajectory collapse is reproducible across additional trajectory sets.**

For each of the 19 additional trajectory sets, each containing 2,000 trajectories, we quantified how many exactly distinct total-vortex-core-size trajectories remained within the backward-time window $[1, \tau]$, where $t = 1$ denotes the timestep immediately preceding formation of the final configuration and larger $\tau$ includes progressively earlier timesteps in the run. We counted two trajectories as distinct if their total-vortex-core-size time series differed at even a single timestep within the window. Thin gray curves show the 19 sets individually, and the dark red curve shows their mean. In every set of 2000 trajectories, the number of distinct trajectories remained high for most of the run and then decreased sharply only at late times, demonstrating that the collapse in trajectory diversity shown in Fig. 5e is not specific to the single set of 2000 trajectories used there but is a reproducible property of the system.

We independently verified that every full trajectory in the complete dataset was distinct from every other full trajectory. The late-time reduction in the number of exact trajectory types therefore does not arise from duplication in the dataset. Instead, it reflects a genuine convergence of many different full histories onto a much smaller number of terminal scalar trajectories. This figure therefore shows that the final stage of self-organization is highly constrained even though the earlier dynamics remains highly diverse.

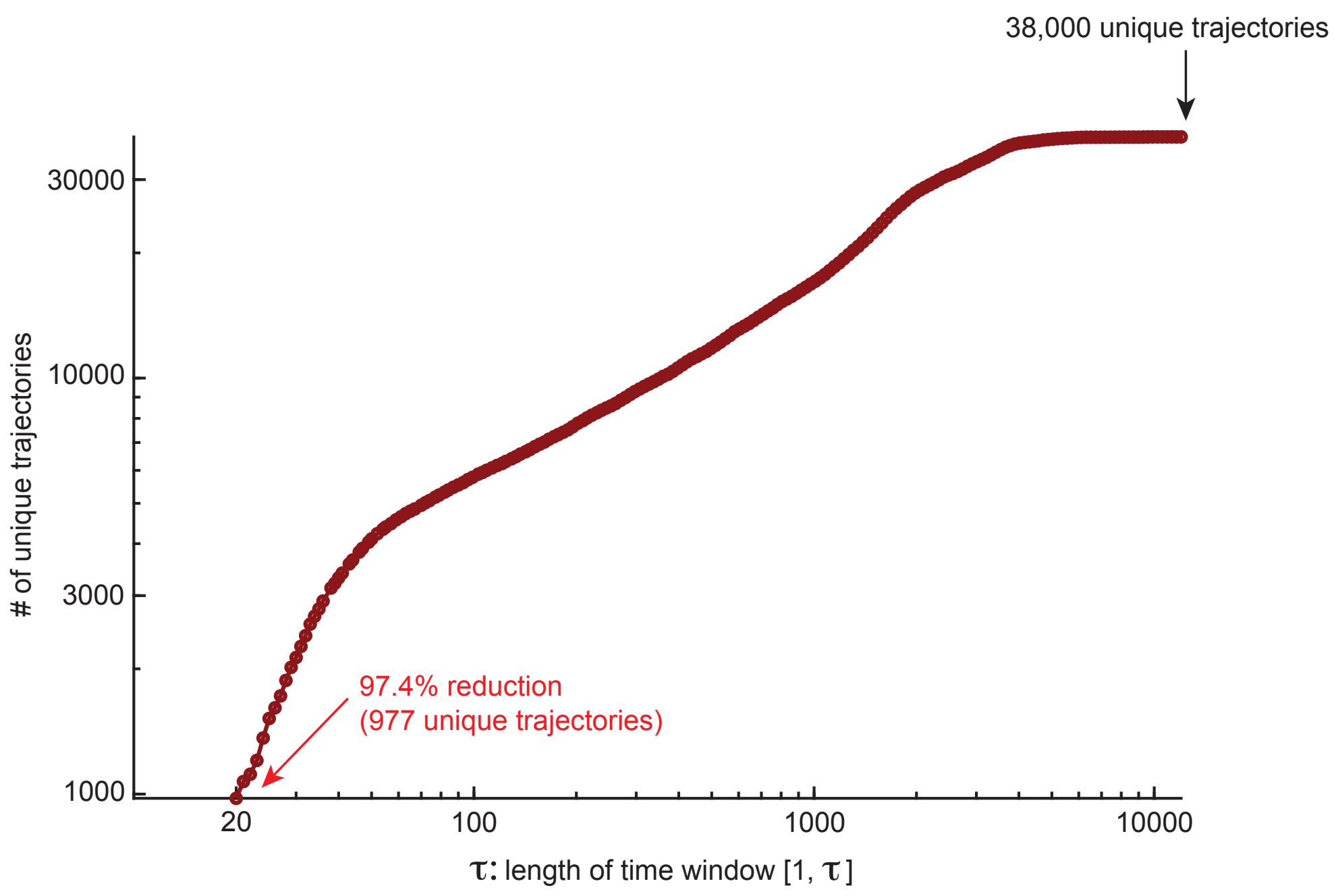

**Supplementary Figure 27: Exact trajectory diversity collapses strongly in the pooled dataset.** We combined 19 complete sets of trajectories–each containing 2000 trajectories–into one pooled dataset containing 38,000 trajectories and repeated the unique-trajectory analysis on the combined set. For each backward-time window $[1, \tau]$, where $t = 1$ denotes the timestep immediately preceding final-pattern formation, we counted how many *exactly distinct* total-vortex-core-size trajectories remained. We counted two trajectories as distinct if their total-vortex-core-size time series differed at even a single timestep within the window. At large $\tau$, the number of unique trajectories rose to 38,000, showing that every full trajectory in the pooled dataset was distinct. Yet by $\tau = 20$, these 38,000 distinct full trajectories had collapsed to only 977 exact late-time scalar trajectories, corresponding to a 97.4% reduction in trajectory diversity. Thus, the late-time convergence shown in Fig. 5e does not depend on analyzing one set of trajectories at a time. Instead, it persists even when all trajectories are pooled together, showing that many globally distinct full histories converge onto a much smaller number of terminal trajectories.

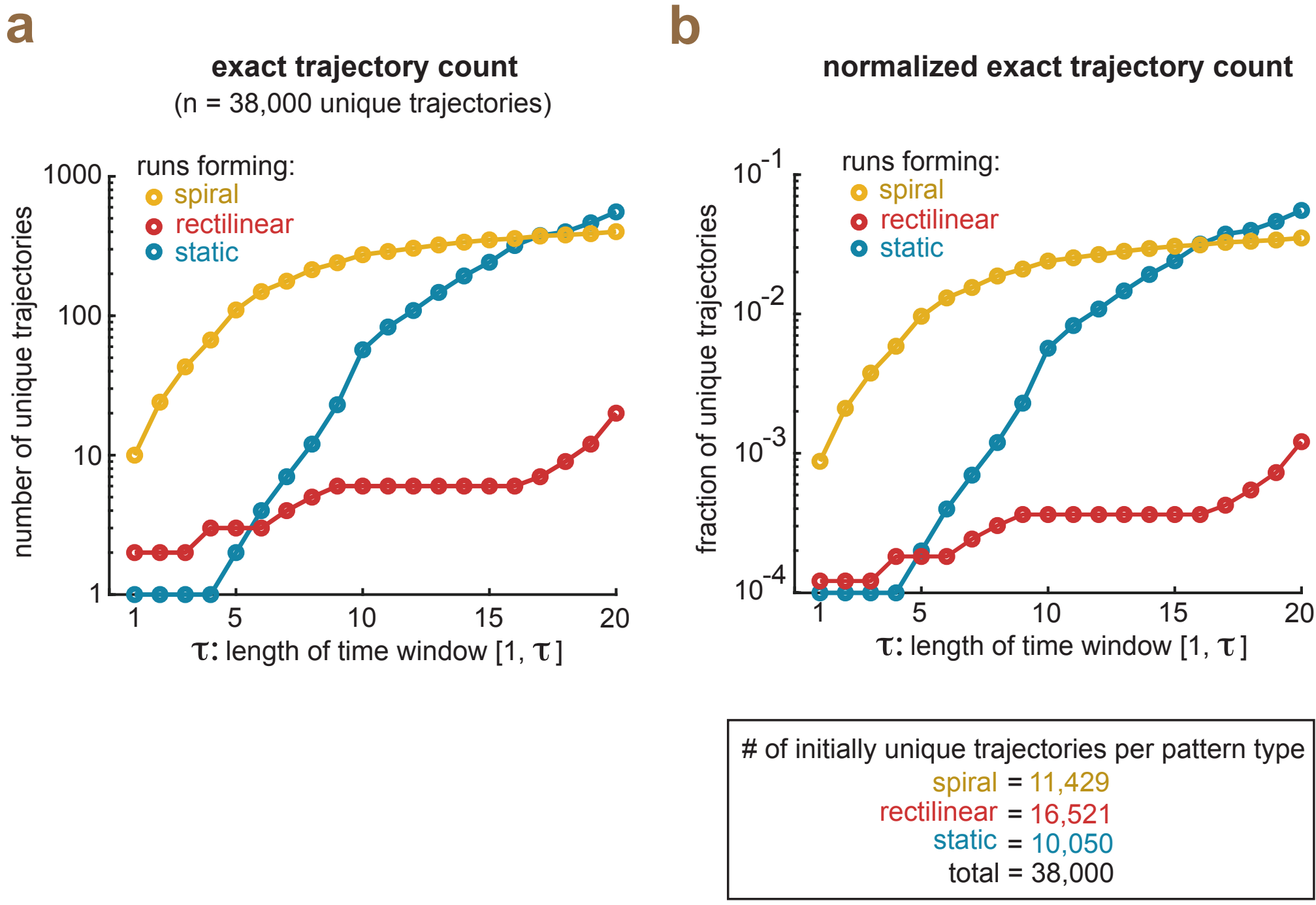

**Supplementary Figure 28: Spiral-destined trajectories remain more diverse than annihilation-bound trajectories even at the final timestep ($\tau$ = 1).**

We extended the pooled unique-trajectory analysis of Supplementary fig. 27 to very small backward-time windows, using all 38,000 trajectories from the 19 trajectory sets not used for the main Fig. 5 analysis. For each trajectory, we considered the scalar observable used in Fig. 5: the total vortex-core size at each backward-time step. Two trajectories were counted as identical within the window [1, $\tau$] if their scalar time series matched exactly at every timestep in that window.

**(a)** Number of exact scalar trajectories as a function of $\tau$, broken down by final pattern type. Static-destined runs collapsed most strongly, reaching a single exact terminal scalar sequence for $\tau \leq 4$. Rectilinear-wave runs also collapsed strongly, remaining confined to only a few exact terminal scalar sequences over the same range. By contrast, spiral-destined runs remained much more diverse even at $\tau$ = 1, indicating that spiral formation did not converge onto a comparably compressed set of terminal scalar trajectories.

**(b)** Same analysis after normalizing by the number of trajectories in each final-pattern class, showing that the asymmetry remained after accounting for unequal class sizes ($n_{\text{static}}$ = 10,050, $n_{\text{spiral}}$ = 11,429, $n_{\text{rectilinear}}$ = 16,521). Thus, the greater late-time heterogeneity of spiral-destined runs was not an artifact of class abundance.

***(caption continues)***

**Supplementary Figure 28 *(continued)*:**

Together, these results show that the asymmetry reported in the main text extends all the way to the final pre-pattern timestep: annihilation-bound outcomes collapsed onto highly constrained terminal scalar trajectories, whereas spiral-destined outcomes remained substantially more heterogeneous.

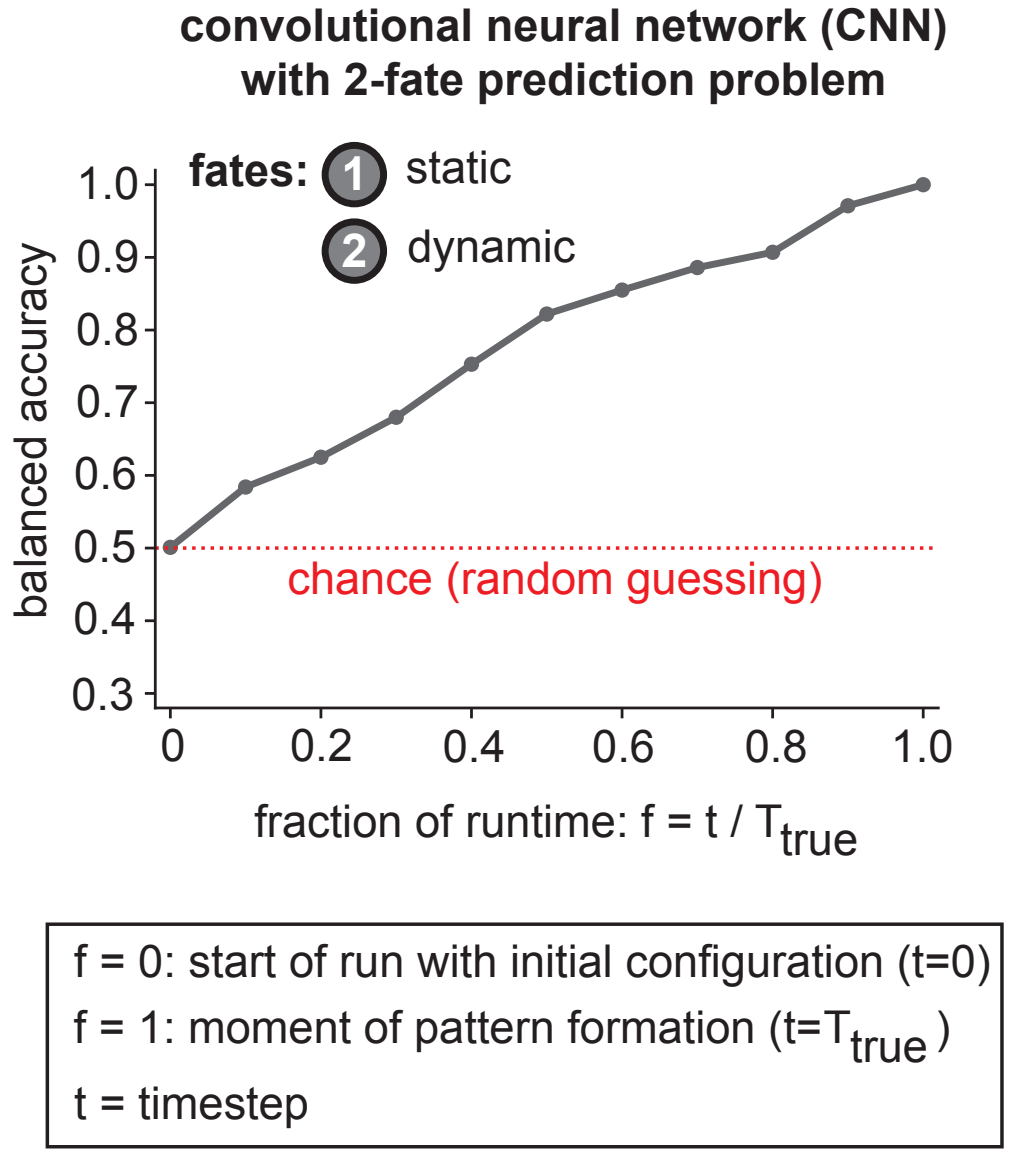

**Supplementary Figure 29: Fate prediction improves as time passes (i.e., as the system evolves towards pattern formation).**

Balanced accuracy of a convolutional neural network (CNN) predicting a binary fate — static or non-static (dynamic) — from a snapshot of lattice configuration taken at timestep $t = f \cdot T_{true}$, where $T_{true}$ is the timestep at which the final pattern forms (see Supplementary note 7); $f$ is the fraction, $t/T_{true}$, between 0 and 1. $t = 0$ is when the system begins with the maximally disordered initial configuration. The CNN on this binary task used at the initial configuration in Supplementary note 2 — static versus dynamic (rectilinear or spiral wave) — rises from chance (1/2) to one as time progresses from $t = 0$ to $T_{true}$. Extending our investigation of machine-learning on initial-configuration (Supplementary note 2) to machine-learning on later snapshots of the lattice configuration (this plot; see Supplementary note 7) revealed that there is no practically extractable predictive signal at $f = 0$, but that it later appears and becomes more predictive along the trajectory.

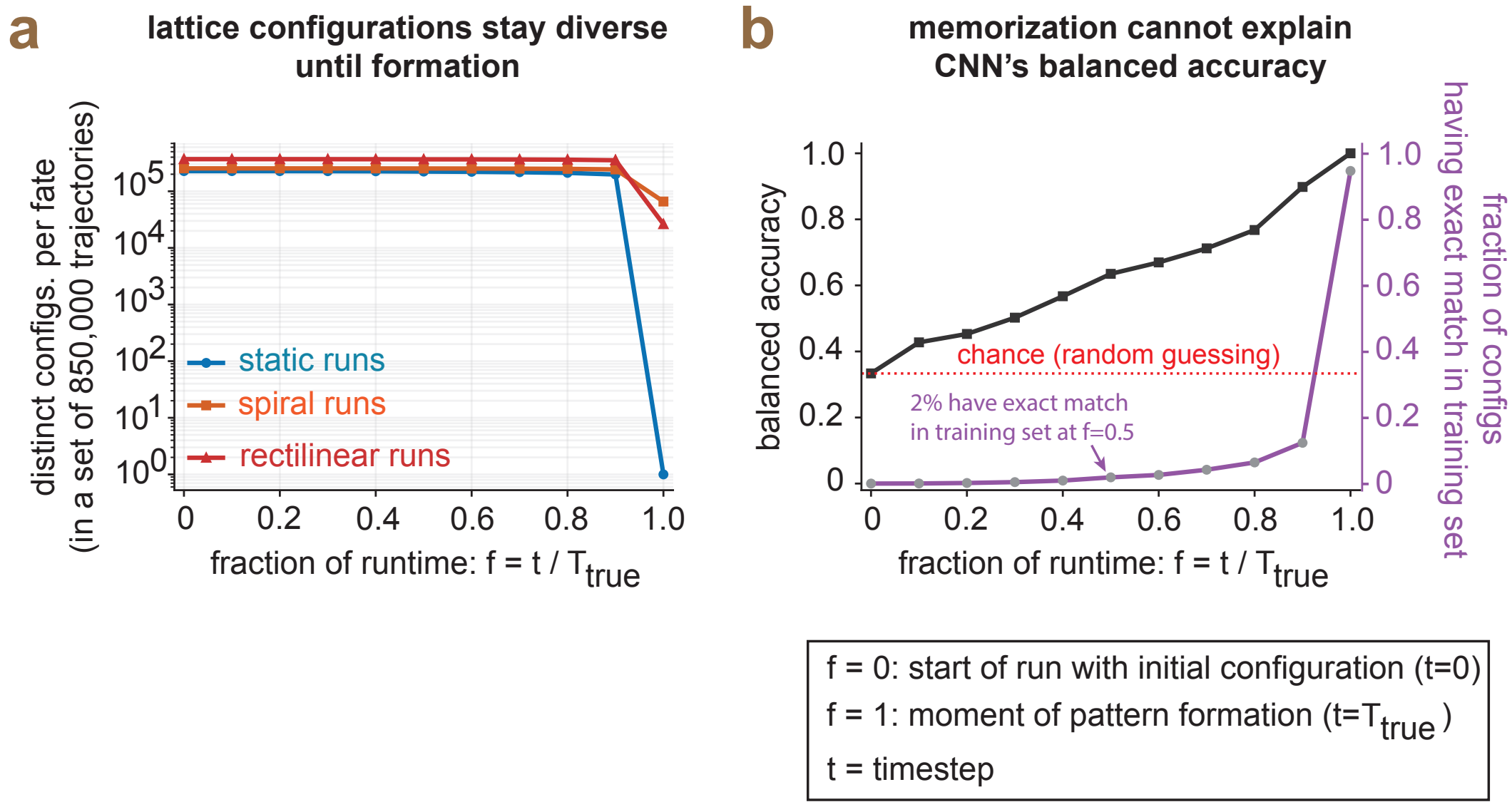

**Supplementary Figure 30: CNN generalizes – it does not memorize configurations seen in training set – to guess more accurately from later snapshots of lattice.**
In the reduced representation of vortex-core-profiles, trajectories of the same fate progressively merge. Therefore, one might worry that the convolutional neural network (CNN) succeeds on later (time-evolved) configurations only because it recognizes configurations it had already seen in training. The results shown here rule out this possibility.

**(a)** The number of distinct lattice configurations per fate, within the training pool of 850,000 trajectories, as a function of timestep $t = f \cdot T_{true}$, where $T_{true}$ is the timestep at which the final pattern forms (see Supplementary note 7); $f$ is the fraction, $t/T_{true}$, between 0 and 1. $t = 0$ is when the system begins with the maximally disordered initial configuration. Plotted as a function of $f$. Virtually throughout the entire trajectory, all three fates retain hundreds of thousands of distinct configurations. Only at the moment of pattern formation do they converge, with static collapsing to a single shared configuration – this is because there is only one static configuration (all cells becoming red) – while spiral-wave fate and, to a lesser degree, rectilinear-wave fate remain spread over tens of thousands of distinct configurations.

**(b)** The fraction of held-out lattice snapshots that have an exact cell-by-cell match anywhere in the training pool of trajectories (purple curve) stays near 2% through the first half of the trajectory (about 1.9% at $f = 0.5$) and rises steeply only near the moment of pattern formation. In contrast, the CNN's balanced accuracy on the full configuration (black curve) is far higher throughout (about 0.64 at $f = 0.5$, chance = 1/3). A pure memorization (lookup) strategy could label at most a few percent of mid-trajectory snapshots, so the CNN must be reading structure within configurations it has never seen.

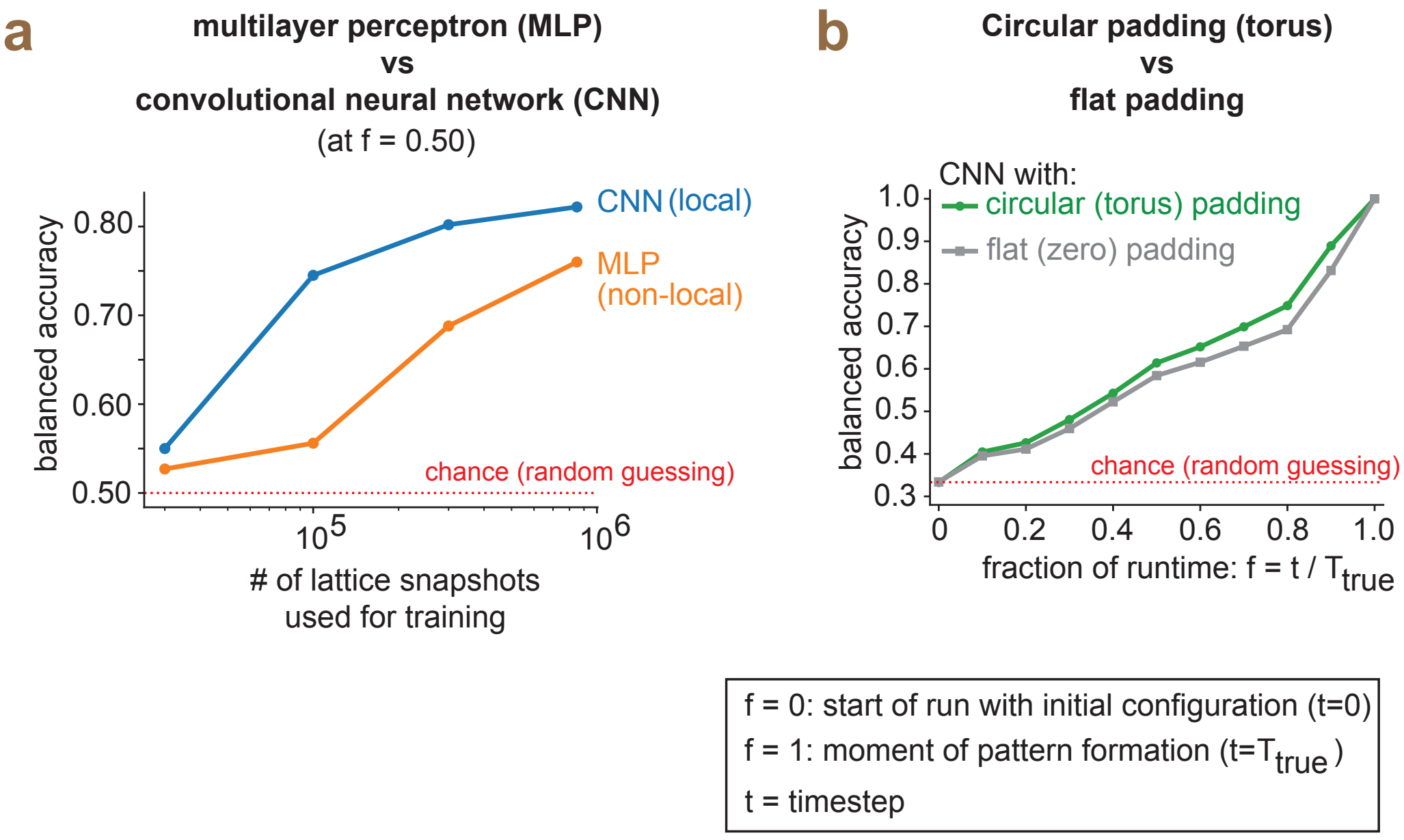

**Supplementary Figure 31: CNN locality improves sample efficiency, while explicit toroidal padding provides only a modest benefit.**

**(a) Locality.** We compared two learning models, convolutional neural network (CNN) and a fully connected multilayer perceptron (MLP). The two differ in their network architectures and capabilities: CNN encodes nearest-neighbor locality whereas MLP does not encode locality. At the midpoint of each trajectory ($f = 0.5$), we found that CNN consistently accomplished a higher balanced accuracy with much less training data (i.e., lattice snapshots at $f = 0.5$) than MLP did. For example, reaching a balanced accuracy of 0.75 required about $10^5$ training configurations whereas MLP required 8.5 x $10^5$ configurations (i.e., 8.5 times more training data). This comparison shows that a learning model that exploits spatial locality learns the predictive signal far more cheaply, which it can do only if the prediction-aiding signal lives in local spatial arrangements of cells. Here, both curves are for the binary static-versus-dynamic task, where the learning-curve comparison was run.

**(b) Torus vs flat boundary.** CNN trained and tested with two types of padding: circular padding (green curve) and flat (zero) padding (grey curve). Circular padding explicitly tells the CNN that the lattice forms a torus due to the periodic boundary condition. Flat (zero) padding does not tell CNN about the periodic boundary condition. Balanced accuracy plotted as a function of timestep $t = f \cdot T_{true}$, where $T_{true}$ is the timestep at which the final pattern forms (see Supplementary note 7); $f$ is the fraction, $t/T_{true}$, between 0 and 1. $t = 0$ is when the system begins with the maximally disordered initial configuration.

***(caption continues)***

**Supplementary Figure 31** ***(continued)*****:** Plotted as a function of $f$. The two curves nearly match in this plot, showing that flat padding already captures most of the prediction-enabling signal. Circular padding adds only a small, though consistent, improvement that grows toward formation (i.e., green curve is consistently above the grey curve across $f$). Knowing that the lattice is a torus is therefore helpful but not necessary to read the local structure.

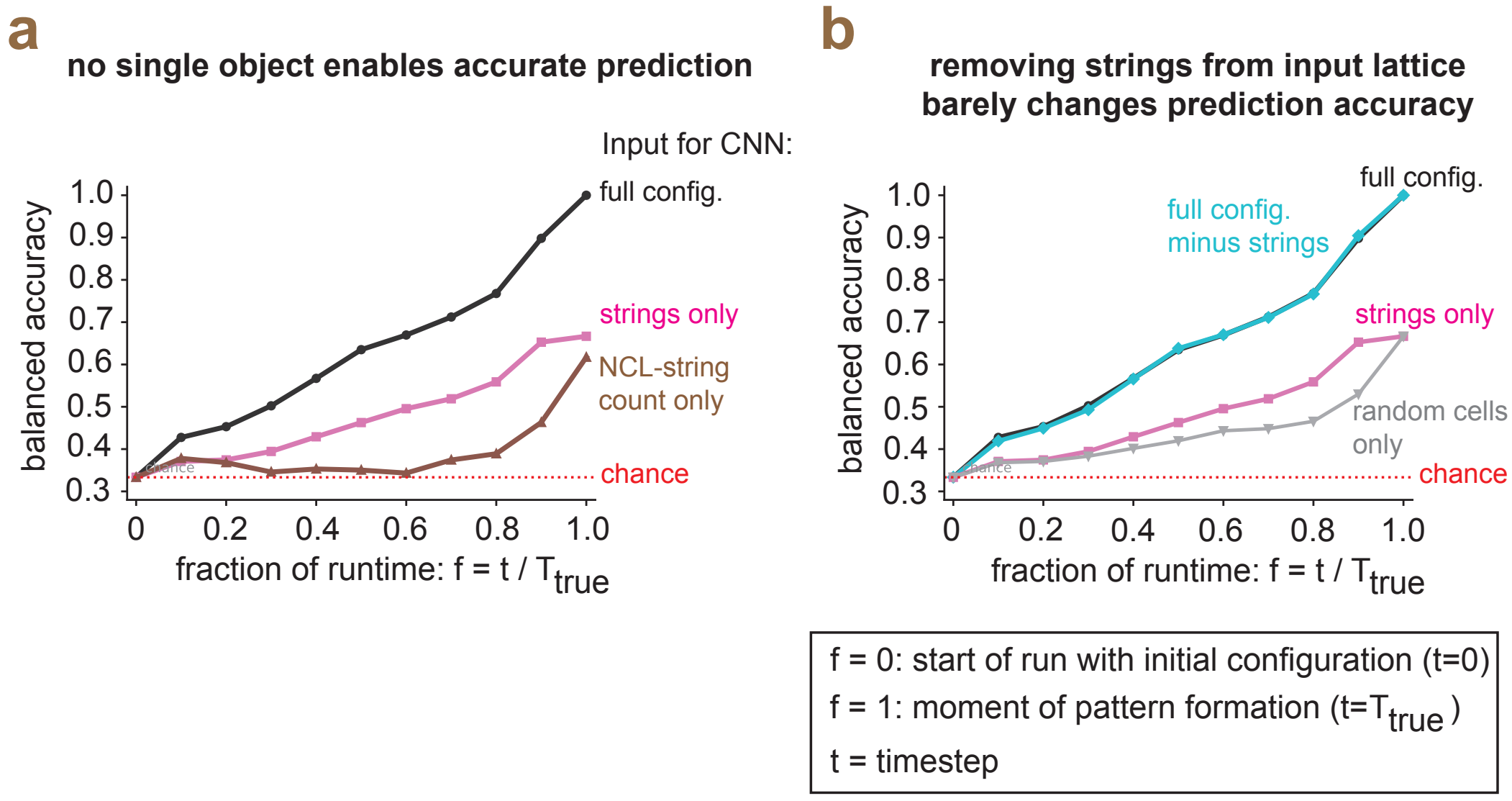

**Supplementary Figure 32: None of the objects identified earlier – strings or NCL-strings – accounts for CNN's gradually increasing prediction accuracy.**

**(a) Any single object, on its own, predicts poorly for most of the trajectory.** CNN's balanced accuracy versus $f$ for the full lattice configuration (i.e., snapshot of the lattice showing the phase (state) of every cell at a given time), for the mesh of strings (all same-state paths that connect vortices), and for the number of NCL strings alone (best of a logistic and a boosted-tree predictor). Regarding their balanced accuracy of fate prediction, both object-only predictors sit far below the full configuration (full lattice snapshot) for most of the trajectory. The NCL-string count becomes informative only close to formation, consistent with NCL strings being a late-time signature (Fig. 4f).

**(b) Removing the strings barely changes the accuracy.** The full lattice configuration, the configuration with the string mesh deleted (configuration minus all strings), the string mesh alone, and an equal number of random cells alone. Deleting the strings leaves accuracy essentially unchanged, while the strings alone are weak. These results establish that the prediction-enabling signal is distributed across the configuration rather than confined to the strings.

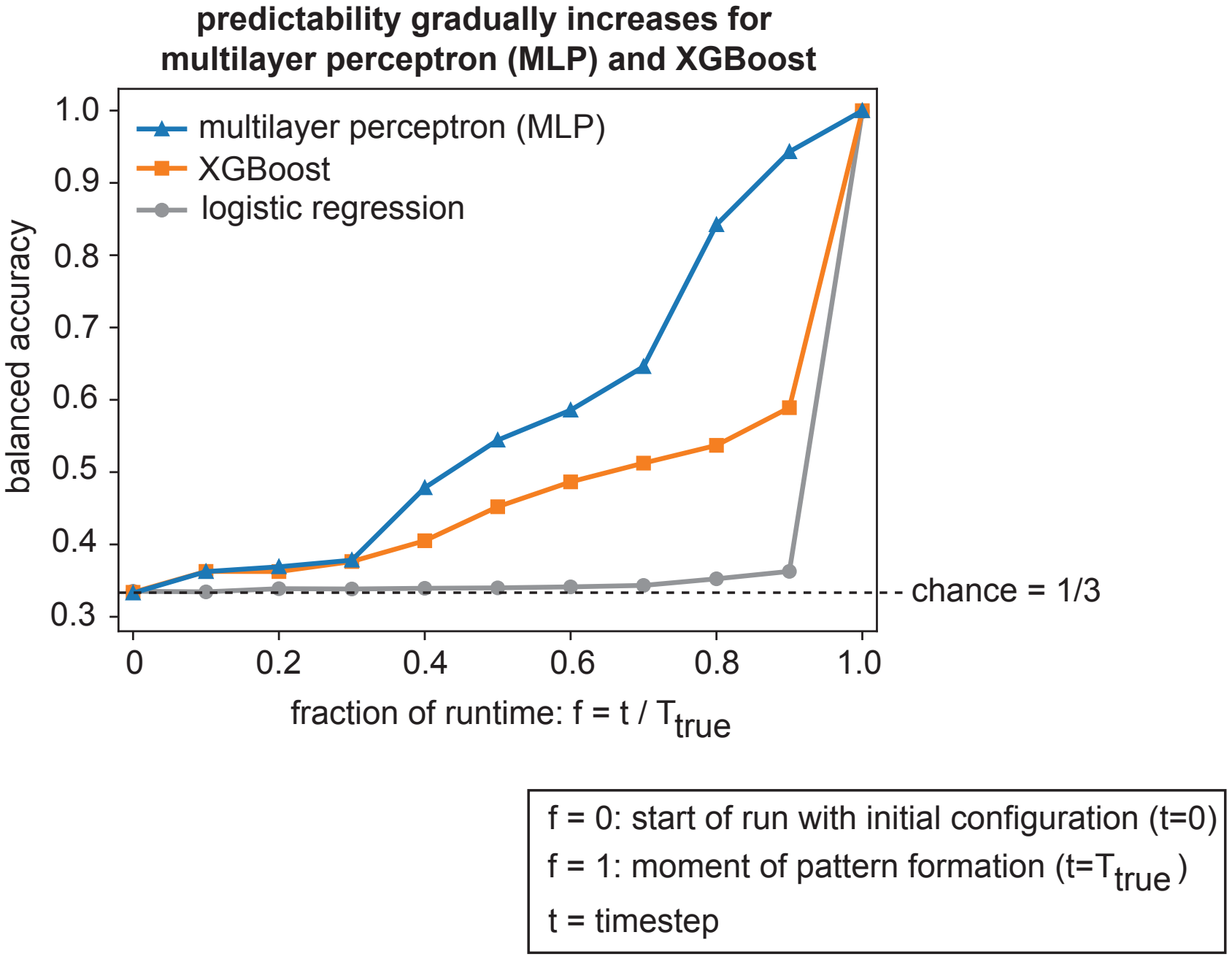

**Supplementary Figure 33: Fully-connected Multilayer perceptron (MLP) and gradient-boosted decision trees (XGBoost) also become more accurate along the trajectory.**

Balanced accuracy for predicting the three-way pattern fate — static, rectilinear wave, or spiral wave — from a single lattice configuration taken along the trajectory at timestep $t = f \cdot T_{true}$, where $T_{true}$ is the timestep at which the final pattern forms (see Supplementary note 7); $f$ is the fraction, $t/T_{true}$, between 0 and 1. $t = 0$ is when the system begins with the maximally disordered initial configuration. Plotted as a function of $f$ for three model classes: logistic regression, gradient-boosted decision trees (XGBoost), and a fully connected multilayer perceptron (MLP). Dotted line marks chance accuracy (1/3). The convolutional neural network trained on the same task is shown in Fig. 6b.

**Data and protocol.** We used the same $10^6$ simulations and the same fixed splits as elsewhere in this work: a training pool of 850,000 configurations, a validation set of 50,000, and a held-out test set of 100,000 trajectories. For each fraction $f$ we extracted, from every trajectory, the single configuration at timestep $t = \text{round}(f\, T_{\text{true}})$. Each cell state was one-hot encoded, giving 784 binary features per lattice. As a proof of principle, each model was trained once, on the full training pool, with a single random seed; error bars are therefore not available. Balanced accuracy is reported on the held-out test set.

***(caption continues)***

**Supplementary Figure 33 *(continued)*:**

**Hyperparameters implemented in Python.** Logistic regression: scikit-learn 1.9.0, lbfgs solver, maximum 2,000 iterations, balanced class weights. XGBoost 3.2.0: multi-class softmax objective, histogram tree method, learning rate 0.05, maximum depth 6, minimum child weight 1, subsample 0.8, column subsample 0.8, L2 regularization 1.0, balanced sample weights, up to 1,200 boosting rounds with early stopping after 60 rounds without improvement in validation multi-class log-loss. MLP: PyTorch 2.9.0, four hidden layers of 2,048, 2,048, 1,024 and 512 units, ReLU activations, dropout 0.25, AdamW (learning rate $10^{-3}$, weight decay $10^{-5}$), batch size 2,048, class-weighted cross-entropy, up to 50 epochs with early stopping after 6 epochs without improvement in validation balanced accuracy and best-epoch weights restored.

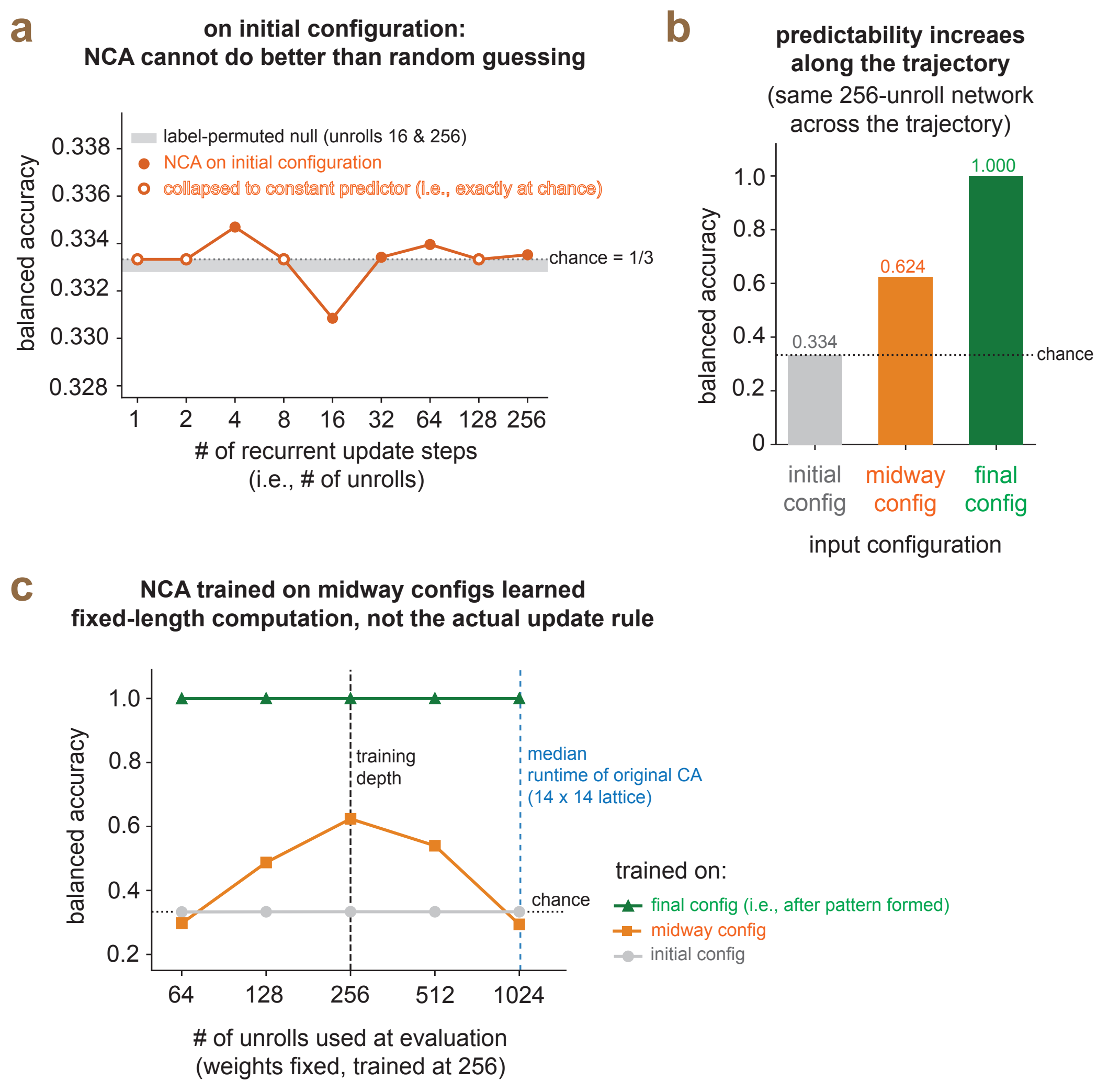

**Supplementary Figure 34: A recurrent neural cellular automaton (NCA) could not perform better than random guessing on initial configuration at any computational depth, and could not learn the actual update rules of the system (our generalized cellular automaton).**

**Main idea:** A neural cellular automaton (NCA) is a learnable cellular automaton. It has the same structure as the generalized cellular automaton (GCA) we study here, except that its update rule is a small neural network whose parameters are fitted to training data instead of being fixed in advance. We built our NCA as follows. We started from the 14 $\times$ 14 lattice and gave every cell a vector of 32 numbers to use as working space, in place of its single discrete state.

***(caption continues)***

**Supplementary Figure 34** ***(continued)*****:**
One update step in NCA then had every cell simultaneously read its own vector and those of its six nearest neighbours, and revise its own 32 numbers according to a learned rule. We used circular padding, so the network obeyed the periodic boundary conditions. We applied the same rule at every cell and at every update step, exactly as a cellular automaton does. After a chosen number of update steps, which we call unrolls, we averaged the vectors over all 196 cells and passed that average to a small classifier that reported one of the three final fates.

Two features of this construction matter for what follows. First, the update rule's weights were shared across every application, so a network that ran for 256 unrolls had exactly as many fitted parameters as one that ran for a single unroll: about 10,600 parameters in total. Unroll depth therefore measures how long the network computes rather than how large it is, and sweeping the depth separates computation from the model capacity. Second, the architecture underlying NCAs is the closest analogue in machine learning to our GCA, which is precisely why we chose to test it. If any learning model were structurally positioned to recover this system's behavior, we reasoned that it would be this one.

We chose the width of 32 numbers per cell. This number is not a property of the system. We picked 32 to give roughly eight times more working space than the automaton's four states while keeping the network deliberately small. We swept the unroll depth but not the width.

**How we trained it.** We trained a separate network from scratch for every unroll count. For a network of $U$ unrolls, we presented a batch of initial configurations, applied the update block $U$ times, pooled over the lattice, classified, and compared the output with the true fate. We then propagated the classification error backwards through all $U$ applications of the update block and summed the resulting gradients into the single shared weight tensor. We repeated this for up to twelve passes over 200,000 initial configurations, used a held-out validation set to choose the stopping epoch, and evaluated on the frozen 100,000-trajectory test set. Chance (random guessing) corresponds to a balanced accuracy of $1/3$. The nine networks in **(a)** are therefore nine independent trainings, not one network run for nine different unrolls.

**(a)** We measured balanced accuracy at $t = 0$ against the number of recurrent update steps (unrolls), from 1 to 256, and found chance throughout. Balanced accuracy spanned only 0.331 to 0.334. The grey band marks the null distribution, which we obtained by retraining the same networks on randomly permuted labels at 16 and at 256 unrolls — the balanced accuracy spans 0.3328 to 0.3333. Every real-label point fell inside or below this band. Open circles mark runs whose balanced accuracy equalled $1/3$ to within $10^{-5}$. Those runs terminated on a constant predictor: validation accuracy never improved at any epoch.

**Supplementary Figure 34 *(continued)*:**
**(b)** We applied the same 256-unroll network at three points along the trajectory. It reached 0.334 on initial configurations, 0.624 at the trajectory midpoint, and 1 on fully formed patterns. The architecture and training budget were therefore adequate. What changed across these three cases was only the configuration we handed the network.

**(c)** We then asked what the NCAs had actually learned. Because the update block shares the learned weights, we could take a trained NCA and re-run it for a different number of update steps without retraining it. Here was our reasoning: if a network had learned a genuine update rule, running it longer could not invalidate it, because a rule does not depend on how many times it is applied — the system would simply evolve further. If a network had instead fitted a computation of fixed length, like an assembly line whose product emerges at a particular station, then stopping early would yield an unfinished result and running past the end would ”spoil” it. All three networks shown were trained at 256 unrolls and re-evaluated at 64, 128, 256, 512 and 1,024 unrolls using the same weights. We found that the network trained on fully formed patterns was entirely insensitive to evaluation depth: it achieved a balanced accuracy of 1 at 128, 256, 512 and 1,024 unrolls, and a balanced accuracy of 0.9998 at 64. Depth-robustness was therefore attainable by NCA, and this curve is the control for the diagnostic itself, because it shows that degradation away from the training depth is not an unavoidable consequence of changing the unroll count. We note that this curve does not by itself establish that the fully-formed-pattern network learned an update rule. Insensitivity to depth follows necessarily from having learned a rule, but it does not imply one: a network that has learned merely to preserve a configuration and read it out is equally insensitive to depth, and since a fully formed pattern already sits in its attractor, that is the simpler account. We therefore draw our conclusion about rule-learning from the midpoint network, where a collapse away from the training depth excludes any rule that could be iterated. That network behaved in the opposite way. It peaked sharply at its training depth, at 0.624, and fell away in both directions, to 0.297 at 64 unrolls and to 0.294 at 1,024 unrolls, which is below chance. We conclude that it had fitted a computation of fixed length rather than learning the automaton’s update rule. The network trained at $t = 0$ remained at chance at every evaluation depth, including 1,024 unrolls, which was comparable to the median runtime of the original CA on a $14 \times 14$ lattice.

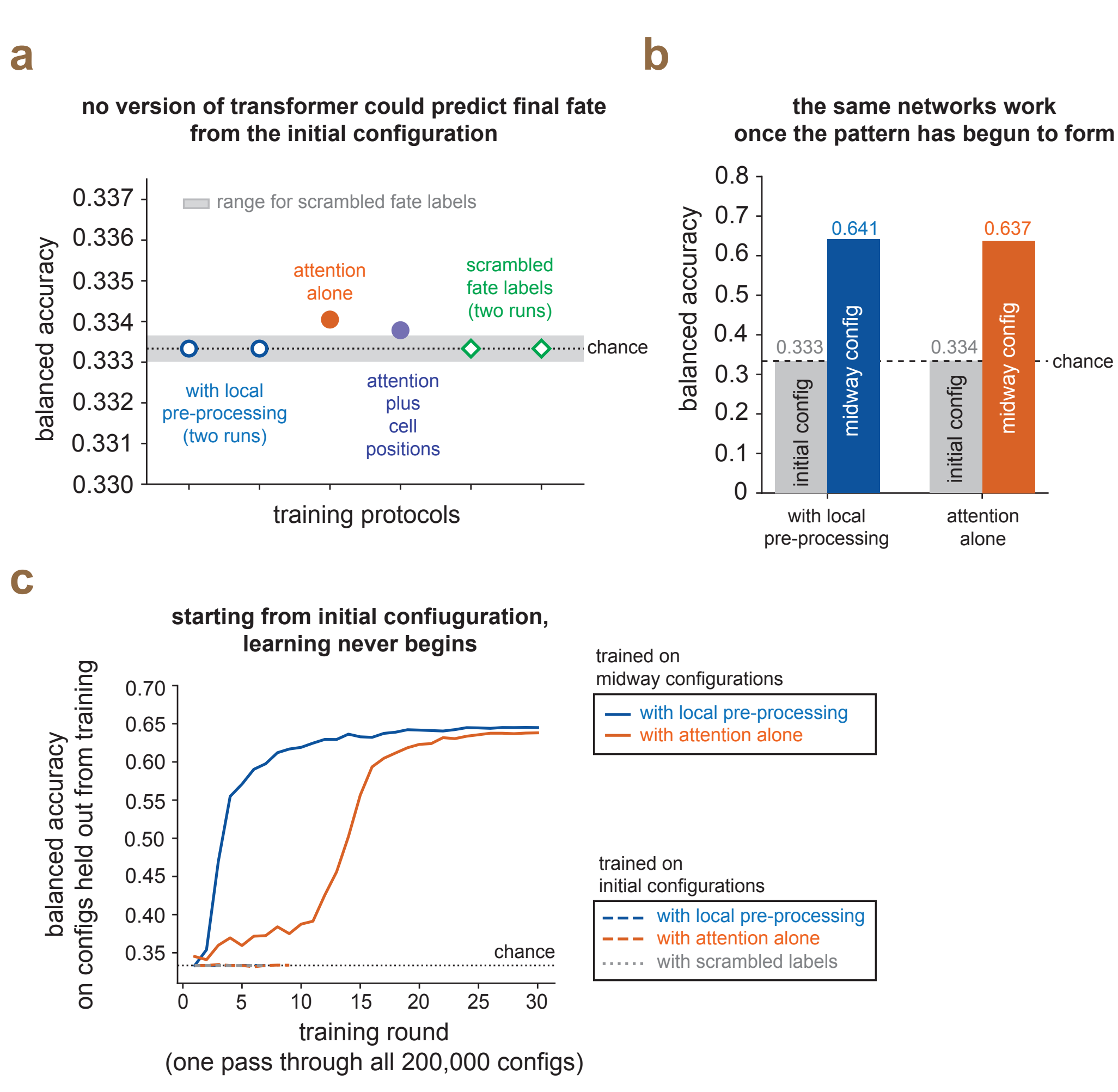

**Supplementary Figure 35: A transformer extracts nothing from the initial configuration.**

**Main idea:** A transformer is a neural network that processes a collection of items by letting every item consult every other item. Its central mechanism is called attention. For each item, the network forms a weighted combination of information drawn from all the other items, and it computes those weights from the items' own contents. The network therefore learns what to look at, instead of having that decided in advance by whoever built it. Attention allows the model (transformer) to relate cells that are arbitrarily far apart on a lattice. In this sense, transformers are distinct in their architecture from the convolutional networks (CNNs) we used. A convolutional layer lets each cell see only its immediate neighbors, so information from a distant cell can reach it only after many layers have passed that information along.
***(caption continues)***

**Supplementary Figure 35** ***(continued)***:
In a transformer, every cell can consult every other cell within a single layer, and which cells it consults is learned rather than prescribed.

**Rationale for testing transformers:** Every model we had previously applied to the initial configuration carried a built-in restriction on what it could look at. The tabular models, like log regression and XG-Boost, treat each cell separately. The convolutional networks and the neural cellular automaton read local neighborhoods and can reach more distant neighbors only by increasing the depth. A transformer removes those restrictions. We reasoned that if the final fate were encoded in some relationship among cells far apart on the lattice, rather than in local structure, then this architecture is better positioned than all the others we examined to find it. A transformer also imposes fewer assumptions than any other model we tested, so a failure here would rule out a correspondingly broad range of possibilities.

**How we adapted the transformer to our lattice:** We treated each of the 196 cells ($14 \times 14$ lattice) as one item. We then made one change specific to our system. In a standard transformer, the network must discover from data how its items are arranged relative to one another. Here the arrangement is already known: the cells sit on a triangular lattice with periodic boundaries. We therefore made the attention between any two cells depend only on their separation across the torus. This has a consequence we wanted. The network's answer does not change if the entire configuration is shifted wholesale across the lattice, which matches the fact that the fate itself would not change under such a shift. We built three versions, differing in how much spatial information we handed the network outright: (1) attention alone; (2) attention with each cell's absolute coordinates supplied as well; and (3) attention preceded by two convolutional layers, which would give the network ready-made local neighborhood features before any attention occurs. We trained the third version twice from different random starting points, to confirm that the outcome did not depend on where training began.

**How we trained the transformer:** We showed the network an initial configuration and asked it to name the fate. We compared its answer with the truth and adjusted its roughly 1.05 million parameters to reduce the error. We repeated this for up to thirty passes over 200,000 initial configurations, used a held-out set of configurations to decide when to stop, and then evaluated the network once on the frozen 100,000-trajectory test set. Balanced accuracy weights the three fates equally, so random guessing scores $1/3$. As a reference, we also repeated the entire procedure after randomly scrambling the fate labels, which destroyed any relationship between configuration and fate by construction. Those runs told us what our training procedure would produce when there is certainly nothing to find.

**Supplementary Figure 35 *(continued)*:**
**(a)** We found that no version of the network read any fate-predictive structures from the initial configuration. All four versions achieved a balanced accuracy between 0.3333 and 0.3340, and the two runs on scrambled labels both had a balanced accuracy of 0.3333 — exactly chance. Real and scrambled runs were therefore indistinguishable. Open symbols mark networks that ended up giving the same answer for every configuration they were shown, which is what a network does when it has found no usable signal at all. Two of the four real-label runs did this, as did both scrambled-label runs. We obtained the same outcome when we varied the number of attention layers: two, four and eight layers all gave exactly 0.3333 (chance exactly) as the balanced accuracy.

**(b)** We then gave the same networks, with the same training budget, lattice configurations taken from midway along each trajectory instead of from its start. Both versions succeeded — their balanced accuracies reached 0.641 and 0.637. For comparison, a convolutional network had a balanced accuracy of 0.635 on the same data. The networks and the training procedure were therefore adequate. What differs between the two bars in each pair was only which configuration we fed the network.

**(c)** We recorded the balanced accuracy on held-out configurations after every pass through the training data. Starting from configurations taken midway along the trajectory, accuracy rose steadily over tens of passes. Starting from initial configurations, the balanced accuracy never left $1/3$ at any point in training, and it was indistinguishable from the run on scrambled labels. Learning did not proceed slowly or partially. It never began.

**Interpretation.** We emphasise that the transformer did not fail as a model. It matched or exceeded every other architecture we tested at every later point on the trajectory — its balanced accuracy reached 0.641 midway and 0.9998 on fully formed patterns. It failed only in the specific sense that it extracted nothing from the initial configuration, and panel **(c)** shows that the failure was total rather than partial. Because the same network, with the same parameters and the same training, succeeded later in the same trajectory, the difference lies in the configuration presented to it rather than in the network. And because a transformer places the fewest restrictions of any model we tested on which relationships among cells it may use, its failure argues that the fate is not practically extractable from the initial configurations in any relationship, local or long-range, that a learned attention pattern could recover.

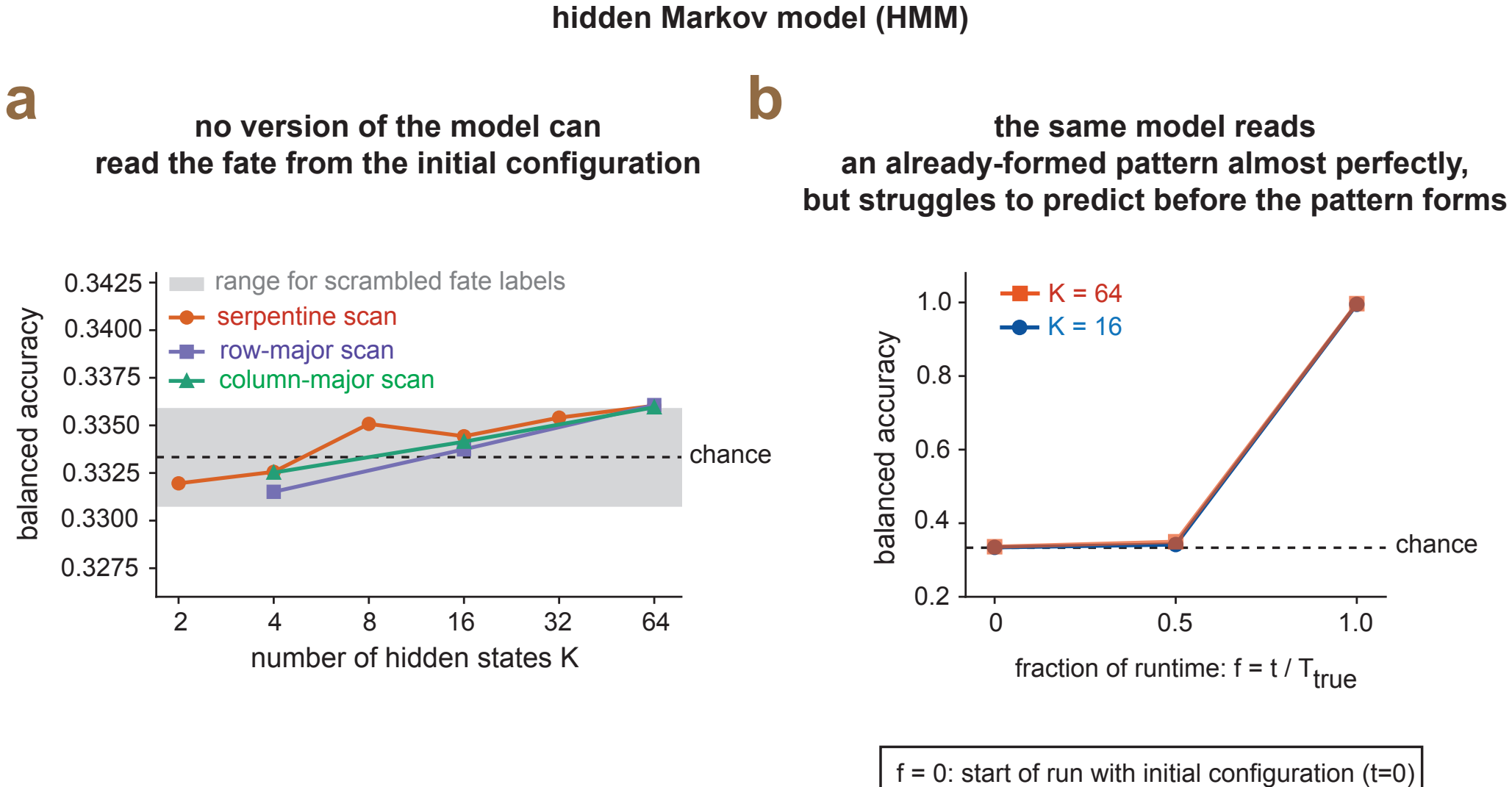

**Supplementary Figure 36: A hidden Markov model could not predict the final fate from the initial configurations, at any model size and in any reading direction.**

**Main idea:** A hidden Markov model (HMM) describes a sequence of observations by supposing that the system producing them passes through a chain of internal states that we cannot see directly. At each position in this sequence, the system occupies one of a fixed number of the hidden states, then emits an observation drawn from a probability distribution attached to that state, and then it moves to the next state according to fixed transition probabilities. Fitting the model to data means estimating three things at once: the transition probabilities between hidden states, the emission probabilities from each state, and the distribution of the starting state. We did this using the Baum–Welch algorithm, a standard iterative procedure that repeatedly infers which hidden states most likely produced the observed sequences and then updates the probabilities accordingly. We chose the number of hidden states, which we call $K$, rather than have it be learned. The number of hidden states sets how many distinct internal situations the model is permitted to distinguish, and it is therefore the natural measure of how elaborate the model is.

**Rationale:** A HMM fills a gap in our sweep of machine-learning models thus far. Namely, every other model we applied to the initial configuration has so far been discriminative: they attempted to learn a direct mapping from a lattice configuration to a fate label. In contrast, a HMM works the other way round: it is generative in that if we fit one model per fate, so that each learns to describe what configurations belonging to that fate tend to look like, and we then classify a new configuration by asking which of the three fate models makes that configuration most probable.

***(caption continues)***

**Supplementary Figure 36** ***(continued)*****:**
This is a structurally different route to the same question, and the HMM is the only generative model in our study.

**Implementation details:** HMM describes sequences, whereas our configurations are two-dimensional. We therefore had to read each $14 \times 14$ lattice out as a sequence of 196 symbols, and the order in which we read the cells was a choice. We used three reading orders: (1) a row-major raster scan, reading each row left to right in turn; (2) a serpentine scan, which reversed every second row so that consecutive symbols in the sequence were always neighbors on the lattice; and (3) a column-major scan, reading down columns instead of across rows. We fitted one model per fate on 15,000 training sequences of that fate, classified each test configuration by whichever fate model assigned it the higher log-likelihood, and used equal prior probabilities for the three fates so that the decision rule matched balanced accuracy as our reported metric. We evaluated every model on the frozen 100,000-trajectory test set. Balanced accuracy for random guessing (i.e., chance) was $1/3$.

**(a)** We swept both free choices: the number of hidden states from 2 to 64, and all three reading orders. We found that neither helped. Across every combination, balanced accuracy at the initial configuration ranged only from 0.332 to 0.336. As a reference, we repeated the entire fitting procedure after randomly scrambling the fate labels, which destroyed any relationship between a configuration and its fate by construction. For those runs, shown as the grey band, balanced accuracy spanned 0.331 to 0.336. Every real-label result therefore fell inside the range produced when there was certainly nothing to find. Making the model 32 times more elaborate changed nothing, and neither did changing which direction we read the lattice in.

**(b)** We then applied the same models to configurations taken from later in the trajectory. On fully formed patterns, they succeeded almost completely: the balanced accuracy reached 0.995 with 16 hidden states and 0.997 with 64. The approach was therefore capable of reading a fate off a lattice configuration once the pattern had already formed, and the failure in panel **(a)** does not reflect a broken procedure. Midway to pattern formation, however, the same models had the balanced accuracy reach only 0.342 and 0.350.

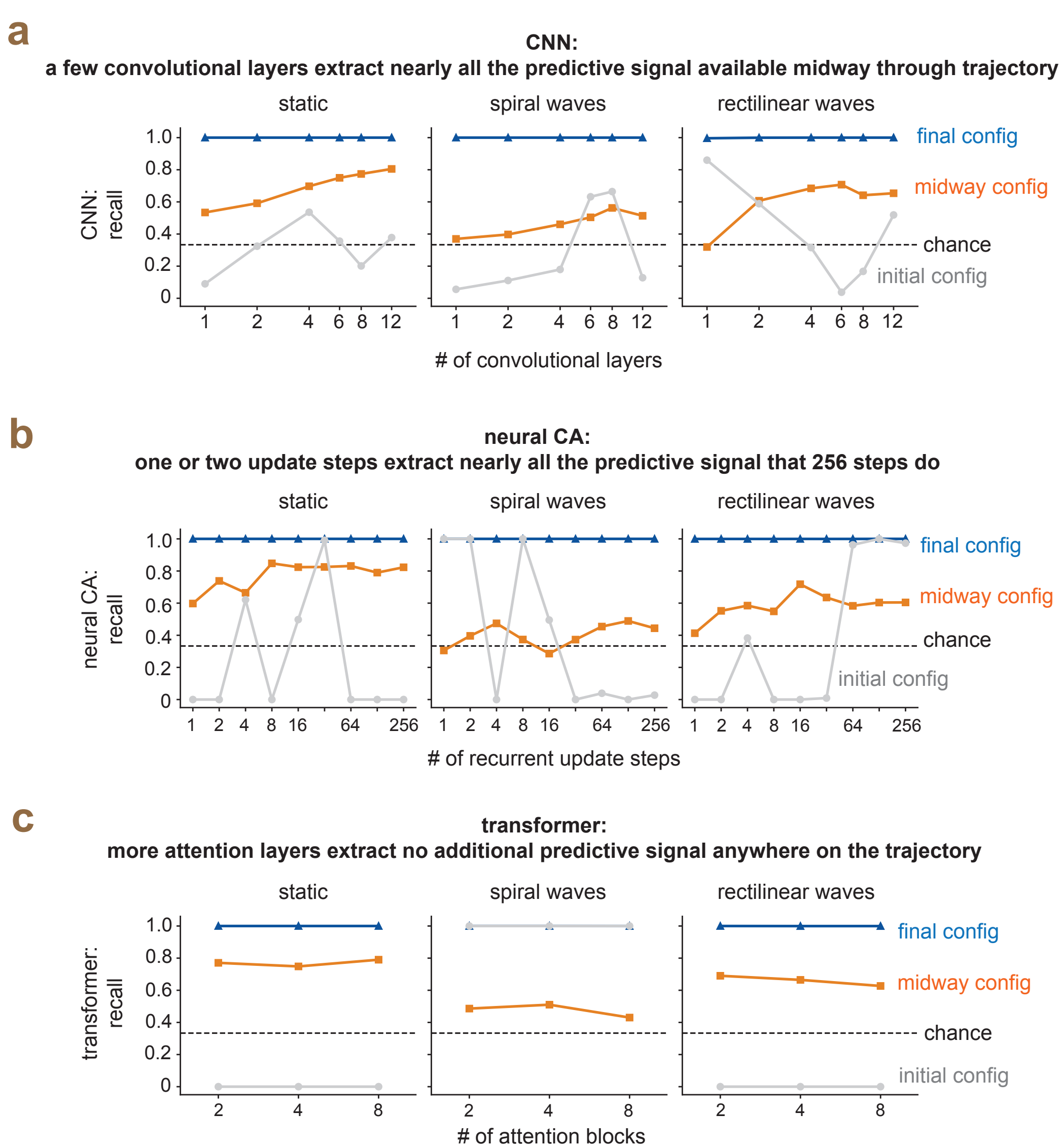

**Supplementary Figure 37: No fate is practically predictable from the initial configuration, and the prediction accuracy attained later in the trajectory was obtained with very little computation.**

**Rationale:** Two questions motivated the analysis shown here. First, a balanced accuracy is an average over the three fates, weighing the incidence of each fate equally — this accounts for the fact that the three fates do not arise with equal probability if we start the system with a maximally random configuration. So a machine-learning model that had learned one fate well and the other two not at all could still report a modest overall score as the balanced accuracy.

***(caption continues)***

**Supplementary Figure 37 *(continued)*:**
We therefore decomposed every model into its per-fate recall (i.e., accuracy on each fate: given $N$ configurations that all have the same fate, what percentage of those does the ML model correctly identify?). Second, and more importantly, we asked whether giving a machine-learning model more computation changed what it could extract. Each of the three architectures we tested here had a natural way to be made to compute for longer, and in each case we swept it. For the convolutional network (CNN), we varied the number of convolutional layers, each of which lets information travel one further step across the lattice. For the recurrent neural cellular automaton (NCA), we varied the number of times its single learned update rule was applied before the answer was read out (see Supplementary fig. 34). For the transformer, we varied the number of attention layers, each of which lets every cell consult every other cell once more (see Supplementary fig. 35). Every model was trained on 200,000 lattice configurations and evaluated on the frozen 100,000-trajectory test set. We repeated the whole sweep at three points along the trajectory: at the initial configuration, midway to pattern formation, and at the fully formed pattern. Chance recall was $1/3$.

**(a)** We found that a CNN extracted nearly all of the available midway accuracy within a few convolutional layers, and that further depth added little. Recall for rectilinear-wave fate almost doubled between one layer and two, from 0.319 to 0.607, and then stayed between 0.64 and 0.71 out to twelve layers. Recall for spiral-wave fate rose gradually to 0.562 at eight layers and did not improve beyond that. Recall for the static fate climbed most steadily, from 0.534 to 0.805, but its rate of improvement fell steadily with depth. Depth therefore saturated, and it did so early. When given fully formed patterns, a single convolutional layer already recovered every fate almost perfectly (1.0, 1.0, and 0.996), and additional layers added nothing. When given the initial configurations, the recalls fluctuated without trend and never rose above chance in a way that survived a change of depth.

**(b)** The NCA saturated even faster. Its balanced accuracy midway rose from 0.439 at a single update step to 0.562 at two steps, and then gained only a further 0.06 over the remaining 254 steps, ending at 0.624 after 256. Recall for static configurations jumped from 0.598 to 0.738 between one step and two and then varied around 0.82 with no further trend. Recall for spiral-wave fate never improved at all across the entire range, wandering between 0.29 and 0.49. When given the fully formed pattern as an input, a single update step already recovered all three fates perfectly, and 256 steps did no better. A 256-fold increase in computation therefore bought almost nothing midway and nothing whatever at either end of the trajectory.

**Supplementary Figure 37 *(continued)*:**

**(c)** For the transformer, additional attention layers changed nothing anywhere. Midway, balanced accuracy was 0.649 with two layers, 0.641 with four and 0.616 with eight, so accuracy fell slightly as the network was made deeper. When given the fully formed pattern as the input, all three fates were recovered perfectly at every depth. At the initial configuration, every depth returned a recall of exactly 1.000 for spirals and exactly 0.000 for the other two fates.

**Why the initial-configuration curves behave as they do.** These curves require explanation, because at first sight a recall of 1.000 looks like a success. But it is the opposite. When a model finds no usable signal, the lowest-cost solution available to it is to ignore its input entirely and name the same fate every time. A model that always answers "spiral" achieves a recall of 1 for spirals, because every genuine spiral is indeed called a spiral, and a recall of 0 for the other two fates. Its balanced accuracy is then $(1 + 0 + 0)/3 = 1/3$ exactly, which is why several of these runs reported 0.333333 with no rounding. The transformer did this at every depth. The NCA did it too, but it is the clearest demonstration that the choice of fate is arbitrary: it named spirals at one, two and eight update steps, static configurations at 32 steps, and rectilinear patterns at 64, 128 and 256 steps. The CNN did not collapse so completely, and instead spread its predictions unevenly across the three fates in a pattern that changed with depth and with random seed. Both behaviors are the same phenomenon. A useful way to see this is that whenever a model's output carries no dependence on its input, its three recalls must sum to one, because the model is simply distributing a fixed set of guesses among the three fates. A perfect model instead has all three recalls equal to one, summing to three. Averaged over depths, the three recalls summed to 1.02 for the convolutional network, 1.00 for the neural cellular automaton and 1.00 for the transformer at the initial configuration; to roughly 1.7 to 1.9 midway; and to 3.00 for all three architectures at the fully formed pattern.

**Conclusion:** These results establish that learning depth was never the limiting factor at any point on the trajectory. At the fully formed pattern, one convolutional layer or one update step already sufficed. Midway of the trajecotry, essentially all of the attainable accuracy arrived within the first one or two operations, and depth beyond that yielded at most marginal gains and sometimes small losses. At the initial configuration, no amount of computation helped: not twelve convolutional layers, not 256 recurrent update steps, not eight attention layers. The ceiling observed midway was therefore not a limit on the computation available to these models but a limit on what their input makes available to them. Spiral-wave fate was consistently the hardest fate, reaching only 0.562 for the convolutional network and 0.489 for the neural cellular automaton midway, while static configurations reached 0.805 and 0.848 respectively.

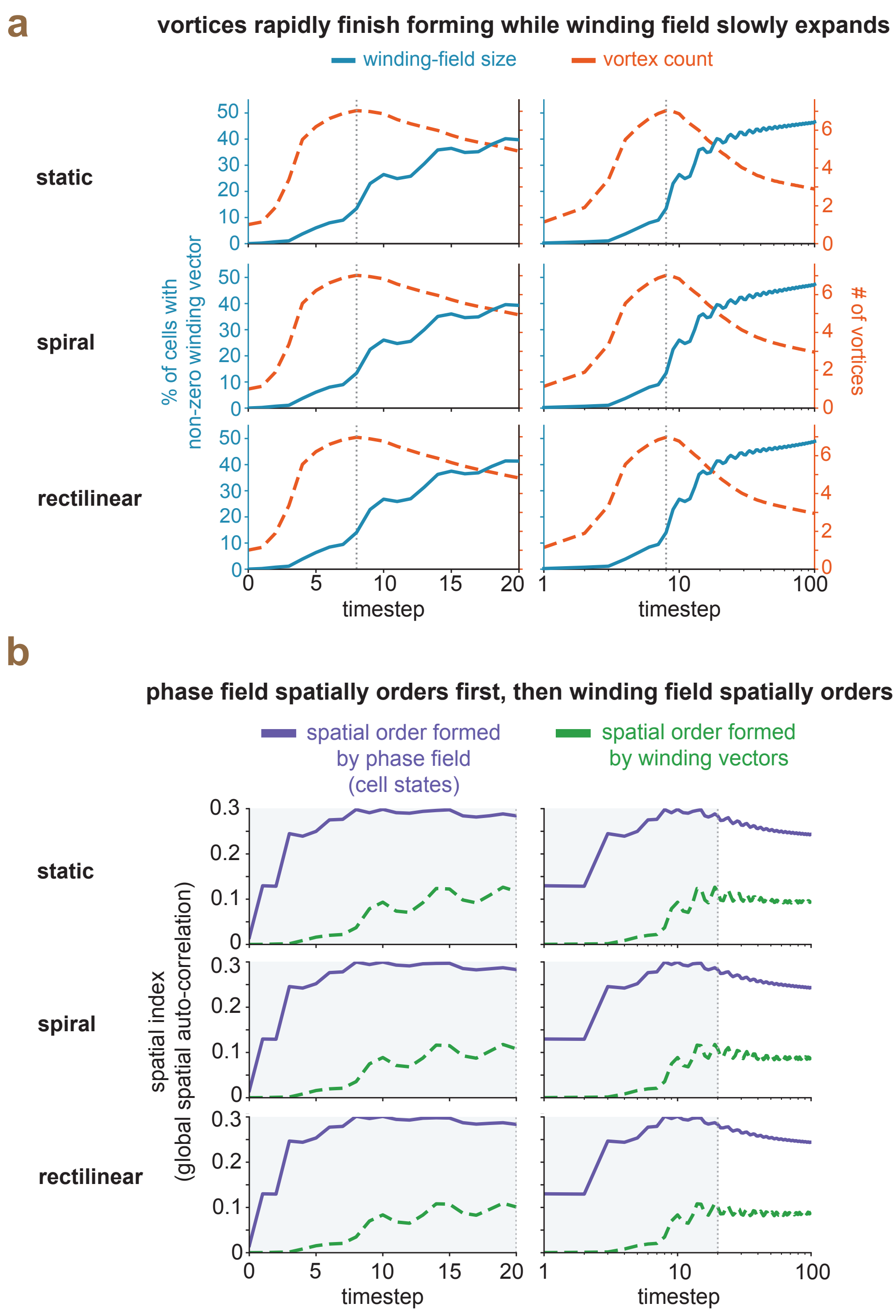

**Supplementary Figure 38: Vortices finish forming while winding field is switching on, and phase field spatially orders first before winding field begins to spatially order.**

Early-time dynamics of three quantities, averaged over $\sim 10^6$ trajectories on a $14 \times 14$ lattice, and separated by eventual fate (static, spiral, rectilinear).

***(caption continues)***

**Supplementary Figure 38 *(continued)*: (a)** Percentage of cells with a non-zero winding vector (teal, left axis) and the mean number of vortices (orange, right axis) versus timestep $t$, shown on a linear axis over $t$ = 0–20 (left) and a logarithmic axis over $t$ = 1–100 (right); the dotted line marks the vortex-count peak near $t \approx 8$. The vortices finishing forming first: their abundance rapidly rises to a peak of $\sim$7 by $t \approx 8$ and then declines by pairwise annihilation, while the winding field switches on slowly during and after the vortices form, its cell fraction still $\sim$0.13 at the vortex peak and climbing toward an early plateau ($\sim$0.4) over $t \approx$ 15–20.

**(b)** Spatial order of the phase field (cell state) and of the winding field over the same time-window, quantified by the spatial index (i.e., weighted Moran's $I$ whose weights are the pairwise cell–cell communication strengths, so that more distant cell pairs contribute less (Supplementary note 1). A representative field component is plotted for each ('gene 1' for the cell-state field and $w_x$ for the winding field. These plots show that the phsae field orders first—its spatial index $I$ rises steeply and nearly saturates within the first $\sim$10 timesteps—before the winding field's spatial index departs from zero.

Together, (a) and (b) place a temporal ordering on the earliest dynamics: the vortices finish forming while the winding field starts and then continues to slowly expand. Phase-field ordering, which yields the vortices, is almost finished when the winding field starts to form. The three fates are nearly indistinguishable in this early time-window.

**Averaged over million trajectories, winding-field dynamics segregate in fate-dependent manner, especially as trajectories approach pattern formation**

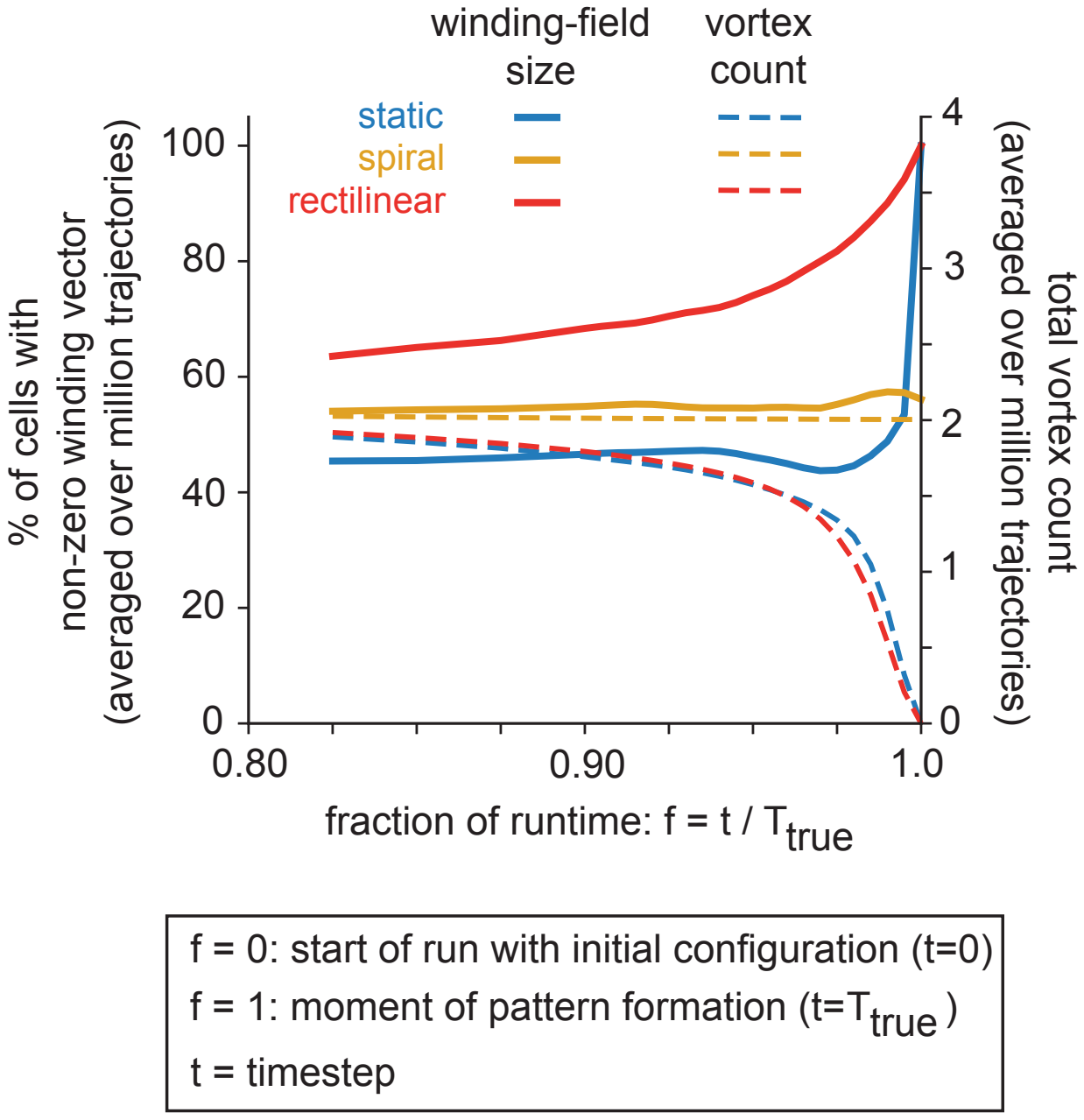

**Supplementary Figure 39: On average, winding-field size and its expansion rate are fate-dependent and separate markedly as the trajectories approach the moment of pattern formation.**

Averaged over $\sim 10^6$ trajectories on a $14 \times 14$ lattice, and separated by eventual fate (static, spiral, rectilinear). Percentage of cells with a non-zero winding vector (solid, left axis) and the mean number of vortices (dashed, right axis) versus scaled time $f = t/T_{\text{true}}$, on a linear axis over the final part of the trajectory ($f$ = 0.8–1.0). For static and rectilinear-wave fates, the winding field expands sharply to $\sim$100% of the lattice just as the last vortices annihilate (vortex count $\rightarrow$ 0). For spiral waves, it instead levels off near 56% and one $+1/-1$ vortex pair survives until the end.

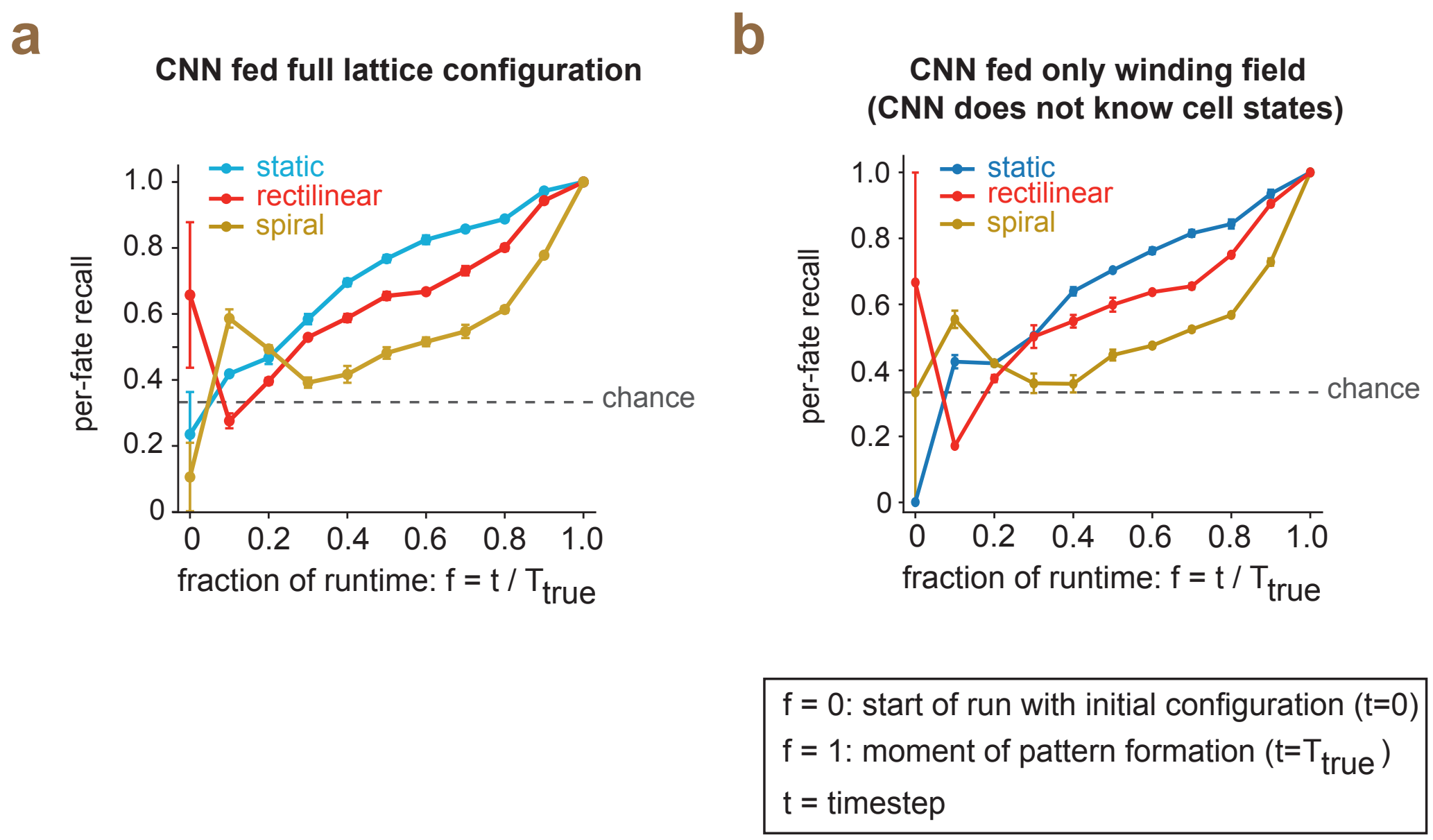

**Supplementary Figure 40: Winding field as an input enables performance that nearly matches the performance on full lattice snapshot (all cell states) as input for CNN. Spiral-wave fate is the hardest to predict and the last to resolve.**

**(a) Per-fate recall versus *f* for CNN that was fed the full lattice snapshot – state of every cell at a particular timestep – as an input.** The pattern type of static- and rectilinear-wave-bound trajectories become legible relatively early. Spiral-wave-bound trajectories remain difficult to predict for much of the trajectory – the recall for these trajectories remain close to chance – and rise steeply only near the pattern-formation moment.

Each point is the mean of three independent training runs (different random seeds), each trained on 850,000 configurations and evaluated on a held-out test set of 100,000 trajectories (26,784 static; 29,812 spiral; 43,404 rectilinear). Error bars are the standard error of the mean over the three runs. Dashed line marks chance accuracy (1/3).

The large error bars at 0% of runtime reflect that, when no predictive structure is yet accessible, individual training runs collapse onto different single fates: across the three runs the static recall ranged from 0.00 to 0.44 while the rectilinear recall ranged from 0.25 to 1.00. The balanced accuracy at this timepoint is nonetheless pinned at chance (Fig. 6b), so the apparent differences among fates at 0% of runtime carry no predictive information.

*(caption continues)*

**Supplementary Figure 40 *(continued)*: (b) Per-fate recall versus $f$ for CNN that was fed the winding field alone, without any cell states.** Graph here look virtually identical to (a) – this and Fig. 6f establish that the winding field nearly matches the performance on snapshots of the full lattice (all cell states) for CNN. The same ordering of curves as in (a) appears here, so the prediction asymmetry is a property of the topological structure and not of the raw lattice configuration. Consistent with this asymmetry, at formation, spiral-wave outcomes remain spread over far more distinct configurations than static or rectilinear outcomes (Supplementary fig. 30). This per-fate asymmetry matches the late-time clustering (Fig. 5) and the NCL-string signature (Fig. 4): vortex-annihilation-bound fates (static and rectilinear-wave fates) are legible (predictable) earlier; spiral-wave fate is only legible at the end of the trajectory, just before the spiral wave forms.

Points are means of three independent training runs (different random seeds), each trained on 850,000 configurations and evaluated on a held-out test set of 100,000 trajectories; error bars are ± s.e.m. Dashed line marks chance (1/3). At f = 0 the winding field is essentially zero everywhere, and individual runs collapse onto a single fate (two onto rectilinear, one onto spiral, none onto static), which is why the per-fate recalls there are widely scattered and carry no predictive information.

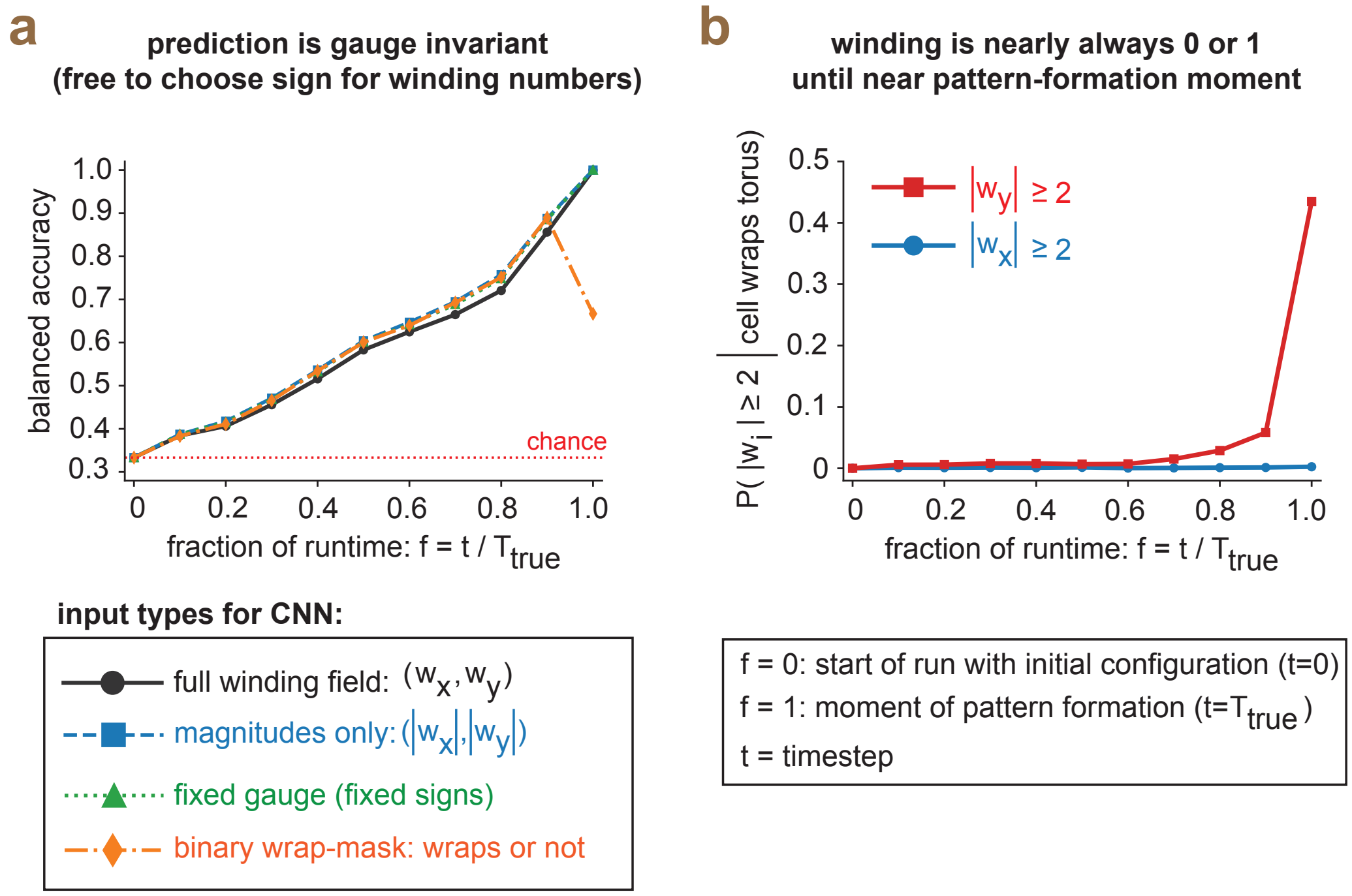

**Supplementary Figure 41: Winding-field-based prediction is gauge invariant, and the wrapping mask – whether a connected region of alike-state cells wraps the torus or not – carries essentially all of the winding field's predictive content (data shown for 14 $\times$ 14 lattice).**

**(a) Balanced accuracy versus time for four encodings of the winding field.** At each timestep $t = f \cdot T_{true}$, we computed the winding field from the lattice configuration at that timestep. A convolutional neural network (CNN) then predicted the three-way fate (static, rectilinear wave, or spiral wave) from that winding field. Here $T_{true}$ is the timestep at which the final pattern forms (see Supplementary note 7), and $f = t/T_{true}$ is the fraction of the runtime, between 0 and 1. At $f = 0$ the system is in its maximally disordered initial configuration. At $f = 1$ the pattern has formed.

We fed the CNN the winding field in four encodings: (1) the full winding field $(w_x, w_y)$; (2) magnitudes only $(|w_x|, |w_y|)$; (3) a gauge-fixed form with a fixed sign convention (for example, left and up positive, right and down negative); and (4) a single binary channel marking only whether each cell's region wraps the torus at all (the wrap mask).

Two results followed. First, the full field, the magnitudes, and the gauge-fixed form yielded essentially the same balanced accuracy at every timestep, including at the moment of pattern formation. Hence, the sign of the winding is only a convention, set by the direction in which one traverses a loop on the torus, so it does not matter for prediction. This is the gauge invariance.

***(caption continues)***

**Supplementary Figure 41 *(continued)*:** Second, the binary wrap mask matched the other three encodings throughout the transient, up to the moment just before the pattern forms. Therefore, throughout the predictive time-window, the winding field can be replaced by one bit per cell: whether the cell belongs to a region that wraps the torus, yes or no. The wrap mask falls below the other encodings only at the final step, $f = 1$. That is where winding magnitudes of two or more become common (panel b), and the one-bit mask cannot represent them. At $f = 1$ the pattern has already formed, so this gap lies outside the time-window in which prediction is the goal.

**(b) Why one bit is enough during the transient.** Among cells whose region wraps the torus, the fraction with winding magnitude of two or more stays tiny for almost the entire trajectory. It is well under 1% for $w_x$, and under about 6% for $w_y$ up to $f = 0.9$. For both components, then, the winding is almost always 0 or 1, so the binary wrap mask discards almost nothing that is present. Winding magnitudes above one become common only as the pattern forms. This is why the wrap mask loses accuracy only at $f = 1$ in panel (a).

The small gap between $w_x$ (blue) and $w_y$ (red) is immaterial. Both components are almost always 0 or 1, so the conclusion is the same for either one.

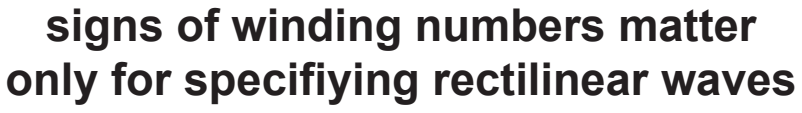

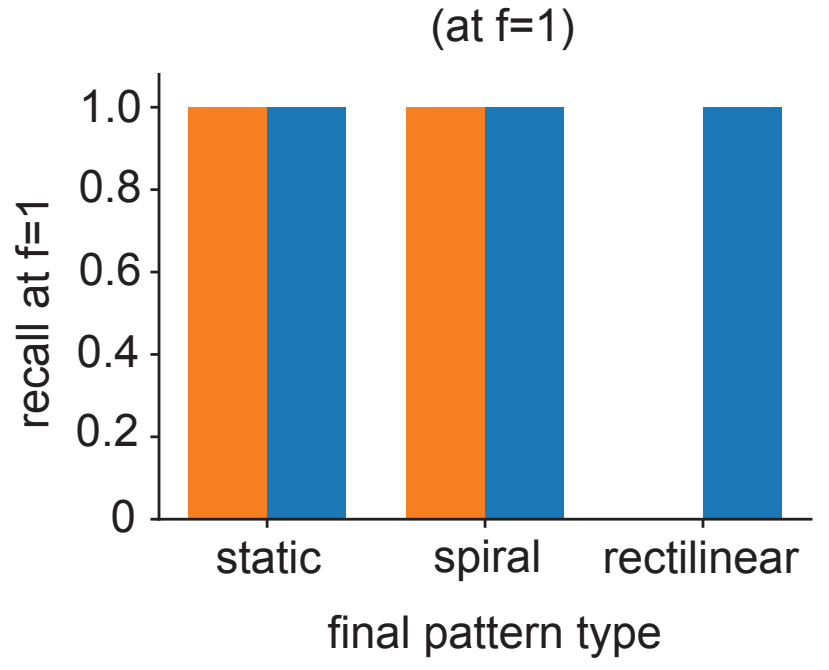

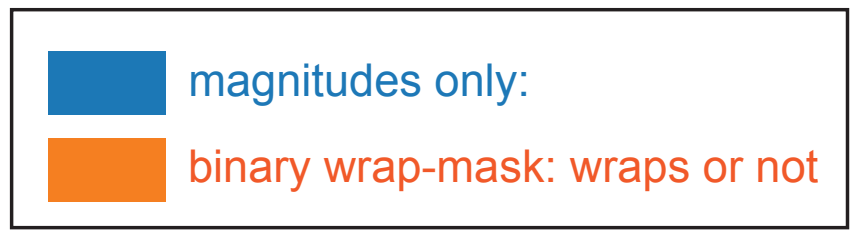

**Supplementary Figure 42: Winding magnitude matters only at the moment of pattern formation, and only for rectilinear-wave-forming trajectories.**

**Per-fate recall at $f = 1$ for the two inputs.** At each timestep $t = f \cdot T_{true}$, we computed the winding field from the lattice configuration at that timestep. A convolutional neural network (CNN) then predicted the three-way fate (static, rectilinear wave, or spiral wave) from that winding field. Here $T_{true}$ is the timestep at which the final pattern forms (see Supplementary note 7), and $f = t/T_{true}$ is the fraction of the runtime, between 0 and 1. At $f = 0$ the system is in its maximally disordered initial configuration. At $f = 1$ the pattern has formed.

The binary wrap-mask classifies static and spiral-wave outcomes perfectly but fails on rectilinear-wave fate (the recall near zero). Restoring the winding magnitude recovers recall to one for rectilinear-wave fate. At the moment of pattern formation, the rectilinear-wave trajectory is distinguished from the other-fate trajectories, not by whether its regions wrap the lattice, but by how many times. The number of wraps is flattened in the bare mask. This is the only place in the analysis where anything beyond the wrap pattern matters.

# Supplementary Notes

## Supplementary Note 1: Cellular automaton describing secrete-and-sense cells

We study a cellular automaton (CA) composed of identical secrete-and-sense cells arranged on a triangular lattice. Each cell occupies one of four discrete states corresponding to the gene-expression levels of two genes. These genes control the secretion rates of two diffusible molecules released into the surrounding three–dimensional space and sensed through surface receptors. Each gene can be either ON (high secretion rate) or OFF (low secretion rate), producing four possible combinations. For visualization we represent these four states as directional phase vectors (right, up, left, and down), which later allow the system to be interpreted as a discrete phase field (in Fig. 2a).

At every timestep, each cell senses the quasi–steady-state concentrations of both molecules produced by itself and by all other cells. The cell then updates its gene-expression state synchronously with all other cells according to threshold response functions. We adopt the quasi–steady-state approximation because molecular diffusion is assumed to occur much faster than the cellular response time required to switch gene expression states. Under this separation of timescales, concentration fields relax rapidly relative to cellular updates, allowing the automaton to operate on discrete timesteps.

To derive the update rule, we begin with the simpler case of a single molecule and then generalize to two interacting molecules. In earlier studies, we examined CA with one molecule [Ref. 1,2]. We had also previously introduced the two-molecule CA—which we study here—and revealed how the molecules must regulate each other's secretion in order for a disordered field of cells to robustly self-organize into dynamical spatial configurations (rectilinear and spiral waves) [Ref. 3]. The current work is about understanding the self-organization dynamics itself.

### A. Prelude: cellular automaton with one molecule

Consider a single diffusible molecule secreted by a spherical cell of radius $R$. The concentration field $c(r, t)$ surrounding the cell obeys the diffusion–degradation equation

$$\frac{\partial c}{\partial t} = D\nabla^2 c - \gamma c + \frac{\eta}{4\pi R^2}\delta(r - R) \tag{S1}$$

where $D$ is the diffusion constant, $\gamma$ is the degradation rate, $\eta$ is the secretion rate of the molecule, and $\delta(\cdot)$ is the Dirac delta function. The secretion rate depends on the internal gene-expression state of the cell. With the boundary condition $\lim_{r\to\infty} c(r) = 0$, the steady-state solution is

$$c(r) = \frac{c_R R}{r} e^{(R-r)/\lambda} \tag{S2}$$

where

$$\lambda = \sqrt{D/\gamma}$$

is the diffusion length and

$$c_R \equiv \frac{\eta\gamma}{4\pi R\lambda(\lambda + R)}.$$

The secretion rate takes one of two values depending on the gene-expression state

$$\eta = \begin{cases} C_{\mathrm{OFF}}, & \text{OFF cell} \\ C_{\mathrm{ON}}, & \text{ON cell} \end{cases} \tag{S3}$$

We express all concentrations in units of $C_{\mathrm{OFF}}$ and therefore set $C_{\mathrm{OFF}} = 1$. All concentrations are therefore dimensionless in the automaton dynamics. The steady-state concentration generated by a single cell becomes

$$c(r) = \begin{cases} \frac{R}{r} e^{(R-r)/\lambda}, & \text{OFF cell} \\ \frac{C_{\mathrm{ON}} R}{r} e^{(R-r)/\lambda}, & \text{ON cell.} \end{cases} \tag{S4}$$

The concentration sensed by the cell is the concentration on its surface ($r = R$):

$$c(R) = \begin{cases} 1, & \text{OFF cell} \\ C_{\mathrm{ON}}, & \text{ON cell.} \end{cases} \tag{S5}$$

The above expressions describe an isolated cell. We now consider $N$ cells arranged on a triangular lattice whose spacing is $a_0$. For convenience we express all lengths in units of $a_0$ and therefore set $a_0 = 1$. The cell radius $R$, diffusion length $\lambda$, and intercellular distances $r_{mj}$ are therefore dimensionless.

Let $X_m$ denote the gene-expression state of cell $m$:

$$X_m = \begin{cases} 0, & \text{OFF cell} \\ 1, & \text{ON cell.} \end{cases} \tag{S6}$$

Each cell senses the average concentration on its surface. The concentration sensed by cell $m$ at timestep $t$ is therefore

$$Y_m(t) = (C_{\mathrm{ON}} - 1)X_m(t) + 1 + \sum_{j \neq m} [(C_{\mathrm{ON}} - 1)X_j(t) + 1] f(r_{mj}) \tag{S7}$$

where the summation runs over all other cells and

$$f(r_{mj}) = \frac{1}{4\pi R^2} \int_{\text{surface of cell } m} \frac{R}{r} e^{(R-r)/\lambda}\, dA = \frac{\lambda e^{(R-r_{mj})/\lambda}}{r_{mj}} \sinh(R/\lambda) \tag{S8}$$

represents the average contribution of cell $j$ to the surface concentration sensed by cell $m$.

Each cell responds through a threshold response function. When the sensed concentration exceeds a threshold $K$, secretion switches to the high state; otherwise it remains low. The update rule for the one-molecule automaton is therefore

$$X_m(t+1) = \begin{cases} 0, & Y_m(t) < K \\ 1, & Y_m(t) > K. \end{cases} \tag{S9}$$

**Spatial correlation measure.** At $t = 0$ we assign each $X_m$ randomly, producing a maximally disordered configuration in which spatial correlations are negligible. We quantify spatial correlations using a spatial index

$$I \equiv \frac{N}{\sum_m \sum_{j \neq m} f(r_{mj})} \frac{\sum_m \sum_{j \neq m} f(r_{mj})(X_m - \langle X \rangle)(X_j - \langle X \rangle)}{\sum_j (X_j - \langle X \rangle)^2}. \tag{S10}$$

This quantity is a modified form of Moran's $I$, a standard measure of spatial autocorrelation in spatial statistics [Ref. 4]. We previously introduced this modified form and studied its properties [Ref. 1,2].

## B. Two-molecule cellular automaton

We now extend the model to two diffusible molecules. Each cell can independently express each gene, producing two binary variables, $X_m^{(1)}$ and $X_m^{(2)}$. A cell's state is then a 2-tuple:

$$(X_m^{(1)}, X_m^{(2)}).$$

The concentration of molecule $z$ sensed by cell $m$ is

$$Y_m^{(z)} = (C_{\text{ON}}^{(z)} - 1)X_m^{(z)} + 1 + \sum_{j \neq m} [(C_{\text{ON}}^{(z)} - 1)X_j^{(z)} + 1] f^{(z)}(r_{mj}) \tag{S11}$$

where $z = 1, 2$.

We define the secretion level

$$C_m^{(z)} = \begin{cases} 1, & \text{OFF state} \\ C_{\text{ON}}^{(z)}, & \text{ON state.} \end{cases} \tag{S12}$$

Note that

$$C_m^{(z)} = (C_{\text{ON}}^{(z)} - 1)X_m^{(z)} + 1.$$

The diffusion kernel becomes

$$f^{(z)}(r_{mj}) = \frac{\lambda^{(z)} e^{(R - r_{mj})/\lambda^{(z)}}}{r_{mj}} \sinh(R/\lambda^{(z)}). \tag{S13}$$

Using this notation the sensed concentration can be written compactly as

$$Y_m^{(z)}(t) = \sum_{j=1}^{N} f_{mj}^{(z)} C_j^{(z)}(t) \tag{S14}$$

with

$$f_{mj}^{(z)} = \begin{cases} f^{(z)}(r_{mj}), & m \neq j \\ 1, & m = j. \end{cases} \tag{S15}$$

The two molecules diffuse independently but regulate each other's secretion through gene regulation.

## Interaction topology

We encode regulatory interactions using an interaction matrix

$$M = \begin{pmatrix} M_{11} & M_{12} \\ M_{21} & M_{22} \end{pmatrix}. \tag{S16}$$

Each entry specifies how molecule *b* regulates secretion of molecule *a*:

$$M_{ab} = \begin{cases} 1, & \text{molecule } b \text{ activates secretion of } a \\ -1, & \text{molecule } b \text{ inhibits secretion of } a \\ 0, & \text{molecule } b \text{ does not regulate } a. \end{cases} \tag{S17}$$

Thresholds are stored in a matrix

$$K = \begin{pmatrix} K_{11} & K_{12} \\ K_{21} & K_{22} \end{pmatrix}. \tag{S18}$$

The response function for cell *m* is

$$g_m^{(ab)}(Y_m^{(b)}) = \begin{cases} \theta(Y_m^{(b)} - K_{ab}), & M_{ab} = 1 \\ \theta(K_{ab} - Y_m^{(b)}), & M_{ab} = -1 \\ 1, & M_{ab} = 0 \end{cases} \tag{S19}$$

where $\theta(x)$ is the Heaviside step function.

For compact notation we write

$$g_m^{(ab)}(Y_m^{(b)}) = \theta((Y_m^{(b)} - K_{ab})M_{ab}) \tag{S20}$$

with the convention that $g_m^{(ab)} = 1$ whenever $M_{ab} = 0$.

The update rule for the two-molecule cellular automaton becomes

$$X_m^{(a)}(t+1) = \prod_{b=1}^{2} g_m^{(ab)}(Y_m^{(b)}(t)). \tag{S21}$$

All cells update their states simultaneously at each timestep, so the values $X_m^{(a)}(t+1)$ are computed from the configuration at time $t$ before any updates are applied.

## Initialization of the cellular automaton

The initial configuration of the lattice specifies the binary gene-expression states of every cell for the two genes controlling secretion of the two diffusible molecules. For a lattice containing $N$ cells, the internal state of cell $m$ at $t = 0$ is defined by the pair

$$(X_m^{(1)}, X_m^{(2)}) \in \{0, 1\}^2,$$

where $X_m^{(z)} = 1$ denotes the ON (high secretion) state for molecule $z$ and $X_m^{(z)} = 0$ denotes the OFF (low secretion) state.

Two quantities characterize the spatial statistics of the initial condition for each gene $z$:

- the fraction of ON cells

$$p_z = \frac{1}{N} \sum_{m=1}^{N} X_m^{(z)},$$

- the spatial index

$$I_z = \frac{N}{\sum_m \sum_{j \neq m} f(r_{mj})} \frac{\sum_m \sum_{j \neq m} f(r_{mj})(X_m^{(z)} - \langle X^{(z)} \rangle)(X_j^{(z)} - \langle X^{(z)} \rangle)}{\sum_j (X_j^{(z)} - \langle X^{(z)} \rangle)^2},$$

  which is a modified form of Moran's $I$ that measures spatial correlations between cellular states. We previously examined how this metric changes during self-organization dynamics [Ref. 3].

We generate initial configurations by specifying target values for $p_z$ and $I_z$ and constructing a configuration consistent with those values.

### General initialization algorithm

The initialization procedure implemented in the simulation code proceeds independently for the two genes.

For gene $z$, we first fix the number of ON cells to be $Np_z$. A binary vector containing $Np_z$ ON states and $N(1-p_z)$ OFF states is randomly permuted across the lattice to generate a provisional configuration.

The algorithm then iteratively adjusts the spatial arrangement of ON and OFF cells to obtain a configuration whose spatial index lies within a specified interval

$$I_z^{\text{target}} \leq I_z \leq I_z^{\text{target}} + \Delta I.$$

At each iteration, one ON cell and one OFF cell are randomly selected and their states are swapped. The Moran index of the new configuration is recalculated, and the swap is accepted if it moves the configuration toward the target interval or otherwise improves the value of $I_z$. This procedure is repeated until the desired interval is reached or a maximum number of iterations is exceeded.

The same algorithm is applied independently to the two genes, producing an $N \times 2$ binary matrix describing the gene-expression state of every cell. The four possible combinations

$$(0,0),\ (0,1),\ (1,0),\ (1,1)$$

are mapped to the four discrete cellular states used in the automaton. Because the two genes are initialized independently, weak correlations between the resulting four-state variables can arise because the two genes are initialized independently while sharing the same spatial lattice.

### Initialization used in this study

In the simulations reported in this work we focus on maximally disordered initial conditions. Specifically, we set

$$p_1 \approx 0.5, \qquad p_2 \approx 0.5$$

and choose

$$I_1 \approx 0, \qquad I_2 \approx 0.$$

Setting the Moran index to zero ensures that the ON and OFF states of each gene are spatially uncorrelated. Intuitively, when half of the cells are ON and half are OFF, there is no systematic tendency for neighboring cells to share the same state or to alternate between states, resulting in a configuration with negligible spatial correlations. The lattice therefore begins in a highly disordered state in which ON and OFF cells are approximately randomly mixed throughout the lattice.

Because both genes are initialized in this manner and independently of one another, the four cellular states occur in roughly equal proportions across the lattice, producing an effectively random distribution of the four states while maintaining negligible spatial correlations for each gene individually. These initial conditions correspond to the disordered starting configurations shown in Fig. 1c, from which the cellular automaton self-organizes over time into static patterns, traveling waves, or spiral waves [Ref. 3].

## Relation to discrete reaction–diffusion systems

The cellular automaton described above can be interpreted as a discrete analogue of a reaction–diffusion system with nonlocal coupling. In classical reaction–diffusion models, continuous concentration fields evolve according to coupled partial differential equations in which reaction terms encode intracellular regulatory dynamics and diffusion terms spread signals in space. In contrast, our automaton replaces continuous fields with discrete agents whose internal states represent gene-expression levels.

Diffusion enters through the steady-state diffusion kernel $f(r_{mj})$, obtained from the Green's function of the diffusion–degradation equation after averaging the emitted and sensed concentrations over the surfaces of the emitting and receiving cells. The automaton therefore corresponds to a nonlocal reaction–diffusion system in which intracellular regulatory dynamics are discretized while diffusion-mediated coupling between cells is retained through the interaction kernel. Because the diffusion kernel decays approximately exponentially with distance, cells interact with many lattice neighbors rather than only nearest neighbors. For the parameter values used in this study ($R = 0.2$ and diffusion lengths $\lambda^{(1)} = 1.0$, $\lambda^{(2)} = 1.2$ in units of the lattice spacing), the effective communication range spans several lattice shells. Nearest neighbors contribute the largest signal, but second-nearest neighbors contribute roughly 20% as much per cell, and third-nearest neighbors still contribute about 10–15%. Contributions become small only beyond approximately the fourth or fifth lattice shell. Thus each cell integrates signals originating from a spatially extended neighborhood rather than from purely local interactions.

To illustrate the spatial extent of interactions, Table S1 lists the relative contributions of successive neighbor shells for the parameters used in this study. These values confirm that the effective communication range spans several lattice spacings, placing the model in an intermediate regime between strictly local cellular automata and fully global coupling.

| Neighbor shell | Distance $r$ | Number of cells | Mol. 1 contribution | Mol. 2 contribution |
|---|---|---|---|---|
| 1st (nearest) | 1 | 6 | 100% | 100% |
| 2nd | $\sqrt{3}$ | 6 | 19% | 23% |
| 3rd | 2 | 6 | 11% | 14% |
| 4th | $\sqrt{7}$ | 12 | 6% | 10% |
| 5th | 3 | 6 | 2% | 3% |

**Table 1:** Relative contributions of successive lattice shells to the concentration sensed by a cell for the parameter values used in this study ($R = 0.2$, $\lambda^{(1)} = 1.0$, $\lambda^{(2)} = 1.2$). Contributions are normalized to the total signal from the nearest-neighbor shell. Signals from multiple lattice shells therefore contribute appreciably to the sensed concentration.

### Interaction matrix studied in this work

Our previous systematic search identified five interaction matrices capable of generating dynamic spatial patterns such as rectilinear waves and spiral waves from initially disordered configurations [Ref. 3].

In this work we focus on the matrix

$$M^* = \begin{pmatrix} 1 & 1 \\ -1 & 0 \end{pmatrix}. \tag{S22}$$

This topology corresponds to a regulatory circuit in which molecule 1 self-activates and represses molecule 2, while molecule 2 activates molecule 1 and does not regulate itself.

We previously found that there are four other interaction matrices capable of producing dynamic spatial patterns [Ref. 3]:

$$\begin{pmatrix} 0 & -1 \\ 1 & 1 \end{pmatrix}, \quad \begin{pmatrix} 1 & -1 \\ 1 & -1 \end{pmatrix}, \quad \begin{pmatrix} 1 & -1 \\ 1 & 1 \end{pmatrix}, \quad \begin{pmatrix} -1 & -1 \\ 1 & 1 \end{pmatrix}. \tag{S23}$$

All five matrices share two key features required for the emergence of dynamic spatial patterns from disorder:

1. the off-diagonal elements must contain both activation and inhibition (+1 and $-1$), and 2. at least one self-activation must be present.

Because swapping molecule labels ($1 \leftrightarrow 2$) leaves the dynamics unchanged, we fix $M_{12} = 1$ and $M_{21} =$

$-1$ without loss of generality.

Throughout this work we therefore study the cellular automaton defined by the interaction matrix $M^*$.

# Supplementary Note 2: Machine-learning analysis of predictability from initial configurations

## 2.1 Overview and rationale

A schematic overview of the full machine-learning (ML) analysis pipeline appears in Supplementary Fig. 4.

Despite the deterministic update rules of the cellular automaton (CA), we discovered that we could not practically predict the final macroscopic fate of the system (static versus dynamic patterning) from the initial configuration alone. We subjected the CA to an extensive and deliberately conservative machine-learning analysis.

Our goal was not to identify an optimal predictive model, but rather to test whether a broad class of supervised learning approaches could extract predictive signal from the initial configuration when provided with very large datasets. We therefore designed the analysis around three guiding principles:

1. **Make the prediction task as easy as possible.** We reduced the problem to binary classification: predicting whether a random initial configuration evolves into a static configuration (label = 1) or into any dynamic pattern (rectilinear or spiral waves; label = 0). This task is strictly easier than predicting the full pattern class.
2. **Eliminate data limitations as a confounding factor.** We generated 1,000,000 independent simulations and evaluated learning curves with up to 850,000 training examples, spanning nearly three orders of magnitude in training set size.
3. **Use a frozen, large test set.** All reported performance metrics were computed on a fixed test set of 100,000 previously unseen initial configurations, ensuring that apparent improvements could not arise from test leakage or overfitting.

As shown below, even under these deliberately favorable conditions, prediction performance saturates at chance-level balanced accuracy and ROC–AUC ($\approx 0.5$), indicating that we detected no practically extractable predictive signal in the initial configuration across the models and data scales examined here.

## 2.2 Dataset generation

Each data point consists of:

- **Input:** a flattened $14 \times 14$ lattice (196 cells), where each cell occupies one of four discrete gene-expression states encoded as integers $\{1, 2, 3, 4\}$.
- **Label:** a binary indicator of the final outcome after full simulation:

$$1 = \text{static configuration}, \qquad 0 = \text{dynamic pattern (rectilinear or spiral wave)}.$$

Initial configurations were sampled uniformly at random, with each cell independently assigned one of the four states with equal probability. We then evolved the CA until it reached its final configuration.

## 2.3 Train / validation / test splits

From the full dataset of 1,000,000 simulations, we constructed the following fixed splits:

- **Training pool:** 850,000 samples
- **Validation set:** 50,000 samples
- **Test set:** 100,000 samples (frozen; never used during training or model selection)

Class balance was consistent across splits, with approximately 26.6% static outcomes and 73.4% dynamic outcomes. All learning curves reported below drew training subsets of increasing size from the same training pool and evaluated performance on the same frozen test set.

## 2.4 Baseline predictors

Before fitting any models, we evaluated trivial baselines: always predicting "dynamic," always predicting "static," and random guessing at the class prior. Because the dataset was class-imbalanced, these baselines achieved high raw accuracy ($\approx 0.73$). However, balanced accuracy equaled 0.5 and ROC–AUC equaled 0.5, establishing that any meaningful predictive model must exceed chance-level balanced accuracy.

## 2.5 Model classes evaluated

We evaluated model classes spanning a wide range of inductive biases and representational capacities:

1. **Logistic regression with one-hot encoding.** Each cell state was encoded as a four-dimensional one-hot vector, yielding a linear classifier over the full lattice.
2. **Extremely Randomized Trees (ExtraTrees).** An ensemble of randomized decision trees capable of capturing nonlinear, high-order interactions among lattice sites.
3. **Histogram-based Gradient Boosting (HGB).** A modern boosted-tree method designed for large structured datasets.
4. **XGBoost.** An aggressively tuned gradient-boosted decision-tree model applied to one-hot encoded initial configurations.
5. **Multilayer perceptron (MLP).** A high-capacity neural network trained on one-hot encoded configurations, with early stopping based on validation ROC–AUC.
6. **Convolutional neural network (CNN).** A spatially structured neural network trained on a two-dimensional, four-channel representation of the lattice.

No manual feature engineering was applied beyond the raw encoding of cell states.

## 2.6 Learning curves and performance saturation

For each model class, we trained on progressively larger subsets of the training pool,

$$N = 10^3,\ 3 \times 10^3,\ 10^4,\ 3 \times 10^4,\ 10^5,\ 3 \times 10^5,\ 8.5 \times 10^5,$$

with three independent random draws at each size. Performance was always evaluated on the same frozen test set.

Across all model classes and training sizes, performance showed the same pattern (Supplementary Fig. 5-9): balanced accuracy $\approx 0.5$, ROC–AUC $\approx 0.5$, and no systematic improvement with increasing data. Increasing the amount of training data by nearly three orders of magnitude therefore failed to reveal detectable predictive signal within the model classes and representations examined here.

## 2.7 Threshold calibration

For classifiers producing continuous scores, we selected decision thresholds on the validation set to maximize balanced accuracy,

$$\text{bal_acc}(t) = \frac{1}{2}\big(\text{TPR}(t) + \text{TNR}(t)\big),$$

and evaluated performance on the frozen test set. Threshold calibration did not improve prediction performance (Supplementary Fig. 6), indicating that the models' score rankings themselves contained no discriminative information about final fate.

## 2.8 Extension to more expressive models

More expressive models—including XGBoost, multilayer perceptrons, and convolutional neural networks—also failed to achieve predictive performance above chance (Supplementary Fig. 7–9). Increasing model expressivity and incorporating spatial inductive biases therefore did not rescue predictability from the initial configuration.

We did not explicitly test permutation-invariant architectures (e.g., DeepSets) or graph neural networks operating on lattice connectivity. However, the model classes we evaluated already span a broad range of inductive biases: tree ensembles operate on unordered feature sets, convolutional networks exploit spatial locality, and fully connected networks can represent arbitrary global functions of the input. All failed to achieve above-chance performance. We therefore do not expect these additional architectures to qualitatively change this outcome, although we cannot exclude the possibility that alternative representations or substantially different model classes could uncover predictive structure not detected here.

A further caveat concerns the CNN's spatial representation. The automaton evolves on a triangular lattice, in which each cell has six nearest neighbors. The CNN instead receives the configuration as a square-grid image and applies square convolutional kernels, whose local receptive field does not match the triangular six-neighbor connectivity. The CNN therefore provides a conservative spatial benchmark rather than an architecture matched to the lattice geometry. This does not weaken our conclusions. A geometry-matched model would only add spatial structure, and the fully connected network, which makes no spatial assumptions and can represent any function of the 196 cells, likewise found no signal. The failure to predict from the initial configuration is therefore not an artifact of the square-grid representation.

## 2.9 Featurewise mutual-information sanity check

As a complementary, model-agnostic test, we quantified the mutual information between each cell's initial four-state identity and the binary fate label. Across all dataset sizes, normalized mutual information collapsed onto the shuffle baseline and decreased toward zero with increasing sample size (Supplementary Fig. 10), consistent with the learning-curve results.

This decrease reflects improved estimation of the true (near-zero) mutual information with larger samples. The small positive values observed at small $N$ arise from finite-sample noise rather than genuine statistical dependence.

These results confirm that ML was unsuccessful not because of insufficient model capacity within the tested model classes, but because, across these analyses, we detected no statistically accessible predictive signal in the initial configuration at the level of individual lattice sites.

## 2.10 Predictability and the limits of representation

A natural question is whether prediction might nevertheless be possible in principle for a finite deterministic system—for example, by constructing a sufficiently large representation that maps each initial configuration to its final outcome.

For the present system, the total number of possible initial configurations is $4^{196} \approx 10^{118}$, vastly exceeding the estimated number of atoms in the observable universe ($\sim 10^{80}$). A brute-force lookup-table strategy is thus physically unrealizable for our system.

Our results, however, point to a limitation beyond brute-force infeasibility. Three independent analyses jointly show that we did not detect statistically extractable predictive structure within the tested representations:

- **Single-cell level (model-agnostic):** Featurewise mutual-information analysis shows that no individual lattice site carries detectable information about the final outcome beyond chance (Supplementary Fig. 10).
- **Local structure:** Convolutional neural networks, which can exploit spatially local patterns, could not extract predictive signal (Supplementary Fig. 7–9).
- **Global structure:** Fully connected neural networks, capable of representing arbitrary global functions of the input, likewise did not achieve performance above chance.

Together, these results argue against the existence of any simple or readily learnable static representation of the mapping from initial configurations to final outcomes among the broad classes we examined.

For generic finite deterministic systems with small state spaces, a lookup table is both constructible and learnable. Our system, even at the modest size of a $14 \times 14$ lattice, already exceeds the regime where such enumeration is physically realizable, removing this fallback entirely. The ML and NMI analyses therefore operate without recourse to exhaustive lookup, and their failure to detect predictive structure indicates that any predictive information present in the initial configuration is not organized in a form accessible to the model classes, representations, and data scales examined here.

## 2.11 Summary

We tested whether supervised learning models could predict the final fate of the cellular automaton from the initial configuration alone using 1,000,000 independent simulations. Under deliberately favorable conditions—binary classification, massive datasets, diverse model families, and frozen evaluation—all models performed at chance level (balanced accuracy $\approx 0.5$, ROC–AUC $\approx 0.5$). Model-agnostic mutual-information analysis likewise detected no statistically accessible predictive signal at the level of individual sites.

Together, these results show that, across the model classes, representations, and data scales examined here, predictive structure is not accessible from the initial configuration to static predictors operating on that configuration alone.

## Supplementary Note 3: Accuracy of the vortex detection algorithm

To quantify the reliability of our vortex-detection algorithm, we analyzed rare single-timestep artifacts that occasionally appeared in our vortex-count trajectories. These artifacts arose from transient ambiguities in contour labeling when vortex cores underwent rapid local rearrangements between successive timesteps. Such events manifested as a brief one-step increase in the total vortex count that disappeared in the following timestep. Here we refer to these events as *blips*.

### Definition and detection of blips

We defined a blip as a single-timestep fluctuation in the total vortex number $n_{\text{total}}(t)$ satisfying

$$n(t) = n(t-1) + 1 \qquad \text{and} \qquad n(t+1) = n(t-1).$$

This pattern corresponded to a transient appearance of an additional vortex that immediately disappeared in the next timestep.

To determine the nature of these events, we decomposed the vortex count into its three components (+1, $-1$, and 0 vortices) and classified each blip according to which component changed. Nearly all detected blips corresponded to the temporary appearance of a 0-vortex, meaning that the counts of +1 and $-1$ vortices remained unchanged during the event. These events therefore do not affect the dynamics of charged vortices, which determine the long-time behavior of the system.

We examined blips only after the early vortex-creation phase of each run (analysis beginning at $t \geq 150$ timesteps), thereby avoiding the initial transient dynamics in which vortex number changed rapidly.

### Measured blip rates

We quantified blip statistics across 1,000 independent runs for each lattice size ($14 \times 14$, $16 \times 16$, and $18 \times 18$) and for each final pattern type (static configurations, rectilinear waves, and spiral waves). Across all conditions, blips occurred extremely rarely.

For example, on a $14 \times 14$ lattice producing spiral waves, we detected 416 blips across 1,000 runs, corresponding to a mean rate of $2.1 \times 10^{-5}$ blips per timestep per run. Static runs on the same lattice showed even lower rates ($7.2 \times 10^{-6}$ per timestep), and rectilinear-wave runs showed similar values ($2.5 \times 10^{-5}$ per timestep). We observed comparable rates for larger lattices. On $16 \times 16$ grids the mean blip rate ranged from $8.8 \times 10^{-6}$ to $7.9 \times 10^{-5}$ per timestep depending on final pattern type, while on $18 \times 18$ grids it ranged from $3.5 \times 10^{-6}$ to $7.9 \times 10^{-5}$ per timestep.

Across all simulations, more than 99% of detected blips corresponded to transient 0 vortices rather than charged vortices.

Summary of vortex-count blip statistics across lattice sizes and final pattern types.

| Grid size | Final pattern | Mean blip rate per timestep | Mean blips per run |
|---|---|---|---|
| $14 \times 14$ | Static | $7.2 \times 10^{-6}$ | 0.143 |
| $14 \times 14$ | Rectilinear wave | $2.45 \times 10^{-5}$ | 0.487 |
| $14 \times 14$ | Spiral wave | $2.10 \times 10^{-5}$ | 0.416 |
| $16 \times 16$ | Static | $8.8 \times 10^{-6}$ | 0.174 |
| $16 \times 16$ | Rectilinear wave | $7.89 \times 10^{-5}$ | 1.567 |
| $16 \times 16$ | Spiral wave | $5.67 \times 10^{-5}$ | 1.126 |
| $18 \times 18$ | Static | $3.46 \times 10^{-6}$ | 0.086 |
| $18 \times 18$ | Rectilinear wave | $7.09 \times 10^{-5}$ | 1.761 |
| $18 \times 18$ | Spiral wave | $7.86 \times 10^{-5}$ | 1.952 |

### Impact on vortex dynamics

These measurements demonstrate that the vortex-detection algorithm is highly reliable. The mean blip rate across all conditions remained below $10^{-4}$ events per timestep, meaning that fewer than one timestep in ten thousand would show a transient counting artifact. Moreover, because nearly all blips involve only neutral (0) vortices and persist for a single timestep, they do not alter the dynamics of charged vortices or the pairwise annihilation events that determine the staircase-like decline in vortex number.

For all analyses presented in the main text and supplementary figures, we removed these artifacts using a simple one-step filter that replaced the transient increase with the surrounding value. Because blips occurred extremely rarely and affected only neutral vortices, this filtering step did not alter the qualitative or quantitative conclusions of the study.

As a reassurance, we had an additional internal consistency check in the form of topological charge conservation, imposed by the periodic boundary conditions of the lattice. Because the system is topologically equivalent to a torus, vortices can only be created or annihilated in $+1/-1$ pairs (Fig. 4c & Supplementary Note 5), implying that the total topological charge must remain exactly zero at all times:

$$\sum_i W_i = 0.$$

Throughout all simulations analyzed in this study, this constraint was satisfied at every timestep (Supplementary Fig. 17). We never observed any violation of global charge neutrality, even during vortex creation and annihilation events. This agreement provides an independent validation that the vortex-identification algorithm correctly detects and classifies vortices and their charges.

Taken together, these results confirm that the vortex-identification procedure accurately tracks the topological structures governing the dynamics of the cellular automaton.

## Supplementary Note 4: Approximating vortices as Brownian particles

### 4.1 Motivation

Tracking individual vortices revealed that charged vortices move through the lattice and annihilate upon encounter with an oppositely charged partner. This observation suggested that the many-body dynamics of vortices might admit a coarse-grained description in which vortices behave as effective particles undergoing stochastic motion.

We therefore asked whether the collective vortex dynamics observed in the cellular automaton could be approximated by a minimal particle model consisting of diffusing particles that annihilate upon encounter in the lattice sizes we examined. The goal of this analysis was not to derive the motion of vortices from the microscopic update rules, but rather to test whether a simple stochastic model could reproduce the statistical structure of vortex lifetimes and annihilation events observed in the automaton for the lattice sizes we examined here.

### 4.2 Empirical characterization of vortex motion

The cellular automaton simulations produce, at each timestep, the spatial coordinates of vortex cores and their associated topological charges. Using the vortex-tracking algorithm described in Methods, we extracted the centroid position of each vortex core over time.

For a given vortex pair, we defined the scalar distance

$$r(t) = \|\mathbf{x}_{+}(t) - \mathbf{x}_{-}(t)\|$$

between the centroids of a +1 vortex and its nearest $-1$ counterpart.

We then computed the discrete-time increments

$$\Delta r(t) = r(t+1) - r(t).$$

The distribution of these increments, aggregated across many trajectories and timesteps, was approximately symmetric and well described by a normal distribution centered near zero. This indicates that vortex motion lacks a detectable drift component and is consistent with unbiased diffusive motion over the timescales examined. Individual vortex centroid displacements were likewise consistent with zero-mean diffusion, with no detectable directional persistence.

We also observed intermittent pauses in vortex movement, during which the centroid position remained unchanged for several consecutive timesteps. The probability distribution of these waiting intervals decayed rapidly, with the vast majority lasting fewer than four timesteps.

Together, these observations motivated approximating vortex motion as an effective diffusion process.

## 4.3 Brownian particle model

To test whether a diffusive particle model could reproduce the statistics of vortex annihilation observed in the automaton, we constructed a minimal stochastic model consisting of charged particles undergoing Brownian motion in two dimensions.

The motion of each particle was modeled by the stochastic differential equations

$$dX_t = \sigma_x \, dW_t^{(x)}, \qquad dY_t = \sigma_y \, dW_t^{(y)},$$

where $W_t$ denotes a Wiener process and $\sigma_x$ and $\sigma_y$ are diffusion parameters.

Because the lattice geometry is isotropic at large scales, we set

$$\sigma_x = \sigma_y.$$

We generated particle trajectories using the discrete-time Euler–Maruyama scheme that corresponded to these stochastic differential equations. At each timestep, we updated the particle position according to

$$X_{t+1} = X_t + \sigma_x \eta_x, \qquad Y_{t+1} = Y_t + \sigma_y \eta_y,$$

where $\eta_x$ and $\eta_y$ are independent samples from a standard normal distribution.

## 4.4 Initialization and annihilation rule

Each simulation began with $2N$ particles placed at random positions within a square domain corresponding to the automaton lattice. Half of the particles were assigned charge +1 and the other half charge $-1$.

Particles evolved independently according to the Brownian dynamics described above. Whenever two oppositely charged particles approached closer than a threshold distance $r_{\text{threshold}}$, we removed the pair from the system, mimicking the annihilation of vortex pairs observed in the cellular automaton.

We repeated this procedure until either all particles had annihilated or a predefined maximum simulation time was reached.

### 4.5 Parameter estimation

We inferred the parameters of the Brownian model directly from the vortex trajectories we extracted from the cellular automaton.

First, we used the distribution of vortex velocity increments to estimate the effective diffusion coefficient. Because the change in pairwise distance over one timestep contained a large fraction of zero increments—this arose due to waiting intervals—we estimated the velocity distribution using differences taken over four timesteps, motivated by each cell behaving as a nearly perfect oscillator that cycles through the four cell states (Supplementary Fig. 1). At this timescale, nearly all vortices had moved at least once.

Second, we estimated the annihilation threshold distance $r_{\text{threshold}}$ from the distribution of pairwise distances at the moment of vortex annihilation in the cellular automaton runs.

Initial particle positions were drawn from a uniform distribution, which reproduced the observed distribution of initial vortex separations.

### 4.6 Comparison with cellular automaton dynamics

We compared the Brownian particle model with the cellular automaton by measuring the time required for the final vortex pair to annihilate.

Across the lattice sizes examined, the Brownian particle model reproduced the same qualitative staircase structure in vortex-pair number observed in the cellular automaton (Fig. 3c) and generated annihilation times of the same order of magnitude as those observed in the automaton. The predicted and observed average run times were strongly correlated (Fig. 3d).

Importantly, we stress that the purpose of this model was not to provide an exact microscopic description of vortex motion. Rather, it demonstrates that the statistical structure of vortex annihilation dynamics can be captured by a minimal coarse-grained model in which vortices behave as diffusing particles that annihilate upon encounter.

## 4.7 Interpretation

These results suggest that our deterministic CA rapidly organizes its microscopic dynamics into an effective defect gas consisting of charged vortices undergoing diffusive motion. In this coarse-grained picture, the long termination times observed in the automaton arise from diffusion-limited encounters between oppositely charged vortices.

While the particle model does not capture all details of the automaton dynamics, it provides a simple physical framework for understanding the temporal structure of vortex annihilation and the emergence of long-lived trajectories during pattern formation.

# Supplementary Note 5: Topological Charge Neutrality on a Triangular Lattice with Periodic Boundary Conditions

## Overview

Here we prove that total plaquette circulation vanishes on a triangular lattice with periodic boundary conditions, which, for the contour-based detection scheme used here, implies vortex-winding neutrality. This is not a property of our specific cellular automaton (CA) dynamics but a topological constraint on any discrete phase field defined on a toroidal lattice. This result follows from discrete edge-cancellation, which is the standard lattice-analogue of Stokes' theorem: the total circulation of the phase field over all triangular plaquettes must sum to zero. Because the winding number assigned to each vortex in the CA corresponds to the circulation of the phase field around its contour, the total vortex winding in our CA also vanishes in practice. A direct corollary of this is that vortex creation and annihilation events must preserve total winding, forbidding the isolated appearance or disappearance of a single non-zero-winding vortex—we indeed experimentally observed this in our CA runs (Supplementary Fig. 17). This discrete topological charge neutrality is the lattice analogue, in spirit, of the same constraint that governs vortex-antivortex pair production in XY models and quantized vortices in superfluids and superconductors (Ref. 5-8).

## 1 Setup and Definitions

### 1.1 Lattice

Let $\Lambda$ be a triangular lattice of $N = L \times L$ sites with periodic boundary conditions in both spatial directions. The periodic boundary conditions make $\Lambda$ topologically equivalent to a torus $\mathbb{T}^2$. Each site $i \in \Lambda$ carries a discrete phase state $\phi_i \in \{1, 2, 3, 4\}$, corresponding to the four phases $\theta_i \in \{0, \pi/2, \pi, 3\pi/2\}$ via $\theta_i = (\phi_i - 1)\pi/2$.

### 1.2 Directed edges and plaquettes

Each directed edge $(i \to j)$ in the lattice (between nearest-neighbor sites $i$ and $j$) carries a phase difference:

$$\Delta_{ij} = \theta_j - \theta_i \pmod{2\pi}, \tag{S24}$$

taken in the range $(-\pi, \pi]$. For our four states, the raw difference $\theta_j - \theta_i$ takes values in $\{-3\pi/2, -\pi, -\pi/2, 0, \pi/2, \pi, 3\pi/2\}$. Reducing modulo $2\pi$ to $(-\pi, \pi]$, the values $\pm 3\pi/2$ map to $\mp\pi/2$, so the reduced differences lie in $\{-\pi/2, 0, \pi/2, \pi\}$.

For each undirected edge $\{i, j\}$, we choose an antisymmetric representative satisfying $\Delta_{ji} = -\Delta_{ij}$, with $\Delta_{ij}$ taking values in $\{-\pi, -\pi/2, 0, \pi/2, \pi\}$. For the generic cases, the representative lies in $(-\pi, \pi]$. For edges whose phase difference is congruent to $\pi$ (mod $2\pi$), we assign the sign according to a fixed orientation of the edge, ensuring $\Delta_{ji} = -\Delta_{ij}$ without exception. This antisymmetry is the only property we use in the proof below.

Each triangular face (plaquette) $f$ of the lattice is a triangle with three vertices $(i, j, k)$ listed in counter-clockwise order. We define the discrete circulation around plaquette $f$ as:

$$\Omega_f = \Delta_{ij} + \Delta_{jk} + \Delta_{ki}, \tag{S25}$$

where $\Delta_{ij}$ is the reduced (signed) phase difference along each directed edge.

### 1.3 Winding number of a vortex contour

**Definition 1** (Winding number). *For each labeled vortex core, we define its* contour *as the closed loop of cells immediately surrounding the core, traversed in counter-clockwise order and returning to the starting cell. The winding number of the vortex is:*

$$W = \frac{1}{2\pi} \sum_{contour} \Delta\phi, \tag{S26}$$

*where each $\Delta\phi$ is the branch-cut-corrected phase difference between consecutive contour cells, lying in $(-\pi, \pi]$.*

**Lemma 1** (Integer winding). *The winding number $W$ of any vortex contour is an integer.*

*Proof.* The contour is a closed loop of $n$ cells $c_0, c_1, \dots, c_{n-1}, c_n = c_0$, returning to the starting cell $c_0$. Each branch-cut-corrected step $\Delta\phi_i$ satisfies

$$\Delta\phi_i = (\theta_{i+1} - \theta_i) + 2\pi m_i, \quad m_i \in \mathbb{Z}, \tag{S27}$$

where $m_i$ is the integer chosen such that $\Delta\phi_i \in (-\pi, \pi]$. Summing over all $n$ steps, we have:

$$\sum_{i=0}^{n-1} \Delta\phi_i = (\theta_n - \theta_0) + 2\pi \sum_{i=0}^{n-1} m_i = 2\pi k, \quad k \in \mathbb{Z}, \tag{S28}$$

since $\theta_n = \theta_0$ (i.e., the loop closes) and $k = \sum_i m_i \in \mathbb{Z}$. Therefore $W = k \in \mathbb{Z}$ for every vortex contour, regardless of the specific phase values on the lattice or the CA update rules. This is simply the discrete analogue of the standard result that the winding number of any closed curve is an integer. □

### 1.4 Topological charge of a vortex

**Definition 2** (Topological charge)**.** *Vortices are classified by the sign (chirality) of their winding number* $W \in \mathbb{Z}$*:*

- $W \geq +1$*: classified as a* $+1$ *vortex (counter-clockwise winding),*
- $W \leq -1$*: classified as a* $-1$ *vortex (clockwise winding),*
- $W = 0$*: classified as a* $0$ *vortex.*

*The* topological charge $q$ *of a vortex is its chirality label* $\in \{-1, 0, +1\}$*. In the rarely observed cases where* $|W| \geq 2$*, the vortex is classified by the sign of* $W$ *and counted as a single* $\pm 1$ *vortex. The charge* $q$ *therefore reflects chirality, not winding magnitude.*

**Remark 1** (Relationship between plaquette circulation and contour winding number)**.** *The plaquette circulation* $\Omega_f = \Delta_{ij} + \Delta_{jk} + \Delta_{ki}$ *defined in Section 1.2 is the local, single-triangle contribution to the total circulation. A vortex core in our CA consists of a connected region of cells, each sharing a* $\pi$*-phase difference with at least one nearest neighbor; the minimal such core comprises two adjacent cells. The contour is the closed loop of cells at distance one from any core cell (Section 1.3). For the vortex cores identified in our CA, the contour winding* $W_v$ *equals the sum of plaquette circulations* $\Omega_f/2\pi$ *over all triangular plaquettes enclosed by the contour, by the same edge-cancellation argument we used in Theorem 1: all interior edges of the enclosed region cancel pairwise, leaving only the boundary contributions that define the contour.*

## 2 Main Result

**Theorem 1** (Total plaquette circulation vanishes on the torus)**.** *For any configuration of phase states* $\{\phi_i\}_{i\in\Lambda}$ *on the triangular lattice* $\Lambda$ *with periodic boundary conditions:*

$$\sum_{f\in\mathcal{F}} \Omega_f \;=\; 0, \tag{S29}$$

*where* $\mathcal{F}$ *denotes the set of all triangular plaquettes of* $\Lambda$*.*

*Proof.* We prove this by edge-cancellation on the torus.

**Step 1: Expressing the total circulation as a sum over directed edges.**

The total circulation over all plaquettes is:

$$\sum_{f\in\mathcal{F}} \Omega_f \;=\; \sum_{f\in\mathcal{F}} \sum_{(i\to j)\in\partial f} \Delta_{ij}, \tag{S30}$$

where $\partial f$ denotes the three directed edges of plaquette $f$ in counter-clockwise order.

**Step 2: Each edge is traversed in opposite directions by its two adjacent plaquettes.**

Every edge $e$ in the triangular lattice is shared by exactly two plaquettes $f^+$ and $f^-$. When both plaquettes are oriented counter-clockwise, edge $e$ is traversed as $(i \to j)$ in $f^+$ and as $(j \to i)$ in $f^-$. By the antisymmetry convention of Section 1.2, $\Delta_{ji} = -\Delta_{ij}$ for every edge. The contributions of $e$ to the total sum therefore cancel:

$$\Delta_{ij} \;+\; \Delta_{ji} \;=\; 0. \tag{S31}$$

**Step 3: Periodic boundary conditions ensure every edge is shared.**

On a lattice with open boundary conditions, boundary edges appear in only one plaquette, leaving non-canceling contributions. The periodic boundary conditions identify opposite edges of the fundamental domain, making the lattice topologically a torus on which every edge is shared by exactly two plaquettes. Therefore all contributions cancel:

$$\sum_{f\in\mathcal{F}} \Omega_f \;=\; \sum_{\text{edges } e} \left(\Delta_{ij}^{(f^+)} + \Delta_{ji}^{(f^-)}\right) \;=\; 0. \tag{S32}$$

□

**Remark 2** (Connection to vortex winding numbers)**.** *In our vortex-identification scheme, we computed the winding number $W_v$ assigned to each detected vortex as the discrete phase circulation along its contour—the closed loop of cells immediately surrounding the connected core region (Section 1.3). This contour circulation corresponds to the sum of plaquette circulations $\Omega_f$ enclosed by the contour, by the same edge-cancellation logic used in the Theorem: interior edges of the enclosed region cancel pairwise, leaving only the boundary contributions. Thus, for the vortex contours in the CA, the vanishing of total plaquette circulation established by the Theorem implies that the total vortex winding vanishes in all CA runs: $\sum_v W_v = 0$.*

## 3 Corollaries

**Corollary 1** (Winding-neutral creation and annihilation)**.** *In the CA dynamics, any creation or annihilation event must preserve* $\sum_v W_v = 0$ *(Supplementary Fig. 17). That is, the winding numbers of all vortices involved in any single event must sum to zero. Isolated creation or annihilation of a single vortex with* $W_v \neq 0$ *is topologically forbidden.*

*In the overwhelmingly dominant case in which the participating vortices had* $W_v = \pm 1$*, the simplest winding-neutral elementary events were:*

1. *simultaneous creation of one* $+1$ *and one* $-1$ *vortex,*
2. *simultaneous annihilation of one* $+1$ *and one* $-1$ *vortex, or*
3. *conversion of a* $(+1)$–$(-1)$ *pair into a* $0$*-vortex.*

*Proof.* By the Theorem, total plaquette circulation vanishes at every timestep $t$ for any instantaneous configuration $\{\phi_i(t)\}$, independently of the dynamics. By the contour-based winding assignment used in our CA (Remark 2), the total vortex winding therefore vanishes at every timestep: $\sum_v W_v(t) = 0$ for all $t$ (Supplementary Fig. 17). Any discrete change between consecutive timesteps must satisfy $\Delta\left(\sum_v W_v\right) = 0$, requiring the winding numbers of all vortices involved in the event to sum to zero. The elementary event types listed above are the simplest winding-neutral events observed in our CA runs when $|W_v| = 1$. □

**Remark 3** (Equal counts of +1 and $-1$ vortices)**.** *Because rare contours with* $|W_v| \leq 2$ *occurred, zero total winding does not by itself require equal numbers of the sign-classified +1 and -1 vortices. Nevertheless, we directly observed* $N_+ = N_-$ *at every timestep in every simulation (Supplementary Fig. 17). Thus, total integer-winding neutrality is the exact topological result, whereas equality of the operational charge counts is an empirical result.*

## 4 Discussion and Generality

**Remark 4** (Generality beyond our CA)**.** *Theorem 1 is purely topological: it requires only (i) a discrete phase field on a lattice, (ii) a local definition of vortex charge as a plaquette circulation with the anti-symmetric edge convention, and (iii) periodic boundary conditions. The CA update rules play no role in the proof. Hence, the same theorem holds for any CA, any initial condition, and any number of discrete phase states, as long as the lattice is toroidal.*

**Remark 5** (Relation to simulation observations)**.** *We verified that* $\sum_v W_v = 0$ *at all timesteps across tens of thousands of CA trajectories (Supplementary Fig. 17). This is not a coincidence or a property specific to our CA's update rules: vanishing total plaquette circulation,* $\sum_f \Omega_f = 0$*, is an* exact topological invariant *of the toroidal lattice, proved by the Theorem. In our CA runs, the vortex winding obeys the same neutrality constraint through the contour construction described in Remark 2, confirming that the topological identity is correctly reflected in our vortex-detection scheme.*

**Remark 6** (Connection to discrete Stokes' theorem)**.** *The proof above is an instance of the discrete Stokes' theorem. The antisymmetric edge differences* $\Delta_{ij}$ *define a discrete 1-cochain on the cell complex of the torus. The plaquette circulations* $\Omega_f$ *are the values of its discrete coboundary (a 2-cochain). The total sum* $\sum_f \Omega_f$ *is the evaluation of this coboundary on the fundamental 2-cycle* $[\mathbb{T}^2]$*. Since the torus has no boundary (*$\partial[\mathbb{T}^2] = \emptyset$*), discrete Stokes gives* $\sum_f \Omega_f = 0$ *directly, without reference to curvature or index theory.*

# Supplementary Note 6: Unique-trajectory and graph-based clustering reveal late-time fate separation

## Overview

We showed that once only the final vortex pair remains, the dynamics of non-contractible-loop (NCL) strings predict whether the run will end in a static configuration, a rectilinear wave, or a spiral wave. That analysis identifies the specific topological structures that carry fate-dependent predictive signature. Having uncovered this mechanism, we then asked a much coarser question: if we discard the full topological description and retains only a single scalar observable at each timestep—the total vortex core size—does a fate-dependent predictive signature still emerge, and if so, when?

What we discovered was striking: for most of each run, trajectories leading to different final outcomes remained effectively indistinguishable at this coarse level. Only very late in each trajectory did they collapse into a small number of stereotyped terminal behaviors that segregated strongly by their final pattern type. We established this using two complementary approaches: (1) a unique-trajectory analysis, which counts how many trajectories are exactly identical within a backward-time window, and (2) a *graph-based clustering* analysis, which groups trajectories whose scalar time series are sufficiently similar under a physically motivated distance threshold.

These two methods imposed different notions of similarity and therefore yielded different numbers. At $\tau = 20$, the unique-trajectory method found 180 distinct exact-match terminal profiles, whereas the graph-based method merged these into 23 connected components aligned with final pattern fate. These are not contradictory results. They are two views of the same late-time compression, differing only in the stringency of the similarity criterion.

## 1. Trajectory representation and backward-time alignment

Each cellular automaton (CA) run terminates upon reaching a final configuration, which is either static, rectilinear-wave, or spiral-wave. Because different runs have different durations, we aligned all trajectories by indexing time *backward* from the terminal event. Thus, $t = 1$ denotes the timestep immediately preceding the final configuration, $t = 2$ denotes two timesteps before the final configuration, and so on. Under this convention, the same column index always refers to the same temporal distance from the end of the run, regardless of total run length.

For each run, we recorded the total vortex core size as a function of backward time and stored it as a row vector of length 12,000, the maximum trajectory length in the analyzed set. Runs shorter than 12,000 timesteps therefore had undefined entries beyond their actual duration. For computational convenience, these missing values were padded before analysis so that all trajectories could be represented in a common matrix format. This choice primarily affects very long backward-time windows; the late-time windows emphasized in Fig. 5 contain actual data for virtually all analyzed trajectories.

This backward-time alignment made it possible to ask the central question addressed in Fig. 5: do trajectories that share a pattern fate also share their final approach to that pattern fate, even when pattern fate is assessed using only a scalar summary of vortex activity and without reference to individual vortices, strings, or NCLs?

## 2. Unique-trajectory analysis

### 2.1. Exact identity within a backward-time window

To ask how many distinct dynamical routes the system follows as it approaches its terminal state, we selected a backward-time window $[1, \tau]$ for varying values of $\tau \leq 12{,}000$. For any given $\tau$, we extracted from each run the sub-vector spanning columns 1 through $\tau$ and counted how many unique row patterns existed.

Two runs were called identical within $[1, \tau]$ if their total-vortex-core-size time series matched exactly at every column in that window. In the implementation, we performed this comparison directly on aligned row vectors after we padded shorter trajectories to a common length. This was therefore an intentionally strict criterion: even a difference of one cell in vortex core size at a single timestep placed two runs into different classes.

### 2.2. Late-time collapse of trajectory diversity

We swept $\tau$ from large values down toward 1 and monitored the number of unique trajectories. At the widest window, $\tau$ = 12,000, nearly every run was distinct, reflecting the enormous diversity of paths the system can take from disordered initial conditions. As the window was shortened toward the terminal event, however, this diversity collapsed sharply. Strikingly, trajectories that had spent thousands of timesteps on entirely different paths were now converging onto the same terminal scalar sequences. By $\tau$ = 20—the final 20 timesteps before the final pattern formed—the 2,000 trajectories had collapsed to only 180 unique sequences Fig. 5e, a 91% reduction.

This compression was not gradual: the number of distinct trajectories remained high for most of the run and dropped steeply only near the end. Thus, although the early and intermediate portions of the dynamics were highly diverse, the final stage of the run was confined to a surprisingly small number of exact scalar trajectories.

## 3. Graph-based clustering

### 3.1. Motivation

The unique-trajectory analysis demanded exact agreement at the level of vortex-core-area trajectories and therefore treated even very small deviations between two trajectories as distinct. We therefore asked whether a broader version of the same collapse appears when near-identical trajectories are grouped together. To do so, we constructed a graph-based clustering method based on pairwise Euclidean distances between scalar time series.

### 3.2. Distance between trajectories and proximity graph

For a given backward-time window $[1, \tau]$, we represented each trajectory by its truncated scalar time series

$$\mathbf{x}_i^{(\tau)} = \left(x_i(1), x_i(2), \dots, x_i(\tau)\right),$$

where $x_i(t)$ is the total vortex core size of run $i$ at backward time $t$. We then defined the distance between trajectories $i$ and $j$ by the Euclidean norm

$$d(i, j) = \left\| \mathbf{x}_i^{(\tau)} - \mathbf{x}_j^{(\tau)} \right\|_2 .$$

As in the unique-trajectory analysis, we embedded all trajectories into a common matrix format before we computed their pairwise distances. Accordingly, at sufficiently long backward-time windows, the distance metric was sensitive both to differences in vortex-core dynamics and to differences in total run length. This did not affect the late-time results emphasized in Fig. 5, where the analyzed backward-time windows lay well within the duration of virtually all trajectories.

Two trajectories were connected by an edge in a proximity graph if and only if

$$d(i, j) \le d_{\text{thr}}(\tau),$$

where we set $\delta = 1$ cell, so that

$$d_{\text{thr}}(\tau) = \sqrt{\delta^2 \tau} = \sqrt{\tau}.$$

**Physical interpretation of the threshold.** This choice corresponds to allowing an average difference of approximately one vortex-core cell per timestep over the chosen window. Equivalently,

$$\frac{d(i,j)}{\sqrt{\tau}} \leq 1$$

means that the root-mean-square difference per timestep does not exceed one cell in total vortex core size. We adopted this value after testing alternatives and found that it provided a useful balance: larger thresholds merged clearly dissimilar trajectories, whereas smaller thresholds fragmented terminal behaviors that were visibly related.

### 3.3. Connected components as trajectory clusters

With the proximity graph constructed, we identified its connected components. Each connected component constituted one cluster; its members were runs whose scalar time series lay within the chosen tolerance of one another, possibly through chains of pairwise connections. This is therefore a more permissive notion of grouping than exact identity.

For each component we recorded its size and the distribution of final pattern types among its member trajectories. We also tracked three summary quantities across backward-time windows: the total number of connected components, the fraction of trajectories belonging to singleton components, and the size of the largest connected component. These quantities allowed us to distinguish trivial fragmentation into isolated trajectories from the emergence of genuinely collective late-time clusters.

### 3.4. Isolated nodes collapse into late-time fate-aligned clusters

Sweeping $\tau$ from large values downward revealed a systematic transition. At large $\tau$ (for example, $\tau$ = 12,000), most nodes remained isolated: nearly every trajectory was too dissimilar from all others for meaningful clusters to form (Fig. 5c, right). The graph was therefore dominated by singletons.

As $\tau$ decreased toward the terminal event, trajectories began to link up. This merging did not occur smoothly. Cluster sizes remained constant over extended intervals and then increased abruptly as $\tau$ crossed specific values (Fig. 5d). By $\tau$ = 20, the 2,000 trajectories had collapsed into just 23 well-separated connected components (Fig. 5c, left). Repeating the analysis across 19 independent trajectory sets yielded the same qualitative picture in every case (Supplementary Fig. 25).

What emerged was striking: at large $\tau$, a large fraction of trajectories remained isolated singletons, so similarity-based prediction was not yet possible for much of the ensemble. As $\tau$ decreased, trajecto-ries were progressively absorbed into larger connected components. Across the additional trajectory sets, the total number of connected components fell reproducibly, the fraction of trajectories in singleton components dropped toward zero, and the largest connected component grew in a stepwise manner (Supplementary Fig. 25). Thus, although fate remained unreadable for most of the trajectory, the final 20 timesteps contained sufficient dynamical signature to group trajectories into late-time clusters aligned with the three final pattern types.

Because this collapse occurs at very short backward-time windows, it reflects genuine convergence of late-stage dynamics rather than any artifact of trajectory padding at long times.

## 4. Relationship between the two analyses

The two methods quantified related but distinct aspects. The unique-trajectory analysis counted the number of exactly distinct scalar time series in a given backward-time window. The graph-based method counted the number of nearby groups under a tolerance of approximately one cell per timestep. Accordingly, the latter number was smaller: at $\tau = 20$, the 180 exact terminal scalar trajectories were merged by the graph-based method into 23 larger connected components aligned with final pattern fate.

These are not contradictory results. The unique-trajectory view revealed that late-time dynamics became highly compressed even under the most stringent possible criterion. The graph-based view revealed that when small deviations were tolerated, those compressed archetypes organized into a still smaller set of late-time groups aligned with final pattern fate. Both views therefore support the same conclusion: the CA's evolutionary paths remain effectively entangled for most of the run and segregate by fate only near the very end.

## 5. Interpretation

These analyses establish that one cannot determine the pattern fate of a given CA trajectory from this scalar observable for most of its duration. For thousands or tens of thousands of timesteps, trajectories leading to different final outcomes remained too fragmented for similarity-based prediction at the level of total vortex core size. A large fraction of trajectories remained isolated or belonged only to very small groups. Only in the final tens of timesteps did this observable acquire fate-dependent predictive structures.

This conclusion is notable precisely because the analysis here used none of the detailed topological information we developed. It ignored individual vortices, NCL strings, and their dynamics, and instead reduced each trajectory to a single number at each timestep. Even at this maximally coarse level, the same central conclusion emerged: in this deterministic system, predictive structure about macroscopic fate is not evident in the initial condition and in most of the trajectory, and crystallizes only late in the self-organization process. That fate-dependent predictive structure survives even this aggressive compression underscores how robustly it is encoded in the late-time dynamics—and how it is practically unextractable everywhere before.

# Supplementary Note 7: Machine-learning analysis of predictability across the trajectory

## 7.1 Overview and relation to Supplementary Note 2

In Supplementary Note 2 we asked a simple question: Can we predict the final fate of the CA from its *initial* configuration alone? We trained and tested several supervised learning models (i.e., models trained on initial configurations labeled with their final fate), using up to $8.5 \times 10^5$ initial configurations. Their balanced accuracy did not rise above 0.5, the value expected from random guessing in the binary task of predicting a static or dynamic final pattern.

Here we ask a more general question. Suppose we instead show a learning model one lattice configuration — a snapshot taken from *later* in a trajectory. When does the fate become predictable? What structure in the configuration makes that prediction possible?

One distinction runs through everything that follows. Because the dynamics are deterministic, the initial configuration already fixes the final fate. We therefore ask whether the learning models tested can infer that fate directly from a single snapshot, without explicitly tracing the dynamics. We found that the dynamics progressively make fate legible by constructing a **winding field** that is nearly zero at the start of a trajectory.

In fact, we found a consistent picture:

- Prediction accuracy (balanced accuracy) is at chance at the initial configuration and rises steadily along the trajectory, reaching near-perfect prediction at the moment the pattern forms.
- Spatial organization matters: a network with built-in nearest-neighbor locality learns the predictive signal with about one-eighth as many training examples as a network without that built-in structure.
- Exact lookup of configurations seen during training does not explain the rise. Moreover, no previously identified object on its own — the vortex-connecting string mesh or the non-contractible-loop strings — predicts well across most of the trajectory.
- Most generally, an emergent topological field — the winding field — records only how each region of same-state cells wraps the torus. A model given only this field predicts almost as well as one shown the entire lattice.
- The winding field is nearly zero at the start and develops over the same short window in which vortices form.

- Prediction changes little when we remove the sign convention from the winding field. Through the transient, the field reduces to a single bit per cell: whether or not the cell's region wraps the torus.
- Prediction is hardest for the spiral fate, which becomes accurately predictable only near formation, echoing the asymmetry found in the clustering analysis (Fig. 5) and the non-contractible-loop strings (Fig. 4).

The machine-learning pipeline used here extends the initial-configuration pipeline, shown schematically in Supplementary Fig. 4, to single configurations sampled later in a trajectory. Unless stated otherwise, every learning curve for the models mentioned below shows the mean of three independent training runs. The runs differ only in their random initialization, or seed. We trained each model on up to $8.5 \times 10^5$ configurations and evaluated it on a frozen, held-out test set: trajectories set aside in advance and never used during training or model selection, exactly as we did for testing a slew of learning models on the initial configuration (see Supplementary Note 2).

### 7.2 Trajectory time: the two clocks $T_{\text{first}}$ and $T_{\text{true}}$

Because the CA is deterministic and its state space is finite ($4^{196}$ possible configurations), every trajectory must eventually repeat a configuration, and from then on it cycles forever. This gives two natural markers of "when the pattern forms."

- $T_{\text{first}}$, the **first-recurrence time**: the first timestep at which the lattice configuration exactly equals a configuration visited earlier in the same trajectory. From this step onward the trajectory simply repeats.
- $T_{\text{true}}$, the **formation time**: the first timestep at which the trajectory enters its terminal cycle, that is, the moment the final pattern first appears. Equivalently, $T_{\text{true}} = T_{\text{first}} - P$, where $P$ is the period of the terminal cycle.

The two coincide (up to one step) for static outcomes, which have period $P = 1$. For rectilinear and spiral waves, whose terminal cycles have longer periods, $T_{\text{first}}$ overshoots $T_{\text{true}}$ by one full period. Throughout the rest of this paper we normalized trajectory time by $T_{\text{first}}$. In this Note, and in Fig. 6, we instead normalize by $T_{\text{true}}$ and write

$$f = \frac{t}{T_{\text{true}}},$$

so that $f = 0$ is the initial configuration and $f = 1$ is the moment of pattern formation for every trajectory, regardless of its cycle period. This makes the fraction $f$ directly comparable across the three fates.

### 7.3 Prediction task, data, and training

**Task.** Given a single lattice configuration sampled at fraction $f$ of a trajectory, predict which of the three fates the trajectory will reach: a static configuration, a rectilinear wave, or a spiral wave. This is harder than the binary static-versus-dynamic task of Supplementary Note 2, because the model must now separate all three outcomes.

**Input.** The $14 \times 14$ lattice (196 cells), each cell in one of four states. For the convolutional network the lattice is presented as a four-channel image, one channel per state, which preserves the two-dimensional arrangement of the cells.

**Data and splits.** We used the same $10^6$ independent simulations as in Supplementary Note 2, with the same fixed splits: a training pool of $8.5 \times 10^5$ configurations, a validation set of $5 \times 10^4$ used only to tune and stop training, and a frozen test set never used during training or model selection. For each fraction $f$ we extracted, from every trajectory, the single configuration at timestep $t = \text{round}(f\, T_{\text{true}})$. The three fates are unequally represented (about 27% static, 30% spiral, 43% rectilinear).

**Metric.** We report *balanced accuracy*: the average, over the three fates, of the per-fate recall, where the recall of a fate is the fraction of trajectories of that fate that the model labels correctly. Balanced accuracy weights the three fates equally, so it cannot be inflated by simply predicting the most common fate. Chance is $1/3$.

**Models.** Our main learning model was a convolutional neural network (CNN). Its early layers combine each cell only with its neighbors, giving the network a built-in preference for local spatial structure. To test whether this preference helps, we compared the CNN with a multilayer perceptron (MLP): a fully connected network that receives the 196 cells as a fixed, flattened vector. The MLP can learn spatial relationships, but its architecture does not specify which cells are neighbors. For tests of individual objects, we also used two simpler, non-neural classifiers: a linear model and gradient-boosted decision trees (a standard method that combines many small decision trees). We trained all neural networks with early stopping: during training, we monitored balanced accuracy on the validation set and retained the model from the point at which that accuracy stopped improving. This procedure helped limit overfitting.

### 7.4 Prediction accuracy rises along the trajectory

When we fed the CNN progressively later configurations, its balanced accuracy climbed steadily from chance at $f = 0$ to near-perfect prediction at the moment of pattern formation (Supplementary Fig. 29a; Fig. 6b; Table 2). The same rise appeared in the easier binary static-versus-dynamic task of Supple-mentary Note 2 (Supplementary Fig. 29b).

It is worth stating plainly what this rise means. Because the CA is deterministic, the initial configuration already fully determines the fate. The result at $f = 0$ does not mean that the dynamics fail to determine fate; it means that the models tested could do no better than randomly guessing fates directly from the initial snapshots. As the dynamics run, later snapshots become progressively easier to read, and by formation the CNN predicts fate nearly perfectly. Prediction remained genuinely difficult through much of the trajectory. At the midpoint, balanced accuracy was only about 0.64, well above chance but far from reliable, and the spiral fate contributed most of the remaining error (Section 7.13).

| $f = t/T_{\text{true}}$ | balanced accuracy | recall (static) | recall (spiral) | recall (rectilinear) |
|---|---|---|---|---|
| 0.0 | 0.33 | 0.24 | 0.11 | 0.66 |
| 0.1 | 0.43 | 0.42 | 0.59 | 0.28 |
| 0.2 | 0.45 | 0.47 | 0.49 | 0.40 |
| 0.3 | 0.50 | 0.58 | 0.39 | 0.53 |
| 0.4 | 0.57 | 0.70 | 0.42 | 0.59 |
| 0.5 | 0.64 | 0.77 | 0.48 | 0.66 |
| 0.6 | 0.67 | 0.82 | 0.52 | 0.67 |
| 0.7 | 0.71 | 0.86 | 0.55 | 0.73 |
| 0.8 | 0.77 | 0.89 | 0.61 | 0.80 |
| 0.9 | 0.90 | 0.97 | 0.78 | 0.94 |
| 1.0 | 1.00 | 1.00 | 1.00 | 1.00 |

**Table 2:** Balanced accuracy and per-fate recall of the full-configuration CNN as a function of trajectory fraction $f$. Values are means over three seeds; chance balanced accuracy is $1/3$. Individual per-fate recalls are non-monotonic early in the transient, although balanced accuracy increases smoothly.

### 7.5 Exact configuration lookup does not explain the rise

Because trajectories of the same fate progressively merge, the CNN might succeed simply by recognizing configurations it has already seen during training. Exact configuration lookup cannot explain its success. At the midpoint of the trajectories ($f = 0.5$), only about 2% of the held-out test configurations had an exact, cell-by-cell match anywhere in the training set, yet the CNN correctly predicted the fate of far more test configurations (Supplementary Fig. 30). The CNN therefore predicted many configurations that it had never encountered exactly. The same figure shows that the number of distinct configurations per fate shrinks over time: static trajectories collapse toward a single shared configuration, whereas spirals remain spread over many distinct configurations. This difference in diversity may help explain why spirals are the hardest fate to predict.

### 7.6 Spatial architecture improves sample efficiency

To test whether locality helps, we compared how each network's accuracy grew as it received more training data (i.e., its learning curve). At the midpoint of each trajectory, the CNN reached with only about $10^5$ training examples the balanced accuracy that the MLP reached with about $8.5 \times 10^5$ — MLP required roughly eight times more data (Supplementary Fig. 31a; Table 3). The CNN therefore learned the predictive signal much more efficiently. This sample-efficiency advantage supports the presence of spatial organization that enables prediction. We carried out this learning-curve comparison on the binary task, where the difference was cleanest.

| training examples | MLP (no locality) | CNN (local) |
|---|---|---|
| $3 \times 10^4$ | 0.53 | 0.55 |
| $1 \times 10^5$ | 0.56 | 0.75 |
| $3 \times 10^5$ | 0.69 | 0.80 |
| $8.5 \times 10^5$ | 0.76 | 0.82 |

**Table 3:** Balanced accuracy at $f = 0.5$ versus training-set size for the CNN and the MLP in the binary static-versus-dynamic task. The CNN with $10^5$ examples matches the MLP with $8.5 \times 10^5$.

### 7.7 The CNN learns most of the signal without explicit toroidal padding

We next asked whether the network needs an explicit representation of the toroidal boundary. A convolution near the edge of the lattice needs values from just beyond the edge. Circular padding supplies them by wrapping around to the opposite side, as required by the torus. Flat (zero) padding instead fills them with zeros and treats the edge as a hard boundary. Using a single-state (red) input, we compared the two. With flat padding, the CNN already captured most of the predictive signal across the trajectory. Circular padding added a small but consistent improvement that grew toward formation (a few percent, reaching roughly six percentage points near $f = 0.9$; Supplementary Fig. 31b). Therefore, explicitly encoding the toroidal boundary therefore helps but is not necessary for the CNN to learn most of the signal.

## 7.8 Previously identified objects do not predict well across most of the trajectory

We then tested objects that we had already characterized. Supplying either the vortex-connecting string mesh or the number of non-contractible-loop (NCL) strings on its own predicted fate poorly for most of the trajectory. The NCL-string count became informative only close to the moment of pattern formation, consistent with NCL strings being a late-time signature (Supplementary Fig. 32a). Conversely, deleting the entire string mesh from the full configuration left accuracy essentially unchanged, whereas the mesh alone predicted weakly (Supplementary Fig. 32b). The string mesh and NCL-string count therefore do not account for prediction across the trajectory. We next sought a more general structure.

## 7.9 An emergent topological winding field carries the predictability

We found that structure by relabeling each cell by a single topological property. Cells that share a state form connected regions on the lattice. On a torus, such a region can either close on itself locally or wrap all the way around — once, twice, or more — in each of the two directions. We labeled every cell by the pair $(w_x, w_y)$ counting how many times its region wraps horizontally and vertically, and by nothing else. We call this per-cell label the **winding field** (Fig. 6d). It keeps only the wrapping topology of each region and discards the cell's state.

Whether a region wraps the torus is a global property: no bounded local patch alone reveals whether the entire region wraps. Providing the winding field therefore makes this global property explicit instead of requiring the CNN to infer it from the raw cell states. Trained on the winding field alone, the CNN recovered about 90% of its full-configuration balanced accuracy at every fraction (about 88% early in the trajectory, rising to about 96% near formation) (Fig. 6f; Table 4). Adding the full vortex description — the size, location, and chirality of every vortex — to the winding field changed accuracy only slightly. A network told only how each region wraps the torus predicts almost as well as one shown the entire lattice. This is our most direct evidence that much of the predictive structure is topological. The winding field is the most general topological representation we identified; NCL strings form a sparse, late-time instance of this topology.

## 7.10 The winding field is built by the self-organization dynamics

To determine whether the dynamics construct the winding field, we tracked a set of order parameters (spatial index) forward in time over about $10^6$ trajectories. The spatial index is a single number that summarizes how ordered the lattice is. In the initial configuration, almost no region wraps the torus, so the winding field is almost entirely zero (Supplementary Fig. 38). The fraction of cells carrying a nonzero winding then switches on sharply between roughly $t = 5$ and $t = 20$: it is negligible through $t = 3$ and reaches about 0.4 by $t = 20$, close to its early-transient plateau. Vortices form spontaneously during this same short window. The dynamics thus construct the topological representation that carries most of the predictive signal.

| $f = t/T_{\text{true}}$ | full configuration | winding field | winding field + vortices |
|---|---|---|---|
| 0.0 | 0.33 | 0.33 | 0.33 |
| 0.1 | 0.43 | 0.38 | 0.40 |
| 0.2 | 0.45 | 0.41 | 0.43 |
| 0.3 | 0.50 | 0.46 | 0.47 |
| 0.4 | 0.57 | 0.52 | 0.53 |
| 0.5 | 0.64 | 0.58 | 0.60 |
| 0.6 | 0.67 | 0.62 | 0.64 |
| 0.7 | 0.71 | 0.66 | 0.69 |
| 0.8 | 0.77 | 0.72 | 0.74 |
| 0.9 | 0.90 | 0.86 | 0.87 |
| 1.0 | 1.00 | 1.00 | 1.00 |

**Table 4:** Balanced accuracy of the CNN given the full configuration, the winding field alone, or the winding field augmented with the full vortex description. The winding field alone recovers most of the full-configuration accuracy at every fraction; the vortex description adds little.

### 7.11 Prediction does not depend on winding sign and reduces to a wrap indicator

We asked which part of the winding field the network actually uses (Supplementary Fig. 41a; Table 5). The sign of a winding number depends on the chosen loop orientation, an orientation or gauge convention. We tried three ways of removing information from the field: keeping only the magnitudes $|w_x|, |w_y|$; fixing the orientation convention; and collapsing the field to a single binary channel that marks only whether each cell's region wraps the torus at all (a wrap mask). These changes had little effect on balanced accuracy through the transient. The model therefore did not need the sign, and a one-bit-per-cell wrap mask was sufficient during the transient.

The reason is structural (Supplementary Fig. 41b). Among cells whose region wraps, the fraction with a winding magnitude of two or more is tiny throughout the transient — well under 1% for $w_x$ and under about 6% for $w_y$ up to $f = 0.9$. A region almost always wraps the torus either zero or one time, so the wrap mask discards almost nothing that is present. Winding magnitudes above one become common only as the pattern forms.

| $f = t/T_{\text{true}}$ | signed $(w_x, w_y)$ | $\lvert w_x \rvert, \lvert w_y \rvert$ | gauge-fixed | wrap mask |
|---|---|---|---|---|
| 0.0 | 0.33 | 0.33 | 0.33 | 0.33 |
| 0.1 | 0.38 | 0.39 | 0.39 | 0.38 |
| 0.2 | 0.41 | 0.42 | 0.42 | 0.41 |
| 0.3 | 0.46 | 0.47 | 0.47 | 0.47 |
| 0.4 | 0.52 | 0.54 | 0.53 | 0.53 |
| 0.5 | 0.58 | 0.60 | 0.60 | 0.60 |
| 0.6 | 0.62 | 0.65 | 0.64 | 0.64 |
| 0.7 | 0.66 | 0.69 | 0.69 | 0.69 |
| 0.8 | 0.72 | 0.76 | 0.75 | 0.75 |
| 0.9 | 0.86 | 0.89 | 0.89 | 0.89 |
| 1.0 | 1.00 | 1.00 | 1.00 | 0.67 |

**Table 5:** Balanced accuracy of the CNN for four encodings of the winding field. Through the transient the four coincide, so the prediction is invariant to the winding sign (a gauge convention) and reduces to a binary wrap indicator. The four agree until formation, where the bare wrap mask alone falls to 2/3 (Section 7.12). Removing the sign does not lower the accuracy, and if anything raises it slightly, consistent with the sign being uninformative.

### 7.12 Winding magnitude matters only at formation, and only for the rectilinear wave

The wrap mask closely tracks the full winding field at every fraction except the last. At the formed pattern ($f = 1$), the mask alone falls to a balanced accuracy of about 2/3, whereas the magnitude-bearing field stays near one (Supplementary Fig. 41; Table 5). The per-fate recalls explain this value. At $f = 1$, the wrap mask classifies static and spiral outcomes perfectly but fails on rectilinear waves (recall near zero), so its balanced accuracy is two-thirds. Restoring the winding magnitude recovers rectilinear recall to one (Supplementary Fig. 42). At formation, the rectilinear wave is distinguished from the other fates not by whether its regions wrap but by how many times they wrap, which the bare mask discards. Among the times and encodings tested, this is the only clear case in which winding magnitude beyond binary wrapping improves prediction.

### 7.13 Spiral waves are the hardest fate to predict and the last to resolve

The difficulty of prediction is not shared equally among the fates (Supplementary Fig. 40; Fig. 6c; Table 2). Static- and rectilinear-bound trajectories become legible progressively. After the non-monotonic early transient, spiral recall remained the lowest of the three from $f = 0.3$ through $f = 0.9$ and rose steeply only close to formation. Because balanced accuracy is the mean of the three per-fate recalls, spiral trajectories contributed most of the mid-trajectory error. Training the CNN on the winding field alone reproduced the same ordering, showing that the asymmetry is present in the topology and not only in the raw configuration. Spiral-wave fate also occupied far more distinct configurations than the other fates (Supplementary Fig. 30); this diversity may help explain why a single snapshot constrains their future less strongly. The per-fate asymmetry matches the one seen in the late-time clustering (Fig. 5) and the NCL strings (Fig. 4). Vortex-annihilation-bound fates (static and rectilinear fates) become legible through progressive topological simplification, whereas spiral-bound trajectories develop a weaker precursor and become accurately predictable only near wave formation.

### 7.14 Summary

Extending the initial-configuration analysis of Supplementary Note 2 forward in time, we found that fate, unreadable from the initial configuration by the models tested, becomes progressively predictable as the dynamics run and reaches near-perfect predictability at formation. A spatially structured model (CNN) learned the signal more efficiently, and exact configuration lookup cannot explain its success. The previously identified vortex-connecting strings and NCL strings do not account for prediction across most of the trajectory. Most generally, an emergent topological winding field — describing how each region of same-state cells wraps the torus — recovered most of the predictive accuracy available from the full lattice. The self-organization dynamics build this field over the first few tens of timesteps. During the transient timesteps, the model did not need the winding sign or even its magnitude: a single bit per cell, recording whether that cell's region wraps, was sufficient for fate prediction. Spiral-wave fate remained the hardest to predict and became accurately predictable only near wave formation. Thus, although the initial configuration completely fixes the final fate, we have shown that it is the self-organization dynamics that construct the topological entities which make the final fate directly legible.

# Supplementary Note 8: Topological self-organization and dynamically constructed predictability

## 8.1 Comparison with the established topological framework

Several ingredients in our paper also appear in the theory of topological defects. For a phase-like order parameter, point defects can be represented by phase winding around a closed contour[9], as in superfluids, superconductors, and two-dimensional magnetic systems[5–8]. Studies have found physical systems in which oppositely charged defects annihilate and thereby produce coarsening[10], while spiral waves associated with phase singularities are well known in oscillatory and excitable media[11].

Our toroidal lattice distinguishes contractible paths from paths that wrap around the torus. This distinction underlies the non-contractible-loop (NCL) strings and the winding field. Crucially, these structures are not conserved: local updates can cut an existing NCL or create a new one by connecting separate regions (Fig. 4e). The winding field also differs from a vortex's winding number. Whereas vortex winding measures phase circulation around a defect core, the winding field assigns each cell a vector describing how its connected same-state region wraps the torus. Established defect theory motivates these descriptions but does not establish a relationship between such dynamically organizing structures and fate predictability. That relationship is the new discovery reported here.

## 8.2. New finding: topological self-organization as a carrier of predictability

Our study uses well-known topological ingredients to ask when an already-determined fate becomes practically predictable. It thus asks whether a non-chaotic deterministic system can be practically unpredictable even when its macroscopic fate is entirely fixed and is not computationally irreducible.

The tested machine-learning models predicted the three fates from initial configurations no more accurately than random guessing. Yet the same model classes became increasingly accurate from progressively later configurations and nearly perfect near the pattern-formation moment. Thus, the dynamics progressively organized fate-predictive structure that became accessible to snapshot-based predictors.

Because fate is fixed throughout a trajectory, the dynamics do not create information about the outcome. Instead, they reorganize the configuration into a form that allows the relevant information to be extracted. The winding field makes this distinction concrete: although it is derived from the configuration and adds no information, it carries nearly all the fate-predictive signal accessible to the convolutional neural network from the complete configuration.

The new finding in our work is therefore not the existence or annihilation of topological defects, but the finding that a dynamically self-organizing topological field carries the emerging predictability of predetermined fates. In this finite, discrete, non-chaotic deterministic system, topological self-organization constructs a collective representation through which an already fixed future becomes practically legible.

### 8.3 Scope of our finding and its mechanisms

Our broader result is an existence statement: in a non-chaotic deterministic system, fate can be fixed before the collective structures that make it practically predictable have formed. Our system establishes this possibility which, we believe, is this study's main contribution. The winding-field mechanism is more specific. It requires connected regions that can wrap around non-contractible directions and therefore applies directly to periodic or otherwise multiply connected domains. In a simply connected open domain, the two winding vectors defined here are unavailable, and defects may enter or leave through the boundary. It is not proposed here as a universal mechanism by which predictability would arise in all systems.

Periodic boundaries define a finite, translationally invariant domain without boundary-induced edge effects. The resulting topological consequences are genuine properties of the model. Related constraints occur in physical and biological systems, although their form depends on the domain and order parameter. For example, topology constrains defects in active nematic vesicles[12], while defects in epithelial monolayers can organize cell extrusion[13]. These examples demonstrate the physical and biological relevance of organized defects, not the occurrence of our specific winding-field mechanism. It would be fruitful to investigate whether other collective structures similarly construct predictability through their dynamics.

## References for Supplementary Notes